\documentclass[11pt]{article}
\usepackage[breaklinks=true,
            colorlinks=true,
            citecolor=black,
            linkcolor=black,
            urlcolor=black]{hyperref}
\usepackage{fullpage}
\usepackage{latexsym,amsmath,amsfonts,amssymb,stmaryrd,amsthm}
\usepackage{url}
\usepackage[square,authoryear,semicolon]{natbib}
\usepackage[nameinlink,capitalize]{cleveref}
\usepackage{manfnt}
\usepackage{todonotes}
\usepackage{mathpartir}
\usepackage{enumitem}
\usepackage{microtype}
\usepackage{xparse}
\usepackage{breakcites}

\theoremstyle{definition}
\newtheorem{definition}{Definition}

\theoremstyle{remark}
\newtheorem{remark}[definition]{Remark}

\theoremstyle{plain}
\newtheorem{lemma}[definition]{Lemma}
\newtheorem{theorem}[definition]{Theorem}
\newtheorem{rul}{Rule}
\Crefname{rul}{Rule}{Rules}

\newcommand{\RedPRL}{\textsc{{\color{red}Red}PRL}}

\newcommand{\G}{\ensuremath{\Gamma}}
\renewcommand{\phi}{\varphi}
\newcommand{\e}{\ensuremath{\varepsilon}}
\newcommand{\eb}{\overline{\e}}
\newcommand{\eq}{\ensuremath{\mathbin{\doteq}}}
\newcommand{\fresh}{\ensuremath{\mathbin{\#}}}
\newcommand{\fd}[1]{\ensuremath{\mathsf{FD}(#1)}}

\newcommand{\oft}[2]{#1\mathbin{:}#2}
\newcommand{\J}{\ensuremath{\mathcal{J}}}
\newcommand{\cube}{\ensuremath{\text{\mancube}}}
\newcommand{\st}[2]{#1 \vert_{(#2)}}

\makeatletter
\def\rightharpoonupfill@{\arrowfill@\relbar\relbar\rightharpoonup}
\newcommand{\overrightharpoonup}{%
\mathpalette{\overarrow@\rightharpoonupfill@}}

\providecommand{\leftsquigarrow}{%
  \mathrel{\mathpalette\reflect@squig\relax}%
}
\newcommand{\reflect@squig}[2]{%
  \reflectbox{$\m@th#1\rightsquigarrow$}%
}
\makeatother

\newcommand{\mkspacer}[1]{\gdef\spacer{\hphantom{#1}}#1}

\NewDocumentCommand\Infer{o m m}{%
  \IfValueTF{#1}
    {\inferrule*[vcenter,right=#1]{#2}{#3}}
    {\inferrule{#2}{#3}}
}

\newcommand{\subst}[3]{\ensuremath{#1 [#2 / #3]}}
\newcommand{\dsubst}[3]{\ensuremath{#1 \langle{#2}/{#3}\rangle}}

\newcommand{\arr}[2]{\ensuremath{#1 \to #2}}
\newcommand{\picl}[3]{\ensuremath{({#1}{:}{#2}) \to #3}}
\newcommand{\lam}[2]{\ensuremath{\lambda{#1}.{#2}}}
\newcommand{\app}[2]{\ensuremath{\mathsf{app}({#1},{#2})}}

\newcommand{\prd}[2]{\ensuremath{#1 \times #2}}
\newcommand{\sigmacl}[3]{\ensuremath{({#1}{:}{#2}) \times #3}}
\newcommand{\pair}[2]{\ensuremath{\langle #1,#2\rangle}}
\newcommand{\fst}[1]{\ensuremath{\mathsf{fst}(#1)}}
\newcommand{\snd}[1]{\ensuremath{\mathsf{snd}(#1)}}

\newcommand{\Eq}[3]{\ensuremath{\mathsf{Eq}_{#1}(#2,#3)}}
\newcommand{\ax}{\ensuremath{\star}}

\newcommand{\Path}[3]{\ensuremath{\mathsf{Path}_{#1}(#2,#3)}}
\newcommand{\dlam}[2]{\ensuremath{\langle #1 \rangle #2}}
\newcommand{\dapp}[2]{\ensuremath{#1 @ #2}}

\newcommand{\Upre}[1][j]{\ensuremath{\mathcal{U}^\mathsf{pre}_{#1}}}
\newcommand{\UKan}[1][j]{\ensuremath{\mathcal{U}^\mathsf{Kan}_{#1}}}
\newcommand{\Ux}[1][j]{\ensuremath{\mathcal{U}^\kappa_{#1}}}
\newcommand{\void}{\ensuremath{\mathsf{void}}}

\newcommand{\nat}{\ensuremath{\mathsf{nat}}}
\newcommand{\z}{\ensuremath{\mathsf{z}}}
\newcommand{\suc}[1]{\ensuremath{\mathsf{s}(#1)}}
\NewDocumentCommand\natrec{g g g}{%
  \ensuremath{\mathsf{natrec}\IfValueT{#1}{(#1;#2,#3)}}}

\newcommand{\bool}{\ensuremath{\mathsf{bool}}}
\newcommand{\wbool}{\ensuremath{\mathsf{wbool}}}
\newcommand{\true}{\ensuremath{\mathsf{true}}}
\newcommand{\false}{\ensuremath{\mathsf{false}}}
\NewDocumentCommand\ifb{G{{}} g g g}{%
  \ensuremath{\mathsf{if}_{#1}
    \IfValueT{#2}{\IfNoValueTF{#3}{({#2})}{({#2};{#3},{#4})}}}}

\newcommand{\C}{\ensuremath{\mathbb{S}^1}}
\newcommand{\base}{\ensuremath{\mathsf{base}}}
\newcommand{\lp}[1]{\ensuremath{\mathsf{loop}_{#1}}}
\NewDocumentCommand\Celim{G{{}} g g g}{%
  \ensuremath{\C\mathsf{\text{-}elim}_{#1}
    \IfValueT{#2}{\IfNoValueTF{#3}{({#2})}{({#2};{#3},{#4})}}}}

\newcommand{\isContr}[1]{\ensuremath{\mathsf{isContr}(#1)}}
\newcommand{\Equiv}[2]{\ensuremath{\mathsf{Equiv}(#1,#2)}}
\NewDocumentCommand\ua{G{{}} g}{%
  \ensuremath{\mathsf{V}_{#1}\IfValueT{#2}{(#2)}}}
\newcommand{\uain}[2]{\ensuremath{\mathsf{Vin}_{#1}(#2)}}
\newcommand{\uaproj}[2]{\ensuremath{\mathsf{Vproj}_{#1}(#2)}}

\newcommand{\etc}[1]{\ensuremath{\overrightharpoonup{#1}}}
\newcommand{\tube}[2]{\ensuremath{#1\hookrightarrow #2}}
\newcommand{\sys}[2]{\etc{\tube{#1}{#2}}}

\NewDocumentCommand\Coe{s G{{}} g g g}{%
  \ensuremath{\mathsf{coe}_{#2}%
  \IfBooleanTF{#1}
    {^{r \rightsquigarrow r'}(M)}
    {\IfValueT{#3}{^{#3 \rightsquigarrow #4}(#5)}}}}

\NewDocumentCommand\NewCompositionOperator{s m m O{\rightsquigarrow} O{y.N_i}}{%
  \IfBooleanTF{#1}
    {\NewDocumentCommand#2{s g g g g}{%
      \IfBooleanTF{##1}
        {\ensuremath{\mathsf{#3}^{r #4 r'}(M;\sys{##2}{#5})}}
        {\IfNoValueTF{##2}
          {\ensuremath{\mathsf{#3}}}
          {\ensuremath{\mathsf{#3}^{##2 #4 ##3}(##4\IfValueT{##5}{;##5})}}}}}
    {\NewDocumentCommand#2{s G{{}} g g g g}{%
      \IfBooleanTF{##1}
        {\ensuremath{\mathsf{#3}_{##2}^{r #4 r'}(M;\sys{##3}{#5})}}
        {\IfNoValueTF{##3}
          {\ensuremath{\mathsf{#3}_{##2}}}
          {\ensuremath{\mathsf{#3}_{##2}^{##3 #4 ##4}(##5\IfValueT{##6}{;##6})}}}}}
}
\NewCompositionOperator{\Hcom}{hcom}
\NewCompositionOperator{\Com}{com}
\NewCompositionOperator{\Ghcom}{ghcom}
\NewCompositionOperator{\Gcom}{gcom}
\NewCompositionOperator*{\Fcom}{fcom}
\NewCompositionOperator*{\Kbox}{box}[\rightsquigarrow][N_i]
\NewCompositionOperator*{\Kcap}{cap}[\leftsquigarrow][y.B_i]

\newcommand{\steps}{\ensuremath{\longmapsto}}
\newcommand{\evals}{\ensuremath{\Downarrow}}
\newcommand{\isvalsym}{\ensuremath{\mathsf{val}}}
\newcommand{\isval}[1]{\ensuremath{#1\ \mathsf{val}}}
\newcommand{\stable}{\cube}
\newcommand{\ssteps}{\ensuremath{\steps_\stable}}
\newcommand{\sisval}[1]{\ensuremath{#1\ \mathsf{val}_\stable}}

\newcommand{\pre}[1]{#1^\mathsf{pre}}
\newcommand{\Kan}[1]{#1^\mathsf{Kan}}
\newcommand{\lift}[1]{#1^\evals}
\newcommand{\PTy}{\ensuremath{\mathsf{PTy}}}
\newcommand{\Coh}{\ensuremath{\mathsf{Coh}}}
\newcommand{\Tm}{\ensuremath{\mathsf{Tm}}}
\NewDocumentCommand\relcts{s m m}{%
  \ensuremath{{#2} \models \IfBooleanTF{#1}{#3}{(#3)}}}

\newcommand{\wftm}[2][\Psi]{\ensuremath{{#2}\ \mathsf{tm}\ [#1]}}
\newcommand{\wfval}[2][\Psi]{\ensuremath{#2\ \mathsf{val}\ [#1]}}
\newcommand{\vper}[1]{\ensuremath{\llbracket #1 \rrbracket}}

\newcommand{\tds}[3]{\ensuremath{{#2} : {#1} \to {#3}}}
\newcommand{\psitd}{\tds{\Psi'}{\psi}{\Psi}}
\newcommand{\td}[2]{\ensuremath{{#1}{#2}}}
\newcommand{\id}[1][\Psi]{\ensuremath{\mathsf{id}_{#1}}}

\newcommand{\wfshape}[1]{{#1}\ \mathsf{valid}}
\newcommand{\ctx}[1]{#1 \gg}
\NewDocumentCommand\judg{O{\Psi} d<> m}
  {#3\ [#1 \IfValueT{#2}{\mid #2}]}
\NewDocumentCommand\wfctx{O{\Psi} d<> m}
  {#3\ \mathsf{ctx}\ [#1 \IfValueT{#2}{\mid #2}]}

\NewDocumentCommand\cwftype{m O{\Psi} d<> m}
  {#4\ \mathsf{type}_\mathsf{#1}\ [#2 \IfValueT{#3}{\mid #3}]}
\NewDocumentCommand\ceqtype{m O{\Psi} d<> m m}
  {#4 \eq #5\ \mathsf{type}_\mathsf{#1}\ [#2 \IfValueT{#3}{\mid #3}]}
\NewDocumentCommand\wftype{m O{\Psi} d<> m m}
  {#4 \gg #5 \ \mathsf{type}_\mathsf{#1}\ [#2 \IfValueT{#3}{\mid #3}]}
\NewDocumentCommand\eqtype{m O{\Psi} d<> m m m}
  {#4 \gg #5 \eq #6\ \mathsf{type}_\mathsf{#1}\ [#2 \IfValueT{#3}{\mid #3}]}

\newcommand{\cwftypep}{\cwftype{pre}}
\newcommand{\ceqtypep}{\ceqtype{pre}}
\newcommand{\wftypep}{\wftype{pre}}
\newcommand{\eqtypep}{\eqtype{pre}}
\newcommand{\cwftypek}{\cwftype{Kan}}
\newcommand{\ceqtypek}{\ceqtype{Kan}}
\newcommand{\wftypek}{\wftype{Kan}}
\newcommand{\eqtypek}{\eqtype{Kan}}
\newcommand{\cwftypex}{\cwftype{\kappa}}
\newcommand{\ceqtypex}{\ceqtype{\kappa}}
\newcommand{\wftypex}{\wftype{\kappa}}
\newcommand{\eqtypex}{\eqtype{\kappa}}

\NewDocumentCommand\coftype{O{\Psi} d<> m m}
  {#3 \in #4\ [#1 \IfValueT{#2}{\mid #2}]}
\NewDocumentCommand\ceqtm{O{\Psi} d<> m m m}
  {#3 \eq #4 \in #5\ [#1 \IfValueT{#2}{\mid #2}]}
\NewDocumentCommand\oftype{O{\Psi} d<> m m m}
  {#3 \gg #4 \in #5\ [#1 \IfValueT{#2}{\mid #2}]}
\NewDocumentCommand\eqtm{O{\Psi} d<> m m m m}
  {#3 \gg #4 \eq #5 \in #6\ [#1 \IfValueT{#2}{\mid #2}]}

\newcommand{\ceqtmtab}[4][\Psi]
  {\ensuremath{{#2} \eq{} & {#3} \in {#4}\ [#1]}}

\title{Computational Higher Type Theory III:\\
Univalent Universes and Exact Equality\nocite{uuee}}

\author{Carlo Angiuli\thanks{\texttt{cangiuli@cs.cmu.edu}}\\Carnegie Mellon University
  \and Kuen-Bang Hou (Favonia)\thanks{\texttt{favonia@ias.edu}}\\Institute for Advanced Study
  \and Robert Harper\thanks{\texttt{rwh@cs.cmu.edu}}\\Carnegie Mellon University}

\date{December, 2017}

\begin{document}

\maketitle

\begin{abstract}
This is the third in a series of papers extending Martin-L\"{o}f's \emph{meaning
explanations} of dependent type theory to a Cartesian cubical realizability
framework that accounts for higher-dimensional types. We extend this framework
to include a cumulative hierarchy of univalent Kan universes of Kan types;
exact equality and other pretypes lacking Kan structure; and a cumulative
hierarchy of pretype universes. As in Parts I and II, the main result is a
\emph{canonicity theorem} stating that closed terms of boolean type evaluate to
either true or false. This establishes the computational interpretation of
Cartesian cubical higher type theory based on cubical programs equipped with a
deterministic operational semantics.
\end{abstract}

\section{Introduction}
\label{sec:intro}

In Parts I and II of this series \citep{ahw2016cubical,ah2016cubicaldep} we
developed \emph{mathematical meaning explanations} for higher-dimensional type
theories with Cartesian cubical structure \citep{ahw2017cubical}. In Part III,
we extend these meaning explanations to support an infinite hierarchy of Kan,
univalent universes \citep{voevodskycmu}.

\paragraph{Mathematical meaning explanations}

We define the judgments of computational higher type theory as dimension-indexed
relations between programs equipped with a deterministic operational semantics.
These relations are cubical analogues of Martin-L\"{o}f's \emph{meaning
explanations} \citep{cmcp} and of the original Nuprl type theory
\citep{constableetalnuprl}, in which types are merely specifications of the
computational behavior of programs. Because types are defined behaviorally, we
trivially obtain the \emph{canonicity} property at every type. (Difficulties
instead lie in checking formation, introduction, and elimination rules. In
contrast, the type theory of \citet{cohen2016cubical} is defined by such rules,
and a separate argument by \citet{hubercanonicity} establishes canonicity.)

\begin{theorem}[Canonicity]
If $M$ is a closed term of type $\bool$, then $M \evals \true$ or $M \evals
\false$.
\end{theorem}

In a sense, our meaning explanations serve as \emph{cubical logical relations},
or a \emph{cubical realizability model}, justifying the rules presented in
\cref{sec:rules}. However, those rules are intended only for reference; the
rules included in the \RedPRL{} proof assistant \citep{redprl} differ
substantially (as described in \cref{sec:rules}).  Moreover, as
$\coftype[x_1,\dots,x_n]{M}{A}$ means that $M$ is a ($n$-dimensional) program
with behavior $A$, programs do not have unique types, nor are typing judgments
decidable.

\paragraph{Cartesian cubes}

Our programs are parametrized by \emph{dimension names} $x, y, \dots$ ranging
over an abstract interval with end points $0$ and $1$. Programs with at most $n$
free dimension names represent $n$-dimensional cubes: points ($n=0$), lines
($n=1$), squares ($n=2$), and so forth. Substituting $\dsubst{}{0}{x}$ or
$\dsubst{}{1}{x}$ yields the left or right face of a cube in dimension $x$;
substituting $\dsubst{}{y}{x}$ yields the $x,y$ diagonal; and weakening by $y$
yields a cube degenerate in the $y$ direction.

The resulting notion of cubes is Cartesian
\citep{licata2014cubical,awodey16cartesian,buchholtz2017}. In contrast, the
\citet{bch} model of type theory has only faces and degeneracies, while the
\citet{cohen2016cubical} type theory uses a de Morgan algebra of cubes with
connections ($x\land y$, $x\lor y$) and reversals ($1-x$) in addition to faces,
diagonals, and degeneracies. The Cartesian notion of cube is appealing because
it results in a \emph{structural} dimension context (with exchange, weakening,
and contraction) and requires no equational reasoning at the dimension level.

\paragraph{Kan operations}

\emph{Kan types} are types equipped with coercion ($\Coe$) and homogeneous
composition ($\Hcom$) operations. If $A$ is a Kan type varying in $x$, the
\emph{coercion} $\Coe*{x.A}$ sends an element $M$ of $\dsubst{A}{r}{x}$ to an
element of $\dsubst{A}{r'}{x}$, such that the coercion is equal to $M$ when
$r=r'$. For example, given a point $M$ in the $\dsubst{}{0}{x}$ side of the type
$A$, written $\coftype[\cdot]{M}{\dsubst{A}{0}{x}}$, we can coerce it to a point
$\Coe{x.A}{0}{1}{M}$ in $\dsubst{A}{1}{x}$, or coerce it to an $x$-line
$\Coe{x.A}{0}{x}{M}$ between $M$ and $\Coe{x.A}{0}{1}{M}$.
\[
\begin{tikzpicture}
  \node (lhs) at (0 , 1) {$M$} ;
  \node (rhs) at (4 , 1) {$\Coe{x.A}{0}{1}{M}$} ;
  \draw (lhs) [->] to node [auto] {$\Coe{x.A}{0}{x}{M}$} (rhs) ;

  \tikzset{shift={(8,0)}}

  \draw (0 , 2) [->] to node [above] {\small $x$} (0.5 , 2) ;
  \draw (0 , 2) [->] to node [left] {\small $y$} (0 , 1.5) ;
  \node (tl) at (1.5 , 2) {$\cdot$} ;
  \node (tr) at (5.5 , 2) {$\cdot$} ;
  \node (bl) at (1.5 , 0) {$\cdot$} ;
  \node (br) at (5.5 , 0) {$\cdot$} ;
  \draw (tl) [->] to node [above] {$M$} (tr) ;
  \draw (tl) [->] to node [left] {$N_0$} (bl) ;
  \draw (tr) [->] to node [right] {$N_1$} (br) ;
  \draw (bl) [->,dashed] to node [below] {$\Hcom{A}{0}{1}{M}{\cdots}$} (br) ;
  \node at (3.5 , 1) {$\Hcom{A}{0}{y}{M}{\cdots}$} ;
\end{tikzpicture}
\]

If $A$ is a Kan type, then \emph{homogeneous composition} in $A$ states that any
open box in $A$ has a composite; for example,
$\Hcom{A}{0}{1}{M}{\tube{x=0}{y.N_0},\tube{x=1}{y.N_1}}$ is the bottom line of
the above square. The cap $M$ is a line on the $\dsubst{}{0}{y}$ side of the
box; $y.N_0$ (resp., $y.N_1$) is a line on the $x=0$ (resp., $x=1$) side of the
box; and the composite is on the $\dsubst{}{1}{y}$ side of the box. Furthermore,
the cap and tubes must be equal where they coincide (the $x=0$ side of $M$ with
the $\dsubst{}{0}{y}$ side of $N_0$), every pair of tubes must be equal where
they coincide (vacuous here, as $x=0$ and $x=1$ are disjoint) and the composite
is equal to the tubes where they coincide (the $x=0$ side of the composite with
the $\dsubst{}{1}{y}$ side of $N_0$). Fillers are the special case in which we
compose to a free dimension name $y$; here,
$\Hcom{A}{0}{y}{M}{\tube{x=0}{y.N_0},\tube{x=1}{y.N_1}}$ is the entire square.

These Kan operations are variants of the uniform Kan conditions first proposed
by \citet{bch}. Notably, \citet{bch} and \citet{cohen2016cubical} combine
coercion and composition into a single heterogeneous composition operation and
do not allow compositions from or to dimension names. Unlike both
\citet{cohen2016cubical} and related work by \citet{licata2014cubical}, we allow
tubes along diagonals ($x=z$), and require every non-trivial box to contain at
least one opposing pair of tubes $x=0$ and $x=1$. The latter restriction
(detailed in \cref{def:valid}) allows us to achieve canonicity for
zero-dimensional elements of the circle and weak booleans.

\paragraph{Pretypes and exact equality}

As in the ``two-level type theories'' of \citet{voevodsky13hts},
\citet{altenkirch16strict}, and \citet{boulier17twolevel}, we allow for
\emph{pretypes} that are not necessarily Kan. In particular, we have types
$\Eq{A}{M}{N}$ of \emph{exact equalities} that internalize (and reflect into)
judgmental equalities $\ceqtm{M}{N}{A}$. Exact equality types are not, in
general, Kan, as one cannot compose exact equalities with non-degenerate lines.
However, unlike in prior two-level type theories, certain exact equality types
\emph{are} Kan (for example, when $A=\nat$; see \cref{sec:future} for a precise
characterization). We write $\cwftypep{A}$ when $A$ is a pretype, and
$\cwftypek{A}$ when $A$ is a Kan type. Pretypes and Kan types are both closed
under most type formers; for example, if $\cwftypex{A}$ and $\cwftypex{B}$ then
$\cwftypex{\arr{A}{B}}$.

\paragraph{Universes and univalence}

We have two cumulative hierarchies of universes $\Upre$ and $\UKan$
internalizing pretypes and Kan types respectively. The Kan universes $\UKan$ are
both Kan and univalent. (See \url{https://git.io/vFjUQ} for a \RedPRL{}-checked
proof of the univalence theorem.) Homogeneous compositions of Kan types are
types whose elements are formal $\Kbox$es of elements of the constituent types.
Every equivalence $E$ between $A$ and $B$ gives rise to the $\ua{x}{A,B,E}$ type
whose $x$-faces are $A$ and $B$; such types are a special case of ``Glue types''
\citep{cohen2016cubical}.

\paragraph{\RedPRL{}}

\RedPRL{} is an interactive proof assistant for computational higher type theory
in the tradition of LCF and Nuprl; the \RedPRL{} logic is principally organized
around dependent refinement rules \citep{spiwack2011,sterling2017}, which are
composed using a simple language of proof tactics. Unlike the inference rules
presented in \cref{sec:rules}, \RedPRL{}'s rules are given in the form of a
goal-oriented sequent calculus which is better-suited for both programming and
automation.

\subsection*{Acknowledgements}

We are greatly indebted to Steve Awodey, Marc Bezem, Evan Cavallo, Daniel
Gratzer, Simon Huber, Dan Licata, Ed Morehouse, Anders M\"ortberg, Jonathan
Sterling, and Todd Wilson for their contributions and advice.

This paper directly continues work previously described in
\citet{ahw2016cubical}, \citet{ah2016cubicaldep}, and \citet{ahw2017cubical},
whose primary antecedents are two-dimensional type theory \citep{lh2dtt}, the
\citet{bch} cubical model of type theory, and the cubical type theories of
\citet{cohen2016cubical} and \citet{licata2014cubical}.

The authors gratefully acknowledge the support of the Air Force Office of
Scientific Research through MURI grant FA9550-15-1-0053.  Any opinions, findings
and conclusions or recommendations expressed in this material are those of the
authors and do not necessarily reflect the views of the AFOSR.
The second author would also like to thank the Isaac Newton Institute for
Mathematical Sciences for its support and hospitality during the program ``Big
Proof'' when part of work on this paper was undertaken. The program was supported by EPSRC
grant number EP/K032208/1.

\newpage
\section{Programming language}
\label{sec:opsem}

The programming language itself has two sorts---dimensions and terms---and
binders for both sorts. Terms are an ordinary untyped lambda calculus with
constructors; dimensions are either dimension constants ($0$ or $1$) or
dimension names ($x,y,\dots$), the latter behaving like nominal constants
\citep{pittsnominal}. Dimensions may appear in terms: for example, $\lp{r}$ is a
term when $r$ is a dimension. The operational semantics is defined on terms that
are closed with respect to term variables but may contain free dimension names.

Dimension names represent generic elements of an abstract interval whose end
points are notated $0$ and $1$. While one may sensibly substitute any dimension
for a dimension name, terms are \emph{not} to be understood solely in terms
of their dimensionally-closed instances (namely, their end points). Rather, a
term's dependence on dimension names is to be understood generically;
geometrically, one might imagine additional unnamed points in the interior of
the abstract interval.

\subsection{Terms}

\begin{align*}
M &:=
\picl{a}{A}{B} \mid
\sigmacl{a}{A}{B} \mid
\Path{x.A}{M}{N} \mid
\Eq{A}{M}{N} \mid
\void \mid
\nat \mid
\bool \\&\mid
\wbool \mid
\C \mid
\Upre \mid
\UKan \mid
\ua{r}{A,B,E} \mid
\uain{r}{M,N} \mid
\uaproj{r}{M,F} \\&\mid
\lam{a}{M} \mid
\app{M}{N} \mid
\pair{M}{N} \mid
\fst{M} \mid
\snd{M} \mid
\dlam{x}{M} \mid
\dapp{M}{r} \mid
\ax \\&\mid
\z \mid
\suc{M} \mid
\natrec{M}{N_1}{n.a.N_2} \mid
\true \mid
\false \mid
\ifb{b.A}{M}{N_1}{N_2} \\&\mid
\base \mid
\lp{r} \mid
\Celim{c.A}{M}{N_1}{x.N_2} \\&\mid
\Coe*{x.A} \mid
\Hcom*{A}{r_i=r_i'} \\&\mid
\Com*{y.A}{r_i=r_i'} \mid
\Fcom*{r_i=r_i'} \\&\mid
\Ghcom*{A}{r_i=r_i'} \mid
\Gcom*{y.A}{r_i=r_i'} \\&\mid
\Kbox*{r_i=r_i'} \mid
\Kcap*{r_i=r_i'}
\end{align*}

We use capital letters like $M$, $N$, and $A$ to denote terms, $r$,
$r'$, $r_i$ to denote dimensions, $x$ to denote dimension names, $\e$
to denote dimension constants ($0$ or $1$), and $\eb$ to denote the
opposite dimension constant of $\e$.  We write $x.-$ for dimension
binders, $a.-$ for term binders, and $\fd{M}$ for the set of dimension
names free in $M$. Additionally, in $\picl{a}{A}{B}$ and $\sigmacl{a}{A}{B}$,
$a$ is bound in $B$. Dimension substitution $\dsubst{M}{r}{x}$ and term
substitution $\subst{M}{N}{a}$ are defined in the usual way.

The final argument of most composition operators is a (possibly empty) list of
triples $(r_i,r_i',y.N_i)$ whose first two components are dimensions, and whose
third is a term (in some cases, with a bound dimension). We write
$\sys{r_i=r_i'}{y.N_i}$ to abbreviate such lists or transformations on such
lists, and $\xi_i$ to abbreviate $r_i=r_i'$ when their identity is irrelevant.

\begin{definition}\label{def:wftm}
We write $\wftm{M}$ when $M$ is a term with no free term variables, and
$\fd{M}\subseteq\Psi$. (Similarly, we write $\wfval{M}$ when $\wftm{M}$ and
$\isval{M}$.)
\end{definition}

\begin{definition}
A total dimension substitution $\psitd$ assigns to each dimension name in
$\Psi$ either $0$, $1$, or a dimension name in $\Psi'$. It follows that if
$\wftm{M}$ then $\wftm[\Psi']{\td{M}{\psi}}$.
\end{definition}

\subsection{Operational semantics}

The following describes a deterministic weak head reduction evaluation strategy
for (term-)closed terms in the form of a transition system with two judgments:
\begin{enumerate}
\item $\isval{M}$, stating that $M$ is a \emph{value}, or
  \emph{canonical form}.
\item $M\steps M'$, stating that $M$ takes \emph{one step of
    evaluation} to $M'$.
\end{enumerate}
These judgments are defined so that if $\isval{M}$, then $M\not\steps$, but the
converse need not be the case. As usual, we write $M\steps^* M'$ to mean that
$M$ transitions to $M'$ in zero or more steps. We say $M$ evaluates to $V$,
written $M \evals V$, when $M\steps^* V$ and $\isval{V}$.

The $\steps$ judgment satisfies two additional conditions. Determinacy implies
that a term has at most one value; dimension preservation states that evaluation
does not introduce new (free) dimension names.
\begin{lemma}[Determinacy]
  If $M\steps M_1$ and $M\steps M_2$, then $M_1 = M_2$.
\end{lemma}

\begin{lemma}[Dimension preservation]
  If $M\steps M'$, then $\fd{M'}\subseteq\fd{M}$.
\end{lemma}

Many rules below are annotated with $\stable$. Those rules define an additional
pair of judgments $\sisval{M}$ and $M\ssteps M'$ by replacing every occurrence
of $\isvalsym$ (resp., $\steps$) in those rules with $\isvalsym_\stable$ (resp.,
$\ssteps$). These rules define the \emph{cubically-stable values} (resp.,
\emph{cubically-stable steps}), characterized by the following property:

\begin{lemma}[Cubical stability]
If $\wftm{M}$, then for any $\psitd$,
\begin{enumerate}
\item if $\sisval{M}$ then $\isval{\td{M}{\psi}}$, and
\item if $M\ssteps M'$ then $\td{M}{\psi}\steps\td{M'}{\psi}$.
\end{enumerate}
\end{lemma}

Cubically-stable values and steps are significant because they are unaffected by
the cubical apparatus. All standard operational semantics rules are
cubically-stable.

\paragraph{Types}

\begin{mathpar}
\Infer[\stable]
  { }
  {\isval{\picl{a}{A}{B}}}
\and
\Infer[\stable]
  { }
  {\isval{\sigmacl{a}{A}{B}}}
\and
\Infer[\stable]
  { }
  {\isval{\Path{x.A}{M}{N}}}
\and
\Infer[\stable]
  { }
  {\isval{\Eq{A}{M}{N}}}
\and
\Infer[\stable]
  { }
  {\isval{\void}}
\and
\Infer[\stable]
  { }
  {\isval{\nat}}
\and
\Infer[\stable]
  { }
  {\isval{\bool}}
\and
\Infer[\stable]
  { }
  {\isval{\wbool}}
\and
\Infer[\stable]
  { }
  {\isval{\C}}
\and
\Infer[\stable]
  { }
  {\isval{\Upre}}
\and
\Infer[\stable]
  { }
  {\isval{\UKan}}
\and
\Infer
  { }
  {\isval{\ua{x}{A,B,E}}}
\and
\Infer[\stable]
  { }
  {\ua{0}{A,B,E} \steps A}
\and
\Infer[\stable]
  { }
  {\ua{1}{A,B,E} \steps B}
\end{mathpar}

\paragraph{Kan operations}

\begin{mathpar}
\Infer[\stable]
  {A\steps A'}
  {\Hcom*{A}{\xi_i} \steps \Hcom*{A'}{\xi_i}}
\and
\Infer[\stable]
  {A\steps A'}
  {\Coe*{x.A} \steps \Coe*{x.A'}}
\and
\Infer[\stable]
  { }
  {\Com*{y.A}{\xi_i} \steps
   \Hcom{\dsubst{A}{r'}{y}}{r}{r'}{\Coe{y.A}{r}{r'}{M}}{\sys{\xi_i}{y.\Coe{y.A}{y}{r'}{N_i}}}}
\and
\Infer[\stable]
  {r = r'}
  {\Fcom*{\xi_i} \steps M}
\and
\Infer
  {r\neq r' \\ r_i\neq r_i'\ (\forall i<j) \\ r_j = r_j'}
  {\Fcom*{r_i=r_i'} \steps \dsubst{N_j}{r'}{y}}
\and
\Infer
  {r\neq r' \\ r_i\neq r_i'\ (\forall i)}
  {\isval{\Fcom*{r_i=r_i'}}}
\and
\Infer[\stable]
  { }
  {\Ghcom{A}{r}{r'}{M}{\cdot} \steps M}
\and
\Infer[\stable]
  {T_\e = \Hcom{A}{r}{z}{M}{
    \tube{s'=\e}{y.N},
    \tube{s'=\eb}{y.\Ghcom{A}{r}{y}{M}{\sys{\xi_i}{y.N_i}}},
    \sys{\xi_i}{y.N_i}}}
  {\Ghcom{A}{r}{r'}{M}{\tube{s=s'}{y.N},\sys{\xi_i}{y.N_i}} \steps \\
   \Hcom{A}{r}{r'}{M}{\sys{s=\e}{z.T_\e},\tube{s=s'}{y.N},\sys{\xi_i}{y.N_i}}}
\and
\Infer[\stable]
  { }
  {\Gcom*{y.A}{\xi_i} \steps
   \Ghcom{\dsubst{A}{r'}{y}}{r}{r'}{\Coe{y.A}{r}{r'}{M}}{\sys{\xi_i}{y.\Coe{y.A}{y}{r'}{N_i}}}}
\end{mathpar}

\paragraph{Dependent function types}

\begin{mathpar}
\Infer[\stable]
  {M \steps M'}
  {\app{M}{N} \steps \app{M'}{N}}
\and
\Infer[\stable]
  { }
  {\app{\lam{a}{M}}{N} \steps \subst{M}{N}{a}}
\and
\Infer[\stable]
  { }
  {\isval{\lam{a}{M}}}
\and
\Infer[\stable]
  { }
  {\Hcom*{\picl{a}{A}{B}}{\xi_i} \steps
   \lam{a}{\Hcom{B}{r}{r'}{\app{M}{a}}{\sys{\xi_i}{y.\app{N_i}{a}}}}}
\and
\Infer[\stable]
  { }
  {\Coe{x.\picl{a}{A}{B}}{r}{r'}{M} \steps
   \lam{a}{\Coe{x.\subst{B}{\Coe{x.A}{r'}{x}{a}}{a}}%
   {r}{r'}{\app{M}{\Coe{x.A}{r'}{r}{a}}}}}
\end{mathpar}

\paragraph{Dependent pair types}

\begin{mathpar}
\Infer[\stable]
  {M \steps M'}
  {\fst{M} \steps \fst{M'}}
\and
\Infer[\stable]
  {M \steps M'}
  {\snd{M} \steps \snd{M'}}
\and
\Infer[\stable]
  { }
  {\isval{\pair{M}{N}}}
\and
\Infer[\stable]
  { }
  {\fst{\pair{M}{N}} \steps M}
\and
\Infer[\stable]
  { }
  {\snd{\pair{M}{N}} \steps N}
\and
\Infer[\stable]
  {F = \Hcom{A}{r}{z}{\fst{M}}{\sys{\xi_i}{y.\fst{N_i}}}}
  {\Hcom*{\sigmacl{a}{A}{B}}{\xi_i}
   \steps \\
   \pair{\Hcom{A}{r}{r'}{\fst{M}}{\sys{\xi_i}{y.\fst{N_i}}}}
        {\Com{z.\subst{B}{F}{a}}{r}{r'}{\snd{M}}{\sys{\xi_i}{y.\snd{N_i}}}}}
\and
\Infer[\stable]
  { }
  {\Coe{x.\sigmacl{a}{A}{B}}{r}{r'}{M} \steps
   \pair{\Coe{x.A}{r}{r'}{\fst{M}}}
        {\Coe{x.\subst{B}{\Coe{x.A}{r}{x}{\fst{M}}}{a}}{r}{r'}{\snd{M}}}}
\end{mathpar}

\paragraph{Path types}

\begin{mathpar}
\Infer[\stable]
  {M \steps M'}
  {\dapp{M}{r} \steps \dapp{M'}{r}}
\and
\Infer[\stable]
  { }
  {\dapp{(\dlam{x}{M})}{r} \steps \dsubst{M}{r}{x}}
\and
\Infer[\stable]
  { }
  {\isval{\dlam{x}{M}}}
\and
\Infer[\stable]
  { }
  {\Hcom*{\Path{x.A}{P_0}{P_1}}{\xi_i} \steps
   \dlam{x}{\Hcom{A}{r}{r'}{\dapp{M}{x}}%
   {\sys{x=\e}{\_.P_\e},\sys{\xi_i}{y.\dapp{N_i}{x}}}}}
\and
\Infer[\stable]
  { }
  {\Coe{y.\Path{x.A}{P_0}{P_1}}{r}{r'}{M} \steps
   \dlam{x}{\Com{y.A}{r}{r'}{\dapp{M}{x}}{\sys{x=\e}{y.P_\e}}}}
\end{mathpar}

\paragraph{Equality types}

\begin{mathpar}
\Infer[\stable]
  { }
  {\isval{\ax}}
\and
\Infer[\stable]
  { }
  {\Hcom*{\Eq{A}{E_0}{E_1}}{\xi_i} \steps \star}
\end{mathpar}

\paragraph{Natural numbers}

\begin{mathpar}
\Infer[\stable]
  { }
  {\isval{\z}}
\and
\Infer[\stable]
  { }
  {\isval{\suc{M}}}
\and
\Infer[\stable]
  {M \steps M'}
  {\natrec{M}{Z}{n.a.S} \steps \natrec{M'}{Z}{n.a.S}}
\and
\Infer[\stable]
  { }
  {\natrec{\z}{Z}{n.a.S} \steps Z}
\and
\Infer[\stable]
  { }
  {\natrec{\suc{M}}{Z}{n.a.S} \steps \subst{\subst{S}{M}{n}}{\natrec{M}{Z}{n.a.S}}{a}}
\and
\Infer[\stable]
  { }
  {\Hcom*{\nat}{\xi_i} \steps M}
\and
\Infer[\stable]
  { }
  {\Coe*{x.\nat} \steps M}
\end{mathpar}

\paragraph{Booleans}

\begin{mathpar}
\Infer[\stable]
  { }
  {\isval{\true}}
\and
\Infer[\stable]
  { }
  {\isval{\false}}
\and
\Infer[\stable]
  {M \steps M'}
  {\ifb{b.A}{M}{T}{F} \steps \ifb{b.A}{M'}{T}{F}}
\and
\Infer[\stable]
  { }
  {\ifb{b.A}{\true}{T}{F} \steps T}
\and
\Infer[\stable]
  { }
  {\ifb{b.A}{\false}{T}{F} \steps F}
\and
\Infer[\stable]
  { }
  {\Hcom*{\bool}{\xi_i} \steps M}
\and
\Infer[\stable]
  { }
  {\Coe*{x.\bool} \steps M}
\end{mathpar}

\paragraph{Weak booleans}

\begin{mathpar}
\Infer[\stable]
  { }
  {\Hcom*{\wbool}{\xi_i} \steps \Fcom*{\xi_i}}
\and
\Infer
  {r\neq r' \\ r_i\neq r_i'\ (\forall i) \\
   H = \Fcom{r}{z}{M}{\sys{r_i=r_i'}{y.N_i}}}
  {\ifb{b.A}{\Fcom*{r_i=r_i'}}{T}{F}
   \steps \\
   \Com{z.\subst{A}{H}{b}}{r}{r'}{\ifb{b.A}{M}{T}{F}}{\sys{r_i=r_i'}{y.\ifb{b.A}{N_i}{T}{F}}}}
\and
\Infer[\stable]
  { }
  {\Coe*{x.\wbool} \steps M}
\end{mathpar}

\paragraph{Circle}

\begin{mathpar}
\Infer[\stable]
  { }
  {\Hcom*{\C}{\xi_i} \steps \Fcom*{\xi_i}}
\and
\Infer[\stable]
  { }
  {\lp{\e} \steps \base}
\and
\Infer[\stable]
  { }
  {\isval{\base}}
\and
\Infer
  { }
  {\isval{\lp{x}}}
\and
\Infer[\stable]
  {M \steps M'}
  {\Celim{c.A}{M}{P}{x.L} \steps \Celim{c.A}{M'}{P}{x.L}}
\and
\Infer[\stable]
  { }
  {\Celim{c.A}{\base}{P}{x.L} \steps P}
\and
\Infer
  { }
  {\Celim{c.A}{\lp{w}}{P}{x.L} \steps \dsubst{L}{w}{x}}
\and
\Infer
  {r \neq r' \\ r_i\neq r_i'\ (\forall i) \\
   F = \Fcom{r}{z}{M}{\sys{r_i=r_i'}{y.N_i}}}
  {\Celim{c.A}{\Fcom*{r_i=r_i'}}{P}{x.L}
   \steps \\
   \Com{z.\subst{A}{F}{c}}{r}{r'}{\Celim{c.A}{M}{P}{x.L}}{\sys{r_i=r_i'}{y.\Celim{c.A}{N_i}{P}{x.L}}}}
\and
\Infer[\stable]
  { }
  {\Coe*{x.\C} \steps M}
\end{mathpar}

\paragraph{Univalence}\

\begin{mathparpagebreakable}
\Infer
  { }
  {\isval{\uain{x}{M,N}}}
\and
\Infer[\stable]
  { }
  {\uain{0}{M,N} \steps M}
\and
\Infer[\stable]
  { }
  {\uain{1}{M,N} \steps N}
\and
\Infer[\stable]
  { }
  {\uaproj{0}{M,F} \steps \app{F}{M}}
\and
\Infer[\stable]
  { }
  {\uaproj{1}{M,F} \steps M}
\and
\Infer
  {M \steps M'}
  {\uaproj{x}{M,F} \steps \uaproj{x}{M',F}}
\and
\Infer
  { }
  {\uaproj{x}{\uain{x}{M,N},F} \steps N}
\and
\Infer
  {O = \Hcom{A}{r}{y}{M}{\sys{\xi_i}{y.N_i}} \\
   \etc{T} =
     \tube{x=0}{y.\app{\fst{E}}{O}},
     \tube{x=1}{y.\Hcom{B}{r}{y}{M}{\sys{\xi_i}{y.N_i}}}}
  {\Hcom*{\ua{x}{A,B,E}}{\xi_i}
   \steps \\
   \uain{x}{\dsubst{O}{r'}{y},
     \Hcom{B}{r}{r'}{\uaproj{x}{M,\fst{E}}}{\sys{\xi_i}{y.\uaproj{x}{N_i,\fst{E}}},\etc{T}}}}
\and
\Infer[\stable]
  { }
  {\Coe{x.\ua{x}{A,B,E}}{0}{r'}{M} \steps
   \uain{r'}{M,\Coe{x.B}{0}{r'}{\app{\fst{\dsubst{E}{0}{x}}}{M}}}}
\and
\Infer[\stable]
  {O = \fst{\app{\snd{\dsubst{E}{r'}{x}}}{\Coe{x.B}{1}{r'}{N}}} \\
   P = \Hcom{\dsubst{B}{r'}{x}}{1}{0}{\Coe{x.B}{1}{r'}{N}}{
    \tube{r'=0}{y.\dapp{\snd{O}}{y}},
    \tube{r'=1}{\_.\Coe{x.B}{1}{r'}{N}}}}
  {\Coe{x.\ua{x}{A,B,E}}{1}{r'}{N} \steps
   \uain{r'}{\fst{O},P}}
\and
\Infer
  {O_\e = \uaproj{w}{\Coe{x.\ua{x}{A,B,E}}{\e}{w}{M},\fst{\dsubst{E}{w}{x}}} \\
   P = \Com{x.B}{y}{x}{\uaproj{y}{M,\fst{\dsubst{E}{y}{x}}}}{\etc{\tube{y=\e}{w.O_\e}}} \\
   Q_\e[a] = \pair%
     {\Coe{y.\dsubst{A}{0}{x}}{\e}{y}{a}}%
     {\dlam{z}{\Com{y.\dsubst{B}{0}{x}}{\e}{y}{\dsubst{\dsubst{P}{0}{x}}{\e}{y}}%
       {\etc{U}}}} \\
   \etc{U} =
     \tube{z=0}{y.\app{\fst{\dsubst{E}{0}{x}}}{\Coe{y.\dsubst{A}{0}{x}}{\e}{y}{a}}},
     \tube{z=1}{y.\dsubst{P}{0}{x}} \\
   R = \dapp{\app{\app{\snd{\app{\snd{\dsubst{E}{0}{x}}}{\dsubst{P}{0}{x}}}}{Q_0[\dsubst{M}{0}{y}]}}%
     {Q_1[\dsubst{(\Coe{x.\ua{x}{A,B,E}}{1}{0}{M})}{1}{y}]}}{y} \\
   \etc{T} =
     \etc{\tube{y=\e}{\_.\dsubst{O_\e}{r'}{w}}},
     \tube{y=r'}{\_.\uaproj{r'}{M,\fst{\dsubst{E}{r'}{x}}}},
     \tube{r'=0}{z.\dapp{\snd{R}}{z}}}
  {\Coe{x.\ua{x}{A,B,E}}{y}{r'}{M} \steps
   \uain{r'}{\fst{R},\Hcom{\dsubst{B}{r'}{x}}{1}{0}{\dsubst{P}{r'}{x}}{\etc{T}}}}
\and
\Infer
  {x\neq y \\
   \etc{T} =
     \tube{x=0}{y.\app{\fst{E}}{\Coe{y.A}{r}{y}{M}}},
     \tube{x=1}{y.\Coe{y.B}{r}{y}{M}}}
  {\Coe{y.\ua{x}{A,B,E}}{r}{r'}{M} \steps
   \uain{x}{\Coe{y.A}{r}{r'}{M},\Com{y.B}{r}{r'}{\uaproj{x}{M,\fst{\dsubst{E}{r}{y}}}}{\etc{T}}}}
\end{mathparpagebreakable}

\paragraph{Universes}\

\begin{mathparpagebreakable}
\Infer[\stable]
  { }
  {\Hcom*{\UKan}{\xi_i} \steps \Fcom*{\xi_i}}
\and
\Infer[\stable]
  { }
  {\Coe*{x.\Ux} \steps M}
\and
\Infer[\stable]
  {r = r'}
  {\Kbox*{\xi_i} \steps M}
\and
\Infer
  {r\neq r' \\ r_i\neq r_i'\ (\forall i<j) \\ r_j = r_j'}
  {\Kbox*{r_i=r_i'} \steps N_j}
\and
\Infer
  {r\neq r' \\ r_i\neq r_i'\ (\forall i)}
  {\isval{\Kbox*{r_i=r_i'}}}
\and
\Infer[\stable]
  {r = r'}
  {\Kcap*{\xi_i} \steps M}
\and
\Infer
  {r\neq r' \\ r_i\neq r_i'\ (\forall i<j) \\ r_j = r_j'}
  {\Kcap*{r_i=r_i'} \steps \Coe{y.B_j}{r'}{r}{M}}
\and
\Infer
  {r\neq r' \\ r_i\neq r_i'\ (\forall i) \\ M \steps M'}
  {\Kcap{r}{r'}{M}{\sys{r_i=r_i'}{y.B_i}} \steps
   \Kcap{r}{r'}{M'}{\sys{r_i=r_i'}{y.B_i}}}
\and
\Infer
  {r\neq r' \\ r_i\neq r_i'\ (\forall i)}
  {\Kcap{r}{r'}{\Kbox*{\xi_i}}{\sys{r_i=r_i'}{y.B_i}} \steps M}
\and
\Infer
  {s\neq s' \\
   s_j\neq s_j'\ (\forall j) \\
   P_j = \Hcom{B_j}{r}{r'}{\Coe{z.B_j}{s'}{z}{M}}{
     \sys{r_i=r_i'}{y.\Coe{z.B_j}{s'}{z}{N_i}}} \\
   F[c] = \Hcom{A}{s'}{z}{\Kcap{s}{s'}{c}{\sys{s_j=s_j'}{z.B_j}}}{
     \sys{s_j=s_j'}{z'.\Coe{z.B_j}{z'}{s}{\Coe{z.B_j}{s'}{z'}{c}}}} \\
   O = \Hcom{A}{r}{r'}{\dsubst{(F[M])}{s}{z}}{\sys{r_i=r_i'}{y.\dsubst{(F[N_i])}{s}{z}}} \\
   Q = \Hcom{A}{s}{s'}{O}{
     \sys{r_i=r_i'}{z.F[\dsubst{N_i}{r'}{y}]},
     \sys{s_j=s_j'}{z.\Coe{z.B_j}{z}{s}{P_j}},
     \tube{r=r'}{z.F[M]}}}
  {\Hcom{\Fcom{s}{s'}{A}{\sys{s_j=s_j'}{z.B_j}}}{r}{r'}{M}{\sys{r_i=r_i'}{y.N_i}} \steps
   \Kbox{s}{s'}{Q}{\sys{s_j=s_j'}{\dsubst{P_j}{s'}{z}}}}
\and
\Infer
  {s\neq s' \\
   s_i\neq s_i'\ (\forall i) \\
   N_i = \Coe{z.B_i}{s'}{z}{\Coe{x.\dsubst{B_i}{s'}{z}}{r}{x}{M}} \\
   O = \dsubst{(\Hcom{A}{s'}{z}{\Kcap{s}{s'}{M}{\sys{s_i=s_i'}{z.B_i}}}{
     \sys{s_i=s_i'}{z.\Coe{z.B_i}{z}{s}{N_i}}})}{r}{x} \\
   P = \Gcom{x.A}{r}{r'}{\dsubst{O}{\dsubst{s}{r}{x}}{z}}{
     \st{\sys{s_i=s_i'}{x.\dsubst{N_i}{s}{z}}}{x\fresh s_i,s_i'},
     \st{\tube{s=s'}{x.\Coe{x.A}{r}{x}{M}}}{x\fresh s,s'}} \\
   Q_k = \Gcom{z.\dsubst{B_k}{r'}{x}}{\dsubst{s}{r'}{x}}{z}{P}{
     \st{\sys{s_i=s_i'}{z.\dsubst{N_i}{r'}{x}}}{x\fresh s_i,s_i'},
     \tube{r=r'}{z.\dsubst{N_k}{r'}{x}}}}
  {\Coe{x.\Fcom{s}{s'}{A}{\sys{s_i=s_i'}{z.B_i}}}{r}{r'}{M} \steps \\
   \dsubst{(\Kbox{s}{s'}{\Hcom{A}{s}{s'}{P}{
     \sys{s_i=s_i'}{z.\Coe{z.B_i}{z}{s}{Q_i}},
     \tube{r=r'}{z.O}}}{\sys{s_i=s_i'}{\dsubst{Q_i}{s'}{z}}})}{r'}{x}}
\end{mathparpagebreakable}

\newpage
\section{Cubical type systems}
\label{sec:typesys}

In this paper, we define the judgments of higher type theory relative to a
\emph{cubical type system}, a family of relations over values in the
previously-described programming language. In this section we describe how to
construct a particular cubical type system that will validate the rules given in
\cref{sec:rules}; this construction is based on similar constructions outlined
by \citet{allen1987types} and \citet{harper1992typesys}.

\begin{definition}
A \emph{candidate cubical type system} is a relation $\tau(\Psi,A_0,B_0,\phi)$
over $\wfval{A_0}$, $\wfval{B_0}$, and binary relations $\phi(M_0,N_0)$ over
$\wfval{M_0}$ and $\wfval{N_0}$.
\end{definition}

For any relation $R$ with value arguments, we define $\lift{R}$ as its
evaluation lifting to terms. For example, $\lift{\tau}(\Psi,A,B,\phi)$ when
there exist $A_0$ and $B_0$ such that $A\evals A_0$, $B\evals B_0$, and
$\tau(\Psi,A_0,B_0,\phi)$.

\begin{definition}
A \emph{$\Psi$-relation} is a family of binary relations $\alpha_\psi(M,N)$
indexed by substitutions $\psitd$, relating $\wftm[\Psi']{M}$ and
$\wftm[\Psi']{N}$. (We will write $\alpha(M,N)$ in place of
$\alpha_{\id}(M,N)$.) We are often interested in \emph{$\Psi$-relations over
values}, which relate only values. If a $\Psi$-relation depends only on the
choice of $\Psi'$ and not $\psi$, we instead call it \emph{context-indexed} and
write $\alpha_{\Psi'}(M,N)$.
\end{definition}

We can precompose any $\Psi$-relation $\alpha$ by a dimension substitution
$\psitd$ to yield a $\Psi'$-relation $(\td{\alpha}{\psi})_{\psi'}(M,N) =
\alpha_{\psi\psi'}(M,N)$. Context-indexed relations are indeed families of
binary relations indexed by contexts $\Psi'$, because the choice of $\Psi$ and
$\psi$ are irrelevant---every $\Psi,\Psi'$ have at least one dimension
substitution between them. We write $R(\Psi')$ for the context-indexed relation
$R$ regarded as a $\Psi'$-relation.

\begin{definition}\label{def:ptyrel}
For any candidate cubical type system $\tau$, the relation
$\PTy(\tau)(\Psi,A,B,\alpha)$ over $\wftm{A}$, $\wftm{B}$, and a $\Psi$-relation
over values $\alpha$ holds if for all $\psitd$ we have
$\lift{\tau}(\Psi',\td{A}{\psi},\td{B}{\psi},\alpha_\psi)$, and for all
$\tds{\Psi_1}{\psi_1}{\Psi}$ and $\tds{\Psi_2}{\psi_2}{\Psi_1}$, we have
$\td{A}{\psi_1} \evals A_1$, $\td{B}{\psi_1} \evals B_1$,
$\lift{\tau}(\Psi_2,\td{A_1}{\psi_2},\td{A}{\psi_1\psi_2},\phi)$,
$\lift{\tau}(\Psi_2,\td{A}{\psi_1\psi_2},\td{A_1}{\psi_2},\phi)$,
$\lift{\tau}(\Psi_2,\td{B_1}{\psi_2},\td{B}{\psi_1\psi_2},\phi)$,
$\lift{\tau}(\Psi_2,\td{B}{\psi_1\psi_2},\td{B_1}{\psi_2},\phi)$, and
$\lift{\tau}(\Psi_2,\td{A_1}{\psi_2},\td{B_1}{\psi_2},\phi)$.
\end{definition}

\begin{definition}\label{def:tmrel}
For any $\Psi$-relation on values $\alpha$, the relation $\Tm(\alpha)(M,N)$ over
$\wftm{M}$ and $\wftm{N}$ holds if for all $\tds{\Psi_1}{\psi_1}{\Psi}$ and
$\tds{\Psi_2}{\psi_2}{\Psi_1}$, we have $\td{M}{\psi_1} \evals M_1$,
$\td{N}{\psi_1} \evals N_1$,
$\lift{\alpha_{\psi_1\psi_2}}(\td{M_1}{\psi_2},\td{M}{\psi_1\psi_2})$,
$\lift{\alpha_{\psi_1\psi_2}}(\td{M}{\psi_1\psi_2},\td{M_1}{\psi_2})$,
$\lift{\alpha_{\psi_1\psi_2}}(\td{N_1}{\psi_2},\td{N}{\psi_1\psi_2})$,
$\lift{\alpha_{\psi_1\psi_2}}(\td{N}{\psi_1\psi_2},\td{N_1}{\psi_2})$, and
$\lift{\alpha_{\psi_1\psi_2}}(\td{M_1}{\psi_2},\td{N_1}{\psi_2})$.
\end{definition}

\begin{definition}
A $\Psi$-relation on values $\alpha$ is \emph{value-coherent}, or
$\Coh(\alpha)$, when for all $\psitd$, if $\alpha_\psi(M_0,N_0)$ then
$\Tm(\td{\alpha}{\psi})(M_0,N_0)$.
\end{definition}

These relations are closed under dimension substitution by construction---for
any $\psitd$, if $\PTy(\tau)(\Psi,A,B,\alpha)$ then
$\PTy(\tau)(\Psi',\td{A}{\psi},\td{B}{\psi},\td{\alpha}{\psi})$, if
$\Tm(\alpha)(M,N)$ then $\Tm(\td{\alpha}{\psi})(\td{M}{\psi},\td{N}{\psi})$, and
if $\Coh(\alpha)$ then $\Coh(\td{\alpha}{\psi})$.

\subsection{Fixed points}

$\Psi$-relations (and context-indexed relations) over values form a complete
lattice when ordered by inclusion. By the Knaster-Tarski fixed point theorem,
any order-preserving operator $F(x)$ on a complete lattice has a least fixed
point $\mu x.F(x)$ that is also its least pre-fixed point
\citep[2.35]{daveypriestleylattices}.

We define the canonical element equality relations of inductive
types---$\mathbb{N}$ for natural numbers, $\mathbb{B}$ for weak booleans, and
$\mathbb{C}$ for the circle---as context-indexed relations (written here as
three-place relations) that are least fixed points of order-preserving
operators:
\begin{align*}
\mathbb{N} &= \mu R.
(\{(\Psi,\z,\z)\}\cup \{(\Psi,\suc{M},\suc{M'}) \mid \Tm(R(\Psi))(M,M')\}) \\
\mathbb{B} &= \mu R.
(\{(\Psi,\true,\true),(\Psi,\false,\false)\} \cup \textsc{FKan}(R)) \\
\mathbb{C} &= \mu R.
(\{(\Psi,\base,\base),((\Psi,x),\lp{x},\lp{x})\} \cup \textsc{FKan}(R))
\end{align*}
where
{\def\spacer{\hphantom{{}={} \{}}
\begin{align*}
\textsc{FKan}(R) &=
\{(\Psi,\Fcom*{r_i=r_i'},\Fcom{r}{r'}{M'}{\sys{r_i=r_i'}{y.N_i'}}) \mid \\&\spacer
(r\neq r') \land
(\forall i. r_i \neq r_i') \land
(\exists i,j. (r_i = r_j) \land (r_i' = 0) \land (r_j' = 1)) \land
\Tm(R(\Psi))(M,M') \\&\spacer
{}\land (\forall i,j,\tds{\Psi'}{\psi}{(\Psi,y)}.
((\td{r_i}{\psi} = \td{r_i'}{\psi}) \land
(\td{r_j}{\psi} = \td{r_j'}{\psi})) \implies
\Tm(R(\Psi'))(\td{N_i}{\psi},\td{N_j'}{\psi})) \\&\spacer
{}\land (\forall i,\psitd.
(\td{r_i}{\psi} = \td{r_i'}{\psi}) \implies
\Tm(R(\Psi'))(\td{\dsubst{N_i}{r}{y}}{\psi},\td{M}{\psi})) \})
\end{align*}}%
The operators $\Tm$ and $\textsc{FKan}$ are order-preserving because they
only use their argument relations in positive positions.

Similarly, candidate cubical type systems form a complete lattice, and we define
a sequence of candidate cubical type systems as least fixed points of
order-preserving operators, using the following auxiliary definitions for each
type former:

{\def\spacer{\hphantom{{}={}\{}}
\allowdisplaybreaks
\begin{align*}
\textsc{Fun}(\tau) &= \{
(\Psi,\picl{a}{A}{B},\picl{a}{A'}{B'},\phi) \mid \\&\spacer
\exists \alpha,\beta^{(-,-,-)}. \PTy(\tau)(\Psi,A,A',\alpha)
  \land \Coh(\alpha) \\&\spacer
{}\land (\forall\psi,M,M'. \Tm(\td{\alpha}{\psi})(M,M') \implies\\&\spacer\qquad
  \PTy(\tau)(\Psi',\subst{\td{B}{\psi}}{M}{a},
  \subst{\td{B'}{\psi}}{M'}{a},\beta^{\psi,M,M'})
  \land \Coh(\beta^{\psi,M,M'})) \\&\spacer
{}\land (\phi = \{(\lam{a}{N},\lam{a}{N'}) \mid
  \forall\psi,M,M'. \Tm(\td{\alpha}{\psi})(M,M') \implies \\&\spacer\qquad
  \Tm(\beta^{\psi,M,M'})(\subst{\td{N}{\psi}}{M}{a},\subst{\td{N'}{\psi}}{M'}{a})
\}) \} \\
\textsc{Pair}(\tau) &= \{
(\Psi,\sigmacl{a}{A}{B},\sigmacl{a}{A'}{B'},\phi) \mid \\&\spacer
\exists \alpha,\beta^{(-,-,-)}. \PTy(\tau)(\Psi,A,A',\alpha)
  \land \Coh(\alpha) \\&\spacer
{}\land (\forall\psi,M,M'. \Tm(\td{\alpha}{\psi})(M,M') \implies\\&\spacer\qquad
  \PTy(\tau)(\Psi',\subst{\td{B}{\psi}}{M}{a},
    \subst{\td{B'}{\psi}}{M'}{a},\beta^{\psi,M,M'})
  \land \Coh(\beta^{\psi,M,M'})) \\&\spacer
{}\land (\phi = \{(\pair{M}{N},\pair{M'}{N'}) \mid
  \Tm(\alpha)(M,M') \land \Tm(\beta^{\id[\Psi],M,M'})(N,N')
\}) \} \\
\textsc{Path}(\tau) &= \{
(\Psi,\Path{x.A}{P_0}{P_1},\Path{x.A'}{P_0'}{P_1'},\phi) \mid \\&\spacer
\exists \alpha. \PTy(\tau)((\Psi,x),A,A',\alpha) \land \Coh(\alpha)
  \land (\forall\e. \Tm(\dsubst{\alpha}{\e}{x})(P_\e,P_\e')) \\&\spacer
{}\land (\phi = \{(\dlam{x}{M},\dlam{x}{M'}) \mid
  \Tm(\alpha)(M,M') \land
  (\forall\e. \Tm(\dsubst{\alpha}{\e}{x})(\dsubst{M}{\e}{x},P_\e))
\}) \} \\
\textsc{Eq}(\tau) &= \{
(\Psi,\Eq{A}{M}{N},\Eq{A'}{M'}{N'},\phi) \mid \\&\spacer
\exists \alpha. \PTy(\tau)(\Psi,A,A',\alpha) \land \Coh(\alpha)
  \land \Tm(\alpha)(M,M') \land \Tm(\alpha)(N,N') \\&\spacer
{}\land (\phi = \{(\ax,\ax) \mid \Tm(\alpha)(M,N) \}) \} \\
\textsc{V}(\tau) &= \{
((\Psi,x),\ua{x}{A,B,E},\ua{x}{A',B',E'},\phi) \mid \\&\spacer
\exists \beta,\alpha^{(-)},\eta^{(-)}.
  \PTy(\tau)((\Psi,x),B,B',\beta) \land \Coh(\beta) \\&\spacer
{}\land (\forall\psi.(\td{x}{\psi} = 0) \implies
  \PTy(\tau)(\Psi',\td{A}{\psi},\td{A'}{\psi},\alpha^\psi)
  \land \Coh(\alpha^\psi) \\&\spacer\qquad
{}\land \PTy(\tau)(\Psi',\Equiv{\td{A}{\psi}}{\td{B}{\psi}},
  \Equiv{\td{A}{\psi}}{\td{B}{\psi}},\eta^\psi)
  \land \Tm(\eta^\psi)(\td{E}{\psi},\td{E'}{\psi})) \\&\spacer
{}\land (\phi = \{(\uain{x}{M,N},\uain{x}{M',N'}) \mid
  \Tm(\beta)(N,N') \land (\forall\psi. (\td{x}{\psi} = 0) \implies \\&\spacer\qquad
\Tm(\alpha^\psi)(\td{M}{\psi},\td{M'}{\psi}) \land
\Tm(\td{\beta}{\psi})(\app{\fst{\td{E}{\psi}}}{\td{M}{\psi}},\td{N}{\psi}) )\} \} \\
\textsc{Fcom}(\tau) &= \{
(\Psi,\Fcom{r}{r'}{A}{\sys{r_i=r_i'}{y.B_i}},
  \Fcom{r}{r'}{A'}{\sys{r_i=r_i'}{y.B_i'}}) \mid \\&\spacer
\exists \alpha,\beta^{(-,-,-)}.
  (r\neq r') \land
  (\forall i. r_i \neq r_i') \land
  (\exists i,j. (r_i = r_j) \land (r_i' = 0) \land (r_j' = 1)) \\&\spacer
{}\land \PTy(\tau)(\Psi,A,A',\alpha) \land \Coh(\alpha) \\&\spacer
{}\land (\forall i,j,\tds{\Psi'}{\psi}{(\Psi,y)}.
  ((\td{r_i}{\psi} = \td{r_i'}{\psi}) \land
  (\td{r_j}{\psi} = \td{r_j'}{\psi})) \implies \\&\spacer\qquad
\PTy(\tau)(\Psi',\td{B_i}{\psi},\td{B_j'}{\psi},\beta^{\psi,i,j})
  \land \Coh(\beta^{\psi,i,j})) \\&\spacer
{}\land (\forall i,\psi.
  (\td{r_i}{\psi} = \td{r_i'}{\psi}) \implies
  \PTy(\tau)(\Psi',\td{\dsubst{B_i}{r}{y}}{\psi},\td{A}{\psi},\_)) \}) \\&\spacer
{}\land (\phi = \{(\Kbox{r}{r'}{M}{\sys{r_i=r_i'}{N_i}},
  \Kbox{r}{r'}{M'}{\sys{r_i=r_i'}{N_i'}}) \mid \Tm(\alpha)(M,M') \\&\spacer\qquad
{}\land (\forall i,j,\psi.
  ((\td{r_i}{\psi} = \td{r_i'}{\psi}) \land
  (\td{r_j}{\psi} = \td{r_j'}{\psi})) \implies
  \Tm(\dsubst{\beta^{\psi,i,j}}{\td{r'}{\psi}}{y})
    (\td{N_i}{\psi},\td{N_j'}{\psi})) \\&\spacer\qquad
{}\land (\forall i,\psi.
  (\td{r_i}{\psi} = \td{r_i'}{\psi}) \implies
  \Tm(\td{\alpha}{\psi})(\td{M}{\psi},
    \Coe{y.\td{B_i}{\psi}}{\td{r'}{\psi}}{\td{r}{\psi}}{\td{N_i}{\psi}})) \}) \} \\
\textsc{Void} &= \{
(\Psi,\void,\void, \{\}) \} \\
\textsc{Nat} &= \{
(\Psi,\nat,\nat,\mathbb{N}_\Psi) \} \\
\textsc{Bool} &= \{
(\Psi,\bool,\bool,\{(\true,\true),(\false,\false)\}) \} \\
\textsc{WB} &= \{
(\Psi,\wbool,\wbool,\mathbb{B}_\Psi) \} \\
\textsc{Circ} &= \{
(\Psi,\C,\C,\mathbb{C}_\Psi) \} \\
\textsc{UPre}(\nu) &= \{
(\Psi,\Upre[j],\Upre[j],\phi) \mid \nu(\Psi,\Upre[j],\Upre[j],\phi) \} \\
\textsc{UKan}(\nu) &= \{
(\Psi,\UKan[j],\UKan[j],\phi) \mid \nu(\Psi,\UKan[j],\UKan[j],\phi) \}
\end{align*}}
In the $\textsc{V}$ case, and for the remainder of this paper, we use the
abbreviations
\begin{align*}
\isContr{C} &:= \prd{C}{(\picl{c}{C}{\picl{c'}{C}{\Path{\_.C}{c}{c'}}})} \\
\Equiv{A}{B} &:=
\sigmacl{f}{\arr{A}{B}}{(\picl{b}{B}{\isContr{\sigmacl{a}{A}{\Path{\_.B}{\app{f}{a}}{b}}}})}.
\end{align*}

For candidate cubical type systems $\nu,\sigma,\tau$, define
\begin{align*}
P(\nu,\sigma,\tau) ={}&
\textsc{Fun}(\tau) \cup
\textsc{Pair}(\tau) \cup
\textsc{Path}(\tau) \cup
\textsc{Eq}(\tau) \cup
\textsc{V}(\tau) \cup
\textsc{Fcom}(\sigma) \\& {}\cup
\textsc{Void} \cup
\textsc{Nat} \cup
\textsc{Bool} \cup
\textsc{WB} \cup
\textsc{Circ} \cup
\textsc{UPre}(\nu) \cup
\textsc{UKan}(\nu) \\
K(\nu,\sigma) ={}&
\textsc{Fun}(\sigma) \cup
\textsc{Pair}(\sigma) \cup
\textsc{Path}(\sigma) \cup
\textsc{V}(\sigma) \cup
\textsc{Fcom}(\sigma) \\& {}\cup
\textsc{Void} \cup
\textsc{Nat} \cup
\textsc{Bool} \cup
\textsc{WB} \cup
\textsc{Circ} \cup
\textsc{UKan}(\nu)
\end{align*}
The operator $P$ includes $\textsc{Eq}$ and $\textsc{UPre}$ while $K$ does not;
furthermore, in $P$ only $\textsc{Fcom}$ varies in $\sigma$. The operators $P$
and $K$ are order-preserving in all arguments because $\PTy$ and each type
operator only use their argument in strictly positive positions.

\begin{lemma}\label{lem:lfp-misc}
In any complete lattice,
\begin{enumerate}
\item If $F(x)$ and $G(x)$ are order-preserving and $F(x)\subseteq G(x)$ for all
$x$, then $\mu x.F(x) \subseteq \mu x.G(x)$.

\item If $F(x,y)$ and $G(x,y)$ are order-preserving and $F(x,y)\subseteq G(x,y)$
whenever $x\subseteq y$, then $\mu_f\subseteq\mu_g$ where $(\mu_f,\mu_g) = \mu
(x,y). (F(x,y),G(x,y))$.
\end{enumerate}
\end{lemma}
\begin{proof}
For part (1), $\mu x.G(x)$ is a pre-fixed point of $F$ because $F(\mu
x.G(x))\subseteq G(\mu x.G(x)) = \mu x.G(x)$. But $\mu x.F(x)$ is the least
such, so $\mu x.F(x) \subseteq \mu x.G(x)$.

For part (2), let $\mu_{\cap} = \mu_f \cap \mu_g$.
Note $(\mu_{\cap}, \mu_g)$ is a pre-fixed point of $(x,y)\mapsto (F(x,y), G(x,y))$
because, by assumption and $(\mu_f,\mu_g)$ being a fixed point,
$F(\mu_{\cap}, \mu_g) \subseteq F(\mu_f, \mu_g) = \mu_f$ and
$F(\mu_{\cap}, \mu_g) \subseteq G(\mu_{\cap}, \mu_g) \subseteq G(\mu_f,\mu_g) = \mu_g$.
This implies $(\mu_f, \mu_g) \subseteq (\mu_{\cap}, \mu_g)$ and thus $\mu_f \subseteq \mu_g$.
\end{proof}

\begin{lemma}
Let $\pre\mu(\nu,\sigma) = \mu\tau. P(\nu,\sigma,\tau)$ and let $\Kan\mu(\nu) =
\mu\sigma. K(\nu,\sigma)$. Then $\pre\mu(\nu,\sigma)$ and $\Kan\mu(\nu)$ are
order-preserving and $\Kan\mu(\nu)\subseteq\pre\mu(\nu,\Kan\mu(\nu))$ for all
$\nu$.
\end{lemma}
\begin{proof}
Part (1) is immediate by part (1) of \cref{lem:lfp-misc}, because
whenever $\nu\subseteq\nu'$ and $\sigma\subseteq\sigma'$,
$P(\nu,\sigma,-)\subseteq P(\nu',\sigma',-)$ and $K(\nu,-)\subseteq K(\nu',-)$.
For part (2), a theorem of \citet{Bekic1984} on simultaneous fixed points
implies $(\Kan\mu(\nu),\pre\mu(\nu,\Kan\mu(\nu))) = \mu (\sigma,\tau).
(K(\nu,\sigma),P(\nu,\sigma,\tau))$. Because each type operator is
order-preserving, $K(\nu,\sigma)\subseteq P(\nu,\sigma,\tau)$ whenever
$\sigma\subseteq\tau$. The result follows by part (2) of \cref{lem:lfp-misc}.
\end{proof}

We mutually define three sequences of candidate cubical type systems:
$\nu_{i+1}$ containing $i$ universes,
$\pre{\tau_{i+1}}$ containing the pretypes in a system with $i$ universes, and
$\Kan{\tau_{i+1}}$ containing the Kan types in a system with $i$ universes:
{\def\spacer{\hphantom{{}={}}}
\begin{align*}
\nu_0 &= \emptyset \\
\nu_n &= \{ (\Psi,\Ux[j],\Ux[j],\phi) \mid (j<n) \land
    (\phi = \{(A_0,B_0) \mid \tau^\kappa_j(\Psi,A_0,B_0,\_)\}) \} \\
\pre{\tau_n} &= \pre\mu(\nu_n,\Kan\mu(\nu_n)) \\
\Kan{\tau_n} &= \Kan\mu(\nu_n) \\
\nu_\omega &= \{ (\Psi,\Ux[j],\Ux[j],\phi) \mid
    \phi = \{(A_0,B_0) \mid \tau^\kappa_j(\Psi,A_0,B_0,\_)\} \} \\
\pre{\tau_\omega} &= \pre\mu(\nu_\omega,\Kan\mu(\nu_\omega)) \\
\Kan{\tau_\omega} &= \Kan\mu(\nu_\omega) \\
\end{align*}}
Observe that $\nu_n\subseteq\nu_{n+i}$, $\nu_n\subseteq\nu_\omega$,
$\tau_n^\kappa\subseteq\tau_{n+i}^\kappa$, $\tau_n^\kappa\subseteq\tau_\omega^\kappa$,
$\Kan{\tau_n}\subseteq\pre{\tau_n}$, and $\Kan{\tau_\omega}\subseteq\pre{\tau_\omega}$.

\subsection{Cubical type systems}

In the remainder of this paper, we consider only candidate cubical type systems
satisfying a number of additional conditions:

\begin{definition}
A \emph{cubical type system} is a candidate cubical type system $\tau$
satisfying:
\begin{description}[align=right,labelwidth=3.5cm]
\item[Functionality.] If $\tau(\Psi,A_0,B_0,\phi)$ and $\tau(\Psi,A_0,B_0,\phi')$
then $\phi=\phi'$.
\item[PER-valuation.] If $\tau(\Psi,A_0,B_0,\phi)$ then $\phi$ is symmetric and
transitive.
\item[Symmetry.] If $\tau(\Psi,A_0,B_0,\phi)$ then $\tau(\Psi,B_0,A_0,\phi)$.
\item[Transitivity.] If $\tau(\Psi,A_0,B_0,\phi)$ and $\tau(\Psi,B_0,C_0,\phi)$
then $\tau(\Psi,A_0,C_0,\phi)$.
\item[Value-coherence.] If $\tau(\Psi,A_0,B_0,\phi)$ then
$\PTy(\tau)(\Psi,A_0,B_0,\alpha)$ for some $\alpha$.
\end{description}
\end{definition}

If $\tau$ is a cubical type system, then $\PTy(\tau)$ is functional, symmetric,
transitive, and $\Psi$-PER-valued in the above senses. If $\alpha$ is a
$\Psi$-PER, then every $\td{\alpha}{\psi}$ is a $\Psi'$-PER, and $\Tm(\alpha)$
is a PER.

\begin{lemma}\label{lem:cts-cts}
If $\nu,\sigma$ are cubical type systems, then $\Kan\mu(\nu)$ and
$\pre\mu(\nu,\sigma)$ are cubical type systems.
\end{lemma}
\begin{proof}
Because the operators $\textsc{Fun}$, $\textsc{Pair}$\dots are disjoint, we can
check them individually in each case. We describe the proof for
$\pre\mu(\nu,\sigma)$; the proof for $\Kan\mu(\nu)$ follows analogously.

\begin{enumerate}
\item \emph{Functionality.}

Define a candidate cubical type system $\Phi = \{(\Psi,A_0,B_0,\phi) \mid
\forall\phi'. \pre\mu(\nu,\sigma)(\Psi,A_0,B_0,\phi') \implies (\phi=\phi')\}$.
Let us show that $\Phi$ is a pre-fixed point of $P(\nu,\sigma,-)$ (that is,
$P(\nu,\sigma,\Phi)\subseteq\Phi$). Because $\pre\mu(\nu,\sigma)$ is the least
pre-fixed point, it will follow that $\pre\mu(\nu,\sigma)\subseteq\Phi$, and
that $\pre\mu(\nu,\sigma)$ is functional.

Assume that $\textsc{Fun}(\Phi)(\Psi,\picl{a}{A}{B},\picl{a}{A'}{B'},\phi)$.
Thus $\PTy(\Phi)(\Psi,A,A',\alpha)$, and in particular, for all $\psitd$,
$\lift{\pre\mu(\nu,\sigma)}(\Psi',\td{A}{\psi},\td{A'}{\psi},\phi')$ implies
$\alpha_\psi = \phi'$, so $\alpha$ is unique in $\pre\mu(\nu,\sigma)$ when it
exists. Similarly, each $\beta^{(-,-,-)}$ is unique in $\pre\mu(\nu,\sigma)$
when it exists. The relation $\phi$ is determined uniquely by $\alpha$ and
$\beta^{(-,-,-)}$. Now let us show
$\Phi(\Psi,\picl{a}{A}{B},\picl{a}{A'}{B'},\phi)$, that is, assume
$\pre\mu(\nu,\sigma)(\Psi,\picl{a}{A}{B},\picl{a}{A'}{B'},\phi')$ and show
$\phi=\phi'$. It follows that $\PTy(\pre\mu(\nu,\sigma))(\Psi,A,A',\alpha')$ for
some $\alpha'$, and similarly for some family $\beta'$, but $\alpha=\alpha'$ and
each $\beta=\beta'$. Because $\phi'$ is defined using the same $\alpha$ and
$\beta^{(-,-,-)}$ as $\phi$, we conclude $\phi=\phi'$. Other cases are similar;
for $\textsc{Fcom},\textsc{UPre},\textsc{UKan}$ we use that $\nu,\sigma$ are
functional.

\item \emph{PER-valuation.}

Define $\Phi = \{(\Psi,A_0,B_0,\phi) \mid \text{$\phi$ is a PER}\}$, and show
that $\Phi$ is a pre-fixed point of $P(\nu,\sigma,-)$. It follows that
$\pre\mu(\nu,\sigma)$ is PER-valued, by $\pre\mu(\nu,\sigma)\subseteq\Phi$.

Assume that $\textsc{Fun}(\Phi)(\Psi,\picl{a}{A}{B},\picl{a}{A'}{B'},\phi)$.
Then $\PTy(\Phi)(\Psi,A,A',\alpha)$, and in particular, for all $\psitd$,
$\lift{\Phi}(\Psi',\td{A}{\psi},\td{A'}{\psi},\alpha_\psi)$, so each
$\alpha_\psi$ is a PER. Similarly, each $\beta^{\psi,M,M'}_{\psi'}$ is a PER.
Now we must show $\Phi(\Psi,\picl{a}{A}{B},\picl{a}{A'}{B'},\phi)$. The relation
$\phi$ is a PER because $\Tm(\alpha\psi)$ and $\Tm(\beta^{\psi,M,M'})$ are PERs,
because $\alpha_\psi$ and $\beta^{\psi,M,M'}_{\psi'}$ are PERs. Most cases
proceed in this fashion. For $\textsc{Nat}$, $\textsc{WB}$, and
$\textsc{Circ}$ we show that $\mathbb{N}$, $\mathbb{B}$, and $\mathbb{C}$ are
symmetric and transitive at each dimension (employing the same strategy as in
parts (3--4)); for $\textsc{Fcom},\textsc{UPre},\textsc{UKan}$ we use that
$\sigma,\nu$ are PER-valued.

\item \emph{Symmetry.}

Define $\Phi = \{(\Psi,A_0,B_0,\phi) \mid
\pre\mu(\nu,\sigma)(\Psi,B_0,A_0,\phi)\}$. Let us show that $\Phi$ is a pre-fixed point
of $P(\nu,\sigma,-)$. It will follow that $\pre\mu(\nu,\sigma)$ is symmetric, by
$\pre\mu(\nu,\sigma)\subseteq\Phi$.

Assume that $\textsc{Fun}(\Phi)(\Psi,\picl{a}{A}{B},\picl{a}{A'}{B'},\phi)$.
Then $\PTy(\Phi)(\Psi,A,A',\alpha)$ and $\Coh(\alpha)$, and thus
$\lift{\pre\mu(\nu,\sigma)}(\Psi',\td{A'}{\psi},\td{A}{\psi},\alpha_\psi)$,
$\td{A}{\psi_1} \evals A_1$, $\td{A'}{\psi_1} \evals A_1'$, and
$\lift{\pre\mu(\nu,\sigma)}(\Psi_2,-,-,\phi)$ relates
$(\td{A}{\psi_1\psi_2},\td{A_1}{\psi_2})$,
$(\td{A_1}{\psi_2},\td{A}{\psi_1\psi_2})$,
$(\td{A'}{\psi_1\psi_2},\td{A_1'}{\psi_2})$,
$(\td{A_1'}{\psi_2},\td{A'}{\psi_1\psi_2})$, and
$(\td{A_1'}{\psi_2},\td{A_1}{\psi_2})$.
Similar facts hold by virtue of $\PTy(\Phi)(\Psi',\subst{\td{B}{\psi}}{M}{a},
\subst{\td{B'}{\psi}}{M'}{a},\beta^{\psi,M,M'})$ and $\Coh(\beta^{\psi,M,M'})$.
We must show $\Phi(\Psi,\picl{a}{A}{B},\picl{a}{A'}{B'},\phi)$, that is,
$\pre\mu(\nu,\sigma)(\Psi,\picl{a}{A'}{B'},\picl{a}{A}{B},\phi)$.
This requires $\PTy(\pre\mu(\nu,\sigma))(\Psi,A',A,\alpha)$ and $\Coh(\alpha)$,
which follows from the above facts; and also
$\PTy(\pre\mu(\nu,\sigma))(\Psi,\subst{\td{B'}{\psi}}{M}{a},
\subst{\td{B}{\psi}}{M'}{a},\beta^{\psi,M,M'})$ and $\Coh(\beta^{\psi,M,M'})$
whenever $\Tm(\td{\alpha}{\psi})(M,M')$, which follows from the symmetry of
$\Tm(\td{\alpha}{\psi})$ (since each $\alpha_\psi$ is a PER, by (2)), and the
above facts. Other cases are similar; for $\textsc{Fcom}$ we use that $\sigma$
is symmetric.

\item \emph{Transitivity.}

Define $\Phi = \{(\Psi,A_0,B_0,\phi) \mid \forall C_0.
\pre\mu(\nu,\sigma)(\Psi,B_0,C_0,\phi)\implies
\pre\mu(\nu,\sigma)(\Psi,A_0,C_0,\phi)\}$. Let us show that $\Phi$ is a
pre-fixed point of $P(\nu,\sigma,-)$. It will follow that $\pre\mu(\nu,\sigma)$
is transitive, by $\pre\mu(\nu,\sigma)\subseteq\Phi$.

Assume that $\textsc{Fun}(\Phi)(\Psi,\picl{a}{A}{B},\picl{a}{A'}{B'},\phi)$.
Then $\PTy(\Phi)(\Psi,A,A',\alpha)$, and thus if
$\lift{\pre\mu(\nu,\sigma)}(\Psi',\td{A'}{\psi},C_0,\alpha_\psi)$ then
$\lift{\pre\mu(\nu,\sigma)}(\Psi',\td{A}{\psi},C_0,\alpha_\psi)$. Furthermore,
$\td{A}{\psi_1} \evals A_1$, $\td{A'}{\psi_1} \evals A_1'$, and for any $C_0$,
$\lift{\pre\mu(\nu,\sigma)}(\Psi_2,-,-,\phi)$ relates
$(\td{A}{\psi_1\psi_2},C_0)$ if and only if $(\td{A_1}{\psi_2},C_0)$;
$(\td{A'}{\psi_1\psi_2},C_0)$ if and only if $(\td{A_1'}{\psi_2},C_0)$;
and if $(\td{A_1'}{\psi_2},C_0)$ then $(\td{A_1}{\psi_2},C_0)$. Similar facts
hold by virtue of $\PTy(\Phi)(\Psi',\subst{\td{B}{\psi}}{M}{a},
\subst{\td{B'}{\psi}}{M'}{a},\beta^{\psi,M,M'})$.

Now we must show $\Phi(\Psi,\picl{a}{A}{B},\picl{a}{A'}{B'},\phi)$, that is, if
$\pre\mu(\nu,\sigma)(\Psi,\picl{a}{A'}{B'},C_0,\phi)$ then
$\pre\mu(\nu,\sigma)(\Psi,\picl{a}{A}{B},C_0,\phi)$. By inspecting $P$, we see this is
only possible if $C_0 = \picl{a}{A''}{B''}$, in which case
$\pre\mu(\nu,\sigma)(\Psi,\picl{a}{A'}{B'},\picl{a}{A''}{B''},\phi)$. Thus we have
$\PTy(\pre\mu(\nu,\sigma))(\Psi,A',A'',\alpha')$ and $\Coh(\alpha')$, so
$\lift{\pre\mu(\nu,\sigma)}(\Psi',\td{A'}{\psi},\td{A''}{\psi},\alpha'_\psi)$, and by
hypothesis, $\lift{\pre\mu(\nu,\sigma)}(\Psi',\td{A}{\psi},\td{A''}{\psi},\alpha_\psi)$
and $\Coh(\alpha)$. We already know $\td{A}{\psi_1} \evals A_1$,
$\td{A''}{\psi_1} \evals A_1''$,
and that $\lift{\pre\mu(\nu,\sigma)}(\Psi_2,-,-,\phi)$ relates
$(\td{A''}{\psi_1\psi_2},\td{A''_1}{\psi_2})$ and vice versa. By
$(\td{A'_1}{\psi_2},\td{A''_1}{\psi_2})$ and the above, we have
$(\td{A_1}{\psi_2},\td{A''_1}{\psi_2})$. Finally, by
$(\td{A'}{\psi_1\psi_2},\td{A'_1}{\psi_2})$ and transitivity we have
$(\td{A'_1}{\psi_2},\td{A'_1}{\psi_2})$, hence by transitivity and symmetry
$(\td{A'_1}{\psi_2},\td{A_1}{\psi_2})$, and again by transitivity
$(\td{A_1}{\psi_2},\td{A_1}{\psi_2})$; as needed,
$(\td{A_1}{\psi_2},\td{A_0}{\psi_2})$ and vice versa follow by transitivity.
As before, $\PTy(\pre\mu(\nu,\sigma))(\Psi,\subst{\td{B}{\psi}}{M}{a},
\subst{\td{B''}{\psi}}{M'}{a},\beta^{\psi,M,M'})$ and $\Coh(\beta^{\psi,M,M'})$
when $\Tm(\td{\alpha}{\psi})(M,M')$ follows by transitivity of
$\Tm(\td{\alpha}{\psi})$ (since each $\alpha_\psi$ is a PER, by (2)). Other
cases are similar; for $\textsc{Fcom}$ we use that $\sigma$ is transitive.

\item \emph{Value-coherence.}

Define $\Phi = \{(\Psi,A_0,B_0,\phi) \mid
\PTy(\pre\mu(\nu,\sigma))(\Psi,A_0,B_0,\alpha)\}$. Let us show that $\Phi$ is a
pre-fixed point of $P(\nu,\sigma,-)$. The property $P(\nu,\sigma,\Phi)\subseteq\Phi$ holds
trivially for base types $\textsc{Void}$, $\textsc{Nat}$\dots as well as
universes $\textsc{UPre}$ and $\textsc{UKan}$; we check $\textsc{Fun}$
($\textsc{Pair}$, $\textsc{Path}$, and $\textsc{Eq}$ are similar) and
$\textsc{V}$ ($\textsc{Fcom}$ is similar). It will follow that
$\pre\mu(\nu,\sigma)$ is value-coherent, by $\pre\mu(\nu,\sigma)\subseteq\Phi$.

Assume that $\textsc{Fun}(\Phi)(\Psi,\picl{a}{A}{B},\picl{a}{A'}{B'},\phi)$.
Then by $\PTy(\Phi)(\Psi,A,A',\alpha)$ and $\Coh(\alpha)$, we have
$\lift{\Phi}(\Psi',\td{A}{\psi},\td{A'}{\psi},\alpha_\psi)$,
$\td{A}{\psi_1}\evals A_1$, $\td{A'}{\psi_1}\evals A_1'$,
$\lift{\Phi}(\Psi_2,\td{A_1}{\psi_2},\td{A}{\psi_1\psi_2},\phi')$, and so forth.
Note that for values $A_0,B_0$, if $\PTy(\tau)(\Psi,A_0,B_0,\alpha)$ then
$\tau(\Psi,A_0,B_0,\alpha_{\id})$ by definition. Therefore
$\lift{\pre\mu(\nu,\sigma)}(\Psi',\td{A}{\psi},\td{A'}{\psi},\alpha_\psi)$, and
so forth. We get similar facts for each $\Tm(\td{\alpha}{\psi})(M,M')$ by
$\PTy(\Phi)(\Psi',\subst{\td{B}{\psi}}{M}{a},
\subst{\td{B'}{\psi}}{M'}{a},\beta^{\psi,M,M'})$ and $\Coh(\beta^{\psi,M,M'})$.
We must show $\Phi(\Psi,\picl{a}{A}{B},\picl{a}{A'}{B'},\phi')$, that
is, $\PTy(\pre\mu(\nu,\sigma))(\Psi,\picl{a}{A}{B},\picl{a}{A'}{B'},\gamma)$. We
know $\sisval{\picl{a}{A}{B}}$, and by the above,
$\PTy(\pre\mu(\nu,\sigma))(\Psi,A,A',\alpha)$, $\Coh(\alpha)$, and when
$\Tm(\td{\alpha}{\psi})(M,M')$,
$\PTy(\pre\mu(\nu,\sigma))(\Psi',\subst{\td{B}{\psi}}{M}{a},
\subst{\td{B'}{\psi}}{M'}{a},\beta^{\psi,M,M'})$ and $\Coh(\beta^{\psi,M,M'})$.
The result holds because $\PTy$, $\Tm$, and $\Coh$ are closed under dimension
substitution.

The $\textsc{V}$ case is mostly similar, but not all instances of
$\ua{x}{A,B,E}$ have the same head constructor. Repeating the previous argument,
by $\textsc{V}(\Phi)(\Psi,\ua{x}{A,B,E},\ua{x}{A',B',E'})$ we have that
$\PTy(\pre\mu(\nu,\sigma))(\Psi,B,B',\beta)$ and for all $\psi$ with $\td{x}{\psi} =
0$, $\PTy(\pre\mu(\nu,\sigma))(\Psi',\td{A}{\psi},\td{A'}{\psi},\alpha^\psi)$. However,
in order to prove $\PTy(\pre\mu(\nu,\sigma))(\Psi,\ua{x}{A,B,E},\ua{x}{A',B',E'})$, we
must observe that when $\td{x}{\psi} = 0$,
$\ua{0}{\td{A}{\psi},\td{B}{\psi},\td{E}{\psi}}\steps\td{A}{\psi}$; when
$\td{x}{\psi} = 1$,
$\ua{1}{\td{A}{\psi},\td{B}{\psi},\td{E}{\psi}}\steps\td{B}{\psi}$; and for
every $\psi_1,\psi_2$ the appropriate relations hold in $\pre\mu(\nu,\sigma)$.
See \cref{rul:ua-form-pre} for the full proof, and \cref{lem:fcom-preform} for
the corresponding proof for $\textsc{Fcom}$.
\qedhere
\end{enumerate}
\end{proof}

\begin{theorem}
$\tau^\kappa_n$ and $\tau^\kappa_\omega$ are cubical type systems.
\end{theorem}
\begin{proof}\mbox{}
\paragraph{System $\tau^\kappa_n$.}
Use strong induction on $n$. Clearly $\nu_0$
is a cubical type system; by \cref{lem:cts-cts} so are $\Kan\tau_0$ and
thus $\pre\tau_0$. Suppose $\tau^\kappa_j$ are cubical type systems for $j<n$.
Then $\nu_n$ is a cubical type system: functionality, symmetry,
transitivity, and value-coherence are immediate; PER-valuation follows from the
previous $\tau^\kappa_j$ being cubical type systems. The induction step follows
by \cref{lem:cts-cts}.

\paragraph{System $\tau^\kappa_\omega$.}
Because each $\tau^\kappa_n$ is a cubical type system, so
is $\nu_\omega$ (as before), and so are $\tau^\kappa_\omega$.
\end{proof}

The cubical type systems employed by \citet{ah2016cubicaldep} are equivalent to
candidate cubical type systems satisfying conditions (1--4): define $A_0
\approx^\Psi B_0$ to hold when $\tau(\Psi,A_0,B_0,\phi)$, and $M_0
\approx^\Psi_{A_0} N_0$ when $\phi(M_0,N_0)$. Condition (5) is needed in the
construction of universes.

\newpage
\section{Mathematical meaning explanations}
\label{sec:meanings}

In this section, we finally define the judgments of higher type theory as
relations parametrized by a choice of cubical type system $\tau$. In these
definitions we suppress dependency on $\tau$, but we will write
$\relcts*{\tau}{\judg{\J}}$ to make the choice of $\tau$ explicit.

The presuppositions of a judgment are facts that must be true before one can
even sensibly state that judgment. For example, in \cref{def:ceqtm} below, we
presuppose that $A$ is a pretype when defining what it means to be equal
elements of $A$; if we do not know $A$ to be a pretype, $\vper{A}$ has no
meaning. In every judgment $\judg{\J}$ we will presuppose that the free
dimensions of all terms are contained in $\Psi$.

\subsection{Judgments}

\begin{definition}\label{def:ceqtypep}
The judgment $\ceqtypep{A}{B}$ holds when $\PTy(\tau)(\Psi,A,B,\alpha)$ and
$\Coh(\alpha)$. Whenever $\PTy(\tau)(\Psi,A,B,\alpha)$ the choice of $\alpha$ is
unique and independent of $B$, so we notate it $\vper{A}$.
\end{definition}

\begin{definition}\label{def:ceqtm}
The judgment $\ceqtm{M}{N}{A}$ holds, presupposing $\ceqtypep{A}{A}$, when
$\Tm(\vper{A})(M,N)$.
\end{definition}

If $A$ and $B$ have no free dimensions and $\ceqtypep{A}{B}$, then for any
$\Psi'$, $\lift\tau(\Psi',A,B,\vper{A})$ and $\vper{A}$ is context-indexed; if
$M$, $N$, and $A$ have no free dimensions and $\ceqtm{M}{N}{A}$, then
$\lift{(\vper{A}(\Psi'))}(M,N)$ for all $\Psi'$. Therefore one can regard the
ordinary meaning explanations as an instance of these meaning explanations, in
which all dependency on dimensions trivializes.

We are primarily interested in \emph{Kan types}, pretypes equipped with Kan
operations that implement composition, inversion, etc., of cubes. These Kan
operations are best specified using judgments augmented by \emph{dimension
context restrictions}. We extend the prior judgments to restricted ones:

\begin{definition}\label{def:satisfies}
For any $\Psi$ and set of unoriented equations $\Xi = (r_1=r_1',\dots,r_n=r_n')$
in $\Psi$ (that is, $\fd{\etc{r_i},\etc{r_i'}}\subseteq\Psi$), we say that
$\psitd$ \emph{satisfies} $\Xi$ if $\td{r_i}{\psi} = \td{r_i'}{\psi}$ for each
$i\in [1,n]$.
\end{definition}

\begin{definition}
\label{def:crestricted}
~\begin{enumerate}
\item
The judgment $\ceqtypep<\Xi>{A}{B}$ holds, presupposing
$\fd{\Xi}\subseteq\Psi$, when $\ceqtypep[\Psi']{\td{A}{\psi}}{\td{B}{\psi}}$
for every $\psitd$ satisfying $\Xi$.

\item
The judgment $\ceqtm<\Xi>{M}{N}{A}$ holds, presupposing
$\cwftypep<\Xi>{A}$, when
$\ceqtm[\Psi']{\td{M}{\psi}}{\td{N}{\psi}}{\td{A}{\psi}}$ for every $\psitd$
satisfying $\Xi$.
\end{enumerate}
\end{definition}

\begin{definition}\label{def:valid}
A list of equations $\etc{r_i=r_i'}$ is valid if either $r_i=r_i'$ for some $i$,
or $r_i=r_j$, $r_i'=0$, and $r_j'=1$ for some $i,j$.
\end{definition}

\begin{definition}\label{def:kan}
The judgment $\ceqtypek{A}{B}$ holds, presupposing $\ceqtypep{A}{B}$, when the
following Kan conditions hold for any $\psitd$:
\begin{enumerate}
\item 
If
\begin{enumerate}
\item $\etc{r_i=r_i'}$ is valid,
\item $\ceqtm[\Psi']{M}{M'}{\td{A}{\psi}}$,
\item $\ceqtm[\Psi',y]<r_i=r_i',r_j=r_j'>{N_i}{N_j'}{\td{A}{\psi}}$
for any $i,j$, and
\item $\ceqtm[\Psi']<r_i=r_i'>{\dsubst{N_i}{r}{y}}{M}{\td{A}{\psi}}$
for any $i$,
\end{enumerate}
then
\begin{enumerate}
\item $\ceqtm[\Psi']{\Hcom*{\td{A}{\psi}}{r_i=r_i'}}%
{\Hcom{\td{B}{\psi}}{r}{r'}{M'}{\sys{r_i=r_i'}{y.N_i'}}}{\td{A}{\psi}}$;
\item if $r=r'$ then
$\ceqtm[\Psi']{\Hcom{\td{A}{\psi}}{r}{r}{M}{\sys{r_i=r_i'}{y.N_i}}}{M}{\td{A}{\psi}}$;
and
\item if $r_i = r_i'$ then
$\ceqtm[\Psi']{\Hcom*{\td{A}{\psi}}{r_i=r_i'}}{\dsubst{N_i}{r'}{y}}{\td{A}{\psi}}$.
\end{enumerate}

\item 
If $\Psi' = (\Psi'',x)$ and $\ceqtm[\Psi'']{M}{M'}{\dsubst{\td{A}{\psi}}{r}{x}}$, then
\begin{enumerate}
\item $\ceqtm[\Psi'']{\Coe*{x.\td{A}{\psi}}}{\Coe{x.\td{B}{\psi}}{r}{r'}{M'}}%
{\dsubst{\td{A}{\psi}}{r'}{x}}$; and
\item if $r=r'$ then
$\ceqtm[\Psi'']{\Coe{x.\td{A}{\psi}}{r}{r}{M}}{M}{\dsubst{\td{A}{\psi}}{r}{x}}$.
\end{enumerate}
\end{enumerate}
\end{definition}

We extend the closed judgments to open terms by functionality, that is, an open
pretype (resp., element of a pretype) is an open term that sends equal elements
of the pretypes in the context to equal closed pretypes (resp., elements). The
open judgments are defined simultaneously, stratified by the length of the
context. (We assume the variables $a_1,\dots,a_n$ in a context are distinct.)

\begin{definition}\label{def:wfctx}
We say $\wfctx{(\oft{a_1}{A_1},\dots,\oft{a_n}{A_n})}$ when
\begin{gather*}
\cwftypep{A_1}, \\
\wftypep{\oft{a_1}{A_1}}{A_2}, \dots \\ \text{and}~
\wftypep{\oft{a_1}{A_1},\dots,\oft{a_{n-1}}{A_{n-1}}}{A_n}.
\end{gather*}
\end{definition}

\begin{definition}\label{def:eqtypep}
We say $\eqtypep{\oft{a_1}{A_1},\dots,\oft{a_n}{A_n}}{B}{B'}$,
presupposing \\
$\wfctx{(\oft{a_1}{A_1},\dots,\oft{a_n}{A_n})}$, when for any $\psitd$ and any
\begin{gather*}
\ceqtm[\Psi']{N_1}{N_1'}{\td{A_1}{\psi}}, \\
\ceqtm[\Psi']{N_2}{N_2'}{\subst{\td{A_2}{\psi}}{N_1}{a_1}}, \dots\\\text{and}~
\ceqtm[\Psi']{N_n}{N_n'}
{\subst{\td{A_n}{\psi}}{N_1,\dots,N_{n-1}}{a_1,\dots,a_n}},
\end{gather*}
$\ceqtypep[\Psi']
{\subst{\td{B}{\psi}}{N_1,\dots,N_n}{a_1,\dots,a_n}}
{\subst{\td{B'}{\psi}}{N_1',\dots,N_n'}{a_1,\dots,a_n}}$.
\end{definition}

\begin{definition}\label{def:eqtm}
We say $\eqtm{\oft{a_1}{A_1},\dots,\oft{a_n}{A_n}}{M}{M'}{B}$,
presupposing \\
$\wftypep{\oft{a_1}{A_1},\dots,\oft{a_n}{A_n}}{B}$,
when for any $\psitd$ and any
\begin{gather*}
\ceqtm[\Psi']{N_1}{N_1'}{\td{A_1}{\psi}}, \\
\ceqtm[\Psi']{N_2}{N_2'}{\subst{\td{A_2}{\psi}}{N_1}{a_1}}, \dots\\\text{and}~
\ceqtm[\Psi']{N_n}{N_n'}
{\subst{\td{A_n}{\psi}}{N_1,\dots,N_{n-1}}{a_1,\dots,a_n}},
\end{gather*}
$\ceqtm[\Psi']
{\subst{\td{M}{\psi}}{N_1,\dots,N_n}{a_1,\dots,a_n}}
{\subst{\td{M'}{\psi}}{N_1',\dots,N_n'}{a_1,\dots,a_n}}
{\subst{\td{B}{\psi}}{N_1,\dots,N_n}{a_1,\dots,a_n}}$.
\end{definition}

One should read $[\Psi]$ as extending across the entire judgment, as it
specifies the starting dimension at which to consider not only $B$ and $M$ but
$\G$ as well.
The open judgments, like the closed judgments, are symmetric and transitive.
In particular, if $\eqtypep{\G}{B}{B'}$ then $\wftypep{\G}{B}$.
As a result, the earlier hypotheses of each definition ensure that later
hypotheses are sensible; for example,
$\wfctx{(\oft{a_1}{A_1},\dots,\oft{a_n}{A_n})}$ and
$\coftype[\Psi']{N_1}{\td{A_1}{\psi}}$ ensure that
$\cwftypep[\Psi']{\subst{\td{A_2}{\psi}}{N_1}{a_1}}$.

\begin{definition}\label{def:eqtypek}
We say $\eqtypek{\oft{a_1}{A_1},\dots,\oft{a_n}{A_n}}{B}{B'}$,
presupposing \\
$\eqtypep{\oft{a_1}{A_1},\dots,\oft{a_n}{A_n}}{B}{B'}$,
when for any $\psitd$ and any
\begin{gather*}
\ceqtm[\Psi']{N_1}{N_1'}{\td{A_1}{\psi}}, \\
\ceqtm[\Psi']{N_2}{N_2'}{\subst{\td{A_2}{\psi}}{N_1}{a_1}}, \dots\\\text{and}~
\ceqtm[\Psi']{N_n}{N_n'}
{\subst{\td{A_n}{\psi}}{N_1,\dots,N_{n-1}}{a_1,\dots,a_n}},
\end{gather*}
we have
$\ceqtypek[\Psi']
{\subst{\td{B}{\psi}}{N_1,\dots,N_n}{a_1,\dots,a_n}}
{\subst{\td{B'}{\psi}}{N_1',\dots,N_n'}{a_1,\dots,a_n}}$.
\end{definition}

Finally, the open judgments can also be augmented by context restrictions. In
order to make sense of \cref{def:restricted}, the presuppositions of the open
judgments require them to be closed under dimension substitution, which we will
prove in \cref{lem:td-judgments}.

\begin{definition}
\label{def:restricted}
~\begin{enumerate}
\item
The judgment $\wfctx<\Xi>{\G}$ holds, presupposing $\fd{\Xi}\subseteq\Psi$,
when $\wfctx[\Psi']{\td{\G}{\psi}}$ for every $\psitd$ satisfying $\Xi$.

\item
The judgment $\eqtypep<\Xi>{\G}{B}{B'}$ holds, presupposing
$\wfctx<\Xi>{\G}$, when
$\eqtypep[\Psi']{\td{\G}{\psi}}{\td{B}{\psi}}{\td{B'}{\psi}}$ for every
$\psitd$ satisfying $\Xi$.

\item
The judgment $\eqtm<\Xi>{\G}{M}{M'}{B}$ holds, presupposing
$\wfctx<\Xi>{\G}$ and $\wftypep<\Xi>{\G}{B}$, when
$\eqtm[\Psi']{\td{\G}{\psi}}{\td{M}{\psi}}{\td{M'}{\psi}}{\td{B}{\psi}}$ for
every $\psitd$ satisfying $\Xi$.

\item
The judgment $\eqtypek<\Xi>{\G}{B}{B'}$ holds, presupposing
$\wfctx<\Xi>{\G}$, when
$\eqtypek[\Psi']{\td{\G}{\psi}}{\td{B}{\psi}}{\td{B'}{\psi}}$ for every $\psitd$
satisfying $\Xi$.
\end{enumerate}
\end{definition}

\subsection{Structural properties}

Every judgment is closed under dimension substitution.

\begin{lemma}\label{lem:td-judgments}
For any $\psitd$,
\begin{enumerate}
\item if $\ceqtypep{A}{B}$ then
$\ceqtypep[\Psi']{\td{A}{\psi}}{\td{B}{\psi}}$;
\item if $\ceqtm{M}{N}{A}$ then
$\ceqtm[\Psi']{\td{M}{\psi}}{\td{N}{\psi}}{\td{A}{\psi}}$;
\item if $\ceqtypek{A}{B}$ then $\ceqtypek[\Psi']{\td{A}{\psi}}{\td{B}{\psi}}$;
\item if $\wfctx{\G}$ then $\wfctx[\Psi']{\td{\G}{\psi}}$;
\item if $\eqtypep{\G}{A}{B}$ then
$\eqtypep[\Psi']{\td{\G}{\psi}}{\td{A}{\psi}}{\td{B}{\psi}}$;
\item if $\eqtm{\G}{M}{N}{A}$ then
$\eqtm[\Psi']{\td{\G}{\psi}}{\td{M}{\psi}}{\td{N}{\psi}}{\td{A}{\psi}}$; and
\item if $\eqtypek{\G}{A}{B}$ then
$\eqtypek[\Psi']{\td{\G}{\psi}}{\td{A}{\psi}}{\td{B}{\psi}}$.
\end{enumerate}
\end{lemma}
\begin{proof}
For proposition (1), by $\PTy(\tau)(\Psi,A,B,\alpha)$ we have
$\PTy(\tau)(\Psi',\td{A}{\psi},\td{B}{\psi},\td{\alpha}{\psi})$. We must show
for all $\tds{\Psi''}{\psi'}{\Psi'}$ that $(\td{\alpha}{\psi})_{\psi'}(M_0,N_0)$
implies $\Tm(\td{\alpha}{\psi\psi'})(M_0,N_0)$; this follows from
value-coherence of $\alpha$ at $\psi\psi'$. Propositions (2) and (3) follow from
$\vper{\td{A}{\psi}} = \td{\vper{A}}{\psi}$ and closure of $\Tm$ and the Kan
conditions under dimension substitution.

Propositions (4), (5), and (6) are proven simultaneously by induction on the
length of $\G$. If $\G=\cdot$, then (4) is trivial, and (5) and (6) follow
because the closed judgments are closed under dimension substitution. The
induction steps for all three use all three induction hypotheses. Proposition
(7) follows similarly.
\end{proof}

\begin{lemma}\label{lem:td-judgres}
For any $\psitd$, if $\judg<\Xi>{\J}$ then
$\judg[\Psi']<\td{\Xi}{\psi}>{\td{\J}{\psi}}$.
\end{lemma}
\begin{proof}
We know that $\judg[\Psi']{\td{\J}{\psi}}$ for any $\psitd$ satisfying $\Xi$,
and want to show that $\judg[\Psi'']{\td{\J}{\psi\psi'}}$ for any $\psitd$ and
$\tds{\Psi''}{\psi'}{\Psi'}$ satisfying $\td{\Xi}{\psi}$. It suffices to
show that if $\psi'$ satisfies $\td{\Xi}{\psi}$, then $\psi\psi'$ satisfies
$\Xi$. But these both hold if and only if for each $(r_i=r_i')\in\Xi$,
$\td{r_i}{\psi\psi'} = \td{r_i'}{\psi\psi'}$.
\end{proof}

\begin{remark}
The context-restricted judgments can be thought of as merely a notational
device, because it is possible to systematically translate $\judg<\Xi>{\J}$
into ordinary judgments by case analysis:

\begin{enumerate}
\item All $\psi$ satisfy an empty set of equations, so $\judg<\cdot>{\J}$ if
and only if $\judg[\Psi']{\td{\J}{\psi}}$ for all $\psi$, which by
\cref{lem:td-judgments} holds if and only if $\judg{\J}$.

\item A $\psi$ satisfies $(\Xi,r=r)$ if and only if it satisfies $\Xi$, so
$\judg<\Xi,r=r>{\J}$ if and only if $\judg<\Xi>{\J}$.

\item No $\psi$ satisfies $(\Xi,0=1)$, so $\judg<\Xi,0=1>{\J}$ always.

\item By \cref{lem:td-judgres}, $\judg[\Psi,x]<\Xi,x=r>{\J}$ if and only if
$\judg<\dsubst{\Xi}{r}{x},r=r>{\dsubst{\J}{r}{x}}$, which holds if and only
if $\judg<\dsubst{\Xi}{r}{x}>{\dsubst{\J}{r}{x}}$.
\end{enumerate}
\end{remark}

The open judgments satisfy the \emph{structural rules} of type theory, like
hypothesis and weakening.

\begin{lemma}[Hypothesis]
If $\wfctx{(\G,\oft{a_i}{A_i},\G')}$ then
$\oftype{\G,\oft{a_i}{A_i},\G'}{a_i}{A_i}$.
\end{lemma}
\begin{proof}
We must show for any $\psitd$ and equal elements $N_1,N_1',\dots,N_n,N_n'$ of
the pretypes in $(\td{\G}{\psi},\oft{a_i}{\td{A_i}{\psi}},\td{\G'}{\psi})$, that
$\ceqtm[\Psi']{N_i}{N_i'}{\td{A_i}{\psi}}$. But this is exactly our assumption
about $N_i,N_i'$.
\end{proof}

\begin{lemma}[Weakening]
~\begin{enumerate}
\item If $\eqtypep{\G,\G'}{B}{B'}$ and $\wftypep{\G}{A}$, then
$\eqtypep{\G,\oft{a}{A},\G'}{B}{B'}$.
\item If $\eqtm{\G,\G'}{M}{M'}{B}$ and $\wftypep{\G}{A}$, then
$\eqtm{\G,\oft{a}{A},\G'}{M}{M'}{B}$.
\end{enumerate}
\end{lemma}
\begin{proof}
For the first part, we must show for any $\psitd$ and equal elements
\begin{gather*}
\ceqtm[\Psi']{N_1}{N_1'}{\td{A_1}{\psi}}, \\
\ceqtm[\Psi']{N_2}{N_2'}{\subst{\td{A_2}{\psi}}{N_1}{a_1}}, \dots \\
\ceqtm[\Psi']{N}{N'}{\subst{\td{A}{\psi}}{N_1,\dots}{a_1,\dots}},
\dots\\\text{and}~
\ceqtm[\Psi']{N_n}{N_n'}
{\subst{\td{A_n}{\psi}}{N_1,\dots,N,\dots,N_{n-1}}{a_1,\dots,a,\dots,a_n}},
\end{gather*}
that the corresponding instances of $B,B'$ are equal closed pretypes.
By $\eqtypep{\G,\G'}{B}{B'}$ we know that $a\fresh \G',B,B'$---since the
contained pretypes become closed when substituting for $a_1,\dots,a_n$.
It also gives us
$\ceqtypep[\Psi']
{\subst{\td{B}{\psi}}{N_1,\dots}{a_1,\dots}}
{\subst{\td{B'}{\psi}}{N_1',\dots}{a_1,\dots}}$
which are the desired instances of $B,B'$ because $a\fresh B,B'$.
The second part follows similarly.
\end{proof}

The definition of equal pretypes was chosen to ensure that equal pretypes have
equal elements.

\begin{lemma}\label{lem:ceqtypep-ceqtm}
If $\ceqtypep{A}{B}$ and $\ceqtm{M}{N}{A}$ then $\ceqtm{M}{N}{B}$.
\end{lemma}
\begin{proof}
If $\PTy(\tau)(\Psi,A,B,\alpha)$ then $\PTy(\tau)(\Psi,B,A,\alpha)$; the result
follows by $\vper{A}=\vper{B}$.
\end{proof}

\begin{lemma}
If $\eqtypep{\G}{A}{B}$ and $\eqtm{\G}{M}{N}{A}$ then $\eqtm{\G}{M}{N}{B}$.
\end{lemma}
\begin{proof}
If $\G = (\oft{a_1}{A_1},\dots,\oft{a_n}{A_n})$ then $\eqtm{\G}{M}{N}{A}$ means
that for any $\psitd$ and equal elements $N_1,N_1',\dots,N_n,N_n'$ of the
pretypes in $\td{\G}{\psi}$, the corresponding instances of $M$ and $N$ are
equal in $\subst{\td{A}{\psi}}{N_1,\dots,N_n}{a_1,\dots,a_n}$. But
$\eqtypep{\G}{A}{B}$ implies this pretype is equal to
$\subst{\td{B}{\psi}}{N_1,\dots,N_n}{a_1,\dots,a_n}$, so the result follows by
\cref{lem:ceqtypep-ceqtm}.
\end{proof}

\subsection{Basic lemmas}

The definition of $\PTy(\tau)(\Psi,A,B,\alpha)$ can be simplified when $\tau$ is
a cubical type system: it suffices to check for all
$\tds{\Psi_1}{\psi_1}{\Psi}$ and $\tds{\Psi_2}{\psi_2}{\Psi_1}$ that
$\td{A}{\psi_1} \evals A_1$, $\td{B}{\psi_1} \evals B_1$,
$\lift{\tau}(\Psi_2,\td{A_1}{\psi_2},\td{A}{\psi_1\psi_2},\phi)$,
$\lift{\tau}(\Psi_2,\td{B_1}{\psi_2},\td{B}{\psi_1\psi_2},\phi')$, and
$\lift{\tau}(\Psi_2,\td{A_1}{\psi_2},\td{B_1}{\psi_2},\phi'')$. Then
$\phi=\phi'=\phi''$ and $\alpha$ exists uniquely. The proof uses the observation
that the following permissive form of transitivity holds for any functional PER
$R$: if $R(\Psi,A,B,\alpha)$ and $R(\Psi,B,C,\beta)$ then $R(\Psi,A,C,\alpha)$
and $\alpha=\beta$.

\begin{lemma}\label{lem:pty-evals}
If $\PTy(\tau)(\Psi,A,A,\alpha)$, then $A\evals A_0$ and
$\PTy(\tau)(\Psi,A,A_0,\alpha)$.
\end{lemma}
\begin{proof}
Check for all $\tds{\Psi_1}{\psi_1}{\Psi}$ and $\tds{\Psi_2}{\psi_2}{\Psi_1}$
that $\lift{\tau}(\Psi_2,\td{A}{\psi_1\psi_2},\td{A_0}{\psi_1\psi_2},\phi)$ and
$\lift{\tau}(\Psi_2,\td{A_1}{\psi_2},\td{A_1'}{\psi_2},\phi')$ where
$\td{A}{\psi_1}\evals A_1$ and $\td{A_0}{\psi_1}\evals A_1'$.
The former holds by $\PTy(\tau)(\Psi,A,A,\alpha)$ at the substitutions
$\id$ and $\psi_1\psi_2$. For the latter, $\PTy(\tau)(\Psi,A,A,\alpha)$ at
$\psi_1,\id[\Psi_1]$ proves that $\lift{\tau}(\Psi_1,A_1,\td{A}{\psi_1},\_)$ and
at $\id,\psi_1$ proves $\lift{\tau}(\Psi_1,\td{A_0}{\psi_1},\td{A}{\psi_1},\_)$.
By transitivity, $\tau(\Psi_1,A_1,A_1',\_)$ so
$\PTy(\tau)(\Psi_1,A_1,A_1',\_)$ and thus
$\lift{\tau}(\Psi_2,\td{A_1}{\psi_2},\td{A_1'}{\psi_2},\phi')$ as required.
\end{proof}

\begin{lemma}\label{lem:cwftypep-evals-ceqtypep}
If $\cwftypep{A}$, then $A\evals A_0$ and $\ceqtypep{A}{A_0}$.
\end{lemma}
\begin{proof}
By \cref{lem:pty-evals} we have $\PTy(\tau)(\Psi,A,A_0,\alpha)$; value-coherence
follows by $\cwftypep{A}$.
\end{proof}

\begin{lemma}\label{lem:coftype-ceqtm}
If $\coftype{M}{A}$, $\coftype{N}{A}$, and $\lift{\vper{A}}(M,N)$, then
$\ceqtm{M}{N}{A}$.
\end{lemma}
\begin{proof}
We check for all $\tds{\Psi_1}{\psi_1}{\Psi}$ and $\tds{\Psi_2}{\psi_2}{\Psi_1}$
that
$\lift{\vper{A}}_{\psi_1\psi_2}(\td{M}{\psi_1\psi_2},\td{N}{\psi_1\psi_2})$; the
other needed relations follow from $\coftype{M}{A}$ and $\coftype{N}{A}$. By
$\cwftypep{A}$, $\lift{\vper{A}}(M,N)$ implies $\Tm(\vper{A})(M_0,N_0)$ where
$M\evals M_0$ and $N\evals N_0$, hence
$\lift{\vper{A}}_{\psi_1\psi_2}(\td{M_0}{\psi_1\psi_2},\td{N_0}{\psi_1\psi_2})$.
By $\coftype{M}{A}$ we have
$\lift{\vper{A}}_{\psi_1\psi_2}(\td{M_0}{\psi_1\psi_2},\td{M}{\psi_1\psi_2})$
and similarly for $N$, so the result follows by transitivity.
\end{proof}

\begin{lemma}\label{lem:coftype-evals-ceqtm}
If $\coftype{M}{A}$, then $M\evals M_0$ and $\ceqtm{M}{M_0}{A}$.
\end{lemma}
\begin{proof}
By $\coftype{M}{A}$, $M\evals M_0$ and $\vper{A}(M_0,M_0)$. By $\cwftypep{A}$,
$\coftype{M_0}{A}$, so the result follows by \cref{lem:coftype-ceqtm}.
\end{proof}

\begin{lemma}\label{lem:cwftypek-evals-ceqtypek}
If $\cwftypek{A}$, $\cwftypek{B}$, and for all $\psitd$,
$\ceqtypek[\Psi']{A_\psi}{B_\psi}$ where
$\td{A}{\psi}\evals A_\psi$ and
$\td{B}{\psi}\evals B_\psi$, then
$\ceqtypek{A}{B}$.
\end{lemma}
\begin{proof}
By \cref{lem:cwftypep-evals-ceqtypep} we have
$\ceqtypep[\Psi']{\td{A}{\psi}}{A_\psi}$ and
$\ceqtypep[\Psi']{\td{B}{\psi}}{B_\psi}$ for all $\psitd$;
thus $\ceqtypep[\Psi']{\td{A}{\psi}}{\td{B}{\psi}}$ for all $\psitd$, and it
suffices to establish that if
\begin{enumerate}
\item $\etc{r_i=r_i'}$ is valid,
\item $\ceqtm[\Psi']{M}{M'}{\td{A}{\psi}}$,
\item $\ceqtm[\Psi',y]<r_i=r_i',r_j=r_j'>{N_i}{N_j'}{\td{A}{\psi}}$
for any $i,j$, and
\item $\ceqtm[\Psi']<r_i=r_i'>{\dsubst{N_i}{r}{y}}{M}{\td{A}{\psi}}$
for any $i$,
\end{enumerate}
then
$\ceqtm[\Psi']{\Hcom*{\td{A}{\psi}}{r_i=r_i'}}%
{\Hcom{\td{B}{\psi}}{r}{r'}{M'}{\sys{r_i=r_i'}{y.N_i'}}}{\td{A}{\psi}}$.
We already know both terms are elements of this type (by \cref{def:kan} and
$\ceqtypep[\Psi']{\td{A}{\psi}}{\td{B}{\psi}}$), so by
\cref{lem:coftype-ceqtm} it suffices to check that these terms are related by
$\lift{\vper{\td{A}{\psi}}}$ or equivalently $\lift{\vper{A_\psi}}$. This is
true because $\Hcom{\td{A}{\psi}}\steps^* \Hcom{A_\psi}$,
$\Hcom{\td{B}{\psi}}\steps^* \Hcom{B_\psi}$, and
by $\ceqtypek[\Psi']{A_\psi}{B_\psi}$ and
$\ceqtypep[\Psi']{\td{A}{\psi}}{A_\psi}$,
$\ceqtm[\Psi']{\Hcom{A_\psi}}{\Hcom{B_\psi}}{A_\psi}$. The remaining $\Hcom$
equations of \cref{def:kan} follow by transitivity and
$\cwftypek[\Psi']{A_\psi}$; the $\Coe$ equations follow by a similar argument.
\end{proof}

In order to establish that a term is a pretype or element, one must frequently
reason about the evaluation behavior of its aspects. When all aspects compute in
lockstep, a \emph{head expansion} lemma applies; otherwise one must appeal to
its generalization, \emph{coherent expansion}:

\begin{lemma}\label{lem:cohexp-ceqtypep}
Assume we have $\wftm{A}$ and a family of terms $\{A^{\Psi'}_\psi\}_{\psitd}$
such that for all $\psitd$,
$\ceqtypep[\Psi']{A^{\Psi'}_{\psi}}{\td{(A^{\Psi}_{\id})}{\psi}}$ and
$\td{A}{\psi} \steps^* A^{\Psi'}_\psi$. Then $\ceqtypep{A}{A^\Psi_{\id}}$.
\end{lemma}
\begin{proof}
We must show that for any
$\tds{\Psi_1}{\psi_1}{\Psi}$ and
$\tds{\Psi_2}{\psi_2}{\Psi_1}$,
$\td{A}{\psi_1}\evals A_1$,
$\td{(A^\Psi_{\id})}{\psi_1}\evals A_1'$, and
$\lift{\tau}(\Psi_2,-,-,\_)$ relates
$\td{A_1}{\psi_2}$,
$\td{A}{\psi_1\psi_2}$,
$\td{(A^\Psi_{\id})}{\psi_1\psi_2}$,
and $\td{A'_1}{\psi_2}$.

\begin{enumerate}
\item $\td{A}{\psi_1}\evals A_1$ and
$\lift{\tau}(\Psi_2,\td{A_1}{\psi_2},\td{A}{\psi_1\psi_2},\phi)$.

We know $\td{A}{\psi_1} \steps^* A^{\Psi_1}_{\psi_1}$ and
$\cwftypep[\Psi_1]{A^{\Psi_1}_{\psi_1}}$, so
$\lift{\tau}(\Psi_2,\td{A_1}{\psi_2},\td{(A^{\Psi_1}_{\psi_1})}{\psi_2},\phi)$
where $A^{\Psi_1}_{\psi_1}\evals A_1$.

By $\ceqtypep[\Psi_1]{A^{\Psi_1}_{\psi_1}}{\td{(A^{\Psi}_{\id})}{\psi_1}}$
under $\psi_2$ and
$\ceqtypep[\Psi_2]{\td{(A^{\Psi}_{\id})}{\psi_1\psi_2}}{A^{\Psi_2}_{\psi_1\psi_2}}$,
we have
$\ceqtypep[\Psi_2]{\td{(A^{\Psi_1}_{\psi_1})}{\psi_2}}{A^{\Psi_2}_{\psi_1\psi_2}}$
and thus
$\lift{\tau}(\Psi_2,\td{(A^{\Psi_1}_{\psi_1})}{\psi_2},A^{\Psi_2}_{\psi_1\psi_2},\phi)$.
The result follows by transitivity and
$\td{A}{\psi_1\psi_2} \steps^* A^{\Psi_2}_{\psi_1\psi_2}$.

\item $\lift{\tau}(\Psi_2,\td{A}{\psi_1\psi_2},\td{(A^\Psi_{\id})}{\psi_1\psi_2},\phi')$.

By $\ceqtypep[\Psi_2]{A^{\Psi_2}_{\psi_1\psi_2}}{\td{(A^{\Psi}_{\id})}{\psi_1\psi_2}}$
we have
$\lift{\tau}(\Psi_2,A^{\Psi_2}_{\psi_1\psi_2},\td{(A^\Psi_{\id})}{\psi_1\psi_2},\phi')$;
the result follows by
$\td{A}{\psi_1\psi_2} \steps^* A^{\Psi_2}_{\psi_1\psi_2}$.

\item $\td{(A^{\Psi}_{\id})}{\psi_1}\evals A_1'$ and
$\lift{\tau}(\Psi_2,\td{(A^\Psi_{\id})}{\psi_1\psi_2},\td{A'_1}{\psi_2},\phi'')$.

Follows from $\cwftypep{A^{\Psi}_{\id}}$.
\qedhere
\end{enumerate}
\end{proof}

\begin{lemma}\label{lem:cohexp-ceqtm}
Assume we have $\wftm{M}$, $\cwftypep{A}$, and a family of terms
$\{M^{\Psi'}_\psi\}_{\psitd}$ such that for all $\psitd$,
$\ceqtm[\Psi']
{M^{\Psi'}_{\psi}}
{\td{(M^{\Psi}_{\id})}{\psi}}
{\td{A}{\psi}}$ and
$\td{M}{\psi} \steps^* M^{\Psi'}_\psi$. Then
$\ceqtm{M}{M^\Psi_{\id}}{A}$.
\end{lemma}
\begin{proof}
We must show that for any
$\tds{\Psi_1}{\psi_1}{\Psi}$ and
$\tds{\Psi_2}{\psi_2}{\Psi_1}$,
$\td{M}{\psi_1}\evals M_1$,
$\td{(M^\Psi_{\id})}{\psi_1}\evals M'_1$, and
$\lift{\vper{A}}_{\psi_1\psi_2}$ relates
$\td{M_1}{\psi_2}$,
$\td{M}{\psi_1\psi_2}$,
$\td{(M^\Psi_{\id})}{\psi_1\psi_2}$,
and $\td{M'_1}{\psi_2}$.

\begin{enumerate}
\item $\td{M}{\psi_1}\evals M_1$ and
$\lift{\vper{A}}_{\psi_1\psi_2}(\td{M_1}{\psi_2},\td{M}{\psi_1\psi_2})$.

We know $\td{M}{\psi_1} \steps^* M^{\Psi_1}_{\psi_1}$ and
$\coftype[\Psi_1]{M^{\Psi_1}_{\psi_1}}{\td{A}{\psi_1}}$, so
$\lift{\vper{A}}_{\psi_1\psi_2}(\td{M_1}{\psi_2},\td{(M^{\Psi_1}_{\psi_1})}{\psi_2})$
where $M^{\Psi_1}_{\psi_1}\evals M_1$.
By $\ceqtm[\Psi_1]
{M^{\Psi_1}_{\psi_1}}
{\td{(M^{\Psi}_{\id})}{\psi_1}}
{\td{A}{\psi_1}}$ under $\psi_2$ and
$\ceqtm[\Psi_2]
{\td{(M^{\Psi}_{\id})}{\psi_1\psi_2}}
{M^{\Psi_2}_{\psi_1\psi_2}}
{\td{A}{\psi_1\psi_2}}$, we have
$\ceqtm[\Psi_2]
{\td{(M^{\Psi_1}_{\psi_1})}{\psi_2}}
{M^{\Psi_2}_{\psi_1\psi_2}}
{\td{A}{\psi_1\psi_2}}$ and thus
$\lift{\vper{A}}_{\psi_1\psi_2}(\td{(M^{\Psi_1}_{\psi_1})}{\psi_2},M^{\Psi_2}_{\psi_1\psi_2})$.
The result follows by transitivity and
$\td{M}{\psi_1\psi_2} \steps^* M^{\Psi_2}_{\psi_1\psi_2}$.

\item $\lift{\vper{A}}_{\psi_1\psi_2}(\td{M}{\psi_1\psi_2},\td{(M^\Psi_{\id})}{\psi_1\psi_2})$.

By $\ceqtm[\Psi_2]
{M^{\Psi_2}_{\psi_1\psi_2}}
{\td{(M^{\Psi}_{\id})}{\psi_1\psi_2}}
{\td{A}{\psi_1\psi_2}}$ we have
$\lift{\vper{A}}_{\psi_1\psi_2}(M^{\Psi_2}_{\psi_1\psi_2},\td{(M^\Psi_{\id})}{\psi_1\psi_2})$;
the result follows by
$\td{M}{\psi_1\psi_2} \steps^* M^{\Psi_2}_{\psi_1\psi_2}$.

\item $\td{(M^{\Psi}_{\id})}{\psi_1}\evals M_1'$ and
$\lift{\vper{A}}_{\psi_1\psi_2}(\td{(M^\Psi_{\id})}{\psi_1\psi_2},\td{M'_1}{\psi_2})$.

Follows from $\coftype{M^{\Psi}_{\id}}{A}$.
\qedhere
\end{enumerate}
\end{proof}

\begin{lemma}\label{lem:cohexp-ceqtypek}
Assume we have $\wftm{A}$ and a family of terms $\{A^{\Psi'}_\psi\}_{\psitd}$
such that for all $\psitd$,
$\ceqtypek[\Psi']{A^{\Psi'}_{\psi}}{\td{(A^{\Psi}_{\id})}{\psi}}$ and
$\td{A}{\psi} \steps^* A^{\Psi'}_\psi$. Then $\ceqtypek{A}{A^\Psi_{\id}}$.
\end{lemma}
\begin{proof}
By \cref{lem:cohexp-ceqtypep}, $\ceqtypep{A}{A^\Psi_{\id}}$; it suffices to
establish the conditions in \cref{def:kan}. First, assume $\psitd$,
\begin{enumerate}
\item $\etc{r_i=r_i'}$ is valid,
\item $\ceqtm[\Psi']{M}{M'}{\td{A}{\psi}}$,
\item $\ceqtm[\Psi',y]<r_i=r_i',r_j=r_j'>{N_i}{N_j'}{\td{A}{\psi}}$
for any $i,j$, and
\item $\ceqtm[\Psi']<r_i=r_i'>{\dsubst{N_i}{r}{y}}{M}{\td{A}{\psi}}$
for any $i$,
\end{enumerate}
and show that
$\ceqtm[\Psi']{\Hcom*{\td{A}{\psi}}{r_i=r_i'}}%
{\Hcom{\td{(A^\Psi_{\id})}{\psi}}{r}{r'}{M'}{\sys{r_i=r_i'}{y.N_i'}}}{\td{A}{\psi}}$.
We apply \cref{lem:cohexp-ceqtm} to $\Hcom*{\td{A}{\psi}}{r_i=r_i'}$ and the
family
\[
\{ \Hcom{A^{\Psi''}_{\psi\psi'}}{\td{r}{\psi'}}{\td{r'}{\psi'}}%
{\td{M}{\psi'}}{\sys{\td{r_i}{\psi'}=\td{r_i'}{\psi'}}{y.\td{N_i}{\psi'}}}
\}^{\Psi''}_{\psi'}
\]
at $\cwftypep[\Psi']{\td{A}{\psi}}$. We know $\Hcom{\td{A}{\psi\psi'}}
\steps^* \Hcom{A^{\Psi''}_{\psi\psi'}}$ by $\td{A}{\psi\psi'} \steps^*
A^{\Psi''}_{\psi\psi'}$, and
$\ceqtm[\Psi'']
{\Hcom{A^{\Psi''}_{\psi\psi'}}}
{\Hcom{\td{(A^{\Psi'}_{\psi})}{\psi'}}}
{\td{A}{\psi\psi'}}$ by
$\ceqtypek[\Psi'']{A^{\Psi''}_{\psi\psi'}}{\td{(A^{\Psi'}_{\psi})}{\psi'}}$ and
$\ceqtypep[\Psi'']{A^{\Psi''}_{\psi\psi'}}{\td{A}{\psi\psi'}}$
(both by transitivity through $\td{(A^{\Psi}_{\id})}{\psi\psi'}$).
We conclude that
$\ceqtm[\Psi']{\Hcom{\td{A}{\psi}}}{\Hcom{A^{\Psi'}_{\psi}}}{\td{A}{\psi}}$, and
the desired result follows by
$\ceqtypek[\Psi']{A^{\Psi'}_{\psi}}{\td{(A^{\Psi}_{\id})}{\psi}}$.
The remaining $\Hcom$ equations of \cref{def:kan} follow by transitivity and
$\cwftypek{A^{\Psi}_{\id}}$.

Next, assuming $\tds{(\Psi',x)}{\psi}{\Psi}$ and
$\ceqtm[\Psi']{M}{M'}{\dsubst{\td{A}{\psi}}{r}{x}}$, show that
$\ceqtm[\Psi']{\Coe*{x.\td{A}{\psi}}}
{\Coe{x.\td{(A^\Psi_{\id})}{\psi}}{r}{r'}{M'}}%
{\dsubst{\td{A}{\psi}}{r'}{x}}$.
We apply \cref{lem:cohexp-ceqtm} to $\Coe*{x.\td{A}{\psi}}$ and
$\{ \Coe{x.A^{\Psi''}_{\psi\psi'}}{\td{r}{\psi'}}{\td{r'}{\psi'}}%
{\td{M}{\psi'}} \}^{\Psi''}_{\psi'}$
at $\cwftypep[\Psi']{\dsubst{\td{A}{\psi}}{r'}{x}}$, using the same argument as
before; we conclude that
$\ceqtm[\Psi']{\Coe{x.\td{A}{\psi}}}{\Coe{x.A^{\Psi'}_{\psi}}}{\dsubst{\td{A}{\psi}}{r'}{x}}$,
and the desired result follows by
$\ceqtypek[\Psi',x]{A^{\Psi'}_{\psi}}{\td{(A^{\Psi}_{\id})}{\psi}}$.
The remaining $\Coe$ equation of \cref{def:kan} follows by transitivity and
$\cwftypek{A^{\Psi}_{\id}}$.
\end{proof}

\begin{lemma}[Head expansion]\label{lem:expansion}
~\begin{enumerate}
\item If $\cwftypep{A'}$ and $A\ssteps^* A'$, then $\ceqtypep{A}{A'}$.
\item If $\coftype{M'}{A}$ and $M\ssteps^* M'$, then $\ceqtm{M}{M'}{A}$.
\item If $\cwftypek{A'}$ and $A\ssteps^* A'$, then $\ceqtypek{A}{A'}$.
\end{enumerate}
\end{lemma}
\begin{proof}
~\begin{enumerate}
\item
By \cref{lem:cohexp-ceqtypep} with $A^{\Psi'}_\psi = \td{A'}{\psi}$, because
$\td{A}{\psi}\steps^* \td{A'}{\psi}$ and
$\cwftypep[\Psi']{\td{A'}{\psi}}$ for all $\psi$.

\item
By \cref{lem:cohexp-ceqtm} with $M^{\Psi'}_\psi = \td{M'}{\psi}$, because
$\td{M}{\psi}\steps^* \td{M'}{\psi}$ and
$\coftype[\Psi']{\td{M'}{\psi}}{\td{A}{\psi}}$ for all $\psi$.

\item
By \cref{lem:cohexp-ceqtypek} with $A^{\Psi'}_\psi = \td{A'}{\psi}$, because
$\td{A}{\psi}\steps^* \td{A'}{\psi}$ and
$\cwftypek[\Psi']{\td{A'}{\psi}}$ for all $\psi$.
\qedhere
\end{enumerate}
\end{proof}

The $\Hcom$ operation implements \emph{homogeneous} composition, in the sense
that $A$ must be degenerate in the bound direction of the tubes. We can obtain
\emph{heterogeneous} composition, $\Com$, by combining $\Hcom$ and $\Coe$.

\begin{theorem}\label{thm:com}
If $\ceqtypek[\Psi,y]{A}{B}$,
\begin{enumerate}
\item $\etc{r_i=r_i'}$ is valid,
\item $\ceqtm{M}{M'}{\dsubst{A}{r}{y}}$,
\item $\ceqtm[\Psi,y]<r_i=r_i',r_j=r_j'>{N_i}{N_j'}{A}$ for any $i,j$, and
\item $\ceqtm<r_i=r_i'>{\dsubst{N_i}{r}{y}}{M}{\dsubst{A}{r}{y}}$ for any $i$,
\end{enumerate}
then
\begin{enumerate}
\item
$\ceqtm{\Com*{y.A}{r_i=r_i'}}{\Com{y.B}{r}{r'}{M'}{\sys{r_i=r_i'}{y.N_i'}}}
{\dsubst{A}{r'}{y}}$;
\item if $r=r'$ then
$\ceqtm{\Com{y.A}{r}{r}{M}{\sys{r_i=r_i'}{y.N_i}}}{M}{\dsubst{A}{r}{y}}$; and
\item if $r_i = r_i'$ then
$\ceqtm{\Com*{y.A}{r_i=r_i'}}{\dsubst{N_i}{r'}{y}}{\dsubst{A}{r'}{y}}$.
\end{enumerate}
\end{theorem}
\begin{proof}
For all $\tds{\Psi'}{\psi}{(\Psi,y)}$ satisfying $r_i=r_i'$ and $r_j=r_j'$, we
know $\ceqtm[\Psi']{\td{N_i}{\psi}}{\td{N_j'}{\psi}}{\td{A}{\psi}}$. By
\cref{def:kan},
$\ceqtm[\Psi']
{\td{(\Coe{y.A}{y}{r'}{N_i})}{\psi}}
{\td{(\Coe{y.B}{y}{r'}{N_j'})}{\psi}}
{\td{\dsubst{A}{r'}{y}}{\psi}}$, and therefore
$\ceqtm[\Psi,y]<r_i=r_i',r_j=r_j'>{\Coe{y.A}{y}{r'}{N_i}}{\Coe{y.B}{y}{r'}{N_j'}}{A}$.
By a similar argument we conclude
$\ceqtm<r_i=r_i'>{\dsubst{(\Coe{y.A}{y}{r'}{N_i})}{r}{y}}%
{\Coe{y.A}{r}{r'}{M}}{\dsubst{A}{r'}{y}}$,
and by \cref{def:kan} directly,
$\ceqtm{\Coe{y.A}{r}{r'}{M}}{\Coe{y.B}{r}{r'}{M'}}{\dsubst{A}{r'}{y}}$.
By \cref{def:kan} we conclude
\begin{gather*}
{\Hcom{\dsubst{A}{r'}{y}}{r}{r'}%
  {\Coe{y.A}{r}{r'}{M}}%
  {\sys{r_i=r_i'}{y.\Coe{y.A}{y}{r'}{N_i}}}} \\
\ceqtm{{}}
{\Hcom{\dsubst{B}{r'}{y}}{r}{r'}%
  {\Coe{y.B}{r}{r'}{M'}}%
  {\sys{r_i=r_i'}{y.\Coe{y.B}{y}{r'}{N_i'}}}}
{\dsubst{A}{r'}{y}}.
\end{gather*}
Result (1) follows by \cref{lem:expansion} on each side.

Result (2) follows by \cref{lem:expansion} and, by \cref{def:kan} twice,
\[
\ceqtm
{\Hcom{\dsubst{A}{r'}{y}}{r'}{r'}%
  {\Coe{y.A}{r'}{r'}{M}}%
  {\sys{r_i=r_i'}{y.\Coe{y.A}{y}{r'}{N_i}}}}
{M}
{\dsubst{A}{r'}{y}}.
\]

Result (3) follows by \cref{lem:expansion} and, by \cref{def:kan} twice,
\[
\ceqtm
{\Hcom{\dsubst{A}{r'}{y}}{r}{r'}%
  {\Coe{y.A}{r}{r'}{M}}%
  {\sys{r_i=r_i'}{y.\Coe{y.A}{y}{r'}{N_i}}}}
{\dsubst{N_i}{r'}{y}}
{\dsubst{A}{r'}{y}}.
\qedhere
\]
\end{proof}

\newpage
\section{Types}
\label{sec:types}

In \cref{sec:typesys} we defined two sequences of cubical type systems, and in
\cref{sec:meanings} we defined the judgments of higher type theory relative to
any cubical type system.  In this section we will prove that $\pre\tau_\omega$
validates certain rules, summarized in part in \cref{sec:rules}. For
non-universe connectives, we in fact prove that the rules hold in every
$\tau^\kappa_n$ and $\tau^\kappa_\omega$.

\subsection{Dependent function types}

Let $\tau=\Kan\mu(\nu)$ or $\pre\mu(\nu,\sigma)$ for any cubical type systems
$\nu,\sigma$; in $\tau$,
whenever $\ceqtypep{A}{A'}$,
$\eqtypep{\oft{a}{A}}{B}{B'}$, and
$\phi = \{(\lam{a}{N},\lam{a}{N'}) \mid \eqtm{\oft{a}{A}}{N}{N'}{B}\}$, we have
$\tau(\Psi,\picl{a}{A}{B},\picl{a}{A'}{B'},\phi)$.
Notice that whenever $\ceqtypep{A}{A'}$ and $\eqtypep{\oft{a}{A}}{B}{B'}$,
we have $\PTy(\tau)(\Psi,\picl{a}{A}{B},\picl{a}{A'}{B'},\_)$ because
$\sisval{\picl{a}{A}{B}}$ and judgments are preserved by dimension
substitution.

\begin{lemma}\label{lem:fun-preintro}
If $\eqtm{\oft aA}{M}{M'}{B}$ then
$\Tm(\vper{\picl{a}{A}{B}})(\lam{a}{M},\lam{a}{M'})$.
\end{lemma}
\begin{proof}
By $\sisval{\lam{a}{M}}$, it suffices to check that
$\vper{\picl{a}{A}{B}}_{\psi}(\lam{a}{\td{M}{\psi}},\lam{a}{\td{M'}{\psi}})$
for any $\psitd$; this holds because
$\eqtm[\Psi']{\oft{a}{\td{A}{\psi}}}{\td{M}{\psi}}{\td{M'}{\psi}}{\td{B}{\psi}}$
and $\vper{\picl{a}{A}{B}}_{\psi} =
\vper{\picl{a}{\td{A}{\psi}}{\td{B}{\psi}}}$.
\end{proof}

\begin{rul}[Pretype formation]\label{rul:fun-form-pre}
If $\ceqtypep{A}{A'}$ and
$\eqtypep{\oft aA}{B}{B'}$ then
$\ceqtypep{\picl{a}{A}{B}}{\picl{a}{A'}{B'}}$.
\end{rul}
\begin{proof}
We have $\PTy(\tau)(\Psi,\picl{a}{A}{B},\picl{a}{A'}{B'},\alpha)$, and by
\cref{lem:fun-preintro}, $\Coh(\alpha)$.
\end{proof}

\begin{rul}[Introduction]\label{rul:fun-intro}
If $\eqtm{\oft aA}{M}{M'}{B}$ then
$\ceqtm{\lam{a}{M}}{\lam{a}{M'}}{\picl{a}{A}{B}}$.
\end{rul}
\begin{proof}
Immediate by \cref{lem:fun-preintro,rul:fun-form-pre}.
\end{proof}

\begin{lemma}\label{lem:fun-preelim}
If $\coftype{M}{\picl{a}{A}{B}}$ and $\coftype{N}{A}$ then $M\evals\lam{a}{O}$
and $\ceqtm{\app{M}{N}}{\subst{O}{N}{a}}{\subst{B}{N}{a}}$.
\end{lemma}
\begin{proof}
For any $\psitd$, we know that $\td{M}{\psi}\evals\lam{a}{O_\psi}$ and
$\vper{\picl{a}{A}{B}}_\psi(\lam{a}{\td{O_{\id}}{\psi}},\lam{a}{O_\psi})$, and
therefore
$\eqtm[\Psi']{\oft{a}{\td{A}{\psi}}}{\td{O_{\id}}{\psi}}{O_\psi}{\td{B}{\psi}}$.
We apply coherent expansion to $\app{M}{N}$,
$\cwftypep{\subst{B}{N}{a}}$, and
$\{\subst{O_\psi}{\td{N}{\psi}}{a}\}^{\Psi'}_\psi$, by
$\app{\td{M}{\psi}}{\td{N}{\psi}} \steps^*
\app{\lam{a}{O_\psi}}{\td{N}{\psi}} \steps
\subst{O_\psi}{\td{N}{\psi}}{a}$ and
$\ceqtm[\Psi']
{\subst{O_\psi}{\td{N}{\psi}}{a}}
{\td{(\subst{O_{\id}}{N}{a})}{\psi}}
{\subst{\td{B}{\psi}}{\td{N}{\psi}}{a}}$.
We conclude by \cref{lem:cohexp-ceqtm} that
$\ceqtm{\app{M}{N}}{\subst{O_{\id}}{N}{a}}{\subst{B}{N}{a}}$, as desired.
\end{proof}

\begin{rul}[Elimination]\label{rul:fun-elim}
If $\ceqtm{M}{M'}{\picl{a}{A}{B}}$ and
$\ceqtm{N}{N'}{A}$ then
$\ceqtm{\app{M}{N}}{\app{M'}{N'}}{\subst{B}{N}{a}}$.
\end{rul}
\begin{proof}
By \cref{lem:fun-preelim} we know
$M\evals\lam{a}{O}$,
$M'\evals\lam{a}{O'}$,
$\ceqtm{\app{M}{N}}{\subst{O}{N}{a}}{\subst{B}{N}{a}}$, and
$\ceqtm{\app{M'}{N'}}{\subst{O'}{N'}{a}}{\subst{B}{N'}{a}}$.
By \cref{lem:coftype-evals-ceqtm},
$\ceqtm{M}{\lam{a}{O}}{\picl{a}{A}{B}}$ and
$\ceqtm{M'}{\lam{a}{O'}}{\picl{a}{A}{B}}$, and so by
$\vper{\picl{a}{A}{B}}(\lam{a}{O},\lam{a}{O'})$,
$\eqtm{\oft{a}{A}}{O}{O'}{B}$.
We conclude
$\ceqtm{\subst{O}{N}{a}}{\subst{O'}{N'}{a}}{\subst{B}{N}{a}}$ and
$\ceqtypep{\subst{B}{N}{a}}{\subst{B}{N'}{a}}$, and the result follows by
symmetry, transitivity, and \cref{lem:ceqtypep-ceqtm}.
\end{proof}

\begin{rul}[Eta]\label{rul:fun-eta}
If $\coftype{M}{\picl{a}{A}{B}}$ then
$\ceqtm{M}{\lam{a}{\app{M}{a}}}{\picl{a}{A}{B}}$.
\end{rul}
\begin{proof}
By \cref{lem:coftype-evals-ceqtm},
$M\evals\lam{a}{O}$ and
$\ceqtm{M}{\lam{a}{O}}{\picl{a}{A}{B}}$;
by transitivity and \cref{rul:fun-intro} it suffices to show
$\eqtm{\oft{a}{A}}{O}{\app{M}{a}}{B}$, that is,
for any $\psitd$ and $\ceqtm[\Psi']{N}{N'}{\td{A}{\psi}}$,
$\ceqtm[\Psi']{\subst{\td{O}{\psi}}{N}{a}}{\app{\td{M}{\psi}}{N'}}{\subst{\td{B}{\psi}}{N}{a}}$.
By \cref{lem:fun-preelim},
$\ceqtm[\Psi']{\subst{O_\psi}{N'}{a}}{\app{\td{M}{\psi}}{N'}}{\subst{\td{B}{\psi}}{N'}{a}}$,
where $\td{M}{\psi}\evals\lam{a}{O_\psi}$. The result then follows by
$\ceqtypep[\Psi']{\subst{\td{B}{\psi}}{N}{a}}{\subst{\td{B}{\psi}}{N'}{a}}$
and
$\eqtm[\Psi']{\oft{a}{\td{A}{\psi}}}{\td{O_{\id}}{\psi}}{O_\psi}{\td{B}{\psi}}$,
the latter by
$\vper{\picl{a}{A}{B}}_\psi(\lam{a}{\td{O}{\psi}},\lam{a}{O_\psi})$.
\end{proof}

\begin{rul}[Computation]
If $\oftype{\oft aA}{M}{B}$ and
$\coftype{N}{A}$ then
$\ceqtm{\app{\lam{a}{M}}{N}}{\subst{M}{N}{a}}{\subst{B}{N}{a}}$.
\end{rul}
\begin{proof}
Immediate by $\coftype{\subst{M}{N}{a}}{\subst{B}{N}{a}}$,
$\app{\lam{a}{M}}{N}\ssteps \subst{M}{N}{a}$, and
\cref{lem:expansion}.
\end{proof}

\begin{rul}[Kan type formation]\label{rul:fun-form-kan}
If $\ceqtypek{A}{A'}$ and
$\eqtypek{\oft aA}{B}{B'}$ then
$\ceqtypek{\picl{a}{A}{B}}{\picl{a}{A'}{B'}}$.
\end{rul}
\begin{proof}
By \cref{rul:fun-form-pre}, it suffices to check the five Kan conditions.

($\Hcom$) First, suppose that $\psitd$,
\begin{enumerate}
\item $\etc{\xi_i}=\etc{r_i=r_i'}$ is valid,
\item $\ceqtm[\Psi']{M}{M'}{\picl{a}{\td{A}{\psi}}{\td{B}{\psi}}}$,
\item $\ceqtm[\Psi',y]<r_i=r_i',r_j=r_j'>{N_i}{N_j'}%
{\picl{a}{\td{A}{\psi}}{\td{B}{\psi}}}$ for any $i,j$, and
\item $\ceqtm[\Psi']<r_i=r_i'>{\dsubst{N_i}{r}{y}}{M}%
{\picl{a}{\td{A}{\psi}}{\td{B}{\psi}}}$ for any $i$,
\end{enumerate}
and show
$\ceqtm[\Psi']%
{\Hcom*{\picl{a}{\td{A}{\psi}}{\td{B}{\psi}}}{\xi_i}}%
{\Hcom{\picl{a}{\td{A'}{\psi}}{\td{B'}{\psi}}}{r}{r'}{M'}{\sys{\xi_i}{N_i'}}}%
{\picl{a}{\td{A}{\psi}}{\td{B}{\psi}}}$.
By \cref{lem:expansion} on both sides and \cref{rul:fun-intro}, it suffices to
show
\begin{align*}
\ctx{\oft{a}{\td{A}{\psi}}}{} &
{\Hcom{\td{B}{\psi}}{r}{r'}{\app{M}{a}}{\sys{\xi_i}{y.\app{N_i}{a}}}} \\
\ceqtmtab[\Psi']{}
{\Hcom{\td{B'}{\psi}}{r}{r'}{\app{M'}{a}}{\sys{\xi_i}{y.\app{N_i'}{a}}}}
{\td{B}{\psi}}
\end{align*}
or that for any $\tds{\Psi''}{\psi'}{\Psi'}$ and
$\ceqtm[\Psi'']{N}{N'}{\td{A}{\psi\psi'}}$,
\begin{align*}
&{\Hcom{\subst{\td{B}{\psi\psi'}}{N}{a}}{\td{r}{\psi}}{\td{r'}{\psi}}%
  {\app{\td{M}{\psi'}}{N}}{\sys{\td{\xi_i}{\psi'}}{y.\app{\td{N_i}{\psi'}}{N}}}} \\
\ceqtmtab[\Psi'']{}
{\Hcom{\subst{\td{B'}{\psi\psi'}}{N'}{a}}{\td{r}{\psi}}{\td{r'}{\psi}}%
  {\app{\td{M'}{\psi'}}{N'}}{\sys{\td{\xi_i}{\psi'}}{y.\app{\td{N_i'}{\psi'}}{N'}}}}
{\subst{\td{B}{\psi\psi'}}{N}{a}}.
\end{align*}
By $\eqtypek{\oft aA}{B}{B'}$ we know
$\ceqtypek[\Psi'']{\subst{\td{B}{\psi\psi'}}{N}{a}}{\subst{\td{B'}{\psi\psi'}}{N'}{a}}$,
so the result follows by \cref{def:kan} once we establish
\begin{enumerate}
\item $\etc{\td{r_i}{\psi'} = \td{r_i'}{\psi'}}$ is valid,
\item $\ceqtm[\Psi'']{\app{\td{M}{\psi'}}{N}}{\app{\td{M'}{\psi'}}{N'}}%
{\subst{\td{B}{\psi\psi'}}{N}{a}}$,
\item $\ceqtm[\Psi'',y]%
<\td{r_i}{\psi'}=\td{r_i'}{\psi'},\td{r_j}{\psi'}=\td{r_j'}{\psi'}>%
{\app{\td{N_i}{\psi'}}{N}}{\app{\td{N_j'}{\psi'}}{N'}}%
{\subst{\td{B}{\psi\psi'}}{N}{a}}$ for any $i,j$, and
\item $\ceqtm[\Psi'']<\td{r_i}{\psi'}=\td{r_i'}{\psi'}>%
{\app{\td{\dsubst{N_i}{r}{y}}{\psi'}}{N}}{\app{\td{M}{\psi'}}{N'}}%
{\subst{\td{B}{\psi\psi'}}{N}{a}}$ for any $i$.
\end{enumerate}
These follow from our hypotheses and a context-restricted variant of
\cref{rul:fun-elim}, namely that
if $\ceqtm<\Xi>{M}{M'}{\picl{a}{A}{B}}$ and
$\ceqtm<\Xi>{N}{N'}{A}$ then
$\ceqtm<\Xi>{\app{M}{N}}{\app{M'}{N'}}{\subst{B}{N}{a}}$. (This statement is
easily proven by expanding the definition of context-restricted judgments.)

Next, we must show that if $r=r'$ then
$\ceqtm[\Psi']
{\Hcom*{\picl{a}{\td{A}{\psi}}{\td{B}{\psi}}}{\xi_i}}{M}
{\picl{a}{\td{A}{\psi}}{\td{B}{\psi}}}$.
By \cref{lem:expansion} on the left and \cref{rul:fun-eta} on the right, it
suffices to show that
\[
\ceqtm[\Psi']
{\lam{a}{\Hcom{\td{B}{\psi}}{r}{r'}{\app{M}{a}}{\sys{\xi_i}{y.\app{N_i}{a}}}}}
{\lam{a}{\app{M}{a}}}
{\picl{a}{\td{A}{\psi}}{\td{B}{\psi}}}.
\]
By \cref{rul:fun-intro}, we show that for any $\tds{\Psi''}{\psi'}{\Psi'}$ and
$\ceqtm[\Psi'']{N}{N'}{\td{A}{\psi\psi'}}$,
\[
\ceqtm[\Psi']
{\Hcom{\subst{\td{B}{\psi\psi'}}{N}{a}}{\td{r}{\psi}}{\td{r'}{\psi}}%
  {\app{\td{M}{\psi'}}{N}}{\sys{\td{\xi_i}{\psi'}}{y.\app{\td{N_i}{\psi'}}{N}}}}
{\app{\td{M}{\psi'}}{N'}}
{\subst{\td{B}{\psi\psi'}}{N}{a}}.
\]
By $\cwftypek[\Psi'']{\subst{\td{B}{\psi\psi'}}{N}{a}}$ and $r=r'$ on the left,
it suffices to show
$\ceqtm[\Psi'']{\app{\td{M}{\psi'}}{N}}{\app{\td{M}{\psi'}}{N'}}%
{\subst{\td{B}{\psi\psi'}}{N}{a}}$, which holds by \cref{rul:fun-elim}.

For the final $\Hcom$ property, show that if $r_i=r_i'$ then
$\ceqtm[\Psi']
{\Hcom*{\picl{a}{\td{A}{\psi}}{\td{B}{\psi}}}{\xi_i}}{\dsubst{N_i}{r'}{y}}
{\picl{a}{\td{A}{\psi}}{\td{B}{\psi}}}$. As before, by \cref{lem:expansion} on
the left, \cref{rul:fun-eta} on the right, and \cref{rul:fun-intro}, show that
for any $\tds{\Psi''}{\psi'}{\Psi'}$ and
$\ceqtm[\Psi'']{N}{N'}{\td{A}{\psi\psi'}}$,
\[
\ceqtm[\Psi']
{\Hcom{\subst{\td{B}{\psi\psi'}}{N}{a}}{\td{r}{\psi}}{\td{r'}{\psi}}%
  {\app{\td{M}{\psi'}}{N}}{\sys{\td{\xi_i}{\psi'}}{y.\app{\td{N_i}{\psi'}}{N}}}}
{\app{\td{\dsubst{N_i}{r'}{y}}{\psi'}}{N'}}
{\subst{\td{B}{\psi\psi'}}{N}{a}}.
\]
This follows by $\cwftypek[\Psi'']{\subst{\td{B}{\psi\psi'}}{N}{a}}$ and
$\td{r_i}{\psi'}=\td{r_i'}{\psi'}$ on the left, and \cref{rul:fun-elim}.

($\Coe$) Now, suppose that $\tds{(\Psi',x)}{\psi}{\Psi}$ and
$\ceqtm[\Psi']{M}{M'}%
{\picl{a}{\dsubst{\td{A}{\psi}}{r}{x}}{\dsubst{\td{B}{\psi}}{r}{x}}}$, and show
that
$\ceqtm[\Psi']{\Coe*{x.\picl{a}{\td{A}{\psi}}{\td{B}{\psi}}}}%
{\Coe{x.\picl{a}{\td{A'}{\psi}}{\td{B'}{\psi}}}{r}{r'}{M'}}%
{\picl{a}{\dsubst{\td{A}{\psi}}{r'}{x}}{\dsubst{\td{B}{\psi}}{r'}{x}}}$.
By \cref{lem:expansion} on both sides and \cref{rul:fun-intro}, we must show
that for any $\tds{\Psi''}{\psi'}{\Psi'}$ and
$\ceqtm[\Psi'']{N}{N'}{\dsubst{\td{A}{\psi\psi'}}{\td{r'}{\psi'}}{x}}$,
\begin{align*}
&{\Coe{x.\subst{\td{B}{\psi\psi'}}{\Coe{x.\td{A}{\psi\psi'}}{\td{r'}{\psi'}}{x}{N}}{a}}%
  {\td{r}{\psi'}}{\td{r'}{\psi'}}%
  {\app{\td{M}{\psi'}}{\Coe{x.\td{A}{\psi\psi'}}{\td{r'}{\psi'}}{\td{r}{\psi'}}{N}}}} \\
\ceqtmtab[\Psi'']{}%
{\Coe{x.\subst{\td{B'}{\psi\psi'}}{\Coe{x.\td{A'}{\psi\psi'}}{\td{r'}{\psi'}}{x}{N'}}{a}}%
  {\td{r}{\psi'}}{\td{r'}{\psi'}}%
  {\app{\td{M'}{\psi'}}{\Coe{x.\td{A'}{\psi\psi'}}{\td{r'}{\psi'}}{\td{r}{\psi'}}{N'}}}}%
{\subst{\dsubst{\td{B}{\psi\psi'}}{\td{r'}{\psi'}}{x}}{N}{a}}.
\end{align*}
By $\ceqtypek[\Psi'',x]{\td{A}{\psi\psi'}}{\td{A'}{\psi\psi'}}$, we have
$\ceqtm[\Psi'']%
{\Coe{x.\td{A}{\psi\psi'}}{\td{r'}{\psi'}}{x}{N}}%
{\Coe{x.\td{A'}{\psi\psi'}}{\td{r'}{\psi'}}{x}{N'}}%
{\dsubst{\td{A}{\psi\psi'}}{\td{r'}{\psi'}}{x}}$, and the corresponding instances
of $\td{B}{\psi\psi'}$ and $\td{B'}{\psi\psi'}$ are equal as Kan types.
By \cref{rul:fun-elim} we have
\[
\ceqtm[\Psi'']
{\app{\td{M}{\psi'}}{\Coe{x.\td{A}{\psi\psi'}}{\td{r'}{\psi'}}{\td{r}{\psi'}}{N}}}
{\app{\td{M'}{\psi'}}{\Coe{x.\td{A'}{\psi\psi'}}{\td{r'}{\psi'}}{\td{r}{\psi'}}{N'}}}
{\subst{\dsubst{\td{B}{\psi\psi'}}{\td{r}{\psi'}}{x}}{\Coe{x.\td{A}{\psi\psi'}}{\td{r'}{\psi'}}{\td{r}{\psi'}}{N}}{a}}
\]
so the above $\Coe$ are equal in
$\subst{\dsubst{\td{B}{\psi\psi'}}{\td{r'}{\psi'}}{x}}%
{\Coe{x.\td{A}{\psi\psi'}}{\td{r'}{\psi'}}{\td{r'}{\psi'}}{N}}{a}$. The result
follows by \cref{lem:ceqtypep-ceqtm} and
$\ceqtm[\Psi'']{\Coe{x.\td{A}{\psi\psi'}}{\td{r'}{\psi'}}{\td{r'}{\psi'}}{N}}%
{N}{\dsubst{\td{A}{\psi\psi'}}{\td{r'}{\psi'}}{x}}$.

Finally, show that if $r=r'$ then
$\ceqtm[\Psi']{\Coe*{x.\picl{a}{\td{A}{\psi}}{\td{B}{\psi}}}}{M}%
{\picl{a}{\dsubst{\td{A}{\psi}}{r'}{x}}{\dsubst{\td{B}{\psi}}{r'}{x}}}$.
By \cref{lem:expansion} on the left, \cref{rul:fun-eta} on the right, and
\cref{rul:fun-intro}, it suffices to show that for any
$\tds{\Psi''}{\psi'}{\Psi'}$ and
$\ceqtm[\Psi'']{N}{N'}{\dsubst{\td{A}{\psi\psi'}}{\td{r'}{\psi'}}{x}}$,
\[
\ceqtm[\Psi'']
{\Coe{x.\subst{\td{B}{\psi\psi'}}{\Coe{x.\td{A}{\psi\psi'}}{\td{r'}{\psi'}}{x}{N}}{a}}%
  {\td{r}{\psi'}}{\td{r'}{\psi'}}%
  {\app{\td{M}{\psi'}}{\Coe{x.\td{A}{\psi\psi'}}{\td{r'}{\psi'}}{\td{r}{\psi'}}{N}}}}
{\app{\td{M}{\psi'}}{N'}}
{\subst{\dsubst{\td{B}{\psi\psi'}}{\td{r'}{\psi'}}{x}}{N}{a}}.
\]
By $\td{r}{\psi'}=\td{r'}{\psi'}$, $\cwftypek[\Psi'',x]{\td{A}{\psi\psi'}}$,
\cref{rul:fun-elim}, and
$\cwftypek[\Psi'',x]{\subst{\td{B}{\psi\psi'}}{\Coe{x.\td{A}{\psi\psi'}}{\td{r'}{\psi'}}{x}{N}}{a}}$,
it suffices to show
$\ceqtm[\Psi'']{\app{\td{M}{\psi'}}{N}}{\app{\td{M}{\psi'}}{N'}}%
{\subst{\dsubst{\td{B}{\psi\psi'}}{\td{r'}{\psi'}}{x}}{N}{a}}$, which again
follows by \cref{rul:fun-elim}.
\end{proof}

\subsection{Dependent pair types}

Let $\tau=\Kan\mu(\nu)$ or $\pre\mu(\nu,\sigma)$ for any cubical type systems
$\nu,\sigma$; in $\tau$,
whenever $\ceqtypep{A}{A'}$,
$\eqtypep{\oft{a}{A}}{B}{B'}$, and
$\phi = \{(\pair{M}{N},\pair{M'}{N'}) \mid
\ceqtm{M}{M'}{A} \land \ceqtm{N}{N'}{\subst{B}{M}{a}}\}$, we have
$\tau(\Psi,\sigmacl{a}{A}{B},\sigmacl{a}{A'}{B'},\phi)$.

\begin{rul}[Pretype formation]\label{rul:pair-form-pre}
If $\ceqtypep{A}{A'}$ and $\eqtypep{\oft aA}{B}{B'}$ then
$\ceqtypep{\sigmacl{a}{A}{B}}{\sigmacl{a}{A'}{B'}}$.
\end{rul}
\begin{proof}
We have $\PTy(\tau)(\Psi,\sigmacl{a}{A}{B},\sigmacl{a}{A'}{B'},\_)$ because
$\sisval{\sigmacl{a}{A}{B}}$ and judgments are preserved by dimension
substitution. For $\Coh(\vper{\sigmacl{a}{A}{B}})$, assume
$\vper{\sigmacl{a}{A}{B}}_\psi(\pair{M}{N},\pair{M'}{N'})$.
Then $\ceqtm[\Psi']{M}{M'}{\td{A}{\psi}}$ and
$\ceqtm[\Psi']{N}{N'}{\subst{\td{B}{\psi}}{M}{a}}$; again,
$\sisval{\pair{M}{N}}$ and these judgments are preserved by dimension
substitution, so
$\Tm(\td{\vper{\sigmacl{a}{A}{B}}}{\psi})(\pair{M}{N},\pair{M'}{N'})$.
\end{proof}

\begin{rul}[Introduction]\label{rul:pair-intro}
If $\ceqtm{M}{M'}{A}$ and $\ceqtm{N}{N'}{\subst{B}{M}{a}}$ then
$\ceqtm{\pair{M}{N}}{\pair{M'}{N'}}{\sigmacl{a}{A}{B}}$.
\end{rul}
\begin{proof}
Immediate by \cref{rul:pair-form-pre}.
\end{proof}

\begin{rul}[Elimination]\label{rul:pair-elim}
If $\ceqtm{P}{P'}{\sigmacl{a}{A}{B}}$ then
$\ceqtm{\fst{P}}{\fst{P'}}{A}$ and
$\ceqtm{\snd{P}}{\snd{P'}}{\subst{B}{\fst{P}}{a}}$.
\end{rul}
\begin{proof}
For any $\psitd$, $\td{P}{\psi}\evals\pair{M_\psi}{N_\psi}$,
$\coftype[\Psi']{M_\psi}{\td{A}{\psi}}$, and
$\coftype[\Psi']{N_\psi}{\subst{\td{B}{\psi}}{M_\psi}{a}}$. For part (1), apply
coherent expansion to $\fst{P}$ with family $\{M_\psi\}^{\Psi'}_\psi$; then
$\ceqtm[\Psi']{\td{(M_{\id})}{\psi}}{M_\psi}{\td{A}{\psi}}$ by
$\coftype{P}{\sigmacl{a}{A}{B}}$ at $\id,\psi$. By \cref{lem:cohexp-ceqtm},
$\ceqtm{\fst{P}}{M_{\id}}{A}$, and part (1) follows by
$\ceqtm{M_{\id}}{M'_{\id}}{A}$ and a symmetric argument on the right side.

For part (2), apply coherent expansion to $\snd{P}$ with family
$\{N_\psi\}^{\Psi'}_\psi$. We have $\ceqtm[\Psi']{\td{(N_{\id})}{\psi}}{N_\psi}%
{\subst{\td{B}{\psi}}{\td{(M_{\id})}{\psi}}{a}}$ by
$\coftype{P}{\sigmacl{a}{A}{B}}$ at $\id,\psi$, so by \cref{lem:cohexp-ceqtm},
$\ceqtm{\snd{P}}{N_{\id}}{\subst{B}{M_{\id}}{a}}$. Part (2) follows by
$\ceqtypep{\subst{B}{M_{\id}}{a}}{\subst{B}{\fst{P}}{a}}$ (by
$\eqtypep{\oft{a}{A}}{B}{B'}$ and $\ceqtm{M_{\id}}{\fst{P}}{A}$),
$\ceqtm{N_{\id}}{N'_{\id}}{\subst{B}{M_{\id}}{a}}$, and
a symmetric argument on the right side.
\end{proof}

\begin{rul}[Computation]
If $\coftype{M}{A}$ then $\ceqtm{\fst{\pair{M}{N}}}{M}{A}$.
If $\coftype{N}{B}$ then $\ceqtm{\snd{\pair{M}{N}}}{N}{B}$.
\end{rul}
\begin{proof}
Immediate by \cref{lem:expansion}.
\end{proof}

\begin{rul}[Eta]\label{rul:pair-eta}
If $\coftype{P}{\sigmacl{a}{A}{B}}$ then
$\ceqtm{P}{\pair{\fst{P}}{\snd{P}}}{\sigmacl{a}{A}{B}}$.
\end{rul}
\begin{proof}
By \cref{lem:coftype-evals-ceqtm}, $P\evals\pair{M}{N}$,
$\ceqtm{P}{\pair{M}{N}}{\sigmacl{a}{A}{B}}$, $\coftype{M}{A}$, and
$\coftype{N}{\subst{B}{M}{a}}$. By \cref{rul:pair-intro,lem:coftype-ceqtm} and
transitivity, we show $\lift{\vper{A}}(M,\fst{P})$ and
$\lift{\vper{\subst{B}{M}{a}}}(N,\snd{P})$. This is immediate by
$\fst{P}\steps^*\fst{\pair{M}{N}}\steps M$ and
$\snd{P}\steps^*\snd{\pair{M}{N}}\steps N$.
\end{proof}

\begin{rul}[Kan type formation]
If $\ceqtypek{A}{A'}$ and $\eqtypek{\oft aA}{B}{B'}$ then
$\ceqtypek{\sigmacl{a}{A}{B}}{\sigmacl{a}{A'}{B'}}$.
\end{rul}
\begin{proof}
It suffices to check the five Kan conditions.

($\Hcom$) First, suppose that $\psitd$,
\begin{enumerate}
\item $\etc{r_i=r_i'}$ is valid,
\item $\ceqtm[\Psi']{M}{M'}{\sigmacl{a}{\td{A}{\psi}}{\td{B}{\psi}}}$,
\item $\ceqtm[\Psi',y]<r_i=r_i',r_j=r_j'>{N_i}{N_j'}{\sigmacl{a}{\td{A}{\psi}}{\td{B}{\psi}}}$
for any $i,j$, and
\item $\ceqtm[\Psi']<r_i=r_i'>{\dsubst{N_i}{r}{y}}{M}{\sigmacl{a}{\td{A}{\psi}}{\td{B}{\psi}}}$
for any $i$,
\end{enumerate}
and show
$\ceqtm[\Psi']{\Hcom*{\sigmacl{a}{\td{A}{\psi}}{\td{B}{\psi}}}{\xi_i}}%
{\Hcom{\sigmacl{a}{\td{A'}{\psi}}{\td{B'}{\psi}}}{r}{r'}{M'}{\sys{\xi_i}{y.N_i'}}}%
{\sigmacl{a}{\td{A}{\psi}}{\td{B}{\psi}}}$. By \cref{lem:expansion} on both
sides and \cref{rul:pair-intro}, it suffices to show (the binary version of)
\begin{gather*}
\coftype[\Psi']
{\Hcom{\td{A}{\psi}}{r}{r'}{\fst{M}}{\sys{\xi_i}{y.\fst{N_i}}}}
{\td{A}{\psi}} \\
\coftype[\Psi']
{\Com{z.\subst{\td{B}{\psi}}{F}{a}}{r}{r'}{\snd{M}}{\sys{\xi_i}{y.\snd{N_i}}}}
{\subst{\td{B}{\psi}}{\Hcom{\td{A}{\psi}}}{a}} \\
\text{where}\ F={\Hcom{\td{A}{\psi}}{r}{z}{\fst{M}}{\sys{\xi_i}{y.\fst{N_i}}}}.
\end{gather*}
We have $\coftype[\Psi']{\Hcom{\td{A}{\psi}}}{\td{A}{\psi}}$ and
$\coftype[\Psi',z]{F}{\td{A}{\psi}}$ by $\cwftypek{A}$ and \cref{rul:pair-elim}.
We show $\coftype[\Psi']{\Com{z.\subst{\td{B}{\psi}}{F}{a}}}%
{\subst{\td{B}{\psi}}{\Hcom{\td{A}{\psi}}}{a}}$ by \cref{thm:com},
observing that $\cwftypek[\Psi',z]{\subst{\td{B}{\psi}}{F}{a}}$,
$\dsubst{F}{r'}{z} = \Hcom{\td{A}{\psi}}$,
\begin{enumerate}
\item $\coftype[\Psi']{\snd{M}}{\subst{\td{B}{\psi}}{\dsubst{F}{r}{z}}{a}}$
by $\ceqtm[\Psi']{\dsubst{F}{r}{z}}{\fst{M}}{\td{A}{\psi}}$ and
\cref{rul:pair-elim},
\item $\ceqtm[\Psi',y]<r_i=r_i',r_j=r_j'>{N_i}{N_j}%
{\subst{\td{B}{\psi}}{\dsubst{F}{y}{z}}{a}}$ by
$\ceqtm[\Psi',y]<r_i=r_i'>{\dsubst{F}{y}{z}}{\fst{N_i}}{\td{A}{\psi}}$ and
\cref{rul:pair-elim}, and
\item $\ceqtm[\Psi']<r_i=r_i'>{\snd{\dsubst{N_i}{r}{y}}}{\snd{M}}%
{\subst{\td{B}{\psi}}{\dsubst{F}{r}{z}}{a}}$
by $\ceqtm[\Psi']{\dsubst{F}{r}{z}}{\fst{M}}{\td{A}{\psi}}$ and
\cref{rul:pair-elim}.
\end{enumerate}

Next, we must show that if $r=r'$ then
$\ceqtm[\Psi']{\Hcom{\sigmacl{a}{\td{A}{\psi}}{\td{B}{\psi}}}}{M}%
{\sigmacl{a}{\td{A}{\psi}}{\td{B}{\psi}}}$. By \cref{lem:expansion},
$\ceqtm[\Psi']{\Hcom{\sigmacl{a}{\td{A}{\psi}}{\td{B}{\psi}}}}%
{\pair{\Hcom{\td{A}{\psi}}}{\Com{z.\subst{\td{B}{\psi}}{F}{a}}}}%
{\sigmacl{a}{\td{A}{\psi}}{\td{B}{\psi}}}$. By \cref{def:kan,thm:com},
$\ceqtm[\Psi']{\Hcom{\td{A}{\psi}}}{\fst{M}}{\td{A}{\psi}}$,
$\ceqtm[\Psi']{\Com{z.\subst{\td{B}{\psi}}{F}{a}}}{\snd{M}}%
{\subst{\td{B}{\psi}}{\dsubst{F}{r}{z}}{a}}$, and
$\ceqtypek[\Psi']{\subst{\td{B}{\psi}}{\dsubst{F}{r}{z}}{a}}
{\subst{\td{B}{\psi}}{\fst{M}}{a}}$. The result follows by \cref{rul:pair-eta}.

For the final $\Hcom$ property, show that if $r_i=r_i'$ then
$\ceqtm[\Psi']{\Hcom{\sigmacl{a}{\td{A}{\psi}}{\td{B}{\psi}}}}{\dsubst{N_i}{r'}{y}}%
{\sigmacl{a}{\td{A}{\psi}}{\td{B}{\psi}}}$. The result follows by
$\ceqtm[\Psi']{\Hcom{\td{A}{\psi}}}{\fst{\dsubst{N_i}{r'}{y}}}{\td{A}{\psi}}$,
$\ceqtm[\Psi']{\Com{z.\subst{\td{B}{\psi}}{F}{a}}}{\snd{\dsubst{N_i}{r'}{y}}}%
{\subst{\td{B}{\psi}}{\dsubst{F}{r'}{z}}{a}}$, and
$\ceqtypek[\Psi']{\subst{\td{B}{\psi}}{\dsubst{F}{r'}{z}}{a}}
{\subst{\td{B}{\psi}}{\fst{\dsubst{N_i}{r'}{y}}}{a}}$.

($\Coe$) Now, suppose that $\tds{(\Psi',x)}{\psi}{\Psi}$ and
$\ceqtm[\Psi']{M}{M'}{\dsubst{(\sigmacl{a}{\td{A}{\psi}}{\td{B}{\psi}})}{r}{x}}$,
and show $\ceqtm[\Psi']{\Coe*{x.\sigmacl{a}{\td{A}{\psi}}{\td{B}{\psi}}}}%
{\Coe{x.\sigmacl{a}{\td{A'}{\psi}}{\td{B'}{\psi}}}{r}{r'}{M'}}%
{\dsubst{(\sigmacl{a}{\td{A}{\psi}}{\td{B}{\psi}})}{r'}{x}}$. By
\cref{lem:expansion} and \cref{rul:pair-intro}, it suffices to show (the binary
version of)
\begin{gather*}
\coftype[\Psi']{\Coe{x.\td{A}{\psi}}{r}{r'}{\fst{M}}}{\dsubst{\td{A}{\psi}}{r'}{x}} \\
\coftype[\Psi']{\Coe{x.\subst{\td{B}{\psi}}{\Coe{x.\td{A}{\psi}}{r}{x}{\fst{M}}}{a}}{r}{r'}{\snd{M}}}%
{\subst{\dsubst{\td{B}{\psi}}{r'}{x}}{\Coe{x.\td{A}{\psi}}{r}{r'}{\fst{M}}}{a}}
\end{gather*}
We know that $\coftype[\Psi']{\Coe{x.\td{A}{\psi}}{r}{r'}{\fst{M}}}%
{\dsubst{\td{A}{\psi}}{r'}{x}}$ and
$\cwftypek[\Psi',x]{\subst{\td{B}{\psi}}{\Coe{x.\td{A}{\psi}}{r}{x}{\fst{M}}}{a}}$
by $\cwftypek[\Psi',x]{\td{A}{\psi}}$,
$\wftypek[\Psi',x]{\oft{a}{\td{A}{\psi}}}{\td{B}{\psi}}$, and
\cref{rul:pair-elim}. We also know that
$\coftype[\Psi']{\snd{M}}{\subst{\dsubst{\td{B}{\psi}}{r}{x}}{\fst{M}}{a}}$ and
$\ceqtm[\Psi']{\dsubst{(\Coe{x.\td{A}{\psi}}{r}{x}{\fst{M}})}{r}{x}}%
{\fst{M}}{\dsubst{A}{r}{x}}$, so
$\coftype[\Psi']{\Coe{x.\subst{\td{B}{\psi}}{\dots}{a}}}%
{\subst{\dsubst{\td{B}{\psi}}{r'}{x}}{\Coe{x.\td{A}{\psi}}{r}{r'}{\fst{M}}}{a}}$
and the result follows.

Finally, show that if $r=r'$ then
$\ceqtm[\Psi']{\Coe{x.\sigmacl{a}{\td{A}{\psi}}{\td{B}{\psi}}}{r}{r}{M}}{M}%
{\dsubst{(\sigmacl{a}{\td{A}{\psi}}{\td{B}{\psi}})}{r}{x}}$. By
\cref{lem:expansion,rul:pair-intro,rul:pair-eta}, this follows from
$\ceqtm[\Psi']{\Coe{x.\td{A}{\psi}}}{\fst{M}}{\dsubst{\td{A}{\psi}}{r}{x}}$
and $\ceqtm[\Psi']{\Coe{x.\subst{\td{B}{\psi}}{\dots}{a}}}{\snd{M}}%
{\subst{\dsubst{\td{B}{\psi}}{r}{x}}{\fst{M}}{a}}$.
\end{proof}

\subsection{Path types}

Let $\tau=\Kan\mu(\nu)$ or $\pre\mu(\nu,\sigma)$ for any cubical type systems
$\nu,\sigma$; in $\tau$,
whenever $\ceqtypep[\Psi,x]{A}{A'}$,
$\ceqtm{P_\e}{P_\e'}{\dsubst{A}{\e}{x}}$ for $\e\in\{0,1\}$, and
$\phi = \{(\dlam{x}{M},\dlam{x}{M'}) \mid
\ceqtm[\Psi,x]{M}{M'}{A} \land
\forall\e. (\ceqtm{\dsubst{M}{\e}{x}}{P_\e}{\dsubst{A}{\e}{x}})\}$, we have
$\tau(\Psi,\Path{x.A}{P_0}{P_1},\Path{x.A'}{P_0'}{P_1'},\phi)$.

\begin{rul}[Pretype formation]\label{rul:path-form-pre}
If $\ceqtypep[\Psi,x]{A}{A'}$ and
$\ceqtm{P_\e}{P_\e'}{\dsubst{A}{\e}{x}}$ for $\e\in\{0,1\}$, then
$\ceqtypep{\Path{x.A}{P_0}{P_1}}{\Path{x.A'}{P_0'}{P_1'}}$.
\end{rul}
\begin{proof}
We have $\PTy(\tau)(\Psi,\Path{x.A}{P_0}{P_1},\Path{x.A'}{P_0'}{P_1'},\_)$
because $\sisval{\Path{x.A}{P_0}{P_1}}$ and judgments are preserved by dimension
substitution. To show $\Coh(\vper{\Path{x.A}{P_0}{P_1}})$, suppose that
$\vper{\Path{x.A}{P_0}{P_1}}(\dlam{x}{M},\dlam{x}{M'})$. Then
$\ceqtm[\Psi,x]{M}{M'}{A}$ and
$\ceqtm{\dsubst{M}{\e}{x}}{P_\e}{\dsubst{A}{\e}{x}}$, so
$\ceqtm[\Psi',x]{\td{M}{\psi}}{\td{M'}{\psi}}{\td{A}{\psi}}$ and
$\ceqtm[\Psi']{\dsubst{\td{M}{\psi}}{\e}{x}}{\td{P_\e}{\psi}}{\dsubst{\td{A}{\psi}}{\e}{x}}$
for any $\psitd$, so by $\sisval{\dlam{x}{M}}$,
$\Tm(\td{\vper{\Path{x.A}{P_0}{P_1}}}{\psi})(\dlam{x}{M},\dlam{x}{M'})$.
\end{proof}

\begin{rul}[Introduction]\label{rul:path-intro}
If $\ceqtm[\Psi,x]{M}{M'}{A}$ and
$\ceqtm{\dsubst{M}{\e}{x}}{P_\e}{\dsubst{A}{\e}{x}}$ for $\e\in\{0,1\}$, then
$\ceqtm{\dlam{x}{M}}{\dlam{x}{M'}}{\Path{x.A}{P_0}{P_1}}$.
\end{rul}
\begin{proof}
Then $\vper{\Path{x.A}{P_0}{P_1}}(\dlam{x}{M},\dlam{x}{M'})$, so the result
follows by $\Coh(\vper{\Path{x.A}{P_0}{P_1}})$.
\end{proof}

\begin{rul}[Elimination]\label{rul:path-elim}
~\begin{enumerate}
\item If $\ceqtm{M}{M'}{\Path{x.A}{P_0}{P_1}}$ then
$\ceqtm{\dapp{M}{r}}{\dapp{M'}{r}}{\dsubst{A}{r}{x}}$.
\item If $\coftype{M}{\Path{x.A}{P_0}{P_1}}$ then
$\ceqtm{\dapp{M}{\e}}{P_\e}{\dsubst{A}{\e}{x}}$.
\end{enumerate}
\end{rul}
\begin{proof}
Apply coherent expansion to $\dapp{M}{r}$ with family
$\{ \dsubst{M_\psi}{\td{r}{\psi}}{x} \mid \td{M}{\psi}\evals \dlam{x}{M_\psi}
\}^{\Psi'}_\psi$. By $\coftype{M}{\Path{x.A}{P_0}{P_1}}$ at $\id,\psi$ we know
$\ceqtm[\Psi',x]{\td{(M_{\id})}{\psi}}{M_\psi}{\td{A}{\psi}}$, so
$\ceqtm[\Psi']{\dsubst{\td{(M_{\id})}{\psi}}{\td{r}{\psi}}{x}}%
{\dsubst{M_\psi}{\td{r}{\psi}}{x}}{\td{\dsubst{A}{r}{x}}{\psi}}$.
Thus by \cref{lem:cohexp-ceqtm},
$\ceqtm{\dapp{M}{r}}{\dsubst{M_{\id}}{r}{x}}{\dsubst{A}{r}{x}}$; part (1)
follows by the same argument on the right side and
$\ceqtm[\Psi,x]{M_{\id}}{M'_{\id}}{A}$.
Part (2) follows from
$\ceqtm{\dapp{M}{\e}}{\dsubst{M_{\id}}{\e}{x}}{\dsubst{A}{\e}{x}}$ and
$\ceqtm{\dsubst{M_{\id}}{\e}{x}}{P_\e}{\dsubst{A}{\e}{x}}$.
\end{proof}

\begin{rul}[Computation]\label{rul:path-comp}
If $\coftype[\Psi,x]{M}{A}$ then
$\ceqtm{\dapp{(\dlam{x}{M})}{r}}{\dsubst{M}{r}{x}}{\dsubst{A}{r}{x}}$.
\end{rul}
\begin{proof}
Immediate by $\dapp{(\dlam{x}{M})}{r}\ssteps\dsubst{M}{r}{x}$,
$\coftype{\dsubst{M}{r}{x}}{\dsubst{A}{r}{x}}$, and \cref{lem:expansion}.
\end{proof}

\begin{rul}[Eta]\label{rul:path-eta}
If $\coftype{M}{\Path{x.A}{P_0}{P_1}}$ then
$\ceqtm{M}{\dlam{x}{(\dapp{M}{x})}}{\Path{x.A}{P_0}{P_1}}$.
\end{rul}
\begin{proof}
By \cref{lem:coftype-evals-ceqtm}, $M\evals\dlam{x}{N}$ and
$\ceqtm{M}{\dlam{x}{N}}{\Path{x.A}{P_0}{P_1}}$. By \cref{rul:path-elim},
$\ceqtm[\Psi,x]{\dapp{M}{x}}{\dapp{(\dlam{x}{N})}{x}}{A}$, so by
\cref{lem:expansion} on the right, $\ceqtm[\Psi,x]{\dapp{M}{x}}{N}{A}$.
By \cref{rul:path-intro},
$\ceqtm{\dlam{x}{(\dapp{M}{x})}}{\dlam{x}{N}}{\Path{x.A}{P_0}{P_1}}$, and the
result follows by transitivity.
\end{proof}

\begin{rul}[Kan type formation]
If $\ceqtypek[\Psi,x]{A}{A'}$ and
$\ceqtm{P_\e}{P_\e'}{\dsubst{A}{\e}{x}}$ for $\e\in\{0,1\}$, then
$\ceqtypek{\Path{x.A}{P_0}{P_1}}{\Path{x.A'}{P_0'}{P_1'}}$.
\end{rul}
\begin{proof}
It suffices to check the five Kan conditions.

($\Hcom$) First, suppose that $\psitd$,
\begin{enumerate}
\item $\etc{\xi_i}=\etc{r_i=r_i'}$ is valid,
\item $\ceqtm[\Psi']{M}{M'}{\Path{x.\td{A}{\psi}}{\td{P_0}{\psi}}{\td{P_1}{\psi}}}$,
\item $\ceqtm[\Psi',y]<r_i=r_i',r_j=r_j'>{N_i}{N_j'}%
{\Path{x.\td{A}{\psi}}{\td{P_0}{\psi}}{\td{P_1}{\psi}}}$ for any $i,j$, and
\item $\ceqtm[\Psi']<r_i=r_i'>{\dsubst{N_i}{r}{y}}{M}%
{\Path{x.\td{A}{\psi}}{\td{P_0}{\psi}}{\td{P_1}{\psi}}}$ for any $i$,
\end{enumerate}
and show the equality
$\ceqtm[\Psi']%
{\Hcom*{\td{(\Path{x.A}{P_0}{P_1})}{\psi}}{\xi_i}}%
{\Hcom{\td{(\Path{x.A'}{P_0'}{P_1'})}{\psi}}{r}{r'}{M'}{\sys{\xi_i}{N_i'}}}%
{\td{(\Path{x.A}{P_0}{P_1})}{\psi}}$. By \cref{lem:expansion,rul:path-intro} on
both sides it suffices to show
\begin{align*}
&{\Hcom{\td{A}{\psi}}{r}{r'}{\dapp{M}{x}}{\sys{x=\e}{\_.\td{P_\e}{\psi}},\sys{\xi_i}{y.\dapp{N_i}{x}}}} \\
\ceqtmtab[\Psi',x]{}%
{\Hcom{\td{A'}{\psi}}{r}{r'}{\dapp{M'}{x}}{\sys{x=\e}{\_.\td{P_\e'}{\psi}},\sys{\xi_i}{y.\dapp{N_i'}{x}}}}%
{\td{A}{\psi}}
\end{align*}
and
$\ceqtm[\Psi']{\dsubst{(\Hcom{\td{A}{\psi}})}{\e}{x}}{\td{P_\e}{\psi}}{\dsubst{\td{A}{\psi}}{\e}{x}}$.
By our hypotheses and \cref{rul:path-elim},
\begin{enumerate}
\item $\ceqtm[\Psi',x]{\dapp{M}{x}}{\dapp{M'}{x}}{\td{A}{\psi}}$,

\item $\ceqtm[\Psi',x]<x=\e>{\td{P_\e}{\psi}}{\td{P_\e'}{\psi}}{\td{A}{\psi}}$ and
$\ceqtm[\Psi',x]<x=\e>{\td{P_\e}{\psi}}{\dapp{M}{x}}{\td{A}{\psi}}$,

\item
$\ceqtm[\Psi',x,y]<r_i=r_i',r_j=r_j'>{\dapp{N_i}{x}}{\dapp{N_j'}{x}}{\td{A}{\psi}}$,
$\ceqtm[\Psi',x,y]<r_i=r_i',x=\e>{\dapp{N_i}{x}}{\td{P_\e'}{\psi}}{\td{A}{\psi}}$, and
$\ceqtm[\Psi',x]<r_i=r_i'>{\dapp{\dsubst{N_i}{r}{y}}{x}}{\dapp{M}{x}}{\td{A}{\psi}}$,
\end{enumerate}
and so by \cref{def:kan},
$\ceqtm[\Psi',x]{\Hcom{\td{A}{\psi}}}{\Hcom{\td{A'}{\psi}}}{\td{A}{\psi}}$ and
$\ceqtm{\dsubst{(\Hcom{\td{A}{\psi}})}{\e}{x}}{\td{P_\e}{\psi}}{\td{A}{\psi}}$.

Next, show if $r=r'$ then
$\ceqtm[\Psi']{\Hcom*{\td{(\Path{x.A}{P_0}{P_1})}{\psi}}{\xi_i}}{M}%
{\td{(\Path{x.A}{P_0}{P_1})}{\psi}}$. By \cref{rul:path-intro,def:kan} the left
side equals $\dlam{x}{(\dapp{M}{x})}$, and \cref{rul:path-eta} completes this
part.

Finally, if $r_i=r_i'$ then
$\ceqtm[\Psi']{\Hcom*{\td{(\Path{x.A}{P_0}{P_1})}{\psi}}{\xi_i}}{\dsubst{N_i}{r'}{y}}%
{\td{(\Path{x.A}{P_0}{P_1})}{\psi}}$. By \cref{rul:path-intro,def:kan} the left
side equals $\dlam{x}{(\dapp{\dsubst{N_i}{r'}{y}}{x})}$, and \cref{rul:path-eta}
completes this part.

($\Coe$) Now, suppose that $\tds{(\Psi',y)}{\psi}{\Psi}$ and
$\ceqtm[\Psi']{M}{M'}%
{\dsubst{\td{({\Path{x.A}{P_0}{P_1}})}{\psi}}{r}{y}}$, and show
that
$\ceqtm[\Psi']{\Coe*{y.\td{({\Path{x.A}{P_0}{P_1}})}{\psi}}}%
{\Coe{y.\td{({\Path{x.A'}{P_0'}{P_1'}})}{\psi}}{r}{r'}{M'}}%
{\dsubst{\td{({\Path{x.A}{P_0}{P_1}})}{\psi}}{r'}{y}}$.
By \cref{lem:expansion} on both sides and \cref{rul:path-intro}, we show
\[
\ceqtm[\Psi',x]
{\Com{y.\td{A}{\psi}}{r}{r'}{\dapp{M}{x}}{\sys{x=\e}{y.\td{P_\e}{\psi}}}}
{\Com{y.\td{A'}{\psi}}{r}{r'}{\dapp{M'}{x}}{\sys{x=\e}{y.\td{P_\e'}{\psi}}}}
{\dsubst{\td{A}{\psi}}{r'}{y}}
\]
and
$\ceqtm[\Psi']
{\dsubst{(\Com{y.\td{A}{\psi}})}{\e}{x}}
{\dsubst{\td{P_\e}{\psi}}{r'}{y}}
{\dsubst{\dsubst{\td{A}{\psi}}{r'}{y}}{\e}{x}}$.
By our hypotheses and \cref{rul:path-elim},
$\ceqtm[\Psi',x]{\dapp{M}{x}}{\dapp{M'}{x}}{\dsubst{\td{A}{\psi}}{r}{y}}$,
$\ceqtm[\Psi',x,y]<x=\e>{\td{P_\e}{\psi}}{\td{P_\e'}{\psi}}{\td{A}{\psi}}$,
and
$\ceqtm[\Psi',x]<x=\e>{\dsubst{\td{P_\e}{\psi}}{r}{y}}{\dapp{M}{x}}{\dsubst{\td{A}{\psi}}{r}{y}}$,
so by \cref{thm:com},
$\ceqtm[\Psi',x]{\Com{y.\td{A}{\psi}}}{\Com{y.\td{A'}{\psi}}}{\dsubst{\td{A}{\psi}}{r'}{y}}$
and
$\ceqtm[\Psi']{\dsubst{(\Com{y.\td{A}{\psi}})}{\e}{x}}{\dsubst{\td{P_\e}{\psi}}{r'}{y}}%
{\dsubst{\dsubst{\td{A}{\psi}}{r'}{y}}{\e}{x}}$.

Finally, show that if $r=r'$ then
$\ceqtm[\Psi']{\Coe*{y.\td{({\Path{x.A}{P_0}{P_1}})}{\psi}}}{M}%
{\dsubst{\td{({\Path{x.A}{P_0}{P_1}})}{\psi}}{r'}{y}}$.
By \cref{rul:path-intro,thm:com} the left side equals $\dlam{x}{(\dapp{M}{x})}$,
and \cref{rul:path-eta} completes the proof.
\end{proof}

\subsection{Equality pretypes}

Let $\tau=\Kan\mu(\nu)$ or $\pre\mu(\nu,\sigma)$ for any cubical type systems
$\nu,\sigma$; in $\tau$, whenever
$\ceqtypep{A}{A'}$,
$\ceqtm{M}{M'}{A}$,
$\ceqtm{N}{N'}{A}$, and
$\phi = \{(\ax,\ax) \mid \ceqtm{M}{N}{A} \}$,
$\tau(\Psi,\Eq{A}{M}{N},\Eq{A'}{M'}{N'},\phi)$.

\begin{rul}[Pretype formation]
If $\ceqtypep{A}{A'}$,
$\ceqtm{M}{M'}{A}$, and
$\ceqtm{N}{N'}{A}$, then
$\ceqtypep{\Eq{A}{M}{N}}{\Eq{A'}{M'}{N'}}$.
\end{rul}
\begin{proof}
We have $\PTy(\tau)(\Psi,\Eq{A}{M}{N},\Eq{A'}{M'}{N'},\vper{\Eq{A}{M}{N}})$
because $\sisval{\Eq{A}{M}{N}}$ and judgments are preserved by dimension
substitution. To show $\Coh(\vper{\Eq{A}{M}{N}})$, suppose that
$\vper{\Eq{A}{M}{N}}_\psi(\ax,\ax)$. Then $\ceqtm{M}{N}{A}$, so
$\ceqtm[\Psi']{\td{M}{\psi}}{\td{N}{\psi}}{\td{A}{\psi}}$ for all $\psitd$, so
$\Tm(\td{\vper{\Eq{A}{M}{N}}}{\psi})(\ax,\ax)$ holds by this and $\sisval{\ax}$.
\end{proof}

\begin{rul}[Introduction]
If $\ceqtm{M}{N}{A}$ then $\coftype{\ax}{\Eq{A}{M}{N}}$.
\end{rul}
\begin{proof}
Then $\vper{\Eq{A}{M}{N}}(\ax,\ax)$, so the result follows by
$\Coh(\vper{\Eq{A}{M}{N}})$.
\end{proof}

\begin{rul}[Elimination]
If $\coftype{E}{\Eq{A}{M}{N}}$ then $\ceqtm{M}{N}{A}$.
\end{rul}
\begin{proof}
Then $\lift{\vper{\Eq{A}{M}{N}}}(E,E)$ so $E\evals\ax$ and $\ceqtm{M}{N}{A}$.
\end{proof}

\begin{rul}[Eta]
If $\coftype{E}{\Eq{A}{M}{N}}$ then $\ceqtm{E}{\ax}{\Eq{A}{M}{N}}$.
\end{rul}
\begin{proof}
Immediate by \cref{lem:coftype-evals-ceqtm}.
\end{proof}

\subsection{Void}

Let $\tau=\Kan\mu(\nu)$ or $\pre\mu(\nu,\sigma)$ for any cubical type systems
$\nu,\sigma$; we have
$\tau(\Psi,\void,\void,\phi)$ for $\phi$ the empty relation. By
$\sisval{\void}$, $\PTy(\tau)(\Psi,\void,\void,\alpha)$ where each
$\alpha_{\Psi'}$ is empty.

\begin{rul}[Pretype formation]
$\cwftypep{\void}$.
\end{rul}
\begin{proof}
We have already observed $\PTy(\tau)(\Psi,\void,\void,\vper{\void})$;
$\Coh(\vper{\void})$ trivially because each $\vper{\void}_{\Psi'}$ is empty.
\end{proof}

\begin{rul}[Elimination]\label{rul:void-elim}
It is never the case that $\coftype{M}{\void}$.
\end{rul}
\begin{proof}
If $\Tm(\vper{\void})(M,M)$ then $\lift{\vper{\void}}_\Psi(M,M)$, but
$\lift{\vper{\void}}_\Psi$ is empty.
\end{proof}

If $\oftype{\G}{M}{\void}$ then it must be impossible to produce elements of
each pretype in $\G$, in which case every (non-context-restricted) judgment
holds under $\G$. In \cref{sec:rules}, we say that if $\coftype{M}{\void}$ then
$\judg{\J}$.

\begin{rul}[Kan type formation]
$\cwftypek{\void}$.
\end{rul}
\begin{proof}
It suffices to check the five Kan conditions. In each condition, we suppose that
$\ceqtm[\Psi']{M}{M'}{\void}$, so by \cref{rul:void-elim} they vacuously hold.
\end{proof}

\subsection{Booleans}

Let $\tau=\Kan\mu(\nu)$ or $\pre\mu(\nu,\sigma)$ for any cubical type systems
$\nu,\sigma$; we have
$\tau(\Psi,\bool,\bool,\phi)$ for $\phi=\{(\true,\true),(\false,\false)\}$.
By $\sisval{\bool}$, $\PTy(\tau)(\Psi,\bool,\bool,\alpha)$ where each
$\alpha_{\Psi'} = \phi$.

\begin{rul}[Pretype formation]
$\cwftypep{\bool}$.
\end{rul}
\begin{proof}
We have already observed $\PTy(\tau)(\Psi,\bool,\bool,\vper{\bool})$; for
$\Coh(\vper{\bool})$ we must show that $\Tm(\vper{\bool})(\true,\true)$ and
$\Tm(\vper{\bool})(\false,\false)$. These hold by
$\sisval{\true}$, $\vper{\bool}_{\Psi'}(\true,\true)$,
$\sisval{\false}$, and $\vper{\bool}_{\Psi'}(\false,\false)$.
\end{proof}

\begin{rul}[Introduction]
$\coftype{\true}{\bool}$ and $\coftype{\false}{\bool}$.
\end{rul}
\begin{proof}
Immediate by $\Coh(\vper{\bool})$.
\end{proof}

\begin{rul}[Computation]\label{rul:bool-comp}
If $\coftype{T}{B}$ then $\ceqtm{\ifb{b.A}{\true}{T}{F}}{T}{B}$.
If $\coftype{F}{B}$ then $\ceqtm{\ifb{b.A}{\false}{T}{F}}{F}{B}$.
\end{rul}
\begin{proof}
Immediate by $\ifb{b.A}{\true}{T}{F}\ssteps T$, $\ifb{b.A}{\false}{T}{F}\ssteps
F$, and \cref{lem:expansion}.
\end{proof}

\begin{rul}[Elimination]\label{rul:bool-elim}
If $\ceqtm{M}{M'}{\bool}$,
$\wftypep{\oft{b}{\bool}}{C}$,
$\ceqtm{T}{T'}{\subst{C}{\true}{b}}$, and
$\ceqtm{F}{F'}{\subst{C}{\false}{b}}$, then
$\ceqtm{\ifb{b.A}{M}{T}{F}}{\ifb{b.A'}{M'}{T'}{F'}}{\subst{C}{M}{b}}$.
\end{rul}
\begin{proof}
Apply coherent expansion to the left side with
$\{\ifb{b.\td{A}{\psi}}{M_\psi}{\td{T}{\psi}}{\td{F}{\psi}} \mid
\td{M}{\psi} \evals M_\psi\}^{\Psi'}_\psi$. We must show
$\ceqtm[\Psi']%
{\ifb{b.\td{A}{\psi}}{M_\psi}{\td{T}{\psi}}{\td{F}{\psi}}}%
{\ifb{b.\td{A}{\psi}}{\td{(M_{\id})}{\psi}}{\td{T}{\psi}}{\td{F}{\psi}}}%
{\subst{\td{C}{\psi}}{\td{M}{\psi}}{b}}$.
Either $M_\psi=\true$ or $M_\psi=\false$. In either case $M_{\id}=M_\psi$
because $\lift{\vper{\bool}}_{\Psi'}(\td{(M_{\id})}{\psi},M_\psi)$ and
$M_{\id}=\true$ or $M_{\id}=\false$. Consider the case $M_\psi=\true$:
we must show
$\coftype[\Psi']{\ifb{b.\td{A}{\psi}}{\true}{\td{T}{\psi}}{\td{F}{\psi}}}%
{\subst{\td{C}{\psi}}{\td{M}{\psi}}{b}}$.
By \cref{lem:coftype-evals-ceqtm} we have
$\ceqtm[\Psi']{\td{M}{\psi}}{\true}{\bool}$ so
$\ceqtypep[\Psi']{\subst{\td{C}{\psi}}{\td{M}{\psi}}{b}}{\subst{\td{C}{\psi}}{\true}{b}}$.
The result follows by \cref{rul:bool-comp} (with
$B=\subst{\td{C}{\psi}}{\true}{b}$). The $M_\psi=\false$ case is symmetric.

We conclude by \cref{lem:cohexp-ceqtm} that
$\ceqtm{\ifb{b.A}{M}{T}{F}}{\ifb{b.A}{M_{\id}}{T}{F}}{\subst{C}{M}{b}}$. By
transitivity, \cref{lem:coftype-evals-ceqtm}, and the same argument on the
right, it suffices to show
$\ceqtm{\ifb{b.A}{M_{\id}}{T}{F}}{\ifb{b.A'}{M'_{\id}}{T'}{F'}}{\subst{C}{M_{\id}}{b}}$.
By $\ceqtm{M}{M'}{\bool}$, either $M_{\id}=M'_{\id}=\true$ or
$M_{\id}=M'_{\id}=\false$, and in either case the result follows by
\cref{rul:bool-comp} on both sides.
\end{proof}

Notice that \cref{rul:bool-elim} places no restrictions on the motives $b.A$ and
$b.A'$; these motives are only relevant in the elimination rule for $\wbool$.

\begin{lemma}\label{lem:bool-discrete}
If $\coftype[\Psi,y]{M}{\bool}$ then
$\ceqtm{\dsubst{M}{r}{y}}{\dsubst{M}{r'}{y}}{\bool}$.
\end{lemma}
\begin{proof}
By $\lift{\vper{\bool}}_{(\Psi,y)}(M,M)$ we know $M\evals\true$ or
$M\evals\false$, so by \cref{lem:coftype-evals-ceqtm} either
$\ceqtm[\Psi,y]{M}{\true}{\bool}$ or $\ceqtm[\Psi,y]{M}{\false}{\bool}$. In the
former case, both $\ceqtm{\dsubst{M}{r}{y}}{\true}{\bool}$ and
$\ceqtm{\dsubst{M}{r'}{y}}{\true}{\bool}$, and similarly in the latter case.
\end{proof}

\begin{rul}[Kan type formation]\label{rul:bool-form-kan}
$\cwftypek{\bool}$.
\end{rul}
\begin{proof}
It suffices to check the five Kan conditions.

($\Hcom$) Suppose that
\begin{enumerate}
\item $\etc{r_i=r_i'}$ is valid,
\item $\ceqtm[\Psi']{M}{M'}{\bool}$,
\item $\ceqtm[\Psi',y]<r_i=r_i',r_j=r_j'>{N_i}{N_j'}{\bool}$ for any $i,j$, and
\item $\ceqtm[\Psi']<r_i=r_i'>{\dsubst{N_i}{r}{y}}{M}{\bool}$ for any $i$,
\end{enumerate}
and show $\ceqtm[\Psi']{\Hcom*{\bool}{r_i=r_i'}}%
{\Hcom{\bool}{r}{r'}{M'}{\sys{r_i=r_i'}{y.N_i'}}}{\bool}$.
This is immediate by \cref{lem:expansion} on both sides, because
$\Hcom*{\bool}{r_i=r_i'}\ssteps M$ and $\ceqtm[\Psi']{M}{M'}{\bool}$.
Similarly, if $r=r'$ it is immediate that
$\ceqtm[\Psi']{\Hcom*{\bool}{r_i=r_i'}}{M}{\bool}$.
Now suppose that $r_i = r_i'$, and show
$\ceqtm[\Psi']{\Hcom*{\bool}{r_i=r_i'}}{\dsubst{N_i}{r'}{y}}{\bool}$.
By \cref{lem:expansion} it suffices to show
$\ceqtm[\Psi']{M}{\dsubst{N_i}{r'}{y}}{\bool}$, which holds by
$\ceqtm[\Psi']{M}{\dsubst{N_i}{r}{y}}{\bool}$ and \cref{lem:bool-discrete}.

($\Coe$) Suppose that $\ceqtm[\Psi']{M}{M'}{\bool}$, and show that
$\ceqtm[\Psi']{\Coe*{x.\bool}}{\Coe{x.\bool}{r}{r'}{M'}}{\bool}$. This is
immediate by \cref{lem:expansion} on both sides, because $\Coe*{x.\bool}\ssteps
M$ and $\ceqtm[\Psi']{M}{M'}{\bool}$. Similarly, if $r=r'$ it is immediate that
$\ceqtm[\Psi']{\Coe*{x.\bool}}{M}{\bool}$.
\end{proof}

\subsection{Natural numbers}

Let $\tau=\Kan\mu(\nu)$ or $\pre\mu(\nu,\sigma)$ for any cubical type systems
$\nu,\sigma$; we have
$\tau(\Psi,\nat,\nat,\mathbb{N}_\Psi)$, where $\mathbb{N}$ is the least
context-indexed relation such that $\mathbb{N}_\Psi(\z,\z)$ and
$\mathbb{N}_\Psi(\suc{M},\suc{M'})$ when $\Tm(\mathbb{N}(\Psi))(M,M')$. By
$\sisval{\nat}$, $\PTy(\tau)(\Psi,\nat,\nat,\mathbb{N}(\Psi))$.

\begin{rul}[Pretype formation]
$\cwftypep{\nat}$.
\end{rul}
\begin{proof}
It suffices to show $\Coh(\vper{\nat})$. We have $\Tm(\vper{\nat})(\z,\z)$ and
$\Tm(\vper{\nat})(\suc{M},\suc{M'})$ when $\Tm(\vper{\nat})(M,M')$ by
$\sisval{\z}$, $\sisval{\suc{M}}$, and
$\Tm(\td{\vper{\nat}}{\psi})(\td{M}{\psi},\td{M'}{\psi})$ for all $\psitd$.
\end{proof}

\begin{rul}[Introduction]
$\coftype{\z}{\nat}$ and if $\ceqtm{M}{M'}{\nat}$ then
$\ceqtm{\suc{M}}{\suc{M'}}{\nat}$.
\end{rul}
\begin{proof}
Immediate by $\Coh(\vper{\nat})$.
\end{proof}

\begin{rul}[Elimination]\label{rul:nat-elim}
If $\wftypep{\oft{n}{\nat}}{A}$,
$\ceqtm{M}{M'}{\nat}$,
$\ceqtm{Z}{Z'}{\subst{A}{\z}{n}}$, and
$\eqtm{\oft{n}{\nat},\oft{a}{A}}{S}{S'}{\subst{A}{\suc{n}}{n}}$, then
$\ceqtm{\natrec{M}{Z}{n.a.S}}{\natrec{M'}{Z'}{n.a.S'}}{\subst{A}{M}{n}}$.
\end{rul}
\begin{proof}
We induct over the definition of $\vper{\nat}$. The equality relation of $\nat$,
$\Tm(\vper{\nat})$, is the lifting of the least pre-fixed point of an
order-preserving operator $N$ on context-indexed relations over values.
Therefore, we prove (1) the elimination rule lifts from values to elements; (2)
the elimination rule holds for values; and thus (3) the elimination rule holds
for elements.

Define $\Phi_\Psi(M_0,M_0')$ to hold
when $\vper{\nat}_\Psi(M_0,M_0')$ and for all
$\wftypep{\oft{n}{\nat}}{A}$,
$\ceqtm{Z}{Z'}{\subst{A}{\z}{n}}$, and
$\eqtm{\oft{n}{\nat},\oft{a}{A}}{S}{S'}{\subst{A}{\suc{n}}{n}}$, we have
$\ceqtm{\natrec{M_0}{Z}{n.a.S}}{\natrec{M_0'}{Z'}{n.a.S'}}{\subst{A}{M_0}{n}}$.
\begin{enumerate}
\item If $\Tm(\Phi(\Psi))(M,M')$ then the elimination rule holds for $M,M'$.

By definition, $\Phi\subseteq\vper{\nat}$, so because $\Tm$ is order-preserving,
$\Tm(\vper{\nat}(\Psi))(M,M')$. Apply coherent expansion to
$\natrec{M}{Z}{n.a.S}$ at $\cwftypep{\subst{A}{M}{n}}$ with
$\{\natrec{M_\psi}{\td{Z}{\psi}}{n.a.\td{S}{\psi}}\mid
\td{M}{\psi}\evals M_\psi\}^{\Psi'}_\psi$.
Then $\coftype[\Psi']{\natrec{M_\psi}{\td{Z}{\psi}}{n.a.\td{S}{\psi}}}%
{\subst{\td{A}{\psi}}{M_\psi}{n}}$ for all $\psitd$ because
$\lift{\Phi}_\Psi(M,M')$ by $\Tm(\Phi(\Psi))(M,M')$. We must show
\[
\ceqtm[\Psi']{\natrec{M_\psi}{\td{Z}{\psi}}{n.a.\td{S}{\psi}}}%
{\natrec{\td{(M_{\id})}{\psi}}{\td{Z}{\psi}}{n.a.\td{S}{\psi}}}%
{\subst{\td{A}{\psi}}{M_\psi}{n}}
\]
but by \cref{lem:coftype-ceqtm} and
$\ceqtm[\Psi']{\td{(M_{\id})}{\psi}}{M_\psi}{\nat}$ it suffices to show these
$\natrec$ are related by $\lift{\vper{\subst{\td{A}{\psi}}{M_\psi}{n}}}$, which
follows from $\lift{\Phi}_{\Psi'}(\td{(M_{\id})}{\psi},M_\psi)$.

\item If $\vper{\nat}_\Psi(M_0,M_0')$ then $\Phi_\Psi(M_0,M_0')$.

We prove that $N(\Phi)\subseteq\Phi$; then $\Phi$ is a pre-fixed point of $N$,
and $\vper{\nat}\subseteq\Phi$ because $\vper{\nat}$ is the least pre-fixed
point of $N$. Suppose $N(\Phi)_\Psi(M_0,M_0')$. There are two cases:
\begin{enumerate}
\item $M_0=M_0'=\z$.

Show $\ceqtm{\natrec{\z}{Z}{n.a.S}}{\natrec{\z}{Z'}{n.a.S'}}{\subst{A}{\z}{n}}$,
which is immediate by $\ceqtm{Z}{Z'}{\subst{A}{\z}{n}}$ and \cref{lem:expansion}
on both sides.

\item $M_0=\suc{M}$, $M_0'=\suc{M'}$, and $\Tm(\Phi(\Psi))(M,M')$.

Show $\ceqtm{\natrec{\suc{M}}{Z}{n.a.S}}{\natrec{\suc{M'}}{Z'}{n.a.S'}}%
{\subst{A}{\suc{M}}{n}}$. By \cref{lem:expansion} on both sides, it suffices to
show
\[
\ceqtm{\subst{\subst{S}{M}{n}}{\natrec{M}{Z}{n.a.S}}{a}}%
{\subst{\subst{S'}{M'}{n}}{\natrec{M'}{Z'}{n.a.S'}}{a}}%
{\subst{A}{\suc{M}}{n}}.
\]
We have $\ceqtm{M}{M'}{\nat}$ and
$\ceqtm{\natrec{M}{Z}{n.a.S}}{\natrec{M'}{Z'}{n.a.S'}}{\subst{A}{M}{n}}$ by
$\Tm(\Phi(\Psi))(M,M')$, so the result follows by
$\eqtm{\oft{n}{\nat},\oft{a}{A}}{S}{S'}{\subst{A}{\suc{n}}{n}}$.
\end{enumerate}

\item Assume $\Tm(\vper{\nat}(\Psi))(M,M')$; $\Tm$ is order-preserving and
$\vper{\nat}\subseteq\Phi$, so $\Tm(\Phi(\Psi))(M,M')$. Thus the elimination
rule holds for $M,M'$, completing the proof.
\qedhere
\end{enumerate}
\end{proof}

\begin{rul}[Computation]
~\begin{enumerate}
\item If $\coftype{Z}{A}$ then $\ceqtm{\natrec{\z}{Z}{n.a.S}}{Z}{A}$.
\item If $\wftypep{\oft{n}{\nat}}{A}$,
$\coftype{M}{\nat}$,
$\coftype{Z}{\subst{A}{\z}{n}}$, and
$\oftype{\oft{n}{\nat},\oft{a}{A}}{S}{\subst{A}{\suc{n}}{n}}$, then
$\ceqtm{\natrec{\suc{M}}{Z}{n.a.S}}%
{\subst{\subst{S}{M}{n}}{\natrec{M}{Z}{n.a.S}}{a}}%
{\subst{A}{\suc{M}}{n}}$.
\end{enumerate}
\end{rul}
\begin{proof}
Part (1) is immediate by \cref{lem:expansion}. For part (2), we have
$\coftype{\natrec{M}{Z}{n.a.S}}{\subst{A}{M}{n}}$ and thus
$\coftype{\subst{\subst{S}{M}{n}}{\natrec{M}{Z}{n.a.S}}{a}}{\subst{A}{\suc{M}}{n}}$
by \cref{rul:nat-elim}, so the result again follows by \cref{lem:expansion}.
\end{proof}

\begin{rul}[Kan type formation]
$\cwftypek{\nat}$.
\end{rul}
\begin{proof}
Identical to \cref{rul:bool-form-kan}.
\end{proof}

\subsection{Circle}
\label{ssec:C}

Let $\tau=\Kan\mu(\nu)$ or $\pre\mu(\nu,\sigma)$ for any cubical type systems
$\nu,\sigma$; we have
$\tau(\Psi,\C,\C,\mathbb{C}_\Psi)$, where $\mathbb{C}$ is the least
context-indexed relation such that:
\begin{enumerate}
\item $\mathbb{C}_\Psi(\base,\base)$,
\item $\mathbb{C}_{(\Psi,x)}(\lp{x},\lp{x})$, and
\item $\mathbb{C}_\Psi(\Fcom*{r_i=r_i'},\Fcom{r}{r'}{M'}{\sys{r_i=r_i'}{y.N_i'}})$
whenever
\begin{enumerate}
\item $r\neq r'$;
$r_i \neq r_i'$ for all $i$;
$r_i = r_j$, $r_i' = 0$, and $r_j' = 1$ for some $i,j$;
\item $\Tm(\mathbb{C}(\Psi))(M,M')$;
\item $\Tm(\mathbb{C}(\Psi'))(\td{N_i}{\psi},\td{N_j'}{\psi})$ for all $i,j$ and
$\tds{\Psi'}{\psi}{(\Psi,y)}$ satisfying $r_i=r_i',r_j=r_j'$; and
\item $\Tm(\mathbb{C}(\Psi'))(\td{\dsubst{N_i}{r}{y}}{\psi},\td{M}{\psi})$ for
all $i,j$ and $\psitd$ satisfying $r_i=r_i'$.
\end{enumerate}
\end{enumerate}
By $\sisval{\C}$ it is immediate that $\PTy(\tau)(\Psi,\C,\C,\mathbb{C}(\Psi))$.

\begin{lemma}\label{lem:C-prekan}
If
\begin{enumerate}
\item $\etc{r_i=r_i'}$ is valid,
\item $\Tm(\vper{\C}(\Psi))(M,M')$,
\item $\Tm(\vper{\C}(\Psi'))(\td{N_i}{\psi},\td{N_j'}{\psi})$ for all $i,j$ and
$\tds{\Psi'}{\psi}{(\Psi,y)}$ satisfying $r_i=r_i',r_j=r_j'$, and
\item $\Tm(\vper{\C}(\Psi'))(\td{\dsubst{N_i}{r}{y}}{\psi},\td{M}{\psi})$ for
all $i,j$ and $\psitd$ satisfying $r_i=r_i'$,
\end{enumerate}
then $\Tm(\vper{\C}(\Psi))(\Fcom*{r_i=r_i'},\Fcom{r}{r'}{M'}{\sys{r_i=r_i'}{y.N_i'}})$.
\end{lemma}
\begin{proof}
Let us abbreviate the above $\Fcom$ terms $L$ and $R$ respectively. Expanding
the definition of $\Tm$, for any $\tds{\Psi_1}{\psi_1}{\Psi}$ and
$\tds{\Psi_2}{\psi_2}{\Psi_1}$ we must show $\td{L}{\psi_1}\evals L_1$,
$\td{R}{\psi_1}\evals R_1$, and
$\lift{\vper{\C}}_{\Psi_2}$ relates
$\td{L_1}{\psi_2}$, $\td{L}{\psi_1\psi_2}$,
$\td{R_1}{\psi_2}$, and $\td{R}{\psi_1\psi_2}$.
We proceed by cases on the first step taken by $\td{L}{\psi_1}$ and
$\td{L}{\psi_1\psi_2}$.

\begin{enumerate}
\item $\td{r}{\psi_1}=\td{r'}{\psi_1}$.

Then $\td{L}{\psi_1}\ssteps \td{M}{\psi_1}$, $\td{R}{\psi_1}\ssteps
\td{M'}{\psi_1}$, and the result follows by $\Tm(\vper{\C}(\Psi))(M,M')$.

\item $\td{r}{\psi_1}\neq\td{r'}{\psi_1}$,
$\td{r_j}{\psi_1}=\td{r_j'}{\psi_1}$
  (where $\td{r_i}{\psi_1}\neq\td{r_i'}{\psi_1}$ for all $i<j$), and
$\td{r}{\psi_1\psi_2}=\td{r'}{\psi_1\psi_2}$.

Then $\td{L}{\psi_1}\steps \td{\dsubst{N_j}{r'}{y}}{\psi_1}$,
$\td{L}{\psi_1\psi_2}\steps \td{M}{\psi_1\psi_2}$,
$\td{R}{\psi_1}\steps \td{\dsubst{N_j'}{r'}{y}}{\psi_1}$, and
$\td{R}{\psi_1\psi_2}\steps \td{M'}{\psi_1\psi_2}$.
Because $\psi_1$ satisfies $r_j=r_j'$, by (3) and (4)
$\Tm(\vper{\C}(\Psi_1,y))(\td{N_j}{\psi_1},\td{N_j'}{\psi_1})$ and
$\Tm(\vper{\C}(\Psi_1))(\td{\dsubst{N_j}{r}{y}}{\psi_1},\td{M}{\psi_1})$.
By the former at $\dsubst{}{\td{r'}{\psi_1}}{y},\psi_2$,
$\lift{\vper{\C}}_{\Psi_2}(\td{\dsubst{N_j}{r'}{y}}{\psi_1\psi_2},\td{L_1}{\psi_2})$
and $\lift{\vper{\C}}_{\Psi_2}(\td{L_1}{\psi_2},\td{R_1}{\psi_2})$.
The latter at $\psi_2,\id[\Psi_2]$ yields
$\lift{\vper{\C}}_{\Psi_2}(\td{\dsubst{N_j}{r}{y}}{\psi_1\psi_2},\td{M}{\psi_1\psi_2})$;
by transitivity and $\td{r}{\psi_1\psi_2}=\td{r'}{\psi_1\psi_2}$ we have
$\lift{\vper{\C}}_{\Psi_2}(\td{L_1}{\psi_2},\td{L}{\psi_1\psi_2})$.
Finally, by $\Tm(\vper{\C}(\Psi))(M,M')$ we have
$\lift{\vper{\C}}_{\Psi_2}(\td{L}{\psi_1\psi_2},\td{R}{\psi_1\psi_2})$.

\item $\td{r}{\psi_1}\neq\td{r'}{\psi_1}$,
$\td{r_i}{\psi_1}=\td{r_i'}{\psi_1}$ (and this is the least such $i$),
$\td{r}{\psi_1\psi_2}\neq\td{r'}{\psi_1\psi_2}$, and
$\td{r_j}{\psi_1\psi_2}=\td{r_j'}{\psi_1\psi_2}$ (and this is the least such $j\leq i$).

Then $\td{L}{\psi_1}\steps \td{\dsubst{N_i}{r'}{y}}{\psi_1}$,
$\td{L}{\psi_1\psi_2}\steps \td{\dsubst{N_j}{r'}{y}}{\psi_1\psi_2}$,
$\td{R}{\psi_1}\steps \td{\dsubst{N_i'}{r'}{y}}{\psi_1}$, and
$\td{R}{\psi_1\psi_2}\steps \td{\dsubst{N_j'}{r'}{y}}{\psi_1\psi_2}$.
In this case, $\td{\dsubst{}{r'}{y}}{\psi_1\psi_2}$ satisfies
$r_i=r_i',r_j=r_j'$, and the result follows because $\Tm(\vper{\C}(\Psi_2))$
relates
$\td{\dsubst{N_i}{r'}{y}}{\psi_1\psi_2}$,
$\td{\dsubst{N_j}{r'}{y}}{\psi_1\psi_2}$,
$\td{\dsubst{N_i'}{r'}{y}}{\psi_1\psi_2}$, and
$\td{\dsubst{N_j'}{r'}{y}}{\psi_1\psi_2}$.

\item $\td{r}{\psi_1}\neq\td{r'}{\psi_1}$,
$\td{r_i}{\psi_1}\neq\td{r_i'}{\psi_1}$ for all $i$, and
$\td{r}{\psi_1\psi_2} = \td{r'}{\psi_1\psi_2}$.

Then $\isval{\td{L}{\psi_1}}$,
$\td{L}{\psi_1\psi_2}\steps \td{M}{\psi_1\psi_2}$,
$\isval{\td{R}{\psi_1}}$, and
$\td{R}{\psi_1\psi_2}\steps \td{M'}{\psi_1\psi_2}$. In this case,
$\td{L_1}{\psi_2} = \td{L}{\psi_1\psi_2}$ and
$\td{R_1}{\psi_2} = \td{R}{\psi_1\psi_2}$, so the result follows by
$\Tm(\vper{\C}(\Psi))(M,M')$.

\item $\td{r}{\psi_1}\neq\td{r'}{\psi_1}$,
$\td{r_i}{\psi_1}\neq\td{r_i'}{\psi_1}$ for all $i$,
$\td{r}{\psi_1\psi_2}\neq\td{r'}{\psi_1\psi_2}$, and
$\td{r_j}{\psi_1\psi_2}=\td{r_j'}{\psi_1\psi_2}$ (the least such $j$).

Then $\isval{\td{L}{\psi_1}}$,
$\td{L}{\psi_1\psi_2}\steps \td{\dsubst{N_j}{r'}{y}}{\psi_1\psi_2}$,
$\isval{\td{R}{\psi_1}}$, and
$\td{R}{\psi_1\psi_2}\steps \td{\dsubst{N_j'}{r'}{y}}{\psi_1\psi_2}$. The result
follows because
$\td{L_1}{\psi_2} = \td{L}{\psi_1\psi_2}$,
$\td{R_1}{\psi_2} = \td{R}{\psi_1\psi_2}$, and
because $\td{\dsubst{}{r'}{y}}{\psi_1\psi_2}$ satisfies $r_j=r_j'$,
$\Tm(\vper{\C}(\Psi_2))%
(\td{\dsubst{N_j}{r'}{y}}{\psi_1\psi_2},\td{\dsubst{N_j'}{r'}{y}}{\psi_1\psi_2})$.

\item $\td{r}{\psi_1}\neq\td{r'}{\psi_1}$,
$\td{r_i}{\psi_1}\neq\td{r_i'}{\psi_1}$ for all $i$, and
$\td{r}{\psi_1\psi_2}\neq\td{r'}{\psi_1\psi_2}$, and
$\td{r_j}{\psi_1\psi_2}\neq\td{r_j'}{\psi_1\psi_2}$ for all $j$.

Then $\isval{\td{L}{\psi_1}}$,
$\isval{\td{L}{\psi_1\psi_2}}$,
$\isval{\td{R}{\psi_1}}$, and
$\isval{\td{R}{\psi_1\psi_2}}$, so it suffices to show
$\vper{\C}_{\Psi_2}(\td{L}{\psi_1\psi_2},\td{R}{\psi_1\psi_2})$.
We know $\etc{\td{r_i}{\psi_1\psi_2}=\td{r_i'}{\psi_1\psi_2}}$ is valid and
$\td{r_i}{\psi_1\psi_2}\neq\td{r_i'}{\psi_1\psi_2}$ for all $i$, so there must
be some $i,j$ for which $\td{r_i}{\psi_1\psi_2} = \td{r_j}{\psi_1\psi_2}$,
$\td{r_i'}{\psi_1\psi_2} = 0$, and $\td{r_j'}{\psi_1\psi_2} = 1$. The result
follows immediately by the third clause of the definition of $\vper{\C}$.
\qedhere
\end{enumerate}
\end{proof}

\begin{rul}[Pretype formation]
$\cwftypep{\C}$.
\end{rul}
\begin{proof}
It remains to show $\Coh(\vper{\C})$. There are three cases:
\begin{enumerate}
\item $\Tm(\vper{\C}(\Psi))(\base,\base)$.

Immediate because $\sisval{\base}$.

\item $\Tm(\vper{\C}(\Psi,x))(\lp{x},\lp{x})$.

Show that if $\tds{\Psi_1}{\psi_1}{(\Psi,x)}$ and
$\tds{\Psi_2}{\psi_2}{\Psi_1}$, $\lp{\td{x}{\psi_1}}\evals M_1$ and
$\lift{\vper{\C}}_{\Psi_2}(\td{M_1}{\psi_2},\lp{\td{x}{\psi_1\psi_2}})$.
If $\td{x}{\psi_1} = \e$ then $M_1=\base$, $\lp{\td{x}{\psi_1\psi_2}}\steps
\base$, and $\vper{\C}_{\Psi_2}(\base,\base)$.
If $\td{x}{\psi_1} = x'$ and $\td{x'}{\psi_2} = \e$, then $M_1=\lp{x'}$,
$\lp{\td{x'}{\psi_2}}\steps \base$, $\lp{\td{x}{\psi_1\psi_2}}\steps \base$, and
$\vper{\C}_{\Psi_2}(\base,\base)$.
Otherwise, $\td{x}{\psi_1} = x'$ and $\td{x'}{\psi_2} = x''$, so $M_1=\lp{x'}$
and $\vper{\C}_{\Psi_2}(\lp{x''},\lp{x''})$.

\item
$\Tm(\vper{\C}(\Psi))(\Fcom*{r_i=r_i'},\Fcom{r}{r'}{M'}{\sys{r_i=r_i'}{y.N_i'}})$
where\dots

This is a special case of \cref{lem:C-prekan}. (Note that $\etc{r_i=r_i'}$ is
valid.)
\qedhere
\end{enumerate}
\end{proof}

\begin{rul}[Introduction]
$\coftype{\base}{\C}$,
$\ceqtm{\lp{\e}}{\base}{\C}$, and
$\coftype{\lp{r}}{\C}$.
\end{rul}
\begin{proof}
The first is a consequence of $\Coh(\vper{\C})$; the second follows by
$\lp{\e}\ssteps \base$ and \cref{lem:expansion}; the third is a consequence of
$\Coh(\vper{\C})$ when $r=x$, and of \cref{lem:expansion} when $r=\e$.
\end{proof}

\begin{rul}[Kan type formation]\label{rul:C-form-kan}
$\cwftypek{\C}$.
\end{rul}
\begin{proof}
It suffices to check the five Kan conditions.

($\Hcom$) First, suppose that
\begin{enumerate}
\item $\etc{r_i=r_i'}$ is valid,
\item $\ceqtm[\Psi']{M}{M'}{\C}$,
\item $\ceqtm[\Psi',y]<r_i=r_i',r_j=r_j'>{N_i}{N_j'}{\C}$ for any $i,j$, and
\item $\ceqtm[\Psi']<r_i=r_i'>{\dsubst{N_i}{r}{y}}{M}{\C}$ for any $i$,
\end{enumerate}
and show $\ceqtm[\Psi']{\Hcom*{\C}{r_i=r_i'}}%
{\Hcom{\C}{r}{r'}{M'}{\sys{r_i=r_i'}{y.N_i'}}}{\C}$.
This is immediate by \cref{lem:expansion} on both sides (because $\Hcom{\C}
\ssteps \Fcom$) and \cref{lem:C-prekan}.

Next, show that if $r=r'$ then $\ceqtm[\Psi']{\Hcom*{\C}{r_i=r_i'}}{M}{\C}$.
This is immediate by $\Hcom*{\C}{r_i=r_i'}\ssteps \Fcom*{r_i=r_i'}\ssteps M$ and
\cref{lem:expansion}.

For the final $\Hcom$ property, show that if $r_i = r_i'$ then
$\ceqtm[\Psi']{\Hcom*{\C}{r_i=r_i'}}{\dsubst{N_i}{r'}{y}}{\C}$. We already know
each side is an element of $\C$, so by \cref{lem:coftype-evals-ceqtm} it
suffices to show
$\lift{\vper{\C}}_{\Psi'}(\Hcom*{\C}{r_i=r_i'},\dsubst{N_i}{r'}{y})$.
If $r=r'$ then $\Hcom \steps^2 M$ and the result follows by
$\ceqtm[\Psi']<r_i=r_i'>{\dsubst{N_i}{r}{y}}{M}{\C}$, because
$\id[\Psi']$ satisfies $r_i=r_i'$. Otherwise, let $r_j=r_j'$ be the first true
equation. Then $\Hcom \steps^2 \dsubst{N_j}{r'}{y}$ and this follows by
$\ceqtm[\Psi',y]<r_i=r_i',r_j=r_j'>{N_i}{N_j}{\C}$.

($\Coe$) Now, suppose that $\ceqtm[\Psi']{M}{M'}{\C}$ and show
$\ceqtm[\Psi']{\Coe*{x.\C}}{\Coe{x.\C}{r}{r'}{M'}}{\C}$. This is immediate by
$\Coe*{x.\C}\ssteps M$ and \cref{lem:expansion} on both sides. Similarly, if
$r=r'$ then $\ceqtm[\Psi']{\Coe*{x.\C}}{M}{\C}$ by \cref{lem:expansion} on the
left.
\end{proof}

\begin{rul}[Computation]\label{rul:C-comp-base}
If $\coftype{P}{B}$ then $\ceqtm{\Celim{c.A}{\base}{P}{x.L}}{P}{B}$.
\end{rul}
\begin{proof}
Immediate by $\Celim{c.A}{\base}{P}{x.L}\ssteps P$ and \cref{lem:expansion}.
\end{proof}

\begin{rul}[Computation]\label{rul:C-comp-loop}
If $\coftype[\Psi,x]{L}{B}$ and
$\ceqtm{\dsubst{L}{\e}{x}}{P}{\dsubst{B}{\e}{x}}$ for $\e\in\{0,1\}$, then
$\ceqtm{\Celim{c.A}{\lp{r}}{P}{x.L}}{\dsubst{L}{r}{x}}{\dsubst{B}{r}{x}}$.
\end{rul}
\begin{proof}
If $r=\e$ then this is immediate by \cref{lem:expansion} and
$\ceqtm{\dsubst{L}{\e}{x}}{P}{\dsubst{B}{\e}{x}}$. If $r=y$ then we apply
coherent expansion to the left side with family
$\{\td{P}{\psi} \mid \td{y}{\psi}=\e\}^{\Psi'}_\psi \cup
\{\dsubst{\td{L}{\psi}}{z}{x} \mid \td{y}{\psi}=z\}^{\Psi'}_\psi$. The
$\id[\Psi]$ element of this family is $\dsubst{L}{y}{x}$;
when $\td{y}{\psi}=\e$ we have
$\ceqtm[\Psi']{\td{\dsubst{L}{y}{x}}{\psi}}{\td{P}{\psi}}%
{\td{\dsubst{B}{y}{x}}{\psi}}$
(by $\td{\dsubst{}{y}{x}}{\psi}=\td{\dsubst{}{\e}{x}}{\psi}$),
and when $\td{y}{\psi}=z$ we have
$\ceqtm[\Psi']{\td{\dsubst{L}{y}{x}}{\psi}}{\dsubst{\td{L}{\psi}}{z}{x}}%
{\td{\dsubst{B}{y}{x}}{\psi}}$
(by $\dsubst{\psi}{z}{x}=\td{\dsubst{}{y}{x}}{\psi}$).
Thus by \cref{lem:cohexp-ceqtm},
$\ceqtm{\Celim{c.A}{\lp{y}}{P}{x.L}}{\dsubst{L}{y}{x}}{\dsubst{B}{y}{x}}$.
\end{proof}

To establish the elimination rule we must induct over the definition of
$\vper{\C}$. As $\vper{\C}$ was defined in \cref{sec:typesys} as the least
pre-fixed point of an order-preserving operator $C$ on context-indexed
relations, we define our induction hypothesis as an auxiliary context-indexed
PER on values $\Phi_\Psi(M_0,M_0')$ that holds when
\begin{enumerate}
\item $\vper{\C}_\Psi(M_0,M_0')$ and
\item whenever $\eqtypek{\oft{c}{\C}}{A}{A'}$,
$\ceqtm{P}{P'}{\subst{A}{\base}{c}}$,
$\ceqtm[\Psi,x]{L}{L'}{\subst{A}{\lp{x}}{c}}$, and
$\ceqtm{\dsubst{L}{\e}{x}}{P}{\subst{A}{\base}{c}}$ for $\e\in\{0,1\}$,
$\ceqtm{\Celim{c.A}{M_0}{P}{x.L}}{\Celim{c.A'}{M_0'}{P'}{x.L'}}{\subst{A}{M_0}{c}}$.
(In other words, the elimination rule holds for $M_0$ and $M_0'$.)
\end{enumerate}

\begin{lemma}\label{lem:C-elim-lift}
If $\Tm(\Phi(\Psi))(M,M')$ then
whenever $\eqtypek{\oft{c}{\C}}{A}{A'}$,
$\ceqtm{P}{P'}{\subst{A}{\base}{c}}$,
$\ceqtm[\Psi,x]{L}{L'}{\subst{A}{\lp{x}}{c}}$, and
$\ceqtm{\dsubst{L}{\e}{x}}{P}{\subst{A}{\base}{c}}$ for $\e\in\{0,1\}$,
$\ceqtm{\Celim{c.A}{M}{P}{x.L}}{\Celim{c.A'}{M'}{P'}{x.L'}}{\subst{A}{M}{c}}$.
\end{lemma}
\begin{proof}
First we apply coherent expansion to the left side with family
$\{ \Celim{c.\td{A}{\psi}}{M_\psi}{\td{P}{\psi}}{x.\td{L}{\psi}} \mid
\td{M}{\psi}\evals M_\psi \}^{\Psi'}_\psi$, by showing that
\[
\ceqtm[\Psi']%
{\Celim{c.\td{A}{\psi}}{M_\psi}{\td{P}{\psi}}{x.\td{L}{\psi}}}%
{\Celim{c.\td{A}{\psi}}{\td{(M_{\id})}{\psi}}{\td{P}{\psi}}{x.\td{L}{\psi}}}%
{\td{(\subst{A}{M}{c})}{\psi}}.
\]
The left side is an element of this type by $\Phi_{\Psi'}(M_\psi,M_\psi)$
and
$\ceqtypek[\Psi']{\subst{\td{A}{\psi}}{M_\psi}{c}}{\subst{\td{A}{\psi}}{\td{M}{\psi}}{c}}$
(by $\ceqtm[\Psi']{M_\psi}{\td{M}{\psi}}{\C}$). The right side is an
element by $\Phi_\Psi(M_{\id},M_{\id})$ and
$\ceqtypek{\subst{A}{M_{\id}}{c}}{\subst{A}{M}{c}}$. The equality follows from
$\td{(M_{\id})}{\psi}\evals M_2$,
$\Phi_{\Psi'}(M_\psi,M_2)$, and
\cref{lem:coftype-evals-ceqtm}. Thus by \cref{lem:cohexp-ceqtm},
$\ceqtm{\Celim{c.A}{M}{P}{x.L}}{\Celim{c.A}{M_{\id}}{P}{x.L}}{\subst{A}{M}{c}}$.

By the same argument on the right side,
$\ceqtypek{\subst{A}{M}{c}}{\subst{A'}{M'}{c}}$ (by $\ceqtm{M}{M'}{\C}$), and
transitivity, it suffices to show
$\ceqtm{\Celim{c.A}{M_{\id}}{P}{x.L}}{\Celim{c.A'}{M'_{\id}}{P'}{x.L'}}{\subst{A}{M}{c}}$;
this is immediate by $\Phi_\Psi(M_{\id},M'_{\id})$ and
$\ceqtypek{\subst{A}{M_{\id}}{c}}{\subst{A}{M}{c}}$.
\end{proof}

\begin{lemma}\label{lem:C-elim-ind}
If ${C(\Phi)}_\Psi(M_0,M_0')$ then $\Phi_\Psi(M_0,M_0')$.
\end{lemma}
\begin{proof}
We must show that $\vper{\C}_\Psi(M_0,M_0')$, and that if
$\eqtypek{\oft{c}{\C}}{A}{A'}$,
$\ceqtm{P}{P'}{\subst{A}{\base}{c}}$,
$\ceqtm[\Psi,x]{L}{L'}{\subst{A}{\lp{x}}{c}}$, and
$\ceqtm{\dsubst{L}{\e}{x}}{P}{\subst{A}{\base}{c}}$ for $\e\in\{0,1\}$, then
$\ceqtm{\Celim{c.A}{M_0}{P}{x.L}}{\Celim{c.A'}{M_0'}{P'}{x.L'}}{\subst{A}{M_0}{c}}$.
There are three cases to consider.
\begin{enumerate}
\item ${C(\Phi)}_\Psi(\base,\base)$.

Then $\vper{\C}_\Psi(\base,\base)$ by definition, and the elimination rule holds
by \cref{rul:C-comp-base} on both sides (with $B=\subst{A}{\base}{c}$) and
$\ceqtm{P}{P'}{\subst{A}{\base}{c}}$.

\item ${C(\Phi)}_{(\Psi,y)}(\lp{y},\lp{y})$.

Then $\vper{\C}_{(\Psi,y)}(\lp{y},\lp{y})$ by definition, and the elimination
rule holds by \cref{rul:C-comp-loop} on both sides (with
$B=\subst{A}{\lp{x}}{c}$ and
$\ceqtypek{\dsubst{\subst{A}{\lp{x}}{c}}{\e}{x}}{\subst{A}{\base}{c}}$) and
$\ceqtm{\dsubst{L}{y}{x}}{\dsubst{L'}{y}{x}}{\subst{A}{\lp{y}}{c}}$.

\item ${C(\Phi)}_\Psi(\Fcom*{r_i=r_i'},\Fcom{r}{r'}{M'}{\sys{r_i=r_i'}{y.N_i'}})$ where
\begin{enumerate}
\item $r\neq r'$; $r_i \neq r_i'$ for all $i$;
$r_i = r_j$, $r_i' = 0$, and $r_j' = 1$ for some $i,j$;
\item $\Tm(\Phi(\Psi))(M,M')$;
\item $\Tm(\Phi(\Psi'))(\td{N_i}{\psi},\td{N_j'}{\psi})$ for all $i,j$ and
$\tds{\Psi'}{\psi}{(\Psi,y)}$ satisfying $r_i=r_i',r_j=r_j'$; and
\item $\Tm(\Phi(\Psi'))(\td{\dsubst{N_i}{r}{y}}{\psi},\td{M}{\psi})$ for
all $i,j$ and $\psitd$ satisfying $r_i=r_i'$.
\end{enumerate}

By construction, $\Phi\subseteq\vper{\C}$, so
$\Tm(\Phi)\subseteq\Tm(\vper{\C})$ and $\vper{\C}_\Psi(\Fcom,\Fcom)$.
By \cref{lem:C-elim-lift} and $\Tm(\Phi(\Psi))(M,M')$,
$\ceqtm{\Celim{c.A}{M}}{\Celim{c.A'}{M'}}{\subst{A}{M}{c}}$.
For all $\psi$ satisfying $r_i=r_i',r_j=r_j'$ we have
$\Tm(\Phi(\Psi'))(\td{N_i}{\psi},\td{N_j'}{\psi})$, so by
\cref{lem:C-elim-lift},
$\ceqtm[\Psi,y]<r_i=r_i',r_j=r_j'>%
{\Celim{c.A}{N_i}}{\Celim{c.A'}{N_j'}}{\subst{A}{N_i}{c}}$.
Similarly, $\ceqtm<r_i=r_i'>{\Celim{c.A}{M}}%
{\Celim{c.A}{\dsubst{N_i}{r}{y}}}{\subst{A}{M}{c}}$.

Apply coherent expansion to the term $\Celim{c.A}{\Fcom*{\xi_i}}{P}{x.L}$ at the
type $\cwftypek{\subst{A}{\Fcom*{\xi_i}}{c}}$ with family:
\[\begin{cases}
\Celim{c.\td{A}{\psi}}{\td{M}{\psi}}{\td{P}{\psi}}{x.\td{L}{\psi}}
& \text{$\td{r}{\psi} = \td{r'}{\psi}$} \\
\Celim{c.\td{A}{\psi}}{\td{\dsubst{N_j}{r'}{y}}{\psi}}{\td{P}{\psi}}{x.\td{L}{\psi}}
& \text{$\td{r}{\psi} \neq \td{r'}{\psi}$,
least $j$ s.t. $\td{r_j}{\psi} = \td{r_j'}{\psi}$} \\
\Com{z.\subst{\td{A}{\psi}}{F}{c}}{\td{r}{\psi}}{\td{r'}{\psi}}%
{\Celim{c.\td{A}{\psi}}{\td{M}{\psi}}{\td{P}{\psi}}{x.\td{L}{\psi}}}%
{\sys{\td{\xi_i}{\psi}}{y.T_i}}
& \text{otherwise} \\
\qquad
F = \Fcom{\td{r}{\psi}}{z}{\td{M}{\psi}}{\sys{\td{\xi_i}{\psi}}{y.\td{N_i}{\psi}}} &\\
\qquad
T_i = \Celim{c.\td{A}{\psi}}{\td{N_i}{\psi}}{\td{P}{\psi}}{x.\td{L}{\psi}} &\\
\end{cases}\]
We must check three equations, noting that $\id$ falls in the third category
above. First:
\[
\ceqtm[\Psi']
{\Com{z.\subst{\td{A}{\psi}}{F}{c}}{\td{r}{\psi}}{\td{r'}{\psi}}%
  {\Celim{c.\td{A}{\psi}}{\td{M}{\psi}}}%
  {\sys{\td{\xi_i}{\psi}}{y.T_i}}}
{\Celim{c.\td{A}{\psi}}{\td{M}{\psi}}}
{\subst{\td{A}{\psi}}{\td{\Fcom}{\psi}}{c}}
\]
when $\td{r}{\psi} = \td{r'}{\psi}$. This follows from \cref{thm:com},
$\subst{\td{A}{\psi}}{\td{\Fcom}{\psi}}{c} =
\dsubst{\subst{\td{A}{\psi}}{F}{c}}{\td{r'}{\psi}}{z}$,
and by \cref{def:kan},
$\ceqtypek[\Psi']{\dsubst{\subst{\td{A}{\psi}}{F}{c}}{\td{r'}{\psi}}{z}}%
{\td{\subst{A}{M}{c}}{\psi}}$ and
$\ceqtypek[\Psi',z]<\td{r_i}{\psi}=\td{r_i'}{\psi}>%
{\subst{\td{A}{\psi}}{F}{c}}{\td{\subst{A}{\dsubst{N_i}{z}{y}}{c}}{\psi}}$.
Next, we must check
\[
\ceqtm[\Psi']
{\Com{z.\subst{\td{A}{\psi}}{F}{c}}{\td{r}{\psi}}{\td{r'}{\psi}}%
  {\Celim{c.\td{A}{\psi}}{\td{M}{\psi}}}%
  {\sys{\td{\xi_i}{\psi}}{y.T_i}}}
{\Celim{c.\td{A}{\psi}}{\td{\dsubst{N_j}{r'}{y}}{\psi}}}
{\subst{\td{A}{\psi}}{\td{\Fcom}{\psi}}{c}}
\]
when $\td{r}{\psi}\neq\td{r'}{\psi}$, $\td{r_j}{\psi}=\td{r_j'}{\psi}$, and
$\td{r_i}{\psi}\neq\td{r_i'}{\psi}$ for $i<j$; again this holds by
\cref{thm:com}. Finally, we must check
\[
\coftype[\Psi']
{\Com{z.\subst{\td{A}{\psi}}{F}{c}}{\td{r}{\psi}}{\td{r'}{\psi}}%
  {\Celim{c.\td{A}{\psi}}{\td{M}{\psi}}}%
  {\sys{\td{\xi_i}{\psi}}{y.T_i}}}
{\subst{\td{A}{\psi}}{\td{\Fcom}{\psi}}{c}}
\]
when $\td{r}{\psi}\neq\td{r'}{\psi}$ and $\td{r_i}{\psi}\neq\td{r_i'}{\psi}$ for
all $i$; again this holds by \cref{thm:com}. Therefore by
\cref{lem:cohexp-ceqtm},
\begin{align*}
& {\Celim{c.A}{\Fcom*{\xi_i}}{P}{x.L}} \\
\ceqtmtab{}
{\Com{z.\subst{\td{A}{\psi}}{\Fcom^{r\rightsquigarrow z}}{c}}{r}{r'}%
  {\Celim{c.A}{M}{P}{x.L}}%
  {\sys{\xi_i}{y.\Celim{c.A}{N_i}{P}{x.L}}}}
{\subst{A}{\Fcom}{c}}.
\end{align*}
By transitivity and a symmetric argument on the right side, it suffices to show
that two $\Com$s are equal, which follows by \cref{thm:com}.
\qedhere
\end{enumerate}
\end{proof}

\begin{rul}[Elimination]\label{rul:C-elim}
If $\ceqtm{M}{M'}{\C}$,
$\eqtypek{\oft{c}{\C}}{A}{A'}$,
$\ceqtm{P}{P'}{\subst{A}{\base}{c}}$,
$\ceqtm[\Psi,x]{L}{L'}{\subst{A}{\lp{x}}{c}}$, and
$\ceqtm{\dsubst{L}{\e}{x}}{P}{\subst{A}{\base}{c}}$ for $\e\in\{0,1\}$, then
$\ceqtm{\Celim{c.A}{M}{P}{x.L}}{\Celim{c.A'}{M'}{P'}{x.L'}}{\subst{A}{M}{c}}$.
\end{rul}
\begin{proof}
\Cref{lem:C-elim-ind} states that $\Phi$ is a pre-fixed point of $C$; because
$\vper{\C}$ is the least pre-fixed point of $C$, $\vper{\C}\subseteq\Phi$, and
therefore $\Tm(\vper{\C})\subseteq\Tm(\Phi)$. We conclude that
$\Tm(\Phi(\Psi))(M,M')$, and the result follows by \cref{lem:C-elim-lift}.
\end{proof}

\subsection{Weak booleans}

Let $\tau=\Kan\mu(\nu)$ or $\pre\mu(\nu,\sigma)$ for any cubical type systems
$\nu,\sigma$; we have
$\tau(\Psi,\wbool,\wbool,\mathbb{B}_\Psi)$, where $\mathbb{B}$ is the least
context-indexed relation such that:
\begin{enumerate}
\item $\mathbb{B}_\Psi(\true,\true)$,
\item $\mathbb{B}_\Psi(\false,\false)$, and
\item $\mathbb{B}_\Psi(\Fcom*{r_i=r_i'},\Fcom{r}{r'}{M'}{\sys{r_i=r_i'}{y.N_i'}})$
whenever
\begin{enumerate}
\item $r\neq r'$;
$r_i \neq r_i'$ for all $i$;
$r_i = r_j$, $r_i' = 0$, and $r_j' = 1$ for some $i,j$;
\item $\Tm(\mathbb{B}(\Psi))(M,M')$;
\item $\Tm(\mathbb{B}(\Psi'))(\td{N_i}{\psi},\td{N_j'}{\psi})$ for all $i,j$ and
$\tds{\Psi'}{\psi}{(\Psi,y)}$ satisfying $r_i=r_i',r_j=r_j'$; and
\item $\Tm(\mathbb{B}(\Psi'))(\td{\dsubst{N_i}{r}{y}}{\psi},\td{M}{\psi})$ for
all $i,j$ and $\psitd$ satisfying $r_i=r_i'$.
\end{enumerate}
\end{enumerate}
By $\sisval{\wbool}$ it is immediate that $\PTy(\tau)(\Psi,\wbool,\wbool,\mathbb{B}(\Psi))$.

We have included $\wbool$ to demonstrate two Kan structures that one may equip
to ordinary inductive types: trivial structure (as in $\bool$) and free
structure (as in $\wbool$, mirroring $\C$). As the $\Fcom$ structure of $\wbool$
is identical to that of $\C$, the proofs in this section are mostly identical to
those in \cref{ssec:C}.

\begin{lemma}\label{lem:wbool-prekan}
If
\begin{enumerate}
\item $\etc{r_i=r_i'}$ is valid,
\item $\Tm(\vper{\wbool}(\Psi))(M,M')$,
\item $\Tm(\vper{\wbool}(\Psi'))(\td{N_i}{\psi},\td{N_j'}{\psi})$ for all $i,j$ and
$\tds{\Psi'}{\psi}{(\Psi,y)}$ satisfying $r_i=r_i',r_j=r_j'$, and
\item $\Tm(\vper{\wbool}(\Psi'))(\td{\dsubst{N_i}{r}{y}}{\psi},\td{M}{\psi})$ for
all $i,j$ and $\psitd$ satisfying $r_i=r_i'$,
\end{enumerate}
then $\Tm(\vper{\wbool}(\Psi))(\Fcom*{r_i=r_i'},\Fcom{r}{r'}{M'}{\sys{r_i=r_i'}{y.N_i'}})$.
\end{lemma}
\begin{proof}
Identical to \cref{lem:C-prekan}.
\end{proof}

\begin{rul}[Pretype formation]
$\cwftypep{\wbool}$.
\end{rul}
\begin{proof}
Show $\Coh(\vper{\wbool})$:
$\Tm(\vper{\wbool}(\Psi))(\true,\true)$ and
$\Tm(\vper{\wbool}(\Psi))(\false,\false)$ because $\sisval{\true}$ and
$\sisval{\false}$, and $\Tm(\vper{\wbool}(\Psi))(\Fcom,\Fcom)$ by
\cref{lem:wbool-prekan}.
\end{proof}

\begin{rul}[Introduction]
If $\ceqtm{M}{M'}{\bool}$ then $\ceqtm{M}{M'}{\wbool}$.
\end{rul}
\begin{proof}
Follows from $\vper{\bool}\subseteq\vper{\wbool}$ and the fact that $\Tm$ is
order-preserving.
\end{proof}

\begin{rul}[Kan type formation]
$\cwftypek{\wbool}$.
\end{rul}
\begin{proof}
Identical to \cref{rul:C-form-kan}.
\end{proof}

We already proved the computation rules in \cref{rul:bool-comp}.  The
elimination rule differs from that of $\bool$, however: the motive $b.A$ must be
Kan, because the eliminator must account and the proof must account for
canonical $\Fcom$ elements of $\wbool$.

\begin{rul}[Elimination]
If $\ceqtm{M}{M'}{\wbool}$,
$\eqtypek{\oft{b}{\wbool}}{A}{A'}$,
$\ceqtm{T}{T'}{\subst{A}{\true}{b}}$, and
$\ceqtm{F}{F'}{\subst{A}{\false}{b}}$, then
$\ceqtm{\ifb{b.A}{M}{T}{F}}{\ifb{b.A'}{M'}{T'}{F'}}{\subst{A}{M}{b}}$.
\end{rul}
\begin{proof}
This proof is analogous to the proof of \cref{rul:C-elim}. First, we define a
context-indexed PER $\Phi_\Psi(M_0,M_0')$ that holds when
$\vper{\wbool}_\Psi(M_0,M_0')$ and the elimination rule is true for $M_0,M_0'$.
Next, we prove that if $\Tm(\Phi(\Psi))(M,M')$ then the elimination rule is true
for $M,M'$. Finally, we prove that $\Phi$ is a pre-fixed point of the operator
defining $\mathbb{B}$. (Here we must check that the elimination rule holds for
$\true$ and $\false$, which are immediate by \cref{rul:bool-comp}.) Therefore
$\Tm(\vper{\wbool})\subseteq\Tm(\Phi)$, so the elimination rule applies to
$\ceqtm{M}{M'}{\wbool}$.
\end{proof}

\subsection{Univalence}

Recall the abbreviations:
\begin{align*}
\isContr{C} &:= \prd{C}{(\picl{c}{C}{\picl{c'}{C}{\Path{\_.C}{c}{c'}}})} \\
\Equiv{A}{B} &:=
\sigmacl{f}{\arr{A}{B}}{(\picl{b}{B}{\isContr{\sigmacl{a}{A}{\Path{\_.B}{\app{f}{a}}{b}}}})}
\end{align*}
Let $\tau=\Kan\mu(\nu)$ or $\pre\mu(\nu,\sigma)$ for any cubical type systems
$\nu,\sigma$; in $\tau$, when
$\ceqtypep[\Psi,x]<x=0>{A}{A'}$,
$\ceqtypep[\Psi,x]{B}{B'}$,
$\ceqtm[\Psi,x]<x=0>{E}{E'}{\Equiv{A}{B}}$, and
$\phi(\uain{x}{M,N},\uain{x}{M',N'})$ for
\begin{enumerate}
\item $\ceqtm[\Psi,x]{N}{N'}{B}$,
\item $\ceqtm[\Psi,x]<x=0>{M}{M'}{A}$, and
\item $\ceqtm[\Psi,x]<x=0>{\app{\fst{E}}{M}}{N}{B}$,
\end{enumerate}
we have $\tau((\Psi,x),\ua{x}{A,B,E},\ua{x}{A',B',E'},\phi)$.

\begin{rul}[Pretype formation]\label{rul:ua-form-pre}
~\begin{enumerate}
\item If $\cwftypep{A}$ then $\ceqtypep{\ua{0}{A,B,E}}{A}$.
\item If $\cwftypep{B}$ then $\ceqtypep{\ua{1}{A,B,E}}{B}$.
\item If $\ceqtypep<r=0>{A}{A'}$,
$\ceqtypep{B}{B'}$, and
$\ceqtm<r=0>{E}{E'}{\Equiv{A}{B}}$, then
$\ceqtypep{\ua{r}{A,B,E}}{\ua{r}{A',B',E'}}$.
\end{enumerate}
\end{rul}
\begin{proof}
Parts (1--2) are immediate by \cref{lem:expansion}. To show part (3), we must
first establish that $\PTy(\tau)(\Psi,\ua{r}{A,B,E},\ua{r}{A',B',E'},\gamma)$,
that is, abbreviating these terms $L$ and $R$, for all
$\tds{\Psi_1}{\psi_1}{\Psi}$ and $\tds{\Psi_2}{\psi_2}{\Psi_1}$,
$\td{L}{\psi_1}\evals L_1$, $\td{R}{\psi_1}\evals R_1$,
$\lift{\tau}(\Psi_2,\td{L_1}{\psi_2},\td{L}{\psi_1\psi_2},\_)$,
$\lift{\tau}(\Psi_2,\td{R_1}{\psi_2},\td{R}{\psi_1\psi_2},\_)$, and
$\lift{\tau}(\Psi_2,\td{L_1}{\psi_2},\td{R_1}{\psi_2},\_)$.
We proceed by cases on the first step taken by $\td{L}{\psi_1}$ and
$\td{L}{\psi_1\psi_2}$.
\begin{enumerate}
\item $\td{r}{\psi_1} = 0$.

Then $\td{L}{\psi_1}\ssteps\td{A}{\psi_1}$,
$\td{R}{\psi_1}\ssteps\td{A'}{\psi_1}$, and the result follows by
$\ceqtypep[\Psi_1]{\td{A}{\psi_1}}{\td{A'}{\psi_1}}$.

\item $\td{r}{\psi_1} = 1$.

Then $\td{L}{\psi_1}\ssteps\td{B}{\psi_1}$,
$\td{R}{\psi_1}\ssteps\td{B'}{\psi_1}$, and the result follows by
$\ceqtypep{B}{B'}$.

\item $\td{r}{\psi_1} = x$ and $\td{r}{\psi_1\psi_2} = 0$.

Then $\isval{\td{L}{\psi_1}}$, $\td{L}{\psi_1\psi_2}\steps\td{A}{\psi_1\psi_2}$,
$\isval{\td{R}{\psi_1}}$, $\td{R}{\psi_1\psi_2}\steps\td{A'}{\psi_1\psi_2}$, and
the result follows by
$\ceqtypep[\Psi_2]{\td{A}{\psi_1\psi_2}}{\td{A'}{\psi_1\psi_2}}$.

\item $\td{r}{\psi_1} = x$ and $\td{r}{\psi_1\psi_2} = 1$.

Then $\isval{\td{L}{\psi_1}}$, $\td{L}{\psi_1\psi_2}\steps\td{B}{\psi_1\psi_2}$,
$\isval{\td{R}{\psi_1}}$, $\td{R}{\psi_1\psi_2}\steps\td{B'}{\psi_1\psi_2}$, and
the result follows by $\ceqtypep{B}{B'}$.

\item $\td{r}{\psi_1} = x$ and $\td{r}{\psi_1\psi_2} = x'$.

Then $\isval{\td{L}{\psi_1}}$, $\isval{\td{L}{\psi_1\psi_2}}$,
$\isval{\td{R}{\psi_1}}$, $\isval{\td{R}{\psi_1\psi_2}}$, and by
$\ceqtypep[\Psi_2]<x'=0>{\td{A}{\psi_1\psi_2}}{\td{A'}{\psi_1\psi_2}}$,
$\ceqtypep[\Psi_2]{\td{B}{\psi_1\psi_2}}{\td{B'}{\psi_1\psi_2}}$, and
$\ceqtm[\Psi_2]<x'=0>{\td{E}{\psi_1\psi_2}}{\td{E'}{\psi_1\psi_2}}%
{\Equiv{\td{A}{\psi_1\psi_2}}{\td{B}{\psi_1\psi_2}}}$, we have
$\tau(\Psi_2,\ua{x'}{\td{A}{\psi_1\psi_2},\td{B}{\psi_1\psi_2},\td{E}{\psi_1\psi_2}},
\ua{x'}{\td{A'}{\psi_1\psi_2},\td{B'}{\psi_1\psi_2},\td{E'}{\psi_1\psi_2}},\_)$.
\end{enumerate}

To complete part (3), we must show $\Coh(\vper{\ua{r}{A,B,E}})$, that is, for
any $\psitd$, if
$\vper{\ua{\td{r}{\psi}}{\td{A}{\psi},\td{B}{\psi},\td{E}{\psi}}}(M_0,N_0)$ then
$\Tm(\vper{\ua{\td{r}{\psi}}{\td{A}{\psi},\td{B}{\psi},\td{E}{\psi}}})(M_0,N_0)$.
If $\td{r}{\psi}=0$ this
follows by
$\vper{\ua{0}{\td{A}{\psi},\td{B}{\psi},\td{E}{\psi}}} = \vper{\td{A}{\psi}}$
and $\Coh(\vper{\td{A}{\psi}})$; if $\td{r}{\psi}=1$ then
this follows by $\vper{\ua{1}{\td{A}{\psi},\td{B}{\psi},\td{E}{\psi}}} =
\vper{\td{B}{\psi}}$ and $\Coh(\vper{\td{B}{\psi}})$. The remaining case is
$\vper{\ua{x}{\td{A}{\psi},\td{B}{\psi},\td{E}{\psi}}}(\uain{x}{M,N},\uain{x}{M',N'})$,
in which
$\ceqtm[\Psi']{N}{N'}{\td{B}{\psi}}$,
$\ceqtm[\Psi']<x=0>{M}{M'}{\td{A}{\psi}}$, and
$\ceqtm[\Psi']<x=0>{\app{\fst{\td{E}{\psi}}}{M}}{N}{\td{B}{\psi}}$.
Again we proceed by cases on the first step taken by the $\psi_1$ and
$\psi_1\psi_2$ instances of the left side.
\begin{enumerate}
\item $\td{x}{\psi_1} = 0$.

Then $\td{L}{\psi_1}\ssteps\td{M}{\psi_1}$,
$\td{R}{\psi_1}\ssteps\td{M'}{\psi_1}$, and the result follows by
$\vper{\ua{0}{\td{A}{\psi\psi_1},\dots}} = \vper{\td{A}{\psi\psi_1}}$ and
$\ceqtm[\Psi_1]{\td{M}{\psi_1}}{\td{M'}{\psi_1}}{\td{A}{\psi\psi_1}}$.

\item $\td{x}{\psi_1} = 1$.

Then $\td{L}{\psi_1}\ssteps\td{N}{\psi_1}$,
$\td{R}{\psi_1}\ssteps\td{N'}{\psi_1}$, and the result follows by
$\vper{\ua{1}{\td{A}{\psi\psi_1},\dots}} = \vper{\td{B}{\psi\psi_1}}$ and
$\ceqtm[\Psi']{N}{N'}{\td{B}{\psi}}$.

\item $\td{x}{\psi_1} = x'$ and $\td{x}{\psi_1\psi_2} = 0$.

Then $\isval{\td{L}{\psi_1}}$, $\td{L}{\psi_1\psi_2}\steps\td{M}{\psi_1\psi_2}$,
$\isval{\td{R}{\psi_1}}$, $\td{R}{\psi_1\psi_2}\steps\td{M'}{\psi_1\psi_2}$, and
the result follows by $\vper{\ua{0}{\td{A}{\psi\psi_1\psi_2},\dots}} =
\vper{\td{A}{\psi\psi_1\psi_2}}$ and
$\ceqtm[\Psi_2]{\td{M}{\psi_1\psi_2}}{\td{M'}{\psi_1\psi_2}}{\td{A}{\psi\psi_1\psi_2}}$.

\item $\td{x}{\psi_1} = x'$ and $\td{x}{\psi_1\psi_2} = 1$.

Then $\isval{\td{L}{\psi_1}}$, $\td{L}{\psi_1\psi_2}\steps\td{N}{\psi_1\psi_2}$,
$\isval{\td{R}{\psi_1}}$, $\td{R}{\psi_1\psi_2}\steps\td{N'}{\psi_1\psi_2}$, and
the result follows by $\vper{\ua{1}{\td{A}{\psi\psi_1\psi_2},\dots}} =
\vper{\td{B}{\psi\psi_1\psi_2}}$ and $\ceqtm[\Psi']{N}{N'}{\td{B}{\psi}}$.

\item $\td{x}{\psi_1} = x'$ and $\td{x}{\psi_1\psi_2} = x''$.

Then $\isval{\td{L}{\psi_1}}$, $\isval{\td{L}{\psi_1\psi_2}}$,
$\isval{\td{R}{\psi_1}}$, $\isval{\td{R}{\psi_1\psi_2}}$, and by
$\ceqtm[\Psi_2]{\td{N}{\psi_1\psi_2}}{\td{N'}{\psi_1\psi_2}}{\td{B}{\psi\psi_1\psi_2}}$,
$\ceqtm[\Psi_2]<x''=0>{\td{M}{\psi_1\psi_2}}{\td{M'}{\psi_1\psi_2}}{\td{A}{\psi\psi_1\psi_2}}$,
and
$\ceqtm[\Psi_2]<x''=0>{\app{\fst{\td{E}{\psi\psi_1\psi_2}}}{\td{M}{\psi_1\psi_2}}}%
{\td{N}{\psi_1\psi_2}}{\td{B}{\psi}}$,
$\vper{\ua{x''}{\td{A}{\psi\psi_1\psi_2},\dots}}
(\uain{x''}{\td{M}{\psi_1\psi_2},\td{N}{\psi_1\psi_2}},
\uain{x''}{\td{M'}{\psi_1\psi_2},\td{N'}{\psi_1\psi_2}})$.
\qedhere
\end{enumerate}
\end{proof}

\begin{rul}[Introduction]\label{rul:ua-intro}
~\begin{enumerate}
\item If $\coftype{M}{A}$ then $\ceqtm{\uain{0}{M,N}}{M}{A}$.
\item If $\coftype{N}{B}$ then $\ceqtm{\uain{1}{M,N}}{N}{B}$.
\item If $\ceqtm<r=0>{M}{M'}{A}$,
$\ceqtm{N}{N'}{B}$,
$\coftype<r=0>{E}{\Equiv{A}{B}}$, and
$\ceqtm<r=0>{\app{\fst{E}}{M}}{N}{B}$, then
$\ceqtm{\uain{r}{M,N}}{\uain{r}{M',N'}}{\ua{r}{A,B,E}}$.
\end{enumerate}
\end{rul}
\begin{proof}
Parts (1--2) are immediate by $\uain{0}{M,N}\ssteps M$,
$\uain{1}{M,N}\ssteps N$, and \cref{lem:expansion}. For part (3), if $r=0$
(resp., $r=1$) the result follows by part (1) (resp., part (2)) and
\cref{rul:ua-form-pre}. If $r=x$ then it follows by $\Coh(\vper{\ua{x}{A,B,E}})$
and the definition of $\vper{\ua{x}{A,B,E}}$.
\end{proof}

\begin{rul}[Elimination]\label{rul:ua-elim}
~\begin{enumerate}
\item If $\coftype{M}{A}$ and $\coftype{F}{\arr{A}{B}}$, then
$\ceqtm{\uaproj{0}{M,F}}{\app{F}{M}}{B}$.
\item If $\coftype{M}{B}$ then $\ceqtm{\uaproj{1}{M,F}}{M}{B}$.
\item If $\ceqtm{M}{M'}{\ua{r}{A,B,E}}$ and
$\ceqtm<r=0>{F}{\fst{E}}{\arr{A}{B}}$, then
$\ceqtm{\uaproj{r}{M,F}}{\uaproj{r}{M',\fst{E}}}{B}$.
\end{enumerate}
\end{rul}
\begin{proof}
Parts (1--2) are immediate by $\uaproj{0}{M,F}\ssteps \app{F}{M}$,
$\uaproj{1}{M,F}\ssteps M$, and \cref{lem:expansion}. For part (3), if $r=0$
(resp., $r=1$) the result follows by part (1) (resp., part (2)),
\cref{rul:fun-elim}, and \cref{rul:ua-form-pre}. If $r=x$ then we apply coherent
expansion to the left side with family
\[\begin{cases}
\app{\td{F}{\psi}}{\td{M}{\psi}} & \text{$\td{x}{\psi}=0$} \\
\td{M}{\psi} & \text{$\td{x}{\psi}=1$} \\
N_\psi & \text{$\td{x}{\psi}=x'$, $\td{M}{\psi}\evals\uain{x'}{O_\psi,N_\psi}$} \\
\end{cases}\]
where $\coftype[\Psi']<x'=0>{O_\psi}{\td{A}{\psi}}$,
$\coftype[\Psi']{N_\psi}{\td{B}{\psi}}$, and
$\ceqtm[\Psi']<x'=0>{\app{\fst{\td{E}{\psi}}}{O_\psi}}{N_\psi}{\td{B}{\psi}}$.
First, show that if $\td{x}{\psi}=0$,
$\ceqtm[\Psi']{\app{\td{F}{\psi}}{\td{M}{\psi}}}{\td{(N_{\id})}{\psi}}{\td{B}{\psi}}$.
By \cref{lem:coftype-evals-ceqtm},
$\ceqtm{M}{\uain{x}{O_{\id},N_{\id}}}{\ua{x}{A,B,E}}$, so by
\cref{rul:ua-intro},
$\ceqtm[\Psi']{\td{M}{\psi}}{\td{(O_{\id})}{\psi}}{\td{A}{\psi}}$.
By assumption,
$\ceqtm[\Psi']{\td{F}{\psi}}{\fst{\td{E}{\psi}}}{\arr{\td{A}{\psi}}{\td{B}{\psi}}}$.
This case is completed by \cref{rul:fun-elim} and
$\ceqtm[\Psi']{\app{\fst{\td{E}{\psi}}}{\td{(O_{\id})}{\psi}}}{\td{(N_{\id})}{\psi}}{\td{B}{\psi}}$.
Next, show that if $\td{x}{\psi}=1$,
$\ceqtm[\Psi']{\td{M}{\psi}}{\td{(N_{\id})}{\psi}}{\td{B}{\psi}}$.
This case is immediate by \cref{rul:ua-intro} and
$\ceqtm{M}{\uain{x}{O_{\id},N_{\id}}}{\ua{x}{A,B,E}}$ under $\psi$.
Finally, show that if $\td{x}{\psi}=x'$,
$\ceqtm[\Psi']{N_{\psi}}{\td{(N_{\id})}{\psi}}{\td{B}{\psi}}$.
By $\coftype{M}{\ua{x}{A,B,E}}$ under $\id,\psi$ we have
$\vper{\ua{x}{A,B,E}}_\psi(\uain{x'}{O_\psi,N_\psi},\uain{x'}{\td{(O_{\id})}{\psi},\td{(N_{\id})}{\psi}})$,
completing this case.

By \cref{lem:cohexp-ceqtm} we conclude $\ceqtm{\uaproj{x}{M,F}}{N_{\id}}{B}$,
and by a symmetric argument, $\ceqtm{\uaproj{x}{M',\fst{E}}}{N'_{\id}}{B}$. We
complete the proof with transitivity and $\ceqtm{N_{\id}}{N'_{\id}}{B}$ by
$\vper{\ua{x}{A,B,E}}(\uain{x}{O_{\id},N_{\id}},\uain{x}{O'_{\id},N'_{\id}})$.
\end{proof}

\begin{rul}[Computation]\label{rul:ua-comp}
If $\coftype<r=0>{M}{A}$,
$\coftype{N}{B}$,
$\coftype<r=0>{F}{\arr{A}{B}}$, and
$\ceqtm<r=0>{\app{F}{M}}{N}{B}$, then
$\ceqtm{\uaproj{r}{\uain{r}{M,N},F}}{N}{B}$.
\end{rul}
\begin{proof}
If $r=0$ then by \cref{lem:expansion} it suffices to show
$\ceqtm{\app{F}{\uain{0}{M,N}}}{N}{B}$; by \cref{rul:fun-elim,rul:ua-intro}
this holds by our hypothesis $\ceqtm{\app{F}{M}}{N}{B}$.
If $r=1$ the result is immediate by \cref{lem:expansion}.
If $r=x$ we apply coherent expansion to the left side with family
\[\begin{cases}
\app{\td{F}{\psi}}{\uain{0}{\td{M}{\psi},\td{N}{\psi}}}
& \text{$\td{x}{\psi}=0$} \\
\td{N}{\psi} & \text{$\td{x}{\psi}=1$ or $\td{x}{\psi}=x'$}
\end{cases}\]
If $\td{x}{\psi}=0$ then
$\ceqtm[\Psi']{\app{\td{F}{\psi}}{\uain{0}{\td{M}{\psi},\td{N}{\psi}}}}%
{\td{N}{\psi}}{\td{B}{\psi}}$ by \cref{rul:fun-elim,rul:ua-intro} and
$\ceqtm<x=0>{\app{F}{M}}{N}{B}$. If $\td{x}{\psi}\neq 0$ then
$\coftype[\Psi']{\td{N}{\psi}}{\td{B}{\psi}}$ and the result follows by
\cref{lem:cohexp-ceqtm}.
\end{proof}

\begin{rul}[Eta]\label{rul:ua-eta}
If $\coftype{N}{\ua{r}{A,B,E}}$ and
$\ceqtm<r=0>{M}{N}{A}$, then
$\ceqtm{\uain{r}{M,\uaproj{r}{N,\fst{E}}}}{N}{\ua{r}{A,B,E}}$.
\end{rul}
\begin{proof}
If $r=0$ or $r=1$ the result is immediate by
\cref{lem:expansion,rul:ua-form-pre}. If $r=x$ then by
\cref{lem:coftype-evals-ceqtm}, $\ceqtm{N}{\uain{x}{M',P'}}{\ua{x}{A,B,E}}$
where $\coftype<x=0>{M'}{A}$, $\coftype{P'}{B}$, and
$\ceqtm<x=0>{\app{\fst{E}}{M'}}{P'}{B}$. By \cref{rul:ua-intro} it suffices
to show that $\ceqtm<x=0>{M}{M'}{A}$,
$\ceqtm{\uaproj{x}{N,\fst{E}}}{P'}{B}$, and
$\ceqtm<x=0>{\app{\fst{E}}{M'}}{P'}{B}$ (which is immediate).
To show $\ceqtm<x=0>{M}{M'}{A}$ it suffices to prove
$\ceqtm<x=0>{N}{M'}{A}$, which follows from
$\ceqtm{N}{\uain{x}{M',P'}}{\ua{x}{A,B,E}}$ and
\cref{rul:ua-form-pre,rul:ua-intro}. To show
$\ceqtm{\uaproj{x}{N,\fst{E}}}{P'}{B}$, by \cref{rul:ua-elim} it suffices to
check $\ceqtm{\uaproj{x}{\uain{x}{M',P'},\fst{E}}}{P'}{B}$, which holds by
\cref{rul:ua-comp}.
\end{proof}

\begin{lemma}\label{lem:ua-hcom}
If $\ceqtypek<x=0>{A}{A'}$,
$\ceqtypek{B}{B'}$,
$\ceqtm<x=0>{E}{E'}{\Equiv{A}{B}}$,

\begin{enumerate}
\item $\etc{\xi_i}=\etc{r_i=r_i'}$ is valid,
\item $\ceqtm{M}{M'}{\ua{x}{A,B,E}}$,
\item $\ceqtm[\Psi,y]<r_i=r_i',r_j=r_j'>{N_i}{N_j'}{\ua{x}{A,B,E}}$
for any $i,j$, and
\item $\ceqtm<r_i=r_i'>{\dsubst{N_i}{r}{y}}{M}{\ua{x}{A,B,E}}$
for any $i$,
\end{enumerate}
then
\begin{enumerate}
\item $\ceqtm{\Hcom*{\ua{x}{A,B,E}}{\xi_i}}%
{\Hcom{\ua{x}{A',B',E'}}{r}{r'}{M'}{\sys{\xi_i}{y.N_i'}}}{\ua{x}{A,B,E}}$;
\item if $r=r'$ then
$\ceqtm{\Hcom{\ua{x}{A,B,E}}{r}{r}{M}{\sys{\xi_i}{y.N_i}}}{M}{\ua{x}{A,B,E}}$;
and
\item if $r_i = r_i'$ then
$\ceqtm{\Hcom*{\ua{x}{A,B,E}}{\xi_i}}{\dsubst{N_i}{r'}{y}}{\ua{x}{A,B,E}}$.
\end{enumerate}
\end{lemma}
\begin{proof}
For part (1), apply coherent expansion to $\Hcom*{\ua{x}{A,B,E}}{\xi_i}$ with family
\[\begin{cases}
\Hcom{\td{A}{\psi}}{\td{r}{\psi}}{\td{r'}{\psi}}%
{\td{M}{\psi}}{\sys{\td{\xi_i}{\psi}}{y.\td{N_i}{\psi}}}
& \text{$\td{x}{\psi}=0$} \\
\Hcom{\td{B}{\psi}}{\td{r}{\psi}}{\td{r'}{\psi}}%
{\td{M}{\psi}}{\sys{\td{\xi_i}{\psi}}{y.\td{N_i}{\psi}}}
& \text{$\td{x}{\psi}=1$} \\
\td{(\uain{x}{\dsubst{O}{r'}{y},
  \Hcom{B}{r}{r'}{\uaproj{x}{M,\fst{E}}}{\etc{T}}})}{\psi}
& \text{$\td{x}{\psi}=x'$} \\
\quad O =
  \Hcom{A}{r}{y}{M}{\sys{\xi_i}{y.N_i}} &\\
\mkspacer{\quad \etc{T} ={}}
  \sys{\xi_i}{y.\uaproj{x}{N_i,\fst{E}}}, &\\
\spacer
  \tube{x=0}{y.\app{\fst{E}}{O}}, &\\
\spacer
  \tube{x=1}{y.\Hcom{B}{r}{y}{M}{\sys{\xi_i}{y.N_i}}}
\end{cases}\]
Consider $\psi=\id$. Using rules for dependent functions, dependent types, and
univalence:
\begin{enumerate}
\item $\coftype[\Psi,y]<x=0>{O}{A}$ and
$\ceqtm<x=0>{\dsubst{O}{r}{y}}{M}{A}$
(by $\ceqtypep<x=0>{\ua{x}{A,B,E}}{A}$).

\item
$\coftype{\uaproj{x}{M,\fst{E}}}{B}$ where
$\ceqtm<x=0>{\uaproj{x}{M,\fst{E}}}{\app{\fst{E}}{M}}{B}$
and $\ceqtm<x=1>{\uaproj{x}{M,\fst{E}}}{M}{B}$.

\item
$\ceqtm[\Psi,y]<r_i=r_i',r_j=r_j'>%
{\uaproj{x}{N_i,\fst{E}}}{\uaproj{x}{N_j,\fst{E}}}{B}$ and
$\ceqtm<r_i=r_i'>{\uaproj{x}{M,\fst{E}}}{\uaproj{x}{\dsubst{N_i}{r}{y},\fst{E}}}{B}$.

\item
$\coftype[\Psi,y]<x=0>{\app{\fst{E}}{O}}{B}$,
$\ceqtm[\Psi,y]<x=0,r_i=r_i'>{\app{\fst{E}}{O}}{\uaproj{x}{N_i,\fst{E}}}{B}$
(both equal $\app{\fst{E}}{N_i}$), and
$\ceqtm<x=0>{\app{\fst{E}}{\dsubst{O}{r}{y}}}{\uaproj{x}{M,\fst{E}}}{B}$
(both equal $\app{\fst{E}}{M}$).

\item
$\coftype[\Psi,y]<x=1>{\Hcom{B}{r}{y}{M}{\sys{\xi_i}{y.N_i}}}{B}$
(by $\ceqtypep<x=1>{\ua{x}{A,B,E}}{B}$),
$\ceqtm[\Psi,y]<x=1,r_i=r_i'>%
{\Hcom{B}{r}{y}{M}{\sys{\xi_i}{y.N_i}}}{\uaproj{x}{N_i,\fst{E}}}{B}$
(both equal $N_i$), and
$\ceqtm[\Psi]<x=1>%
{\Hcom{B}{r}{r}{M}{\sys{\xi_i}{y.N_i}}}{\uaproj{x}{M,\fst{E}}}{B}$
(both equal $M$).

\item By the above,
$\coftype{\Hcom{B}{r}{r'}{\uaproj{x}{M,\fst{E}}}{\etc{T}}}{B}$ and
$\ceqtm<x=0>{\Hcom{B}}{\app{\fst{E}}{\dsubst{O}{r'}{y}}}{B}$, so
$\coftype{\uain{x}{\dsubst{O}{r'}{y},\Hcom{B}{r}{r'}{\uaproj{x}{M,\fst{E}}}{\etc{T}}}}{\ua{x}{A,B,E}}$.
\end{enumerate}

When $\td{x}{\psi}=x'$, coherence is immediate. When $\td{x}{\psi}=0$,
$\ceqtm[\Psi']{\uain{0}{\dsubst{O}{\td{r'}{\psi}}{y},\dots}}{\Hcom{\td{A}{\psi}}}{\td{A}{\psi}}$
as required. When $\td{x}{\psi}=1$,
$\ceqtm[\Psi']{\uain{1}{\dots,\Hcom{\td{B}{\psi}}{\td{r}{\psi}}{\td{r'}{\psi}}{\dots}{\etc{T}}}}%
{\Hcom{\td{B}{\psi}}}{\td{B}{\psi}}$ as required. Therefore
\cref{lem:cohexp-ceqtm} applies, and part (1) follows by repeating this argument
on the right side.

For part (2), show that
$\ceqtm{\uain{x}{\dsubst{O}{r'}{y},\Hcom{B}{r}{r'}{\uaproj{x}{M,\fst{E}}}{\etc{T}}}}%
{M}{\ua{x}{A,B,E}}$ when $r=r'$. By the above,
$\ceqtm{\uain{x}{\dots}}{\uain{x}{M,\uaproj{x}{M,\fst{E}}}}{\ua{x}{A,B,E}}$, so
the result follows by \cref{rul:ua-eta}.

For part (3), show
$\ceqtm{\uain{x}{\dsubst{O}{r'}{y},\Hcom{B}{r}{r'}{\uaproj{x}{M,\fst{E}}}{\etc{T}}}}%
{\dsubst{N_i}{r'}{y}}{\ua{x}{A,B,E}}$ when $r_i=r_i'$. By the above,
$\ceqtm{\uain{x}{\dots}}{\uain{x}{\dsubst{N_i}{r'}{y},\uaproj{x}{\dsubst{N_i}{r'}{y},\fst{E}}}}{\ua{x}{A,B,E}}$,
so the result again follows by \cref{rul:ua-eta}.
\end{proof}

\begin{lemma}\label{lem:ua-coe-xy}
If $\ceqtypek[\Psi,y]<x=0>{A}{A'}$,
$\ceqtypek[\Psi,y]{B}{B'}$,
$\ceqtm[\Psi,y]<x=0>{E}{E'}{\Equiv{A}{B}}$, and
$\ceqtm{M}{M'}{\dsubst{(\ua{x}{A,B,E})}{r}{y}}$ for $x\neq y$, then
$\ceqtm{\Coe{y.\ua{x}{A,B,E}}{r}{r'}{M}}%
{\Coe{y.\ua{x}{A',B',E'}}{r}{r'}{M'}}%
{\dsubst{(\ua{x}{A,B,E})}{r'}{y}}$ and
$\ceqtm{\Coe{y.\ua{x}{A,B,E}}{r}{r}{M}}{M}%
{\dsubst{(\ua{x}{A,B,E})}{r}{y}}$.
\end{lemma}
\begin{proof}
We apply coherent expansion to $\Coe{y.\ua{x}{A,B,E}}{r}{r'}{M}$ with family
\[\begin{cases}
\Coe{y.\td{A}{\psi}}{\td{r}{\psi}}{\td{r'}{\psi}}{\td{M}{\psi}}
&\text{$\td{x}{\psi} = 0$} \\
\Coe{y.\td{B}{\psi}}{\td{r}{\psi}}{\td{r'}{\psi}}{\td{M}{\psi}}
&\text{$\td{x}{\psi} = 1$} \\
\td{(\uain{x}{\Coe{y.A}{r}{r'}{M},\Com{y.B}{r}{r'}{\uaproj{x}{M,\fst{\dsubst{E}{r}{y}}}}{\etc{T}}})}{\psi}
&\text{$\td{x}{\psi} = x'$} \\
\mkspacer{\quad\etc{T}={}}
\tube{x=0}{y.\app{\fst{E}}{\Coe{y.A}{r}{y}{M}}}, &\\
\spacer
\tube{x=1}{y.\Coe{y.B}{r}{y}{M}} &
\end{cases}\]
Consider $\psi=\id$.
\begin{enumerate}
\item $\coftype{\uaproj{x}{M,\fst{\dsubst{E}{r}{y}}}}{\dsubst{B}{r}{y}}$
(by $\coftype{M}{\ua{x}{\dsubst{A}{r}{y},\dots}}$),
$\ceqtm<x=0>{\uaproj{x}{M,\fst{\dsubst{E}{r}{y}}}}{\app{\fst{\dsubst{E}{r}{y}}}{M}}{\dsubst{B}{r}{y}}$,
and $\ceqtm<x=1>{\uaproj{x}{M,\fst{\dsubst{E}{r}{y}}}}{M}{\dsubst{B}{r}{y}}$.

\item $\coftype[\Psi,y]<x=0>{\app{\fst{E}}{\Coe{y.A}{r}{y}{M}}}{B}$ because
$\coftype[\Psi,y]<x=0>{\fst{E}}{\arr{A}{B}}$ and
$\coftype[\Psi,y]<x=0>{\Coe{y.A}{r}{y}{M}}{A}$
(by $\coftype[\Psi]<x=0>{M}{\dsubst{A}{r}{y}}$). Under $\dsubst{}{r}{y}$ this
$\ceqtm<x=0>{{}}{\app{\fst{\dsubst{E}{r}{y}}}{M}}{\dsubst{B}{r}{y}}$.

\item $\coftype[\Psi,y]<x=1>{\Coe{y.B}{r}{y}{M}}{B}$
(by $\coftype[\Psi]<x=1>{M}{\dsubst{B}{r}{y}}$)
and $\ceqtm<x=1>{\Coe{y.B}{r}{r}{M}}{M}{\dsubst{B}{r}{y}}$.

\item Therefore $\coftype{\Com{y.B}}{\dsubst{B}{r'}{y}}$,
$\ceqtm<x=0>{\Com{y.B}}{\app{\fst{\dsubst{E}{r'}{y}}}{\Coe{y.A}{r}{r'}{M}}}{\dsubst{B}{r'}{y}}$,
and $\ceqtm<x=1>{\Com{y.B}}{\Coe{y.B}{r}{r'}{M}}{\dsubst{B}{r'}{y}}$. It
follows that
$\coftype{\uain{x}{\dots}}{\ua{x}{\dsubst{A}{r'}{y},\dsubst{B}{r'}{y},\dsubst{E}{r'}{y}}}$.
\end{enumerate}

When $\td{x}{\psi}=x'$, coherence is immediate. When $\td{x}{\psi}=0$, we have
$\ceqtm[\Psi']%
{\uain{0}{\Coe{y.\td{A}{\psi}}{\td{r}{\psi}}{\td{r'}{\psi}}{\td{M}{\psi}},\dots}}%
{\Coe{y.\td{A}{\psi}}{\td{r}{\psi}}{\td{r'}{\psi}}{\td{M}{\psi}}}%
{\dsubst{\td{A}{\psi}}{\td{r'}{\psi}}{y}}$. When $\td{x}{\psi}=1$,
$\ceqtm[\Psi']{\uain{1}{\dots}}
{\Coe{y.\td{B}{\psi}}{\td{r}{\psi}}{\td{r'}{\psi}}{\td{M}{\psi}}}
{\dsubst{\td{B}{\psi}}{\td{r'}{\psi}}{y}}$.
Therefore \cref{lem:cohexp-ceqtm} applies, and the first part follows by
the same argument on the right side.

For the second part,
$\ceqtm{\Coe{y.\ua{x}{A,B,E}}{r}{r}{M}}%
{\uain{x}{\Coe{y.A}{r}{r}{M},\Com{y.B}{r}{r}{\uaproj{x}{M,\fst{\dsubst{E}{r}{y}}}}{\etc{T}}}}%
{\dsubst{(\ua{x}{A,B,E})}{r}{y}}$, which equals
$\uain{x}{M,\uaproj{x}{M,\fst{\dsubst{E}{r}{y}}}}$ and $M$ by \cref{rul:ua-eta}.
\end{proof}

\begin{lemma}\label{lem:ua-coe-from-0}
If $\ceqtypek[\Psi,x]<x=0>{A}{A'}$,
$\ceqtypek[\Psi,x]{B}{B'}$,
$\ceqtm[\Psi,x]<x=0>{E}{E'}{\Equiv{A}{B}}$, and
$\ceqtm{M}{M'}{\dsubst{(\ua{x}{A,B,E})}{0}{x}}$, then
$\ceqtm{\Coe{x.\ua{x}{A,B,E}}{0}{r'}{M}}%
{\Coe{x.\ua{x}{A',B',E'}}{0}{r'}{M'}}%
{\dsubst{(\ua{x}{A,B,E})}{r'}{x}}$ and
$\ceqtm{\Coe{x.\ua{x}{A,B,E}}{0}{0}{M}}{M}%
{\dsubst{(\ua{x}{A,B,E})}{0}{x}}$.
\end{lemma}
\begin{proof}
By \cref{lem:expansion} on both sides, it suffices to show (the binary version of)
\[
\coftype{\uain{r'}{M,\Coe{x.B}{0}{r'}{\app{\fst{\dsubst{E}{0}{x}}}{M}}}}%
{\dsubst{(\ua{x}{A,B,E})}{r'}{x}}.
\]
By \cref{rul:ua-form-pre}, $\coftype{M}{\dsubst{A}{0}{x}}$, so
$\coftype{\app{\fst{\dsubst{E}{0}{x}}}{M}}{\dsubst{B}{0}{x}}$ and
$\coftype{\Coe{x.B}{0}{r'}{\dots}}{\dsubst{B}{r'}{x}}$.
Then $\coftype<r'=0>{M}{\dsubst{A}{r'}{x}}$ and
$\ceqtm<r'=0>{\Coe{x.B}{0}{r'}{\dots}}%
{\app{\fst{\dsubst{E}{0}{x}}}{M}}%
{\dsubst{B}{r'}{x}}$ so the first part follows by \cref{rul:ua-intro}. When
$r'=0$, $\ceqtm{\uain{0}{M,\dots}}{M}{\dsubst{(\ua{x}{A,B,E})}{0}{x}}$,
completing the second part.
\end{proof}

\begin{lemma}\label{lem:ua-coe-from-1}
If $\ceqtypek[\Psi,x]<x=0>{A}{A'}$,
$\ceqtypek[\Psi,x]{B}{B'}$,
$\ceqtm[\Psi,x]<x=0>{E}{E'}{\Equiv{A}{B}}$, and
$\ceqtm{N}{N'}{\dsubst{(\ua{x}{A,B,E})}{1}{x}}$, then
$\ceqtm{\Coe{x.\ua{x}{A,B,E}}{1}{r'}{N}}%
{\Coe{x.\ua{x}{A',B',E'}}{1}{r'}{N'}}%
{\dsubst{(\ua{x}{A,B,E})}{r'}{x}}$ and
$\ceqtm{\Coe{x.\ua{x}{A,B,E}}{1}{1}{N}}{N}{\dsubst{(\ua{x}{A,B,E})}{1}{x}}$.
\end{lemma}
\begin{proof}
By \cref{lem:expansion} on both sides, it suffices to show (the binary version of)
$\coftype{\uain{r'}{\fst{O},P}}{\dsubst{(\ua{x}{A,B,E})}{r'}{x}}$ where
\begin{gather*}
O = \fst{\app{\snd{\dsubst{E}{r'}{x}}}{\Coe{x.B}{1}{r'}{N}}} \\
P = \Hcom{\dsubst{B}{r'}{x}}{1}{0}{\Coe{x.B}{1}{r'}{N}}{
  \tube{r'=0}{y.\dapp{\snd{O}}{y}},
  \tube{r'=1}{\_.\Coe{x.B}{1}{r'}{N}}}.
\end{gather*}
By \cref{rul:ua-form-pre}, $\coftype{N}{\dsubst{B}{1}{x}}$, so
$\coftype{\Coe{x.B}{1}{r'}{N}}{\dsubst{B}{r'}{x}}$ and
\[
\coftype{O}{\sigmacl{a}{\dsubst{A}{r'}{x}}%
{\Path{\_.\dsubst{B}{r'}{x}}{\app{\fst{\dsubst{E}{r'}{x}}}{a}}{\Coe{x.B}{1}{r'}{N}}}}.
\]
Therefore $\coftype[\Psi,y]<r'=0>{\dapp{\snd{O}}{y}}{\dsubst{B}{r'}{x}}$ and
$\ceqtm<r'=0>{\dapp{\snd{O}}{1}}{\Coe{x.B}{1}{r'}{N}}{\dsubst{B}{r'}{x}}$, so
by $\cwftypek{\dsubst{B}{r'}{x}}$, $\coftype{P}{\dsubst{B}{r'}{x}}$. We also have
$\coftype<r'=0>{\fst{O}}{\dsubst{A}{r'}{x}}$ and
$\ceqtm<r'=0>{\app{\fst{\dsubst{E}{r'}{x}}}{\fst{O}}}{P}{\dsubst{B}{r'}{x}}$
(by
$\ceqtm<r'=0>{\dapp{\snd{O}}{0}}{\app{\fst{\dsubst{E}{r'}{x}}}{\fst{O}}}{\dsubst{B}{r'}{x}}$)
so the first part follows by \cref{rul:ua-intro}. When $r'=1$,
$\ceqtm{\uain{1}{\fst{O},P}}{P}{\dsubst{(\ua{x}{A,B,E})}{1}{x}}$, but $P \eq
\Coe{x.B}{1}{r'}{N} \eq N$, completing the second part.
\end{proof}

\begin{lemma}\label{lem:ua-coe-from-y}
If $\ceqtypek[\Psi,x]<x=0>{A}{A'}$,
$\ceqtypek[\Psi,x]{B}{B'}$,
$\ceqtm[\Psi,x]<x=0>{E}{E'}{\Equiv{A}{B}}$, and
$\ceqtm{M}{M'}{\dsubst{(\ua{x}{A,B,E})}{y}{x}}$, then
$\ceqtm{\Coe{x.\ua{x}{A,B,E}}{y}{r'}{M}}%
{\Coe{x.\ua{x}{A',B',E'}}{y}{r'}{M'}}%
{\dsubst{(\ua{x}{A,B,E})}{r'}{x}}$ and
$\ceqtm{\Coe{x.\ua{x}{A,B,E}}{y}{y}{M}}{M}%
{\dsubst{(\ua{x}{A,B,E})}{y}{x}}$.
\end{lemma}
\begin{proof}
We apply coherent expansion to $\Coe{x.\ua{x}{A,B,E}}{y}{r'}{M}$ with the family
$\Coe{x.\ua{x}{\td{A}{\psi},\td{B}{\psi},\td{E}{\psi}}}{\e}{\td{r'}{\psi}}{\td{M}{\psi}}$
when $\td{y}{\psi} = \e$ and
$\td{(\uain{r'}{\fst{R},\Hcom{\dsubst{B}{r'}{x}}{1}{0}{\dsubst{P}{r'}{x}}{\etc{T}}})}{\psi}$
otherwise, where
\begin{gather*}
O_\e = \uaproj{w}{\Coe{x.\ua{x}{A,B,E}}{\e}{w}{M},\fst{\dsubst{E}{w}{x}}} \\
P = \Com{x.B}{y}{x}{\uaproj{y}{M,\fst{\dsubst{E}{y}{x}}}}{\etc{\tube{y=\e}{w.O_\e}}} \\
Q_\e[a] = \pair%
  {\Coe{y.\dsubst{A}{0}{x}}{\e}{y}{a}}%
  {\dlam{z}{\Com{y.\dsubst{B}{0}{x}}{\e}{y}{\dsubst{\dsubst{P}{0}{x}}{\e}{y}}%
    {\etc{U}}}} \\
\etc{U} =
  \tube{z=0}{y.\app{\fst{\dsubst{E}{0}{x}}}{\Coe{y.\dsubst{A}{0}{x}}{\e}{y}{a}}},
  \tube{z=1}{y.\dsubst{P}{0}{x}} \\
R = \dapp{\app{\app{\snd{\app{\snd{\dsubst{E}{0}{x}}}{\dsubst{P}{0}{x}}}}{Q_0[\dsubst{M}{0}{y}]}}%
  {Q_1[\dsubst{(\Coe{x.\ua{x}{A,B,E}}{1}{0}{M})}{1}{y}]}}{y} \\
\etc{T} =
  \etc{\tube{y=\e}{\_.\dsubst{O_\e}{r'}{w}}},
  \tube{y=r'}{\_.\uaproj{r'}{M,\fst{\dsubst{E}{r'}{x}}}},
  \tube{r'=0}{z.\dapp{\snd{R}}{z}}.
\end{gather*}
Consider $\psi=\id$.
\begin{enumerate}
\item $\coftype[\Psi,w]<y=\e>{O_\e}{\dsubst{B}{w}{x}}$
by $\coftype[\Psi,w]<y=\e>{\Coe{x.\ua{x}{A,B,E}}{\e}{w}{M}}%
{\ua{w}{\dsubst{A}{w}{x},\dots}}$ (by
$\coftype{M}{\ua{y}{\dsubst{A}{y}{x},\dots}}$) and
$\coftype[\Psi,w]<w=0>{\fst{\dsubst{E}{w}{x}}}{\arr{\dsubst{A}{w}{x}}{\dsubst{B}{w}{x}}}$.

\item $\coftype[\Psi,x]{P}{B}$ by
$\coftype{\uaproj{y}{M,\fst{\dsubst{E}{y}{x}}}}{\dsubst{B}{y}{x}}$ and
$\ceqtm<y=\e>{\dsubst{O_\e}{y}{w}}{\uaproj{y}{M,\fst{\dsubst{E}{y}{x}}}}%
{\dsubst{B}{y}{x}}$.

\item Let $C =
\sigmacl{a'}{\dsubst{A}{0}{x}}{\Path{\_.\dsubst{B}{0}{x}}{\app{\fst{\dsubst{E}{0}{x}}}{a'}}{\dsubst{P}{0}{x}}}$.
Then $\coftype{Q_\e[a]}{C}$ for any
$\coftype{a}{\dsubst{\dsubst{A}{0}{x}}{\e}{y}}$ with $y\fresh a$ and
$\ceqtm{\dsubst{\dsubst{P}{0}{x}}{\e}{y}}%
{\app{\fst{\dsubst{\dsubst{E}{0}{x}}{\e}{y}}}{a}}%
{\dsubst{\dsubst{B}{0}{x}}{\e}{y}}$, because
$\coftype{\Coe{y.\dsubst{A}{0}{x}}{\e}{y}{a}}{\dsubst{A}{0}{x}}$ and by
\begin{enumerate}
\item
$\coftype{\dsubst{\dsubst{P}{0}{x}}{\e}{y}}{\dsubst{\dsubst{B}{0}{x}}{\e}{y}}$,
\item
$\coftype{\app{\fst{\dsubst{E}{0}{x}}}{\Coe{y.\dsubst{A}{0}{x}}{\e}{y}{a}}}{\dsubst{B}{0}{x}}$,
\item
$\ceqtm{\dsubst{\dsubst{P}{0}{x}}{\e}{y}}%
{\app{\fst{\dsubst{\dsubst{E}{0}{x}}{\e}{y}}}{\Coe{y.\dsubst{A}{0}{x}}{\e}{\e}{a}}}%
{\dsubst{\dsubst{B}{0}{x}}{\e}{y}}$, and
\item $\coftype{\dsubst{P}{0}{x}}{\dsubst{B}{0}{x}}$,
\end{enumerate}
we have
$\coftype{\dlam{z}{\Com}}%
{\Path{\_.\dsubst{B}{0}{x}}%
  {\app{\fst{\dsubst{E}{0}{x}}}{\Coe{y.\dsubst{A}{0}{x}}{\e}{y}{a}}}%
  {\dsubst{P}{0}{x}}}$.

\item
$\coftype{Q_0[\dsubst{M}{0}{y}]}{C}$ because
$\coftype{\dsubst{M}{0}{y}}{\dsubst{\dsubst{A}{0}{x}}{0}{y}}$ and
$\dsubst{\dsubst{P}{0}{x}}{0}{y}$ $\eq$
$\dsubst{\dsubst{O_0}{0}{w}}{0}{y}$ $\eq$
$\app{\fst{\dsubst{\dsubst{E}{0}{x}}{0}{y}}}{\dsubst{M}{0}{y}}$.

\item
$\coftype{Q_1[\dsubst{(\Coe{x.\ua{x}{A,B,E}}{1}{0}{M})}{1}{y}]}{C}$ because
$\coftype{\dsubst{(\Coe{x.\ua{x}{A,B,E}}{1}{0}{M})}{1}{y}}%
{\dsubst{\dsubst{A}{0}{x}}{1}{y}}$ (by
$\coftype{\dsubst{M}{1}{y}}{\dsubst{\dsubst{B}{1}{x}}{1}{y}}$) and
$\dsubst{\dsubst{P}{0}{x}}{1}{y}$ $\eq$
$\dsubst{\dsubst{O_1}{0}{w}}{1}{y}$ which in turn equals
$\app{\fst{\dsubst{\dsubst{E}{0}{x}}{1}{y}}}{\dsubst{(\Coe{x.\ua{x}{A,B,E}}{1}{0}{M})}{1}{y}}$.

\item $\coftype{R}{C}$ because
$\coftype{\snd{\app{\snd{\dsubst{E}{0}{x}}}{\dsubst{P}{0}{x}}}}%
{(\picl{c}{C}{\picl{c'}{C}{\Path{\_.C}{c}{c'}}})}$ and we further apply this to
$Q_0[\dsubst{M}{0}{y}]$, $Q_1[\dsubst{(\Coe{x.\ua{x}{A,B,E}}{1}{0}{M})}{1}{y}]$,
and $y$.

\item
$\coftype{\Hcom{\dsubst{B}{r'}{x}}{1}{0}{\dsubst{P}{r'}{x}}{\etc{T}}}{\dsubst{B}{r'}{x}}$
because
\begin{enumerate}
\item $\coftype<y=\e>{\dsubst{O_\e}{r'}{w}}{\dsubst{B}{r'}{x}}$,

\item
$\coftype<y=r'>{\uaproj{r'}{M,\fst{\dsubst{E}{r'}{x}}}}{\dsubst{B}{r'}{x}}$,

\item $\coftype[\Psi,z]<r'=0>{\dapp{\snd{R}}{z}}{\dsubst{B}{r'}{x}}$ by
$\coftype[\Psi,z]{\dapp{\snd{R}}{z}}{\dsubst{B}{0}{x}}$,

\item $\coftype{\dsubst{P}{r'}{x}}{\dsubst{B}{r'}{x}}$,

\item $\ceqtm<y=\e>{\dsubst{P}{r'}{x}}{\dsubst{O_\e}{r'}{w}}{\dsubst{B}{r'}{x}}$,

\item $\ceqtm<y=r'>{\dsubst{P}{r'}{x}}%
{\uaproj{r'}{M,\fst{\dsubst{E}{r'}{x}}}}{\dsubst{B}{r'}{x}}$,

\item $\ceqtm<r'=0>{\dsubst{P}{r'}{x}}{\dapp{\snd{R}}{1}}{\dsubst{B}{r'}{x}}$
by $\ceqtm{\dapp{\snd{R}}{1}}{\dsubst{P}{0}{x}}{\dsubst{B}{0}{x}}$,

\item $\ceqtm<y=\e,y=r'>{\dsubst{O_\e}{r'}{w}}%
{\uaproj{r'}{M,\fst{\dsubst{E}{r'}{x}}}}{\dsubst{B}{r'}{x}}$,

\item $\ceqtm[\Psi,z]<y=0,r'=0>{\dsubst{O_0}{r'}{w}}%
{\dapp{\snd{R}}{z}}{\dsubst{B}{r'}{x}}$ by
$\dapp{\snd{R}}{z}$ $\eq$
$\dapp{\snd{Q_0[\dsubst{M}{0}{y}]}}{z}$ $\eq$
$\dapp{(\dlam{z}{\dsubst{\dsubst{P}{0}{x}}{0}{y}})}{z}$ $\eq$
$\dsubst{O_0}{0}{w}$,

\item $\ceqtm[\Psi,z]<y=1,r'=0>{\dsubst{O_1}{r'}{w}}%
{\dapp{\snd{R}}{z}}{\dsubst{B}{r'}{x}}$ because we have
$\dapp{\snd{R}}{z}$ $\eq$
$\dapp{\snd{Q_1[\dsubst{(\Coe{x.\ua{x}{A,B,E}}{1}{0}{M})}{1}{y}]}}{z}$ $\eq$
$\dapp{(\dlam{z}{\dsubst{\dsubst{P}{0}{x}}{1}{y}})}{z}$ $\eq$
$\dsubst{O_1}{0}{w}$, and

\item
$\ceqtm[\Psi,z]<y=r',r'=0>{\uaproj{r'}{M,\fst{\dsubst{E}{r'}{x}}}}%
{\dapp{\snd{R}}{z}}{\dsubst{B}{r'}{x}}$ because
$\dapp{\snd{R}}{z}$ $\eq$
$\dapp{\snd{Q_0[\dsubst{M}{0}{y}]}}{z}$ $\eq$
$\dsubst{\dsubst{P}{0}{x}}{0}{y}$ $\eq$
$\uaproj{y}{M,\fst{\dsubst{E}{y}{x}}}$.
\end{enumerate}

\item
$\coftype{\uain{r'}{\fst{R},\Hcom{\dsubst{B}{r'}{x}}}}{\ua{r'}{\dsubst{A}{r'}{x},\dots}}$
because $\coftype<r'=0>{\fst{R}}{\dsubst{A}{0}{x}}$,
$\coftype{\Hcom{\dsubst{B}{r'}{x}}}{\dsubst{B}{r'}{x}}$, and
$\ceqtm<r'=0>{\app{\fst{\dsubst{E}{r'}{x}}}{\fst{R}}}%
{\Hcom{\dsubst{B}{r'}{x}}}{\dsubst{B}{r'}{x}}$ (by
$\Hcom{\dsubst{B}{r'}{x}}$ $\eq$ $\dapp{\snd{R}}{0}$).
\end{enumerate}

When $\td{y}{\psi}=y'$, coherence is immediate.
When $\td{y}{\psi}=\e$, we prove coherence by \cref{rul:ua-eta}, using
$\ceqtm<y=\e>{\Hcom{\dsubst{B}{r'}{x}}}%
{\uaproj{r'}{\Coe{x.\ua{x}{A,B,E}}{\e}{r'}{M},\fst{\dsubst{E}{r'}{x}}}}%
{\dsubst{B}{r'}{x}}$ (by $\eq$ $\dsubst{O_\e}{r'}{w}$),
$\ceqtm<y=0,r'=0>{\fst{R}}{M}{\dsubst{A}{r'}{x}}$ (by $\eq$
$\fst{Q_0[\dsubst{M}{0}{y}]}$), and
$\ceqtm<y=1,r'=0>{\fst{R}}{\Coe{x.\ua{x}{A,B,E}}{1}{0}{M}}{\dsubst{A}{r'}{x}}$
(by $\eq$ $\fst{Q_1[\dsubst{(\Coe{x.\ua{x}{A,B,E}}{1}{0}{M})}{1}{y}]}$).
Therefore \cref{lem:cohexp-ceqtm} applies, and the first part follows by the
same argument on the right side.

The second part follows by \cref{rul:ua-eta},
$\ceqtm<y=r'>{\Hcom{\dsubst{B}{r'}{x}}}%
{\uaproj{r'}{M,\fst{\dsubst{E}{r'}{x}}}}%
{\dsubst{B}{r'}{x}}$, and
$\ceqtm<y=r',r'=0>{\fst{R}}{M}{\dsubst{A}{r'}{x}}$ (as calculated previously).
\end{proof}

\begin{rul}[Kan type formation]
~\begin{enumerate}
\item If $\cwftypek{A}$ then $\ceqtypek{\ua{0}{A,B,E}}{A}$.
\item If $\cwftypek{B}$ then $\ceqtypek{\ua{1}{A,B,E}}{B}$.
\item If $\ceqtypek<r=0>{A}{A'}$,
$\ceqtypek{B}{B'}$, and
$\ceqtm<r=0>{E}{E'}{\Equiv{A}{B}}$, then
$\ceqtypek{\ua{r}{A,B,E}}{\ua{r}{A',B',E'}}$.
\end{enumerate}
\end{rul}
\begin{proof}
Parts (1--2) follow from \cref{lem:expansion}. For part (3), we check the
Kan conditions.

($\Hcom$) For any $\psitd$, consider a valid composition scenario in
$\ua{\td{r}{\psi}}{\td{A}{\psi},\td{B}{\psi},\td{E}{\psi}}$. If $\td{r}{\psi}=0$
(resp., $1$) then the composition is in $\td{A}{\psi}$ (resp., $\td{B}{\psi}$)
and the $\Hcom$ Kan conditions follow from
$\ceqtypek[\Psi']{\td{A}{\psi}}{\td{A'}{\psi}}$ (resp.,
$\ceqtypek[\Psi']{\td{B}{\psi}}{\td{B'}{\psi}}$). Otherwise, $\td{r}{\psi}=x$ and
the $\Hcom$ Kan conditions follow from \cref{lem:ua-hcom} at
$\ceqtypek[\Psi']<x=0>{\td{A}{\psi}}{\td{A'}{\psi}}$,
$\ceqtypek[\Psi']{\td{B}{\psi}}{\td{B'}{\psi}}$, and
$\ceqtm[\Psi']<x=0>{\td{E}{\psi}}{\td{E'}{\psi}}{\Equiv{\td{A}{\psi}}{\td{B}{\psi}}}$.

($\Coe$) Consider any $\tds{(\Psi',x)}{\psi}{\Psi}$ and
$\ceqtm[\Psi']{M}{M'}{\dsubst{\td{(\ua{r}{A,B,E})}{\psi}}{s}{x}}$. If
$\td{r}{\psi}=0$ (resp., $1$) then the $\Coe$ Kan conditions follow from
$\ceqtypek[\Psi',x]{\td{A}{\psi}}{\td{A'}{\psi}}$
(resp., $\ceqtypek[\Psi',x]{\td{B}{\psi}}{\td{B'}{\psi}}$).
If $\td{r}{\psi}=y\neq x$, then the $\Coe$ Kan conditions follow from
\cref{lem:ua-coe-xy}. Otherwise, $\td{r}{\psi}=x$; the result follows
from \cref{lem:ua-coe-from-0} if $s=0$,
from \cref{lem:ua-coe-from-1} if $s=1$, and
from \cref{lem:ua-coe-from-y} if $s=y$.
\end{proof}

\subsection{Composite types}

Unlike the other type formers, $\Fcom$s are only pretypes when their
constituents are Kan types. (For this reason, in \cref{sec:typesys} we only
close $\pre\tau_i$ under $\Fcom$s of types from $\Kan\tau_i$.) The results of
this section hold in $\tau=\Kan\mu(\nu)$ for any cubical type system $\nu$, and
therefore in each $\pre\tau_i$ as well. In this section, we will say that
$A,\sys{r_i=r_i'}{y.B_i}$ and $A',\sys{r_i=r_i'}{y.B_i'}$ are \emph{(equal) type
compositions $r\rightsquigarrow r'$} whenever:
\begin{enumerate}
\item $\etc{r_i=r_i'}$ is valid,
\item $\ceqtypek{A}{A'}$,
\item $\ceqtypek[\Psi,y]<r_i=r_i',r_j=r_j'>{B_i}{B_j'}$ for any $i,j$, and
\item $\ceqtypek<r_i=r_i'>{\dsubst{B_i}{r}{y}}{A}$ for any $i$.
\end{enumerate}

\begin{lemma}\label{lem:fcom-preform}
If $A,\sys{r_i=r_i'}{y.B_i}$ and $A',\sys{r_i=r_i'}{y.B_i'}$ are equal type
compositions $r\rightsquigarrow r'$, then
\begin{enumerate}
\item $\PTy(\tau)(\Psi,\Fcom{r}{r'}{A}{\sys{r_i=r_i'}{y.B_i}},
\Fcom{r}{r'}{A'}{\sys{r_i=r_i'}{y.B_i'}},\_)$,
\item if $r=r'$ then $\ceqtypek{\Fcom{r}{r}{A}{\sys{r_i=r_i'}{y.B_i}}}{A}$, and
\item if $r_i = r_i'$ then
$\ceqtypek{\Fcom{r}{r'}{A}{\sys{r_i=r_i'}{B_i}}}{\dsubst{B_i}{r'}{y}}$.
\end{enumerate}
\end{lemma}
\begin{proof}
Part (1) is precisely the statement of \cref{lem:C-prekan}, applied to the
context-indexed PER $\{(\Psi,A_0,B_0)\mid\tau(\Psi,A_0,B_0,\_)\}$ instead of
$\vper{\C}(\Psi)$; as the $\Fcom$ structure of these PERs is defined
identically, the same proof applies. Part (2) is immediate by
\cref{lem:expansion}. For part (3), if $r=r'$, the result follows by
\cref{lem:expansion} and $\ceqtypek<r_i=r_i'>{\dsubst{B_i}{r}{y}}{A}$.
Otherwise, there is some least $j$ such that $r_j=r_j'$. Apply coherent
expansion to the left side with family
\[\begin{cases}
\td{A}{\psi} & \text{$\td{r}{\psi}=\td{r'}{\psi}$} \\
\td{\dsubst{B_j}{r'}{y}}{\psi} &
\text{$\td{r}{\psi}\neq\td{r'}{\psi}$, $\td{r_j}{\psi}=\td{r_j'}{\psi}$, and
$\forall k<j,\td{r_k}{\psi}\neq\td{r_k'}{\psi}$}.
\end{cases}\]
If $\td{r}{\psi}=\td{r'}{\psi}$ then
$\ceqtypek[\Psi']{\td{\dsubst{B_j}{r}{y}}{\psi}}{\td{A}{\psi}}$. If
$\td{r}{\psi}\neq\td{r'}{\psi}$, there is some least $k$ such that
$\td{r_k}{\psi}=\td{r_k'}{\psi}$; then
$\ceqtypek[\Psi']{\td{\dsubst{B_j}{r'}{y}}{\psi}}{\td{\dsubst{B_k}{r'}{y}}{\psi}}$.
By \cref{lem:cohexp-ceqtypek},
$\ceqtypek{\Fcom}{\dsubst{B_j}{r'}{y}}$, and part (3) follows by
$\ceqtypek{\dsubst{B_i}{r'}{y}}{\dsubst{B_j}{r'}{y}}$.
\end{proof}

\begin{lemma}\label{lem:fcom-preintro}
If
\begin{enumerate}
\item $A,\sys{r_i=r_i'}{y.B_i}$ is a type composition $r\rightsquigarrow r'$,
\item $\ceqtm{M}{M'}{A}$,
\item $\ceqtm<r_i=r_i',r_j=r_j'>{N_i}{N_j'}{\dsubst{B_i}{r'}{y}}$ for any $i,j$, and
\item $\ceqtm<r_i=r_i'>{\Coe{y.B_i}{r'}{r}{N_i}}{M}{A}$ for any $i$,
\end{enumerate}
then $\Tm(\vper{\Fcom{r}{r'}{A}{\sys{r_i=r_i'}{y.B_i}}})(
\Kbox{r}{r'}{M}{\sys{r_i=r_i'}{N_i}},
\Kbox{r}{r'}{M'}{\sys{r_i=r_i'}{N_i'}})$.
\end{lemma}
\begin{proof}
We focus on the unary case; the binary case follows similarly. For any
$\tds{\Psi_1}{\psi_1}{\Psi}$ and $\tds{\Psi_2}{\psi_2}{\Psi_1}$ we must show
$\td{\Kbox}{\psi_1}\evals X_1$ and
$\lift{\vper{\Fcom{r}{r'}{A}{\sys{r_i=r_i'}{y.B_i}}}}_{\psi_1\psi_2}
(\td{X_1}{\psi_2},\td{\Kbox}{\psi_1\psi_2})$. We proceed by cases on the first
step taken by $\td{\Kbox}{\psi_1}$ and $\td{\Kbox}{\psi_1\psi_2}$.

\begin{enumerate}
\item $\td{r}{\psi_1}=\td{r'}{\psi_1}$.

Then $\td{\Kbox}{\psi_1}\ssteps \td{M}{\psi_1}$,
$\vper{\Fcom}_{\psi_1\psi_2} = \vper{A}_{\psi_1\psi_2}$ by
\cref{lem:fcom-preform}, and
$\lift{\vper{A}}_{\psi_1\psi_2}(\td{X_1}{\psi_2},\td{M}{\psi_1\psi_2})$ by
$\coftype{M}{A}$.

\item $\td{r}{\psi_1}\neq\td{r'}{\psi_1}$,
$\td{r_j}{\psi_1}=\td{r_j'}{\psi_1}$ (where this is the least such $j$), and
$\td{r}{\psi_1\psi_2}=\td{r'}{\psi_1\psi_2}$.

Then $\td{\Kbox}{\psi_1}\steps \td{N_j}{\psi_1}$,
$\td{\Kbox}{\psi_1\psi_2}\steps \td{M}{\psi_1\psi_2}$, and
$\vper{\Fcom}_{\psi_1\psi_2} = \vper{A}_{\psi_1\psi_2}$ by
\cref{lem:fcom-preform}. By
$\ceqtypek[\Psi_2]{\td{\dsubst{B_j}{r'}{y}}{\psi_1\psi_2}}{\td{A}{\psi_1\psi_2}}$
and $\coftype[\Psi_1]{\td{N_j}{\psi_1}}{\td{\dsubst{B_j}{r'}{y}}{\psi_1}}$ at
$\id[\Psi_1],\psi_2$ we have
$\lift{\vper{A}}_{\psi_1\psi_2}(\td{X_1}{\psi_2},\td{N_j}{\psi_1\psi_2})$.
We also have
$\lift{\vper{A}}_{\psi_1\psi_2}(\td{N_j}{\psi_1\psi_2},\td{M}{\psi_1\psi_2})$ by
$\ceqtm[\Psi_2]{\td{(\Coe{y.B_j}{r'}{r}{N_j})}{\psi_1\psi_2}}%
{\td{M}{\psi_1\psi_2}}{\td{A}{\psi_1\psi_2}}$ and
$\ceqtm[\Psi_2]{\td{(\Coe{y.B_j}{r'}{r}{N_j})}{\psi_1\psi_2}}%
{\td{N_j}{\psi_1\psi_2}}{\td{A}{\psi_1\psi_2}}$; the result follows by
transitivity.

\item $\td{r}{\psi_1}\neq\td{r'}{\psi_1}$,
$\td{r_i}{\psi_1}=\td{r_i'}{\psi_1}$ (least such),
$\td{r}{\psi_1\psi_2}\neq\td{r'}{\psi_1\psi_2}$, and
$\td{r_j}{\psi_1\psi_2}=\td{r_j'}{\psi_1\psi_2}$ (least such).

Then $\td{\Kbox}{\psi_1}\steps \td{N_i}{\psi_1}$,
$\td{\Kbox}{\psi_1\psi_2}\steps \td{N_j}{\psi_1\psi_2}$, and
$\vper{\Fcom}_{\psi_1\psi_2} = \vper{\dsubst{B_i}{r'}{y}}_{\psi_1\psi_2}$ by
\cref{lem:fcom-preform}. The result follows by
$\coftype[\Psi_1]{\td{N_i}{\psi_1}}{\td{\dsubst{B_i}{r'}{y}}{\psi_1}}$ and
$\ceqtm[\Psi_2]{\td{N_i}{\psi_1\psi_2}}{\td{N_j}{\psi_1\psi_2}}%
{\td{\dsubst{B_i}{r'}{y}}{\psi_1\psi_2}}$.

\item $\td{r}{\psi_1}\neq\td{r'}{\psi_1}$,
$\td{r_i}{\psi_1}\neq\td{r_i'}{\psi_1}$ for all $i$, and
$\td{r}{\psi_1\psi_2} = \td{r'}{\psi_1\psi_2}$.

Then $\isval{\td{\Kbox}{\psi_1}}$,
$\td{\Kbox}{\psi_1\psi_2}\steps \td{M}{\psi_1\psi_2}$,
$\vper{\Fcom}_{\psi_1\psi_2} = \vper{A}_{\psi_1\psi_2}$ by
\cref{lem:fcom-preform}, and the result follows by $\coftype{M}{A}$.

\item $\td{r}{\psi_1}\neq\td{r'}{\psi_1}$,
$\td{r_i}{\psi_1}\neq\td{r_i'}{\psi_1}$ for all $i$,
$\td{r}{\psi_1\psi_2}\neq\td{r'}{\psi_1\psi_2}$, and
$\td{r_j}{\psi_1\psi_2}=\td{r_j'}{\psi_1\psi_2}$ (the least such $j$).

Then $\isval{\td{\Kbox}{\psi_1}}$,
$\td{\Kbox}{\psi_1\psi_2}\steps \td{N_j}{\psi_1\psi_2}$,
$\vper{\Fcom}_{\psi_1\psi_2} = \vper{\dsubst{B_i}{r'}{y}}_{\psi_1\psi_2}$ by
\cref{lem:fcom-preform}, and the result follows by
$\coftype[\Psi_2]{\td{N_j}{\psi_1\psi_2}}{\td{\dsubst{B_i}{r'}{y}}{\psi_1\psi_2}}$.

\item $\td{r}{\psi_1}\neq\td{r'}{\psi_1}$,
$\td{r_i}{\psi_1}\neq\td{r_i'}{\psi_1}$ for all $i$, and
$\td{r}{\psi_1\psi_2}\neq\td{r'}{\psi_1\psi_2}$, and
$\td{r_j}{\psi_1\psi_2}\neq\td{r_j'}{\psi_1\psi_2}$ for all $j$.

Then $\isval{\td{\Kbox}{\psi_1}}$ and $\isval{\td{\Kbox}{\psi_1\psi_2}}$, and
the result follows by the definition of $\vper{\Fcom}$.
\qedhere
\end{enumerate}
\end{proof}

\begin{rul}[Pretype formation]\label{rul:fcom-form-pre}
If $A,\sys{r_i=r_i'}{y.B_i}$ and $A',\sys{r_i=r_i'}{y.B_i'}$ are equal type
compositions $r\rightsquigarrow r'$, then
\begin{enumerate}
\item $\ceqtypep{\Fcom{r}{r'}{A}{\sys{r_i=r_i'}{y.B_i}}}%
{\Fcom{r}{r'}{A'}{\sys{r_i=r_i'}{y.B_i'}}}$,
\item if $r=r'$ then $\ceqtypep{\Fcom{r}{r}{A}{\sys{r_i=r_i'}{y.B_i}}}{A}$, and
\item if $r_i = r_i'$ then
$\ceqtypep{\Fcom{r}{r'}{A}{\sys{r_i=r_i'}{B_i}}}{\dsubst{B_i}{r'}{y}}$.
\end{enumerate}
\end{rul}
\begin{proof}
For part (1), by \cref{lem:fcom-preform} it suffices to show
$\Coh(\vper{\Fcom})$. Let $\vper{\Fcom}_\psi(M_0,N_0)$ for any $\psitd$. If
$\td{r}{\psi}=\td{r'}{\psi}$ then $\Tm(\td{\vper{\Fcom}}{\psi})(M_0,N_0)$ by
$\td{\vper{\Fcom}}{\psi}=\td{\vper{A}}{\psi}$ and $\Coh(\vper{A})$. Similarly,
if $\td{r_i}{\psi}=\td{r_i'}{\psi}$ for some $i$, then
$\Tm(\td{\vper{\Fcom}}{\psi})(M_0,N_0)$ by
$\td{\vper{\Fcom}}{\psi}=\vper{\td{\dsubst{B_i}{r'}{y}}{\psi}}$ and
$\Coh(\vper{\td{\dsubst{B_i}{r'}{y}}{\psi}})$.
If $\td{r}{\psi}\neq\td{r'}{\psi}$ and $\td{r_i}{\psi}\neq\td{r_i'}{\psi}$, then
$M_0$ and $N_0$ are $\Kbox$es and the result follows by
\cref{lem:fcom-preintro}.

Parts (2--3) are immediate by \cref{lem:fcom-preform}.
\end{proof}

\begin{rul}[Introduction]\label{rul:fcom-intro}
If
\begin{enumerate}
\item $A,\sys{r_i=r_i'}{y.B_i}$ is a type composition $r\rightsquigarrow r'$,
\item $\ceqtm{M}{M'}{A}$,
\item $\ceqtm<r_i=r_i',r_j=r_j'>{N_i}{N_j'}{\dsubst{B_i}{r'}{y}}$ for any $i,j$, and
\item $\ceqtm<r_i=r_i'>{\Coe{y.B_i}{r'}{r}{N_i}}{M}{A}$ for any $i$,
\end{enumerate}
then
\begin{enumerate}
\item $\ceqtm{\Kbox*{r_i=r_i'}}{\Kbox{r}{r'}{M'}{\sys{r_i=r_i'}{N_i'}}}%
{\Fcom{r}{r'}{A}{\sys{r_i=r_i'}{y.B_i}}}$;
\item if $r=r'$ then $\ceqtm{\Kbox*{r_i=r_i'}}{M}{A}$; and
\item if $r_i = r_i'$ then $\ceqtm{\Kbox*{r_i=r_i'}}{N_i}{\dsubst{B_i}{r'}{y}}$.
\end{enumerate}
\end{rul}
\begin{proof}
Part (1) is immediate by \cref{lem:fcom-preintro,rul:fcom-form-pre}; part (2) is
immediate by \cref{lem:expansion}. For part (3), if $r=r'$, the result follows
by \cref{lem:expansion}. Otherwise, there is a least $j$ such that $r_j=r_j'$,
and we apply coherent expansion to the left side with family
\[\begin{cases}
\td{M}{\psi} & \text{$\td{r}{\psi}=\td{r'}{\psi}$} \\
\td{N_k}{\psi} &
\text{$\td{r}{\psi}\neq\td{r'}{\psi}$, $\td{r_k}{\psi}=\td{r_k'}{\psi}$, and
$\forall k'<k,\td{r_{k'}}{\psi}\neq\td{r_{k'}'}{\psi}$}.
\end{cases}\]
If $\td{r}{\psi}=\td{r'}{\psi}$ then
$\ceqtm[\Psi']{\td{M}{\psi}}{\td{N_j}{\psi}}{\td{\dsubst{B_i}{r'}{y}}{\psi}}$ by
$\ceqtm[\Psi']{\td{M}{\psi}}{\td{(\Coe{y.B_j}{r'}{r}{N_j})}{\psi}}{\td{A}{\psi}}$,
$\ceqtm[\Psi']{\td{(\Coe{y.B_j}{r'}{r}{N_j})}{\psi}}{\td{N_j}{\psi}}%
{\td{\dsubst{B_i}{r'}{y}}{\psi}}$, and
$\ceqtypek[\Psi']{\td{\dsubst{B_i}{r'}{y}}{\psi}}{\td{A}{\psi}}$. If
$\td{r}{\psi}\neq\td{r'}{\psi}$ then
$\ceqtm[\Psi']{\td{N_k}{\psi}}{\td{N_j}{\psi}}{\td{\dsubst{B_i}{r'}{y}}{\psi}}$ by
$\ceqtm[\Psi']{\td{N_k}{\psi}}{\td{N_j}{\psi}}{\td{\dsubst{B_j}{r'}{y}}{\psi}}$
and $\ceqtypek[\Psi',y]{\td{B_i}{\psi}}{\td{B_j}{\psi}}$.
Thus by \cref{lem:cohexp-ceqtypek} we have
$\ceqtm{\Fcom}{N_j}{\dsubst{B_i}{r'}{y}}$, and part (3) follows by
$\ceqtm{N_j}{N_i}{\dsubst{B_i}{r'}{y}}$.
\end{proof}

\begin{rul}[Elimination]\label{rul:fcom-elim}
If $A,\sys{r_i=r_i'}{y.B_i}$ and $A',\sys{r_i=r_i'}{y.B_i'}$ are equal type
compositions $r\rightsquigarrow r'$ and
$\ceqtm{M}{M'}{\Fcom{r}{r'}{A}{\sys{r_i=r_i'}{y.B_i}}}$, then
\begin{enumerate}
\item $\ceqtm{\Kcap{r}{r'}{M}{\sys{r_i=r_i'}{y.B_i}}}%
{\Kcap{r}{r'}{M'}{\sys{r_i=r_i'}{y.B_i'}}}{A}$;
\item if $r=r'$ then $\ceqtm{\Kcap{r}{r'}{M}{\sys{r_i=r_i'}{y.B_i}}}{M}{A}$; and
\item if $r_i=r_i'$ then
$\ceqtm{\Kcap{r}{r'}{M}{\sys{r_i=r_i'}{y.B_i}}}{\Coe{y.B_i}{r'}{r}{M}}{A}$.
\end{enumerate}
\end{rul}
\begin{proof}
Part (2) is immediate by \cref{lem:expansion,rul:fcom-form-pre}. For part (3),
if $r=r'$ then the result follows by part (2), $\cwftypek[\Psi,y]{B_i}$, and
$\ceqtypek{\dsubst{B_i}{r}{y}}{A}$. Otherwise, $r\neq r'$ and there is a least
$j$ such that $r_j=r_j'$. Apply coherent expansion to the left side with family
\[\begin{cases}
\td{M}{\psi} & \text{$\td{r}{\psi}=\td{r'}{\psi}$} \\
\Coe{y.\td{B_k}{\psi}}{\td{r'}{\psi}}{\td{r}{\psi}}{\td{M}{\psi}} &
\text{$\td{r}{\psi}\neq\td{r'}{\psi}$, $\td{r_k}{\psi}=\td{r_k'}{\psi}$, and
$\forall i<k,\td{r_i}{\psi}\neq\td{r_i'}{\psi}$}
\end{cases}\]
When $\td{r}{\psi}=\td{r'}{\psi}$,
$\ceqtm[\Psi']{\td{(\Coe{y.B_j}{r'}{r}{M})}{\psi}}{\td{M}{\psi}}{\td{A}{\psi}}$
by $\coftype{M}{\dsubst{B_j}{r'}{y}}$ (by \cref{rul:fcom-form-pre}),
$\cwftypek[\Psi,y]{B_j}$, and
$\ceqtypek[\Psi']{\td{\dsubst{B_j}{r}{y}}{\psi}}{\td{A}{\psi}}$.
When $\td{r}{\psi}\neq\td{r'}{\psi}$ and
$\td{r_k}{\psi}=\td{r_k'}{\psi}$ where $k$ is the least such, we have
$\ceqtm[\Psi']{\td{(\Coe{y.B_j}{r'}{r}{M})}{\psi}}%
{\Coe{y.\td{B_k}{\psi}}{\td{r'}{\psi}}{\td{r}{\psi}}{\td{M}{\psi}}}%
{\td{A}{\psi}}$ by $\ceqtypek[\Psi',y]{\td{B_j}{\psi}}{\td{B_k}{\psi}}$
and $\ceqtypek[\Psi']{\td{\dsubst{B_j}{r}{y}}{\psi}}{\td{A}{\psi}}$.
We conclude that $\ceqtm{\Kcap}{\Coe{y.B_j}{r'}{r}{M}}{A}$ by
\cref{lem:cohexp-ceqtm}, and part (3) follows by
$\ceqtm{\Coe{y.B_j}{r'}{r}{M}}{\Coe{y.B_i}{r'}{r}{M}}{A}$.

For part (1), if $r=r'$ or $r_i=r_i'$ then the result follows by the previous
parts. If $r\neq r'$ and $r_i\neq r_i'$ for all $i$, then for any $\psitd$,
$\ceqtm[\Psi']{\td{M}{\psi}}{\Kbox{r}{r'}{O_\psi}{\sys{\td{\xi_i}{\psi}}{N_{i,\psi}}}}%
{\td{\Fcom}{\psi}}$ by \cref{lem:coftype-evals-ceqtm}. Apply coherent expansion
to the left side with family
\[\begin{cases}
\td{M}{\psi} & \text{$\td{r}{\psi}=\td{r'}{\psi}$} \\
\Coe{y.\td{B_j}{\psi}}{\td{r'}{\psi}}{\td{r}{\psi}}{\td{M}{\psi}} &
\text{$\td{r}{\psi}\neq\td{r'}{\psi}$, $\td{r_j}{\psi}=\td{r_j'}{\psi}$, and
$\forall i<j,\td{r_i}{\psi}\neq\td{r_i'}{\psi}$} \\
O_\psi & \text{$\td{r}{\psi}\neq\td{r'}{\psi}$ and
$\forall i,\td{r_i}{\psi}\neq\td{r_i'}{\psi}$} \\
&\quad\text{where $\td{M}{\psi}\evals
\Kbox{\td{r}{\psi}}{\td{r'}{\psi}}{O_\psi}{\sys{\td{\xi_i}{\psi}}{N_{i,\psi}}}$}.
\end{cases}\]
When $\td{r}{\psi}=\td{r'}{\psi}$,
$\ceqtm[\Psi']{\td{M}{\psi}}{\td{(O_{\id})}{\psi}}{\td{A}{\psi}}$ because
$\ceqtm[\Psi']{\td{M}{\psi}}{\td{\Kbox}{\psi}}{\td{\Fcom}{\psi}}$,
$\ceqtypek[\Psi']{\td{\Fcom}{\psi}}{\td{A}{\psi}}$ (by
\cref{rul:fcom-form-pre}), and
$\ceqtm[\Psi']{\td{\Kbox}{\psi}}{\td{(O_{\id})}{\psi}}{\td{\Fcom}{\psi}}$ (by
\cref{rul:fcom-intro}).
When $\td{r}{\psi}\neq\td{r'}{\psi}$ and $\td{r_j}{\psi}=\td{r_j'}{\psi}$ where
$j$ is the least such, $\ceqtm[\Psi']{\td{(O_{\id})}{\psi}}%
{\Coe{y.\td{B_j}{\psi}}{\td{r'}{\psi}}{\td{r}{\psi}}{\td{M}{\psi}}}{\td{A}{\psi}}$
because
$\ceqtm[\Psi']{\td{M}{\psi}}{\td{(N_{j,\id})}{\psi}}{\td{\dsubst{B_j}{r'}{y}}{\psi}}$
(by \cref{rul:fcom-form-pre,rul:fcom-intro}) and
$\ceqtm<r_j=r_j'>{O_{\id}}{\Coe{y.B_j}{r'}{r}{N_{j,\id}}}{A}$.
When $\td{r}{\psi}\neq\td{r'}{\psi}$ and $\td{r_i}{\psi}\neq\td{r_i'}{\psi}$ for
all $i$, $\ceqtm{\td{(O_{\id})}{\psi}}{O_\psi}{\td{A}{\psi}}$ by
$\vper{\Fcom}(\td{(\Kbox{r}{r'}{O_{\id}}{\sys{\xi_i}{N_{i,\id}}})}{\psi},
\Kbox{\td{r}{\psi}}{\td{r'}{\psi}}{O_\psi}{\sys{\td{\xi_i}{\psi}}{N_{i,\psi}}})$
(by $\coftype{M}{\Fcom}$ at $\id,\psi$). Therefore $\ceqtm{\Kcap}{O_{\id}}{A}$
by \cref{lem:cohexp-ceqtm}, and part (1) follows by a symmetric argument on the
right side.
\end{proof}

\begin{rul}[Computation]\label{rul:fcom-comp}
If
\begin{enumerate}
\item $A,\sys{r_i=r_i'}{y.B_i}$ is a type composition $r\rightsquigarrow r'$,
\item $\ceqtm{M}{M'}{A}$,
\item $\ceqtm<r_i=r_i',r_j=r_j'>{N_i}{N_j'}{\dsubst{B_i}{r'}{y}}$ for any $i,j$, and
\item $\ceqtm<r_i=r_i'>{\Coe{y.B_i}{r'}{r}{N_i}}{M}{A}$ for any $i$,
\end{enumerate}
then
$\ceqtm{\Kcap{r}{r'}{\Kbox{r}{r'}{M}{\sys{r_i=r_i'}{N_i}}}{\sys{r_i=r_i'}{y.B_i}}}{M}{A}$.
\end{rul}
\begin{proof}
By \cref{rul:fcom-intro,rul:fcom-comp}, we know both sides have this type, so it
suffices to show $\lift{\vper{A}}(\Kcap,M)$.
If $r=r'$ then $\Kcap\steps\Kbox\steps M$ and $\lift{\vper{A}}(M,M)$.
If $r\neq r'$ and $r_i=r_i'$ where $i$ is the least such, then
$\Kcap\steps\Coe{y.B_i}{r'}{r}{\Kbox}$, and
$\lift{\vper{A}}(\Coe{y.B_i}{r'}{r}{\Kbox},M)$ by
$\ceqtm{\Kbox}{N_i}{\dsubst{B_i}{r'}{y}}$ and
$\ceqtm{\Coe{y.B_i}{r'}{r}{N_i}}{M}{A}$.
If $r\neq r'$ and $r_i\neq r_i'$ for all $i$, then
$\Kcap\steps M$ and $\lift{\vper{A}}(M,M)$.
\end{proof}

\begin{rul}[Eta]\label{rul:fcom-eta}
If $A,\sys{\xi_i}{y.B_i}$ is a type composition $r\rightsquigarrow r'$ and
$\coftype{M}{\Fcom{r}{r'}{A}{\sys{\xi_i}{y.B_i}}}$, then
$\ceqtm{\Kbox{r}{r'}{\Kcap{r}{r'}{M}{\sys{\xi_i}{y.B_i}}}{\sys{\xi_i}{M}}}%
{M}{\Fcom{r}{r'}{A}{\sys{\xi_i}{y.B_i}}}$.
\end{rul}
\begin{proof}
By $\coftype{\Kcap{r}{r'}{M}{\sys{\xi_i}{y.B_i}}}{A}$ (by \cref{rul:fcom-elim}),
$\coftype<r_i=r_i'>{M}{\dsubst{B_i}{r'}{y}}$ (by \cref{rul:fcom-form-pre}),
$\ceqtm<r_i=r_i'>{\Coe{y.B_i}{r'}{r}{M}}{\Kcap}{A}$ (by \cref{rul:fcom-elim}),
and \cref{rul:fcom-intro}, we have $\coftype{\Kbox}{\Fcom}$. Thus, by
\cref{lem:coftype-ceqtm}, it suffices to show $\lift{\vper{\Fcom}}(\Kbox,M)$.
If $r=r'$ then $\Kbox\steps\Kcap\steps M$ and $\lift{\vper{\Fcom}}(M,M)$.
If $r\neq r'$ and $r_i=r_i'$ for the least such $i$, then $\Kbox\steps M$ and
$\lift{\vper{\Fcom}}(M,M)$. If $r\neq r'$ and $r_i\neq r_i'$ for all $i$, then
$M\evals\Kbox{r}{r'}{O}{\sys{\xi_i}{N_i}}$ and
$\ceqtm{M}{\Kbox{r}{r'}{O}{\sys{\xi_i}{N_i}}}{\Fcom}$. The result follows by
transitivity and \cref{rul:fcom-intro}:
\begin{enumerate}
\item $\ceqtm{\Kcap{r}{r'}{M}{\sys{\xi_i}{y.B_i}}}{O}{A}$ by
\cref{lem:coftype-ceqtm} and $\Kcap\steps^* O$,
\item $\ceqtm<r_i=r_i'>{M}{N_i}{\dsubst{B_i}{r'}{y}}$ by
$\ceqtm{M}{\Kbox{r}{r'}{O}{\sys{\xi_i}{N_i}}}{\Fcom}$ and \cref{rul:fcom-intro},
and
\item $\ceqtm<r_i=r_i'>{\Coe{y.B_i}{r'}{r}{M}}{\Kcap{r}{r'}{M}{\sys{\xi_i}{y.B_i}}}{A}$
by \cref{rul:fcom-elim} as before.
\qedhere
\end{enumerate}
\end{proof}

Our implementation of $\Coe$rcion for $\Fcom$ requires Kan compositions whose
lists of equations might be invalid (in the sense of \cref{def:valid}), although
Kan types are only guaranteed to have compositions for valid lists of equations.
However, we can implement such \emph{generalized} homogeneous compositions
$\Ghcom$ using only ordinary homogeneous compositions $\Hcom$.

\begin{theorem}\label{thm:ghcom}
If $\ceqtypek{A}{B}$,
\begin{enumerate}
\item $\ceqtm{M}{M'}{A}$,
\item $\ceqtm[\Psi,y]<r_i=r_i',r_j=r_j'>{N_i}{N_j'}{A}$ for any $i,j$, and
\item $\ceqtm[\Psi]<r_i=r_i'>{\dsubst{N_i}{r}{y}}{M}{A}$ for any $i$,
\end{enumerate}
then
\begin{enumerate}
\item $\ceqtm{\Ghcom*{A}{r_i=r_i'}}%
{\Ghcom{B}{r}{r'}{M'}{\sys{r_i=r_i'}{y.N_i'}}}{A}$;
\item if $r=r'$ then
$\ceqtm{\Ghcom{A}{r}{r}{M}{\sys{r_i=r_i'}{y.N_i}}}{M}{A}$; and
\item if $r_i = r_i'$ then
$\ceqtm{\Ghcom*{A}{r_i=r_i'}}{\dsubst{N_i}{r'}{y}}{A}$.
\end{enumerate}
\end{theorem}
\begin{proof}
Use induction on the length of $\etc{r_i=r_i'}$. If there are zero tubes, for
part (1) we must show
$\ceqtm{\Ghcom{A}{r}{r'}{M}{\cdot}}{\Ghcom{B}{r}{r'}{M'}{\cdot}}{A}$, which is
immediate by \cref{lem:expansion} on each side. Part (2) is immediate by
\cref{lem:expansion} on the left, and part (3) is impossible without tubes.

Now consider the case
$\Ghcom{A}{r}{r'}{M}{\tube{s=s'}{y.N},\sys{\xi_i}{y.N_i}}$, where we know
$\Ghcom$s with one fewer tube have the desired properties. By
\cref{lem:expansion} we must show (the binary version of)
\begin{gather*}
\coftype{\Hcom{A}{r}{r'}{M}{\sys{s=\e}{z.T_\e},\tube{s=s'}{y.N},\sys{\xi_i}{y.N_i}}}{A}
\\
\text{where}\
T_\e = \Hcom{A}{r}{z}{M}{
  \tube{s'=\e}{y.N},
  \tube{s'=\eb}{y.\Ghcom{A}{r}{y}{M}{\sys{\xi_i}{y.N_i}}},
  \sys{\xi_i}{y.N_i}}.
\end{gather*}
First, show $\coftype[\Psi,z]<s=\e>{T_\e}{A}$ by \cref{def:kan}, noting the
composition is valid by $s'=\e,s'=\eb$,
\begin{enumerate}
\item $\coftype<s=\e>{M}{A}$ by $\coftype{M}{A}$,

\item $\coftype[\Psi,y]<s=\e,s'=\e>{N}{A}$ (by
$\coftype[\Psi,y]<s=s'>{N}{A}$, because $s=s'$ whenever $s=\e,s'=\e$),
$\ceqtm[\Psi,y]<s=\e,s'=\e,r_i=r_i'>{N}{N_i}{A}$ (by
$\ceqtm[\Psi,y]<s=s',r_i=r_i'>{N}{N_i}{A}$), and
$\ceqtm<s=\e,s'=\e>{\dsubst{N}{r}{y}}{M}{A}$ (by
$\ceqtm<s=s'>{\dsubst{N}{r}{y}}{M}{A}$), and

\item
$\coftype[\Psi,y]<s=\e,s'=\eb>{\Ghcom{A}{r}{y}{M}{\sys{\xi_i}{y.N_i}}}{A}$
(by part (1) of the induction hypothesis),
$\ceqtm[\Psi,y]<s=\e,s'=\eb,r_i=r_i'>{\Ghcom{A}}{N_i}{A}$ (by part (3) of the
induction hypothesis), and
$\ceqtm[\Psi,y]<s=\e,s'=\eb>{\dsubst{(\Ghcom{A})}{r}{y}}{M}{A}$ (by part (2)
of the induction hypothesis).
\end{enumerate}
The remaining adjacency conditions are immediate. To check
$\coftype{\Hcom{A}}{A}$ it suffices to observe that
$\coftype[\Psi,z]<s=\e>{T_\e}{A}$ (by the above);
$\ceqtm[\Psi,z]<s=\e,s=s'>{T_\e}{\dsubst{N}{z}{y}}{A}$ (by the $s'=\e$ tube
in $T_\e$);
$\ceqtm[\Psi,z]<s=\e,r_i=r_i'>{T_\e}{\dsubst{N_i}{z}{y}}{A}$ (by the
$r_i=r_i'$ tube in $T_\e$);
$\ceqtm<s=\e>{\dsubst{T_\e}{r}{z}}{M}{A}$ (by $r=\dsubst{z}{r}{z}$ in $T_\e$);
and the $\etc{s=\e}$ tubes ensure the composition is valid. Part (1) follows by
repeating this argument on the right side, and parts (2--3) follow from
\cref{def:kan}.
\end{proof}

\begin{theorem}\label{thm:gcom}
If $\ceqtypek[\Psi,y]{A}{B}$,
\begin{enumerate}
\item $\ceqtm{M}{M'}{\dsubst{A}{r}{y}}$,
\item $\ceqtm[\Psi,y]<r_i=r_i',r_j=r_j'>{N_i}{N_j'}{A}$ for any $i,j$, and
\item $\ceqtm<r_i=r_i'>{\dsubst{N_i}{r}{y}}{M}{\dsubst{A}{r}{y}}$ for any $i$,
\end{enumerate}
then
\begin{enumerate}
\item
$\ceqtm{\Gcom*{y.A}{r_i=r_i'}}{\Gcom{y.B}{r}{r'}{M'}{\sys{r_i=r_i'}{y.N_i'}}}
{\dsubst{A}{r'}{y}}$;
\item if $r=r'$ then
$\ceqtm{\Gcom{y.A}{r}{r}{M}{\sys{r_i=r_i'}{y.N_i}}}{M}{\dsubst{A}{r}{y}}$; and
\item if $r_i = r_i'$ then
$\ceqtm{\Gcom*{y.A}{r_i=r_i'}}{\dsubst{N_i}{r'}{y}}{\dsubst{A}{r'}{y}}$.
\end{enumerate}
\end{theorem}
\begin{proof}
The implementation of $\Gcom$ by $\Ghcom$ and $\Coe$ mirrors exactly the
implementation of $\Com$ by $\Hcom$ and $\Coe$; the proof is thus identical to
that of \cref{thm:com}, appealing to \cref{thm:ghcom} instead of \cref{def:kan}.
\end{proof}

\begin{lemma}\label{lem:fcom-hcom}
If $A,\sys{s_j=s_j'}{z.B_j}$ and $A',\sys{s_j=s_j'}{z.B_j'}$ are equal type
compositions $s\rightsquigarrow s'$ and, letting $\Fcom :=
\Fcom{s}{s'}{A}{\sys{s_j=s_j'}{z.B_j}}$,
\begin{enumerate}
\item $\etc{r_i=r_i'}$ is valid,
\item $\ceqtm{M}{M'}{\Fcom}$,
\item $\ceqtm[\Psi,y]<r_i=r_i',r_{i'}=r_{i'}'>{N_i}{N_{i'}'}{\Fcom}$
for any $i,i'$, and
\item $\ceqtm<r_i=r_i'>{\dsubst{N_i}{r}{y}}{M}{\Fcom}$ for any $i$,
\end{enumerate}
then
\begin{enumerate}
\item $\ceqtm{\Hcom*{\Fcom}{r_i=r_i'}}%
{\Hcom{\Fcom{s}{s'}{A'}{\sys{s_j=s_j'}{z.B_j'}}}{r}{r'}{M'}{\sys{r_i=r_i'}{y.N_i'}}}{\Fcom}$;
\item if $r=r'$ then
$\ceqtm{\Hcom{\Fcom}{r}{r}{M}{\sys{r_i=r_i'}{y.N_i}}}{M}{\Fcom}$; and
\item if $r_i = r_i'$ then
$\ceqtm{\Hcom*{\Fcom}{r_i=r_i'}}{\dsubst{N_i}{r'}{y}}{\Fcom}$.
\end{enumerate}
\end{lemma}
\begin{proof}
If $s=s'$ or $s_j=s_j'$ for some $j$, the results are immediate by parts (2--3)
of \cref{lem:fcom-preform}. Otherwise, $s\neq s'$ and $s_j\neq s_j'$ for all
$j$; apply coherent expansion to $\Hcom{\Fcom}$ with family
\[\begin{cases}
\Hcom{\td{A}{\psi}}{\td{r}{\psi}}{\td{r'}{\psi}}%
{\td{M}{\psi}}{\sys{\td{r_i}{\psi}=\td{r_i'}{\psi}}{y.\td{N_i}{\psi}}}
& \text{$\td{s}{\psi}=\td{s'}{\psi}$} \\
\Hcom{\td{\dsubst{B_j}{s'}{z}}{\psi}}{\td{r}{\psi}}{\td{r'}{\psi}}%
{\td{M}{\psi}}{\sys{\td{r_i}{\psi}=\td{r_i'}{\psi}}{y.\td{N_i}{\psi}}}
& \text{$\td{s}{\psi}\neq\td{s'}{\psi}$,
least $\td{s_j}{\psi}=\td{s_j'}{\psi}$} \\
\td{(\Kbox{s}{s'}{Q}{\sys{s_j=s_j'}{\dsubst{P_j}{s'}{z}}})}{\psi}
& \text{$\td{s}{\psi}\neq\td{s'}{\psi}$,
$\forall j.\td{s_j}{\psi}\neq\td{s_j'}{\psi}$} \\
\quad P_j =
  \Hcom{B_j}{r}{r'}{\Coe{z.B_j}{s'}{z}{M}}%
  {\sys{r_i=r_i'}{y.\Coe{z.B_j}{s'}{z}{N_i}}} &\\
\quad F[c] =
  \Hcom{A}{s'}{z}{\Kcap{s}{s'}{c}{\sys{s_j=s_j'}{z.B_j}}}{\etc{T}} &\\
\quad \etc{T} =
  {\sys{s_j=s_j'}{z'.\Coe{z.B_j}{z'}{s}{\Coe{z.B_j}{s'}{z'}{c}}}} &\\
\quad O =
  \Hcom{A}{r}{r'}{\dsubst{(F[M])}{s}{z}}{\sys{r_i=r_i'}{y.\dsubst{(F[N_i])}{s}{z}}} &\\
\quad Q =
  \Hcom{A}{s}{s'}{O}{
     \sys{r_i=r_i'}{z.F[\dsubst{N_i}{r'}{y}]},\etc{U}} &\\
\quad \etc{U} =
  \sys{s_j=s_j'}{z.\Coe{z.B_j}{z}{s}{P_j}},
  \tube{r=r'}{z.F[M]}
\end{cases}\]
Consider $\psi=\id$.
\begin{enumerate}
\item $\ceqtm[\Psi,z]<s_j=s_j',s_{j'}=s_{j'}'>{P_j}{P_{j'}}{B_j}$ for all
$j,j'$, by
\begin{enumerate}
\item $\ceqtypek[\Psi,z]<s_j=s_j',s_{j'}=s_{j'}'>{B_j}{B_{j'}}$,
\item $\ceqtm[\Psi,z]<s_j=s_j',s_{j'}=s_{j'}'>%
{\Coe{z.B_j}{s'}{z}{M}}{\Coe{z.B_{j'}}{s'}{z}{M}}{B_j}$
by $\ceqtypek<s_j=s_j'>{\Fcom}{\dsubst{B_j}{s'}{z}}$,
\item $\ceqtm[\Psi,z,y]<s_j=s_j',s_{j'}=s_{j'}',r_i=r_i',r_{i'}=r_{i'}'>%
{\Coe{z.B_j}{s'}{z}{N_i}}{\Coe{z.B_{j'}}{s'}{z}{N_{i'}}}{B_j}$ for all
$i,i'$, and
\item
$\ceqtm[\Psi,z]<s_j=s_j',s_{j'}=s_{j'}',r_i=r_i'>%
{\Coe{z.B_j}{s'}{z}{M}}{\Coe{z.B_j}{s'}{z}{\dsubst{N_i}{r}{y}}}{B_j}$ for all
$i$ by
$\ceqtm[\Psi,z]<s_j=s_j',r_i=r_i'>{M}{\dsubst{N_i}{r}{y}}{\dsubst{B_j}{s'}{z}}$.
\end{enumerate}

\item $\ceqtm[\Psi,z]{F[c]}{F[c']}{A}$ for any $\ceqtm{c}{c'}{\Fcom}$, by
\begin{enumerate}
\item $\ceqtm{\Kcap{s}{s'}{c}{\sys{s_j=s_j'}{z.B_j}}}%
{\Kcap{s}{s'}{c'}{\sys{s_j=s_j'}{z.B_j}}}{A}$,
\item $\ceqtm[\Psi,z']<s_j=s_j',s_{j'}=s_{j'}'>%
{\Coe{z.B_j}{z'}{s}{\Coe{z.B_j}{s'}{z'}{c}}}%
{\Coe{z.B_{j'}}{z'}{s}{\Coe{z.B_{j'}}{s'}{z'}{c'}}}{A}$ for all $j,j'$ by
$\ceqtypek<s_j=s_j'>{\Fcom}{\dsubst{B_j}{s'}{z}}$ and
$\ceqtypek<s_j=s_j'>{\dsubst{B_j}{s}{z}}{A}$, and
\item $\ceqtm<s_j=s_j'>%
{\dsubst{(\Coe{z.B_j}{z'}{s}{\Coe{z.B_j}{s'}{z'}{c}})}{s'}{z'}}%
{\Kcap{s}{s'}{c}{\sys{s_j=s_j'}{z.B_j}}}{A}$ for all $j$ because both sides
$\eq$ $\Coe{z.B_j}{s'}{s}{c}$.
\end{enumerate}

\item $\coftype{O}{A}$ by
\begin{enumerate}
\item $\coftype{\dsubst{(F[M])}{s}{z}}{A}$ by $\coftype{M}{\Fcom}$,
\item $\ceqtm[\Psi,y]<r_i=r_i',r_{i'}=r_{i'}'>%
{\dsubst{(F[N_i])}{s}{z}}{\dsubst{(F[N_{i'}])}{s}{z}}{A}$ for all $i,i'$ by
$\ceqtm[\Psi,y]<r_i=r_i',r_{i'}=r_{i'}'>{N_i}{N_{i'}}{\Fcom}$, and
\item $\ceqtm<r_i=r_i'>%
{\dsubst{(F[\dsubst{N_i}{r}{y}])}{s}{z}}{\dsubst{(F[M])}{s}{z}}{A}$ for all $i$
by $\ceqtm<r_i=r_i',r_{i'}=r_{i'}'>{\dsubst{N_i}{r}{y}}{M}{\Fcom}$.
\end{enumerate}

\item $\coftype{Q}{A}$ by
\begin{enumerate}
\item $\ceqtm[\Psi,z]<r_i=r_i',r_{i'}=r_{i'}'>%
{F[\dsubst{N_i}{r'}{y}]}{F[\dsubst{N_{i'}}{r'}{y}]}{A}$ for all $i,i'$ by
$\ceqtm<r_i=r_i',r_{i'}=r_{i'}'>%
{\dsubst{N_i}{r'}{y}}{\dsubst{N_{i'}}{r'}{y}}{\Fcom}$,

\item $\ceqtm[\Psi,z]<s_j=s_j',s_{j'}=s_{j'}'>%
{\Coe{z.B_j}{z}{s}{P_j}}{\Coe{z.B_{j'}}{z}{s}{P_{j'}}}{A}$ by
$\ceqtm[\Psi,z]<s_j=s_j',s_{j'}=s_{j'}'>{P_j}{P_{j'}}{B_j}$,

\item $\coftype[\Psi,z]<r=r'>{F[M]}{A}$ by $\coftype{M}{\Fcom}$,

\item $\coftype{O}{A}$,

\item $\ceqtm<r_i=r_i'>{\dsubst{(F[\dsubst{N_i}{r'}{y}])}{s}{z}}{O}{A}$
for all $i$,

\item $\ceqtm<s_j=s_j'>{\dsubst{(\Coe{z.B_j}{z}{s}{P_j})}{s}{z}}{O}{A}$ for all
$j$, because the left side
$\ceqtm<s_j=s_j'>{\dsubst{(\Coe{z.B_j}{z}{s}{P_j})}{s}{z}}%
{\Hcom{\dsubst{B_j}{s}{z}}{r}{r'}{\Coe{z.B_j}{s'}{s}{M}}%
{\sys{r_i=r_i'}{y.\Coe{z.B_j}{s'}{s}{N_i}}}}{A}$, and this $\eq$ $O$ by
$\ceqtypek<s_j=s_j'>{\dsubst{B_j}{s}{z}}{A}$,
$\ceqtm<s_j=s_j'>{\Coe{z.B_j}{s'}{s}{M}}{\dsubst{(F[M])}{s}{z}}{A}$ (because the
right side $\eq$ $\Coe{z.B_j}{s}{s}{\Coe{z.B_j}{s'}{s}{M}}$), and
$\ceqtm<s_j=s_j',r_i=r_i'>{\Coe{z.B_j}{s'}{s}{N_i}}{\dsubst{(F[N_i])}{s}{z}}{A}$
for all $i$ (because the right side $\eq$
$\Coe{z.B_j}{s}{s}{\Coe{z.B_j}{s'}{s}{N_i}}$),

\item $\ceqtm<r=r'>{\dsubst{(F[M])}{s}{z}}{O}{A}$,

\item $\ceqtm[\Psi,z]<r_i=r_i',s_j=s_j'>%
{F[\dsubst{N_i}{r'}{y}]}{\Coe{z.B_j}{z}{s}{P_j}}{A}$ for all $i,j$ because both
sides $\eq$ $\Coe{z.B_j}{z}{s}{\Coe{z.B_j}{s'}{z}{\dsubst{N_i}{r'}{y}}}$,

\item $\ceqtm[\Psi,z]<r_i=r_i',r=r'>{F[\dsubst{N_i}{r'}{y}]}{F[M]}{A}$
for all $i$ by $\ceqtm<r_i=r_i'>{\dsubst{N_i}{r'}{y}}{M}{\Fcom}$, and

\item $\ceqtm[\Psi,z]<s_j=s_j',r=r'>{\Coe{z.B_j}{z}{s}{P_j}}{F[M]}{A}$
for all $j$ because both sides are $\eq$
$\Coe{z.B_j}{z}{s}{\Coe{z.B_j}{s'}{z}{M}}$.
\end{enumerate}

\item $\coftype{\Kbox{s}{s'}{Q}{\sys{s_j=s_j'}{\dsubst{P_j}{s'}{z}}}}{\Fcom}$ by
$\coftype{Q}{A}$, $\ceqtm<s_j=s_j',s_{j'}=s_{j'}'>%
{\dsubst{P_j}{s'}{z}}{\dsubst{P_{j'}}{s'}{z}}{\dsubst{B_j}{s'}{z}}$ for all
$j,j'$, and $\ceqtm<s_j=s_j'>{\Coe{z.B_j}{s'}{s}{\dsubst{P_j}{s'}{z}}}{Q}{A}$
for all $j$.
\end{enumerate}

When $\td{s}{\psi}\neq\td{s'}{\psi}$ and $\td{s_j}{\psi}\neq\td{s_j'}{\psi}$ for
all $j$, coherence is immediate. When $\td{s}{\psi}=\td{s'}{\psi}$,
$\td{\Kbox}{\psi} \eq \td{Q}{\psi} \eq{}$
$\ceqtm[\Psi']{\td{O}{\psi}}{\Hcom{\td{A}{\psi}}{\td{r}{\psi}}{\td{r'}{\psi}}%
{\td{M}{\psi}}{\sys{\td{\xi_i}{\psi}}{y.\td{N_i}{\psi}}}}{\td{A}{\psi}}$ by
$\ceqtm[\Psi']{\td{\dsubst{(F[M])}{s}{z}}{\psi}}{\td{M}{\psi}}{\td{A}{\psi}}$
and similarly for each tube. When $\td{s}{\psi}\neq\td{s'}{\psi}$ and
$\td{s_j}{\psi}=\td{s_j'}{\psi}$ for the least such $j$,
$\td{\Kbox}{\psi} \eq \td{\dsubst{P_j}{s'}{z}}{\psi} \eq
\td{(\Hcom{\dsubst{B_j}{s'}{z}}{r}{r'}{\Coe{z.B_j}{s'}{s'}{M}}%
{\sys{r_i=r_i'}{y.\Coe{z.B_j}{s'}{s'}{N_i}}})}{\psi}
\eq \td{(\Hcom{\dsubst{B_j}{s'}{z}}{r}{r'}{M}{\sys{r_i=r_i'}{y.N_i}})}{\psi}$.
By \cref{lem:cohexp-ceqtm}, $\ceqtm{\Hcom{\Fcom}}{\Kbox}{\Fcom}$; part (1)
follows by a symmetric argument on the right side.

For part (2), if $r=r'$ then
$Q\eq \ceqtm{\dsubst{(F[M])}{s'}{z}}{\Kcap{s}{s'}{M}{\sys{s_j=s_j'}{z.B_j}}}{A}$
and $\dsubst{P_j}{s'}{z}\eq
\ceqtm<s_j=s_j'>{\Coe{z.B_j}{s'}{s'}{M}}{M}{\dsubst{B_j}{s'}{z}}$ for all $j$,
so $\ceqtm{\Kbox{s}{s'}{Q}{\sys{s_j=s_j'}{\dsubst{P_j}{s'}{z}}}}{M}{\Fcom}$ by
\cref{rul:fcom-eta}, and part (2) follows by transitivity.

For part (3), if $r_i=r_i'$ then
$Q\eq \ceqtm{\dsubst{(F[\dsubst{N_i}{r'}{y}])}{s'}{z}}%
{\Kcap{s}{s'}{\dsubst{N_i}{r'}{y}}{\sys{s_j=s_j'}{z.B_j}}}{A}$
and $\dsubst{P_j}{s'}{z}\eq
\ceqtm<s_j=s_j'>{\Coe{z.B_j}{s'}{s'}{\dsubst{N_i}{r'}{y}}}%
{\dsubst{N_i}{r'}{y}}{\dsubst{B_j}{s'}{z}}$ for all $j$, so
$\ceqtm{\Kbox}{\dsubst{N_i}{r'}{y}}{\Fcom}$ by \cref{rul:fcom-eta}, and part (3)
follows by transitivity.
\end{proof}

\begin{lemma}\label{lem:fcom-coe}
Let $\Fcom := \Fcom{s}{s'}{A}{\sys{s_i=s_i'}{z.B_i}}$. If
\begin{enumerate}
\item $\etc{s_i=s_i'}$ is valid in $(\Psi,x)$,
\item $\ceqtypek[\Psi,x]{A}{A'}$,
\item $\ceqtypek[\Psi,x,z]<s_i=s_i',s_j=s_j'>{B_i}{B_j'}$ for any $i,j$,
\item $\ceqtypek[\Psi,x]<s_i=s_i'>{\dsubst{B_i}{s}{z}}{A}$ for any $i$, and
\item $\ceqtm{M}{M'}{\dsubst{\Fcom}{r}{x}}$,
\end{enumerate}
then
\begin{enumerate}
\item $\ceqtm{\Coe*{x.\Fcom}}%
{\Coe{x.\Fcom{s}{s'}{A'}{\sys{s_i=s_i'}{z.B_i'}}}{r}{r'}{M'}}{\dsubst{\Fcom}{r'}{x}}$; and
\item if $r=r'$ then
$\ceqtm{\Coe*{x.\Fcom}}{M}{\dsubst{\Fcom}{r'}{x}}$.
\end{enumerate}
\end{lemma}
\begin{proof}
If $s=s'$ or $s_i=s_i'$ for some $i$, the results are immediate by parts (2--3)
of \cref{lem:fcom-preform}. Otherwise, $s\neq s'$ and $s_i\neq s_i'$ for all
$i$; apply coherent expansion to $\Coe*{x.\Fcom}$ with family
\[\begin{cases}
\Coe{x.\td{A}{\psi}}{\td{r}{\psi}}{\td{r'}{\psi}}{\td{M}{\psi}}
& \text{$\td{s}{\psi}=\td{s'}{\psi}$} \\
\Coe{x.\td{\dsubst{B_i}{s'}{z}}{\psi}}{\td{r}{\psi}}{\td{r'}{\psi}}{\td{M}{\psi}}
& \text{$\td{s}{\psi}\neq\td{s'}{\psi}$,
least $\td{s_i}{\psi}=\td{s_i'}{\psi}$} \\
\td{\dsubst{(\Kbox{s}{s'}{R}{\sys{\xi_i}{\dsubst{Q_i}{s'}{z}}})}{r'}{x}}{\psi}
& \text{$\td{s}{\psi}\neq\td{s'}{\psi}$,
$\forall i.\td{s_i}{\psi}\neq\td{s_i'}{\psi}$} \\
\quad N_i =
  \Coe{z.B_i}{s'}{z}{\Coe{x.\dsubst{B_i}{s'}{z}}{r}{x}{M}} &\\
\quad O = \dsubst{(\Hcom{A}{s'}{z}{\Kcap{s}{s'}{M}{\sys{\xi_i}{z.B_i}}}{
  \sys{\xi_i}{z.\Coe{z.B_i}{z}{s}{N_i}}})}{r}{x} \\
\quad P = \Gcom{x.A}{r}{r'}{\dsubst{O}{\dsubst{s}{r}{x}}{z}}{
  \st{\sys{\xi_i}{x.\dsubst{N_i}{s}{z}}}{x\fresh\xi_i},T} &\\
\quad T =
  \st{\tube{s=s'}{x.\Coe{x.A}{r}{x}{M}}}{x\fresh s,s'} &\\
\quad Q_k = \Gcom{z.\dsubst{B_k}{r'}{x}}{\dsubst{s}{r'}{x}}{z}{P}{
     \st{\sys{\xi_i}{z.\dsubst{N_i}{r'}{x}}}{x\fresh\xi_i},
     \tube{r=r'}{z.\dsubst{N_k}{r'}{x}}}\hspace{-0.6em} &\\
\quad R =
  \Hcom{A}{s}{s'}{P}{\sys{\xi_i}{z.\Coe{z.B_i}{z}{s}{Q_i}},\tube{r=r'}{z.O}}
\end{cases}\]
Consider $\psi=\id$.

\begin{enumerate}
\item $\ceqtm[\Psi,x,z]<\dsubst{\xi_i}{r}{x},\dsubst{\xi_j}{r}{x}>{N_i}{N_j}{B_i}$
for all $i,j$ by
$\coftype<\dsubst{\xi_i}{r}{x}>{M}{\dsubst{\dsubst{B_i}{s'}{z}}{r}{x}}$
(by $\coftype{M}{\dsubst{\Fcom}{r}{x}}$ and
$\ceqtypek[\Psi,x]<\xi_i>{\Fcom}{\dsubst{B_i}{s'}{z}}$) and
$\ceqtypek[\Psi,x,z]<\xi_i,\xi_j>{B_i}{B_j}$.

\item $\coftype[\Psi,z]{O}{\dsubst{A}{r}{x}}$ by
\begin{enumerate}
\item $\coftype{\dsubst{(\Kcap{s}{s'}{M}{\sys{\xi_i}{z.B_i}})}{r}{x}}{\dsubst{A}{r}{x}}$
by $\coftype{M}{\dsubst{\Fcom}{r}{x}}$,

\item $\ceqtm[\Psi,z]<\dsubst{\xi_i}{r}{x},\dsubst{\xi_j}{r}{x}>%
{\Coe{z.\dsubst{B_i}{r}{x}}{z}{\dsubst{s}{r}{x}}{\dsubst{N_i}{r}{x}}}%
{\Coe{z.\dsubst{B_j}{r}{x}}{z}{\dsubst{s}{r}{x}}{\dsubst{N_j}{r}{x}}}%
{\dsubst{A}{r}{x}}$ for all $i,j$
by $\ceqtypek<\dsubst{\xi_i}{r}{x}>{\dsubst{\dsubst{B_i}{s}{z}}{r}{x}}{\dsubst{A}{r}{x}}$, and

\item $\ceqtm<\dsubst{\xi_i}{r}{x}>%
{\dsubst{(\Kcap{s}{s'}{M}{\sys{\xi_i}{z.B_i}})}{r}{x}}%
{\dsubst{\dsubst{(\Coe{z.B_i}{z}{s}{N_i})}{s'}{z}}{r}{x}}%
{\dsubst{A}{r}{x}}$ for all $i$ by
$\ceqtm<\dsubst{\xi_i}{r}{x}>{\dsubst{\Kcap}{r}{x}}%
{\dsubst{(\Coe{z.B_i}{s'}{s}{M})}{r}{x}}{\dsubst{A}{r}{x}}$ and
$\ceqtm<\dsubst{\xi_i}{r}{x}>{\dsubst{\dsubst{N_i}{s'}{z}}{r}{x}}{M}%
{\dsubst{\dsubst{B_i}{s'}{z}}{r}{x}}$.
\end{enumerate}

\item $\coftype{P}{\dsubst{A}{r'}{x}}$ by
\begin{enumerate}
\item $\coftype{\dsubst{O}{\dsubst{s}{r}{x}}{z}}{\dsubst{A}{r}{x}}$,

\item $\ceqtm[\Psi,x]<\xi_i,\xi_j>{\dsubst{N_i}{s}{z}}{\dsubst{N_j}{s}{z}}{A}$
for all $i,j$ such that $x\fresh\xi_i,\xi_j$ by
$\ceqtm[\Psi,x]<\dsubst{\xi_i}{r}{x},\dsubst{\xi_j}{r}{x}>%
{\dsubst{N_i}{s}{z}}{\dsubst{N_j}{s}{z}}{\dsubst{B_i}{s}{z}}$ and
$\ceqtypek[\Psi,x]<\xi_i>{\dsubst{B_i}{s}{z}}{A}$,

\item $\coftype[\Psi,x]<s=s'>{\Coe{x.A}{r}{x}{M}}{A}$ if $x\fresh s,s'$ by
$\ceqtypek<\dsubst{s}{r}{x}=\dsubst{s'}{r}{x}>{\dsubst{\Fcom}{r}{x}}{\dsubst{A}{r}{x}}$,

\item $\ceqtm<\xi_i>{\dsubst{O}{\dsubst{s}{r}{x}}{z}}%
{\dsubst{\dsubst{N_i}{s}{z}}{r}{x}}{\dsubst{A}{r}{x}}$ for all $i$ such that
$x\fresh\xi_i$ by ${\dsubst{O}{\dsubst{s}{r}{x}}{z}}\eq{}$
$\ceqtm<\dsubst{\xi_i}{r}{x}>{\dsubst{\dsubst{(\Coe{z.B_i}{z}{s}{N_i})}{s}{z}}{r}{x}}%
{\dsubst{\dsubst{N_i}{s}{z}}{r}{x}}{\dsubst{A}{r}{x}}$,

\item $\ceqtm<s=s'>{\dsubst{O}{\dsubst{s}{r}{x}}{z}}%
{\dsubst{(\Coe{x.A}{r}{x}{M})}{r}{x}}{\dsubst{A}{r}{x}}$ if $x\fresh s,s'$ by
$\dsubst{O}{\dsubst{s}{r}{x}}{z} = \dsubst{O}{s}{z} \eq{}
\ceqtm<s=s'>{\dsubst{\Kcap}{r}{x}}{M}{\dsubst{A}{r}{x}}$, and

\item $\ceqtm[\Psi,x]<\xi_i,s=s'>{\dsubst{N_i}{s}{z}}{\Coe{x.A}{r}{x}{M}}{A}$
for all $i$ such that $x\fresh \xi_i,s,s'$ by
$\ceqtm[\Psi,x]<\xi_i,s=s'>{\dsubst{N_i}{s}{z}}%
{\Coe{x.\dsubst{B_i}{s'}{z}}{r}{x}{M}}{\dsubst{B_i}{s'}{z}}$
and $\ceqtypek[\Psi,x]<\xi_i,s=s'>{\dsubst{B_i}{s'}{z}}{A}$.
\end{enumerate}

\item $\ceqtm[\Psi,z]<\dsubst{\xi_k}{r'}{x},\dsubst{\xi_{k'}}{r'}{x}>%
{Q_k}{Q_{k'}}{\dsubst{B_k}{r'}{x}}$ for all $k,k'$ by
\begin{enumerate}
\item $\coftype<\dsubst{\xi_k}{r'}{x}>{P}{\dsubst{\dsubst{B_k}{s}{z}}{r'}{x}}$
by $\ceqtypek[\Psi,x]<\xi_k>{A}{\dsubst{B_k}{s}{z}}$,

\item $\ceqtm[\Psi,z]<\dsubst{\xi_k}{r'}{x},\xi_i,\xi_j>%
{\dsubst{N_i}{r'}{x}}{\dsubst{N_j}{r'}{x}}{\dsubst{B_k}{r'}{x}}$ for all $i,j$
such that $x\fresh\xi_i,\xi_j$ by
$\ceqtm[\Psi,x,z]<\xi_i,\xi_j>{N_i}{N_j}{B_i}$ and
$\ceqtypek[\Psi,x,z]<\xi_i,\xi_k>{B_i}{B_k}$,

\item $\ceqtm[\Psi,z]<\dsubst{\xi_k}{r'}{x},\dsubst{\xi_{k'}}{r'}{x}>%
{\dsubst{N_k}{r'}{x}}{\dsubst{N_{k'}}{r'}{x}}{\dsubst{B_k}{r'}{x}}$,

\item $\ceqtm<\dsubst{\xi_k}{r'}{x},\xi_i>%
{P}{\dsubst{\dsubst{N_i}{s}{z}}{r'}{x}}{\dsubst{\dsubst{B_k}{s}{z}}{r'}{x}}$ for
all $i$ such that $x\fresh\xi_i$ by
$\ceqtm<\xi_i>{P}{\dsubst{\dsubst{N_i}{s}{z}}{r'}{x}}{\dsubst{A}{r'}{x}}$ and
$\ceqtypek<\dsubst{\xi_k}{r'}{x}>{\dsubst{A}{r'}{x}}{\dsubst{\dsubst{B_k}{s}{z}}{r'}{x}}$,

\item $\ceqtm<\dsubst{\xi_k}{r'}{x},r=r'>%
{P}{\dsubst{\dsubst{N_k}{s}{z}}{r'}{x}}{\dsubst{\dsubst{B_k}{s}{z}}{r'}{x}}$
because $P\eq
\ceqtm<\dsubst{\xi_k}{r'}{x},r=r'>{\dsubst{O}{\dsubst{s}{r}{x}}{z}}%
{\dsubst{\dsubst{(\Coe{z.B_k}{z}{s}{N_k})}{\dsubst{s}{r}{x}}{z}}{r}{x}}
{\dsubst{A}{r'}{x}}$, and

\item $\ceqtm[\Psi,z]<\dsubst{\xi_k}{r'}{x},\xi_i,r=r'>%
{\dsubst{N_i}{r'}{x}}{\dsubst{N_k}{r'}{x}}{\dsubst{B_k}{r'}{x}}$ for all $i$
such that $x\fresh\xi_i$.
\end{enumerate}

\item $\coftype{\dsubst{R}{r'}{x}}{\dsubst{A}{r'}{x}}$ by
\begin{enumerate}
\item $\coftype{P}{\dsubst{A}{r'}{x}}$,

\item $\ceqtm[\Psi,z]<\dsubst{\xi_i}{r'}{x},\dsubst{\xi_j}{r'}{x}>%
{\Coe{z.\dsubst{B_i}{r'}{x}}{z}{\dsubst{s}{r'}{x}}{Q_i}}%
{\Coe{z.\dsubst{B_j}{r'}{x}}{z}{\dsubst{s}{r'}{x}}{Q_j}}%
{\dsubst{A}{r'}{x}}$ for all $i,j$ by $\ceqtypek[\Psi,z,x]<\xi_i,\xi_j>{B_i}{B_j}$ and
$\ceqtypek<\dsubst{\xi_i}{r'}{x}>{\dsubst{\dsubst{B_i}{s}{z}}{r'}{x}}{\dsubst{A}{r'}{x}}$,

\item $\coftype[\Psi,z]<r=r'>{O}{\dsubst{A}{r'}{x}}$,

\item $\ceqtm<\dsubst{\xi_i}{r'}{x}>%
{P}{\dsubst{\dsubst{(\Coe{z.B_i}{z}{s}{Q_i})}{s}{z}}{r'}{x}}{\dsubst{A}{r'}{x}}$
for all $i$ by
$\ceqtm<\dsubst{\xi_i}{r'}{x}>{\dsubst{\dsubst{Q_i}{s}{z}}{r'}{x}}{P}%
{\dsubst{\dsubst{B_i}{s}{z}}{r'}{x}}$ and
$\ceqtypek<\dsubst{\xi_i}{r'}{x}>{\dsubst{\dsubst{B_i}{s}{z}}{r'}{x}}{\dsubst{A}{r'}{x}}$,

\item $\ceqtm<r=r'>{P}{\dsubst{\dsubst{O}{s}{z}}{r'}{x}}{\dsubst{A}{r'}{x}}$ by
$\dsubst{\dsubst{O}{s}{z}}{r'}{x} = \dsubst{O}{\dsubst{s}{r'}{x}}{z}$, and

\item $\ceqtm[\Psi,z]<\dsubst{\xi_i}{r'}{x},r=r'>%
{\dsubst{(\Coe{z.B_i}{z}{s}{Q_i})}{r'}{x}}{\dsubst{O}{r'}{x}}{\dsubst{A}{r'}{x}}$
for all $i$ by $\dsubst{O}{r'}{x} =
\ceqtm[\Psi,z]<\dsubst{\xi_i}{r'}{x}>{O}{\dsubst{(\Coe{z.B_i}{z}{s}{N_i})}{r}{x}}{\dsubst{A}{r'}{x}}$
and
$\ceqtm[\Psi,z]<\dsubst{\xi_i}{r'}{x},r=r'>{\dsubst{Q_i}{r'}{x}}{\dsubst{N_i}{r'}{x}}{\dsubst{A}{r'}{x}}$.
\end{enumerate}

\item $\coftype{\Kbox{\dsubst{s}{r'}{x}}{\dsubst{s'}{r'}{x}}{\dsubst{R}{r'}{x}}%
{\sys{\dsubst{\xi_i}{r'}{x}}{\dsubst{Q_i}{\dsubst{s'}{r'}{x}}{z}}}}{\dsubst{\Fcom}{r'}{x}}$ by
\begin{enumerate}
\item $\coftype{\dsubst{R}{r'}{x}}{\dsubst{A}{r'}{x}}$,

\item $\ceqtm<\dsubst{\xi_i}{r'}{x},\dsubst{\xi_j}{r'}{x}>%
{\dsubst{Q_i}{\dsubst{s'}{r'}{x}}{z}}{\dsubst{Q_j}{\dsubst{s'}{r'}{x}}{z}}{\dsubst{\dsubst{B_i}{s'}{z}}{r'}{x}}$
for all $i,j$, and

\item $\ceqtm<\dsubst{\xi_i}{r'}{x}>%
{\dsubst{(\Coe{z.B_i}{s'}{s}{\dsubst{Q_i}{s'}{z}})}{r'}{x}}{\dsubst{R}{r'}{x}}{\dsubst{A}{r'}{x}}$
for all $i$ by $\ceqtm<\dsubst{\xi_i}{r'}{x}>%
{\dsubst{R}{r'}{x}}{\dsubst{\dsubst{(\Coe{z.B_i}{z}{s}{Q_i})}{s'}{z}}{r'}{x}}{\dsubst{A}{r'}{x}}$.
\end{enumerate}
\end{enumerate}

Consider $\psitd$. When $\td{s}{\psi}\neq\td{s'}{\psi}$ and
$\td{s_i}{\psi}\neq\td{s_i'}{\psi}$ for all $i$, coherence is immediate. When
$\td{s}{\psi}=\td{s'}{\psi}$, then by $s\neq s'$, we must have $x\fresh s,s'$
and thus $\td{\dsubst{s}{r'}{x}}{\psi} = \td{\dsubst{s'}{r'}{x}}{\psi}$ also.
Thus $\td{\dsubst{\Kbox}{r'}{x}}{\psi}\eq \td{\dsubst{R}{r'}{x}}{\psi}\eq
\ceqtm[\Psi']{\td{\dsubst{P}{r'}{x}}{\psi}}%
{\td{\dsubst{(\Coe{x.A}{r}{x}{M})}{r'}{x}}{\psi}}{\dsubst{A}{r'}{x}}$ as
required. When $\td{s}{\psi}\neq\td{s'}{\psi}$ and
$\td{s_i}{\psi}=\td{s_i'}{\psi}$ for the least such $i$, again $x\fresh
s_i,s_i'$ and
$\td{\dsubst{\Kbox}{r'}{x}}{\psi}\eq \td{\dsubst{\dsubst{Q_i}{s'}{z}}{r'}{x}}{\psi}\eq
\ceqtm[\Psi']{\td{\dsubst{\dsubst{N_i}{s'}{z}}{r'}{x}}{\psi}}%
{\td{(\Coe{x.\dsubst{B_i}{s'}{z}}{r}{r'}{M})}{\psi}}{\dsubst{A}{r'}{x}}$.
By \cref{lem:cohexp-ceqtm},
$\ceqtm{\Coe*{x.\Fcom}}{\dsubst{\Kbox}{r'}{x}}{\dsubst{\Fcom}{r'}{x}}$; part (1)
follows by a symmetric argument on the right side.

For part (2), if $r=r'$ then
$\dsubst{R}{r'}{x}\eq \ceqtm{\dsubst{\dsubst{O}{s'}{z}}{r'}{x}}%
{\dsubst{(\Kcap{s}{s'}{M}{\sys{\xi_i}{z.B_i}})}{r'}{x}}{\dsubst{A}{r'}{x}}$
and $\dsubst{\dsubst{Q_i}{s'}{z}}{r'}{x}\eq
\ceqtm<\dsubst{\xi_i}{r'}{x}>{\dsubst{\dsubst{N_k}{s'}{z}}{r'}{x}}{M}{\dsubst{\dsubst{B_i}{s'}{z}}{r'}{x}}$
for all $i$, so $\ceqtm{\dsubst{\Kbox}{r'}{x}}{M}{\dsubst{\Fcom}{r'}{x}}$ by
\cref{rul:fcom-eta}, and part (2) follows by transitivity.
\end{proof}

\begin{rul}[Kan type formation]\label{rul:fcom-form-kan}
If $A,\sys{r_i=r_i'}{y.B_i}$ and $A',\sys{r_i=r_i'}{y.B_i'}$ are equal type
compositions $r\rightsquigarrow r'$, then
\begin{enumerate}
\item $\ceqtypek{\Fcom{r}{r'}{A}{\sys{r_i=r_i'}{y.B_i}}}%
{\Fcom{r}{r'}{A'}{\sys{r_i=r_i'}{y.B_i'}}}$,
\item if $r=r'$ then $\ceqtypek{\Fcom{r}{r}{A}{\sys{r_i=r_i'}{y.B_i}}}{A}$, and
\item if $r_i = r_i'$ then
$\ceqtypek{\Fcom{r}{r'}{A}{\sys{r_i=r_i'}{B_i}}}{\dsubst{B_i}{r'}{y}}$.
\end{enumerate}
\end{rul}
\begin{proof}
We already showed parts (2--3) in \cref{lem:fcom-preform}. For part (1), the
$\Hcom$ conditions follow from \cref{lem:fcom-hcom} at $\td{\Fcom}{\psi}$ for
any $\psitd$; the $\Coe$ conditions follow from \cref{lem:fcom-coe} at
$x.\td{\Fcom}{\psi}$ for any $\tds{(\Psi',x)}{\psi}{\Psi}$.
\end{proof}

\subsection{Universes}

Our type theory has two hierarchies of universes, $\Upre$ and $\UKan$,
constructed by two sequences $\pre\tau_j$ and $\Kan\tau_j$ of cubical type
systems. To prove theorems about universe types in the cubical type system
$\pre\tau_\omega$, we must analyze these sequences as constructed in
\cref{sec:typesys}.

\begin{lemma}\label{lem:monotone-judg}
If $\tau,\tau'$ are cubical type systems, $\tau\subseteq\tau'$, and
$\relcts*{\tau}{\J}$ for any judgment $\J$, then $\relcts*{\tau'}{\J}$.
\end{lemma}
\begin{proof}
The result follows by $\PTy(\tau)\subseteq\PTy(\tau')$ and the functionality of
$\tau,\tau'$; the latter ensures that any (pre)type in $\tau$ has no other
meanings in $\tau'$.
\end{proof}

\begin{lemma}\label{lem:pty-ceqtypex}
If $\tau$ is a cubical type system, $\wftm{A}$, $\wftm{B}$, and for all
$\tds{\Psi_1}{\psi_1}{\Psi}$ and
$\tds{\Psi_2}{\psi_2}{\Psi_1}$, we have
$\td{A}{\psi_1}\evals A_1$,
$\td{A_1}{\psi_2}\evals A_2$,
$\td{A}{\psi_1\psi_2}\evals A_{12}$,
$\td{B}{\psi_1}\evals B_1$,
$\td{B_1}{\psi_2}\evals B_2$,
$\td{B}{\psi_1\psi_2}\evals B_{12}$,
$\relcts{\tau}{\ceqtypex[\Psi_2]{A_2}{A_{12}}}$,
$\relcts{\tau}{\ceqtypex[\Psi_2]{B_2}{B_{12}}}$, and
$\relcts{\tau}{\ceqtypex[\Psi_2]{A_2}{B_2}}$, then
$\relcts{\tau}{\ceqtypex{A}{B}}$.
\end{lemma}
\begin{proof}
We apply coherent expansion to $A$ and the family of terms
$\{A^{\Psi'}_\psi \mid \td{A}{\psi}\evals A^{\Psi'}_\psi\}^{\Psi'}_\psi$.
By our hypotheses at $\psi,\id[\Psi']$ and $\id,\id$ we know
$\relcts{\tau}{\cwftypex[\Psi']{A^{\Psi'}_\psi}}$ and
$\relcts{\tau}{\cwftypex[\Psi']{\td{(A^{\Psi}_{\id})}{\psi}}}$;
for any $\tds{\Psi''}{\psi'}{\Psi'}$, our hypotheses at $\psi,\psi'$ and
$\id,\psi\psi'$ show
$\relcts{\tau}{\ceqtypex[\Psi'']{A'}{A^{\Psi''}_{\psi\psi'}}}$ where
$\td{(A^{\Psi'}_\psi)}{\psi'}\evals A'$, and
$\relcts{\tau}{\ceqtypex[\Psi'']{A''}{A^{\Psi''}_{\psi\psi'}}}$ where
$\td{(A^{\Psi}_{\id})}{\psi\psi'}\evals A''$, hence
$\relcts{\tau}{\ceqtypex[\Psi'']{A'}{A''}}$.

If $\kappa=\mathsf{pre}$ then by \cref{lem:cwftypep-evals-ceqtypep},
$\relcts{\tau}{\ceqtypep[\Psi']{\td{(A^\Psi_{\id})}{\psi}}{A_0'}}$ where
$\td{(A^\Psi_{\id})}{\psi}\evals A_0'$; thus we have
$\relcts{\tau}{\ceqtypep[\Psi']{A^{\Psi'}_{\psi}}{\td{(A^\Psi_{\id})}{\psi}}}$
by transitivity, and by \cref{lem:cohexp-ceqtypep},
$\relcts{\tau}{\ceqtypep{A}{A_0}}$ where $A\evals A_0$.
If $\kappa=\mathsf{Kan}$ then by
\cref{lem:cwftypek-evals-ceqtypek},
$\relcts{\tau}{\ceqtypek[\Psi']{A^{\Psi'}_\psi}{\td{(A^{\Psi}_{\id})}{\psi}}}$,
and by \cref{lem:cohexp-ceqtypek}, $\relcts{\tau}{\ceqtypek{A}{A_0}}$ where
$A\evals A_0$. In either case, we repeat the argument for $B$ to obtain
$\relcts{\tau}{\ceqtypex{B}{B_0}}$ where $B\evals B_0$, and the result follows
by symmetry and transitivity.
\end{proof}

\begin{rul}[Pretype formation]\label{rul:U-form-pre}
If $i<j$ or $j=\omega$ then $\relcts{\pre\tau_j}{\cwftypep{\Ux[i]}}$ and
$\relcts{\Kan\tau_j}{\cwftypep{\UKan[i]}}$.
\end{rul}
\begin{proof}
In each case we have $\PTy(\tau^{\kappa'}_j)(\Psi,\Ux[i],\Ux[i],\_)$ by
$\sisval{\Ux[i]}$ and the definition of $\tau^{\kappa'}_j$. For
$\Coh(\vper{\Ux[i]})$, show that if $\vper{\Ux[i]}_{\Psi'}(A_0,B_0)$ then
$\Tm(\vper{\Ux[i]}(\Psi'))(A_0,B_0)$. But $\Tm(\vper{\Ux[i]}(\Psi'))(A,B)$ if
and only if $\PTy(\tau^\kappa_i)(\Psi',A,B,\_)$, so this is immediate by
value-coherence of $\tau^\kappa_i$.
\end{proof}

\begin{rul}[Cumulativity]
If $\relcts{\pre\tau_\omega}{\ceqtm{A}{B}{\Ux[i]}}$ and $i\leq j$ then
$\relcts{\pre\tau_\omega}{\ceqtm{A}{B}{\Ux[j]}}$.
\end{rul}
\begin{proof}
In \cref{sec:typesys} we observed that $\tau^\kappa_i\subseteq\tau^\kappa_j$
whenever $i\leq j$; thus $\vper{\Ux[i]}\subseteq\vper{\Ux[j]}$, and the result
follows because $\Tm$ is order-preserving.
\end{proof}

\begin{lemma}\label{lem:U-preelim}
~\begin{enumerate}
\item If $\relcts{\pre\tau_\omega}{\ceqtm{A}{B}{\UKan[i]}}$ then
$\relcts{\Kan\tau_i}{\ceqtypek{A}{B}}$.
\item If $\relcts{\pre\tau_\omega}{\ceqtm{A}{B}{\Upre[i]}}$ then
$\relcts{\pre\tau_i}{\ceqtypep{A}{B}}$.
\end{enumerate}
\end{lemma}
\begin{proof}
We prove part (1) by strong induction on $i$. For each $i$, define $\Phi =
\{(\Psi,A_0,B_0,\phi) \mid \relcts{\Kan\tau_i}{\ceqtypek{A_0}{B_0}} \}$,
and show $K(\nu_i,\Phi)\subseteq\Phi$. We will conclude
$\Kan\tau_i\subseteq\Phi$ and so $\relcts{\Kan\tau_i}{\ceqtypek{A_0}{B_0}}$
whenever $\vper{\UKan[i]}(A_0,B_0)$; part (1) will follow by
\cref{lem:pty-ceqtypex}.

To establish $K(\nu_i,\Phi)\subseteq\Phi$, we check each type former
independently. Consider the case
$\textsc{Fun}(\Phi)(\Psi,\picl{a}{A}{B},\picl{a}{A'}{B'},\phi)$. Then
$\PTy(\Phi)(\Psi,A,A',\alpha)$, which by \cref{lem:pty-ceqtypex} implies
$\relcts{\Kan\tau_i}{\ceqtypek{A}{A'}}$; similarly,
$\relcts{\Kan\tau_i}{\eqtypek{\oft{a}{A}}{B}{B'}}$. By
\cref{rul:fun-form-kan}, we conclude
$\relcts{\Kan\tau_i}{\ceqtypek{\picl{a}{A}{B}}{\picl{a}{A'}{B'}}}$. The
same argument applies for every type former except for $\textsc{UKan}$, where we
must show $\relcts{\Kan\tau_i}{\cwftypek{\UKan[j]}}$ for every $j<i$. The
$\Coe$ conditions are trivial by $\Coe*{x.\UKan[j]} \ssteps M$; the $\Hcom$
conditions hold by $\Hcom{\UKan[j]} \ssteps \Fcom$,
$\relcts{\Kan\tau_i}{\ceqtm{A}{B}{\UKan[j]}}$ implies
$\relcts{\Kan\tau_i}{\ceqtypek{A}{B}}$ (by induction), and
\cref{rul:fcom-form-kan}.

We prove part (2) directly for all $i$, by establishing
$P(\nu_i,\Kan\tau_i,\Phi)\subseteq\Phi$ for $\Phi = \{(\Psi,A_0,B_0,\phi) \mid
\relcts{\pre\tau_i}{\ceqtypek{A_0}{B_0}} \}$ and appealing to
\cref{lem:pty-ceqtypex}. Most type formers follow the same pattern as above; we
only discuss $\textsc{Fcom}$, $\textsc{UPre}$, and $\textsc{UKan}$. For
$\textsc{Fcom}$, we appeal to part (1) and \cref{rul:fcom-form-pre}, observing
that $\PTy(\Kan\tau_i)(\Psi,A,B,\_)$ if and only if
$\Tm(\vper{\UKan[i]}(\Psi))(A,B)$. For $\textsc{UPre}$ and $\textsc{UKan}$,
$\relcts{\pre\tau_i}{\cwftypep{\Ux[j]}}$ for all $j<i$ is immediate by
\cref{rul:U-form-pre}.
\end{proof}

\begin{rul}[Elimination]\label{rul:U-elim}
If $\relcts{\pre\tau_\omega}{\ceqtm{A}{B}{\Ux[i]}}$ then
$\relcts{\pre\tau_\omega}{\ceqtypex{A}{B}}$.
\end{rul}
\begin{proof}
Immediate by $\tau^\kappa_i\subseteq\pre\tau_\omega$ and
\cref{lem:U-preelim,lem:monotone-judg}.
\end{proof}

\begin{rul}[Introduction]\label{rul:U-intro}
In $\pre\tau_\omega$,
\begin{enumerate}
\item If $\ceqtm{A}{A'}{\Ux}$ and $\eqtm{\oft aA}{B}{B'}{\Ux}$ then
$\ceqtm{\picl{a}{A}{B}}{\picl{a}{A'}{B'}}{\Ux}$.

\item If $\ceqtm{A}{A'}{\Ux}$ and $\eqtm{\oft aA}{B}{B'}{\Ux}$ then
$\ceqtm{\sigmacl{a}{A}{B}}{\sigmacl{a}{A'}{B'}}{\Ux}$.

\item If $\ceqtm[\Psi,x]{A}{A'}{\Ux}$ and
$\ceqtm{P_\e}{P_\e'}{\dsubst{A}{\e}{x}}$ for $\e\in\{0,1\}$ then
$\ceqtm{\Path{x.A}{P_0}{P_1}}{\Path{x.A'}{P_0'}{P_1'}}{\Ux}$.

\item If $\ceqtm{A}{A'}{\Upre}$, $\ceqtm{M}{M'}{A}$, and $\ceqtm{N}{N'}{A}$ then
$\ceqtm{\Eq{A}{M}{N}}{\Eq{A'}{M'}{N'}}{\Upre}$.

\item $\coftype{\void}{\Ux}$.

\item $\coftype{\nat}{\Ux}$.

\item $\coftype{\bool}{\Ux}$.

\item $\coftype{\wbool}{\Ux}$.

\item $\coftype{\C}{\Ux}$.

\item If $\ceqtm<r=0>{A}{A'}{\Ux}$, $\ceqtm{B}{B'}{\Ux}$, and
$\ceqtm<r=0>{E}{E'}{\Equiv{A}{B}}$, then
$\ceqtm{\ua{r}{A,B,E}}{\ua{r}{A',B',E'}}{\Ux}$.

\item If $i<j$ then $\coftype{\Ux[i]}{\Upre[j]}$.

\item If $i<j$ then $\coftype{\UKan[i]}{\UKan[j]}$.
\end{enumerate}
\end{rul}
\begin{proof}
Note that \cref{rul:U-elim} is needed to make sense of these rules; for example,
in part (1), by \cref{rul:U-elim} and
$\relcts{\pre\tau_\omega}{\coftype{A}{\Ux}}$ we conclude
$\relcts{\pre\tau_\omega}{\cwftypex{A}}$, which is a presupposition of
$\relcts{\pre\tau_\omega}{\eqtm{\oft aA}{B}{B'}{\Ux}}$.

For part (1), by $\relcts{\pre\tau_\omega}{\ceqtm{A}{A'}{\Ux}}$ and
\cref{lem:U-preelim}, $\relcts{\tau^\kappa_j}{\ceqtypex{A}{A'}}$; similarly, by
$\relcts{\pre\tau_\omega}{\eqtm{\oft aA}{B}{B'}{\Ux}}$ and
\cref{lem:U-preelim,lem:monotone-judg},
$\relcts{\tau^\kappa_j}{\eqtypex{\oft aA}{B}{B'}}$. By
\cref{rul:fun-form-pre}, we conclude that
$\relcts{\tau^\kappa_j}{\ceqtypep{\picl{a}{A}{B}}{\picl{a}{A'}{B'}}}$, and in
particular, $\PTy(\tau^\kappa_j)(\Psi,\picl{a}{A}{B},\picl{a}{A'}{B'},\_)$.
Therefore $\Tm(\vper{\Ux})(\picl{a}{A}{B},\picl{a}{A'}{B'})$ as needed.
Parts (2--12) follow the same pattern.
\end{proof}

\begin{rul}[Kan type formation]
$\relcts{\pre\tau_\omega}{\cwftypek{\UKan[i]}}$.
\end{rul}
\begin{proof}
By \cref{rul:U-intro},
$\relcts{\pre\tau_\omega}{\coftype{\UKan[i]}{\UKan[i+1]}}$; the result follows
by \cref{rul:U-elim}.
\end{proof}

\begin{rul}[Subsumption]
If $\relcts{\pre\tau_\omega}{\ceqtm{A}{A'}{\UKan[i]}}$ then
$\relcts{\pre\tau_\omega}{\ceqtm{A}{A'}{\Upre[i]}}$.
\end{rul}
\begin{proof}
By $\Kan\tau_i\subseteq\pre\tau_i$ we have
$\vper{\UKan[i]}\subseteq\vper{\Upre[i]}$ and thus
$\Tm(\vper{\UKan[i]})\subseteq\Tm(\vper{\Upre[i]})$.
\end{proof}

\newpage
\section{Rules}
\label{sec:rules}

In this section we collect the rules proven in \cref{sec:meanings,sec:types}
(relative to $\pre\tau_\omega$) for easy reference. Note, however, that these
rules do not constitute our higher type theory, which was defined in
\cref{sec:typesys,sec:meanings} and whose properties were verified in
\cref{sec:types}.
One can settle on a different collection of rules depending on the need.
For example, the \RedPRL{} proof assistant \citep{redprl} based on
this paper uses a sequent calculus rather than natural deduction, judgments
without any presuppositions, and a unified context for dimensions and terms.

For the sake of concision and clarity, we state the following rules in
\emph{local form}, extending them to \emph{global form} by \emph{uniformity},
also called \emph{naturality}. (This format was suggested by
\citet{martin1984intuitionistic}, itself inspired by Gentzen's original concept
of natural deduction.)
While the rules in \cref{sec:types} are stated only for closed terms, the
corresponding generalizations to open-term sequents follow by the definition of
the open judgments, the fact that the introduction and elimination rules respect
equality (proven in \cref{sec:types}), and the fact that all substitutions
commute with term formers.

In the rules below, $\Psi$ and $\Xi$ are unordered sets, and the equations in
$\Xi$ are also unordered. $\J$ stands for any type equality or element equality
judgment, and $\kappa$ for either $\mathsf{pre}$ or $\mathsf{Kan}$. The
$\ssteps$ judgment is the \emph{cubically-stable stepping} relation defined in
\cref{sec:opsem}.

\paragraph{Structural rules}

\begin{mathpar}
\Infer
  {\cwftypex{A}}
  {\oftype{\oft{a}{A}}{a}{A}}
\and
\Infer
  {\judg{\J} \\
   \cwftypex{A}}
  {\ctx{\oft aA} \judg{\J}}
\and
\Infer
  {\judg{\J} \\
   \psitd}
  {\judg[\Psi']{\td{\J}{\psi}}}
\and
\Infer
  {\ceqtypek{A}{A'}}
  {\ceqtypep{A}{A'}}
\and
\Infer
  {\ceqtypex{A}{A'}}
  {\ceqtypex{A'}{A}}
\and
\Infer
  {\ceqtypex{A}{A'} \\
   \ceqtypex{A'}{A''}}
  {\ceqtypex{A}{A''}}
\and
\Infer
  {\ceqtm{M'}{M}{A}}
  {\ceqtm{M}{M'}{A}}
\and
\Infer
  {\ceqtm{M}{M'}{A} \\
   \ceqtm{M'}{M''}{A}}
  {\ceqtm{M}{M''}{A}}
\and
\Infer
  {\ceqtm{M}{M'}{A} \\
   \ceqtypex{A}{A'}}
  {\ceqtm{M}{M'}{A'}}
\and
\Infer
  {\eqtypex{\oft{a}{A}}{B}{B'} \\
   \ceqtm{N}{N'}{A}}
  {\ceqtypex{\subst{B}{N}{a}}{\subst{B'}{N'}{a}}}
\and
\Infer
  {\eqtm{\oft{a}{A}}{M}{M'}{B} \\
   \ceqtm{N}{N'}{A}}
  {\ceqtm{\subst{M}{N}{a}}{\subst{M'}{N'}{a}}{\subst{B}{N}{a}}}
\end{mathpar}

\paragraph{Restriction rules}

\begin{mathpar}
\Infer
  {\judg{\J}}
  {\judg<\cdot>{\J}}
\and
\Infer
  {\judg<\Xi>{\J}}
  {\judg<\Xi,\e=\e>{\J}}
\and
\Infer
  { }
  {\judg<\Xi,\e=\eb>{\J}}
\and
\Infer
  {\judg<\dsubst{\Xi}{r}{x}>{\dsubst{\J}{r}{x}}}
  {\judg[\Psi,x]<\Xi,x=r>{\J}}
\end{mathpar}

\paragraph{Computation rules}

\begin{mathpar}
\Infer
  {\ceqtypex{A'}{B} \\
   A \ssteps A'}
  {\ceqtypex{A}{B}}
\and
\Infer
  {\ceqtm{M'}{N}{A} \\
   M \ssteps M'}
  {\ceqtm{M}{N}{A}}
\end{mathpar}

\paragraph{Kan conditions}\

\begin{mathparpagebreakable}
\Infer
  {r_i = r_j \\
   r_i' = 0 \\
   r_j' = 1}
  {\wfshape{\etc{r_i = r_i'}}}
\and
\Infer
  {r_i = r_i'}
  {\wfshape{\etc{r_i = r_i'}}}
\and
\Infer
  { {\begin{array}{ll}
   &\wfshape{\etc{r_i = r_i'}} \\
   &\ceqtypek{A}{A'} \\
   &\ceqtm{M}{M'}{A} \\
   (\forall i,j) &\ceqtm[\Psi,y]<r_i = r_i',r_j = r_j'>{N_i}{N_j'}{A} \\
   (\forall i)   &\ceqtm<r_i = r_i'>{\dsubst{N_i}{r}{y}}{M}{A}
   \end{array}}}
  {\ceqtm{\Hcom*{A}{r_i=r_i'}}{\Hcom{A'}{r}{r'}{M'}{\sys{r_i=r_i'}{y.N_i'}}}{A}}
\and
\Infer
  { {\begin{array}{ll}
   &\wfshape{\etc{r_i = r_i'}} \\
   &\cwftypek{A} \\
   &\coftype{M}{A} \\
   (\forall i,j) &\ceqtm[\Psi,y]<r_i = r_i',r_j = r_j'>{N_i}{N_j}{A} \\
   (\forall i)   &\ceqtm<r_i = r_i'>{\dsubst{N_i}{r}{y}}{M}{A}
   \end{array}}}
  {\ceqtm{\Hcom{A}{r}{r}{M}{\sys{r_i=r_i'}{y.N_i}}}{M}{A}}
\and
\Infer
  { {\begin{array}{ll}
   & r_i = r_i' \\
   &\cwftypek{A} \\
   &\coftype{M}{A} \\
   (\forall i,j) &\ceqtm[\Psi,y]<r_i = r_i',r_j = r_j'>{N_i}{N_j}{A} \\
   (\forall i)   &\ceqtm<r_i = r_i'>{\dsubst{N_i}{r}{y}}{M}{A}
   \end{array}}}
  {\ceqtm{\Hcom*{A}{r_i=r_i'}}{\dsubst{N_i}{r'}{y}}{A}}
\and
\Infer
  {\ceqtypek[\Psi,x]{A}{A'} \\
   \ceqtm{M}{M'}{\dsubst{A}{r}{x}}}
  {\ceqtm{\Coe*{x.A}}{\Coe{x.A'}{r}{r'}{M'}}{\dsubst{A}{r'}{x}}}
\and
\Infer
  {\cwftypek[\Psi,x]{A} \\
   \coftype{M}{\dsubst{A}{r}{x}}}
  {\ceqtm{\Coe{x.A}{r}{r}{M}}{M}{\dsubst{A}{r}{x}}}
\and
\Infer
  { {\begin{array}{ll}
   &\wfshape{\etc{r_i = r_i'}} \\
   &\ceqtypek[\Psi,y]{A}{A'} \\
   &\ceqtm{M}{M'}{\dsubst{A}{r}{y}} \\
   (\forall i,j) &\ceqtm[\Psi,y]<r_i = r_i',r_j = r_j'>{N_i}{N_j'}{A} \\
   (\forall i)   &\ceqtm<r_i = r_i'>{\dsubst{N_i}{r}{y}}{M}{\dsubst{A}{r}{y}}
   \end{array}}}
  {\ceqtm{\Com*{y.A}{r_i=r_i'}}{\Com{y.A'}{r}{r'}{M'}{\sys{r_i=r_i'}{y.N_i'}}}{\dsubst{A}{r'}{y}}}
\and
\Infer
  { {\begin{array}{ll}
   &\wfshape{\etc{r_i = r_i'}} \\
   &\cwftypek[\Psi,y]{A} \\
   &\coftype{M}{\dsubst{A}{r}{y}} \\
   (\forall i,j) &\ceqtm[\Psi,y]<r_i = r_i',r_j = r_j'>{N_i}{N_j}{A} \\
   (\forall i)   &\ceqtm<r_i = r_i'>{\dsubst{N_i}{r}{y}}{M}{\dsubst{A}{r}{y}}
   \end{array}}}
  {\ceqtm{\Com{y.A}{r}{r}{M}{\sys{r_i=r_i'}{y.N_i}}}{M}{\dsubst{A}{r}{y}}}
\and
\Infer
  { {\begin{array}{ll}
   & r_i = r_i' \\
   &\cwftypek[\Psi,y]{A} \\
   &\coftype{M}{\dsubst{A}{r}{y}} \\
   (\forall i,j) &\ceqtm[\Psi,y]<r_i = r_i',r_j = r_j'>{N_i}{N_j}{A} \\
   (\forall i)   &\ceqtm<r_i = r_i'>{\dsubst{N_i}{r}{y}}{M}{\dsubst{A}{r}{y}}
   \end{array}}}
  {\ceqtm{\Com*{y.A}{r_i=r_i'}}{\dsubst{N_i}{r'}{y}}{\dsubst{A}{r'}{y}}}
\and
\end{mathparpagebreakable}

\paragraph{Dependent function types}

\begin{mathpar}
\Infer
  {\ceqtypex{A}{A'} \\
   \eqtypex{\oft aA}{B}{B'}}
  {\ceqtypex{\picl{a}{A}{B}}{\picl{a}{A'}{B'}}}
\and
\Infer
  {\eqtm{\oft aA}{M}{M'}{B}}
  {\ceqtm{\lam{a}{M}}{\lam{a}{M'}}{\picl{a}{A}{B}}}
\and
\Infer
  {\ceqtm{M}{M'}{\picl{a}{A}{B}} \\
   \ceqtm{N}{N'}{A}}
  {\ceqtm{\app{M}{N}}{\app{M'}{N'}}{\subst{B}{N}{a}}}
\and
\Infer
  {\oftype{\oft aA}{M}{B} \\
   \coftype{N}{A}}
  {\ceqtm{\app{\lam{a}{M}}{N}}{\subst{M}{N}{a}}{\subst{B}{N}{a}}}
\and
\Infer
  {\coftype{M}{\picl{a}{A}{B}}}
  {\ceqtm{M}{\lam{a}{\app{M}{a}}}{\picl{a}{A}{B}}}
\end{mathpar}

\paragraph{Dependent pair types}

\begin{mathpar}
\Infer
  {\ceqtypex{A}{A'} \\
   \eqtypex{\oft aA}{B}{B'}}
  {\ceqtypex{\sigmacl{a}{A}{B}}{\sigmacl{a}{A'}{B'}}}
\and
\Infer
  {\ceqtm{M}{M'}{A} \\
   \ceqtm{N}{N'}{\subst{B}{M}{a}}}
  {\ceqtm{\pair{M}{N}}{\pair{M'}{N'}}{\sigmacl{a}{A}{B}}}
\and
\Infer
  {\ceqtm{P}{P'}{\sigmacl{a}{A}{B}}}
  {\ceqtm{\fst{P}}{\fst{P'}}{A}}
\and
\Infer
  {\ceqtm{P}{P'}{\sigmacl{a}{A}{B}}}
  {\ceqtm{\snd{P}}{\snd{P'}}{\subst{B}{\fst{P}}{a}}}
\and
\Infer
  {\coftype{M}{A}}
  {\ceqtm{\fst{\pair{M}{N}}}{M}{A}}
\and
\Infer
  {\coftype{N}{B}}
  {\ceqtm{\snd{\pair{M}{N}}}{N}{B}}
\and
\Infer
  {\coftype{P}{\sigmacl{a}{A}{B}}}
  {\ceqtm{P}{\pair{\fst{P}}{\snd{P}}}{\sigmacl{a}{A}{B}}}
\end{mathpar}

\paragraph{Path types}

\begin{mathpar}
\Infer
  {\ceqtypex[\Psi,x]{A}{A'} \\
   (\forall\e)\ \ceqtm{P_\e}{P_\e'}{\dsubst{A}{\e}{x}}}
  {\ceqtypex{\Path{x.A}{P_0}{P_1}}{\Path{x.A'}{P_0'}{P_1'}}}
\and
\Infer
  {\ceqtm[\Psi,x]{M}{M'}{A} \\
   (\forall\e)\ \ceqtm{\dsubst{M}{\e}{x}}{P_\e}{\dsubst{A}{\e}{x}}}
  {\ceqtm{\dlam{x}{M}}{\dlam{x}{M'}}{\Path{x.A}{P_0}{P_1}}}
\and
\Infer
  {\ceqtm{M}{M'}{\Path{x.A}{P_0}{P_1}}}
  {\ceqtm{\dapp{M}{r}}{\dapp{M'}{r}}{\dsubst{A}{r}{x}}}
\and
\Infer
  {\coftype{M}{\Path{x.A}{P_0}{P_1}}}
  {\ceqtm{\dapp{M}{\e}}{P_\e}{\dsubst{A}{\e}{x}}}
\and
\Infer
  {\coftype[\Psi,x]{M}{A}}
  {\ceqtm{\dapp{(\dlam{x}{M})}{r}}{\dsubst{M}{r}{x}}{\dsubst{A}{r}{x}}}
\and
\Infer
  {\coftype{M}{\Path{x.A}{P_0}{P_1}}}
  {\ceqtm{M}{\dlam{x}{(\dapp{M}{x})}}{\Path{x.A}{P_0}{P_1}}}
\end{mathpar}

\paragraph{Equality pretypes}

\begin{mathpar}
\Infer
  {\ceqtypep{A}{A'} \\
   \ceqtm{M}{M'}{A} \\
   \ceqtm{N}{N'}{A}}
  {\ceqtypep{\Eq{A}{M}{N}}{\Eq{A'}{M'}{N'}}}
\and
\Infer
  {\ceqtm{M}{N}{A}}
  {\coftype{\ax}{\Eq{A}{M}{N}}}
\and
\Infer
  {\coftype{E}{\Eq{A}{M}{N}}}
  {\ceqtm{M}{N}{A}}
\and
\Infer
  {\coftype{E}{\Eq{A}{M}{N}}}
  {\ceqtm{E}{\ax}{\Eq{A}{M}{N}}}
\end{mathpar}

\paragraph{Void}

\begin{mathpar}
\Infer
  { }
  {\cwftypek{\void}}
\and
\Infer
  {\coftype{M}{\void}}
  {\judg{\J}}
\end{mathpar}

\paragraph{Natural numbers}

\begin{mathpar}
\Infer
  { }
  {\cwftypek{\nat}}
\and
\Infer
  { }
  {\coftype{\z}{\nat}}
\and
\Infer
  {\ceqtm{M}{M'}{\nat}}
  {\ceqtm{\suc{M}}{\suc{M'}}{\nat}}
\and
\Infer
  {\wftypex{\oft{n}{\nat}}{A} \\
   \ceqtm{M}{M'}{\nat} \\
   \ceqtm{Z}{Z'}{\subst{A}{\z}{n}} \\
   \eqtm{\oft{n}{\nat},\oft{a}{A}}{S}{S'}{\subst{A}{\suc{n}}{n}}}
  {\ceqtm{\natrec{M}{Z}{n.a.S}}{\natrec{M'}{Z'}{n.a.S'}}{\subst{A}{M}{n}}}
\and
\Infer
  {\coftype{Z}{A}}
  {\ceqtm{\natrec{\z}{Z}{n.a.S}}{Z}{A}}
\and
\Infer
  {\wftypex{\oft{n}{\nat}}{A} \\
   \coftype{M}{\nat} \\
   \coftype{Z}{\subst{A}{\z}{n}} \\
   \oftype{\oft{n}{\nat},\oft{a}{A}}{S}{\subst{A}{\suc{n}}{n}}}
  {\ceqtm{\natrec{\suc{M}}{Z}{n.a.S}}%
         {\subst{\subst{S}{M}{n}}{\natrec{M}{Z}{n.a.S}}{a}}%
         {\subst{A}{\suc{M}}{n}}}
\end{mathpar}

\paragraph{Booleans}

\begin{mathpar}
\Infer
  { }
  {\cwftypek{\bool}}
\and
\Infer
  { }
  {\coftype{\true}{\bool}}
\and
\Infer
  { }
  {\coftype{\false}{\bool}}
\and
\Infer
  {\wftypep{\oft{b}{\bool}}{C} \\
   \ceqtm{M}{M'}{\bool} \\
   \ceqtm{T}{T'}{\subst{C}{\true}{b}} \\
   \ceqtm{F}{F'}{\subst{C}{\false}{b}}}
  {\ceqtm{\ifb{b.A}{M}{T}{F}}{\ifb{b.A'}{M'}{T'}{F'}}{\subst{C}{M}{b}}}
\and
\Infer
  {\coftype{T}{B}}
  {\ceqtm{\ifb{b.A}{\true}{T}{F}}{T}{B}}
\and
\Infer
  {\coftype{F}{B}}
  {\ceqtm{\ifb{b.A}{\false}{T}{F}}{F}{B}}
\end{mathpar}

\paragraph{Weak Booleans}

\begin{mathpar}
\Infer
  { }
  {\cwftypek{\wbool}}
\and
\Infer
  {\ceqtm{M}{M'}{\bool}}
  {\ceqtm{M}{M'}{\wbool}}
\and
\Infer
  {\eqtypek{\oft{b}{\wbool}}{A}{A'} \\
   \ceqtm{M}{M'}{\wbool} \\
   \ceqtm{T}{T'}{\subst{A}{\true}{b}} \\
   \ceqtm{F}{F'}{\subst{A}{\false}{b}}}
  {\ceqtm{\ifb{b.A}{M}{T}{F}}{\ifb{b.A'}{M'}{T'}{F'}}{\subst{A}{M}{b}}}
\end{mathpar}

\paragraph{Circle}

\begin{mathpar}
\Infer
  { }
  {\cwftypek{\C}}
\and
\Infer
  { }
  {\coftype{\base}{\C}}
\and
\Infer
  { }
  {\coftype{\lp{r}}{\C}}
\and
\Infer
  { }
  {\ceqtm{\lp{\e}}{\base}{\C}}
\and
\Infer
  {\eqtypek{\oft{c}{\C}}{A}{A'} \\
   \ceqtm{M}{M'}{\C} \\
   \ceqtm{P}{P'}{\subst{A}{\base}{c}} \\
   \ceqtm[\Psi,x]{L}{L'}{\subst{A}{\lp{x}}{c}} \\
   (\forall\e)\ \ceqtm{\dsubst{L}{\e}{x}}{P}{\subst{A}{\base}{c}}}
  {\ceqtm{\Celim{c.A}{M}{P}{x.L}}{\Celim{c.A'}{M'}{P'}{x.L'}}{\subst{A}{M}{c}}}
\and
\Infer
  {\coftype{P}{B}}
  {\ceqtm{\Celim{c.A}{\base}{P}{x.L}}{P}{B}}
\and
\Infer
  {\coftype[\Psi,x]{L}{B} \\
   (\forall\e)\ \ceqtm{\dsubst{L}{\e}{x}}{P}{\dsubst{B}{\e}{x}}}
  {\ceqtm{\Celim{c.A}{\lp{r}}{P}{x.L}}{\dsubst{L}{r}{x}}{\dsubst{B}{r}{x}}}
\end{mathpar}

\paragraph{Univalence}\

\begin{mathparpagebreakable}
\isContr{C} := \prd{C}{(\picl{c}{C}{\picl{c'}{C}{\Path{\_.C}{c}{c'}}})}
\and
\Equiv{A}{B} :=
\sigmacl{f}{\arr{A}{B}}{(\picl{b}{B}{\isContr{\sigmacl{a}{A}{\Path{\_.B}{\app{f}{a}}{b}}}})}
\and
\Infer
  {\ceqtypex<r=0>{A}{A'} \\
   \ceqtypex{B}{B'} \\
   \ceqtm<r=0>{E}{E'}{\Equiv{A}{B}}}
  {\ceqtypex{\ua{r}{A,B,E}}{\ua{r}{A',B',E'}}}
\and
\Infer
  {\cwftypex{A}}
  {\ceqtypex{\ua{0}{A,B,E}}{A}}
\and
\Infer
  {\cwftypex{B}}
  {\ceqtypex{\ua{1}{A,B,E}}{B}}
\and
\Infer
  {\ceqtm<r=0>{M}{M'}{A} \\
   \ceqtm{N}{N'}{B} \\
   \coftype<r=0>{E}{\Equiv{A}{B}} \\
   \ceqtm<r=0>{\app{\fst{E}}{M}}{N}{B}}
  {\ceqtm{\uain{r}{M,N}}{\uain{r}{M',N'}}{\ua{r}{A,B,E}}}
\and
\Infer
  {\coftype{M}{A}}
  {\ceqtm{\uain{0}{M,N}}{M}{A}}
\and
\Infer
  {\coftype{N}{B}}
  {\ceqtm{\uain{1}{M,N}}{N}{B}}
\and
\Infer
  {\ceqtm{M}{M'}{\ua{r}{A,B,E}} \\
   \ceqtm<r=0>{F}{\fst{E}}{\arr{A}{B}}}
  {\ceqtm{\uaproj{r}{M,F}}{\uaproj{r}{M',\fst{E}}}{B}}
\and
\Infer
  {\coftype{M}{A} \\
   \coftype{F}{\arr{A}{B}}}
  {\ceqtm{\uaproj{0}{M,F}}{\app{F}{M}}{B}}
\and
\Infer
  {\coftype{M}{B}}
  {\ceqtm{\uaproj{1}{M,F}}{M}{B}}
\and
\Infer
  {\coftype<r=0>{M}{A} \\
   \coftype{N}{B} \\
   \coftype<r=0>{F}{\arr{A}{B}} \\
   \ceqtm<r=0>{\app{F}{M}}{N}{B}}
  {\ceqtm{\uaproj{r}{\uain{r}{M,N},F}}{N}{B}}
\and
\Infer
  {\coftype{N}{\ua{r}{A,B,E}} \\
   \ceqtm<r=0>{M}{N}{A}}
  {\ceqtm{\uain{r}{M,\uaproj{r}{N,\fst{E}}}}{N}{\ua{r}{A,B,E}}}
\end{mathparpagebreakable}

\paragraph{Universes}\

\begin{mathparpagebreakable}
\Infer
  { }
  {\cwftypep{\Upre}}
\and
\Infer
  { }
  {\cwftypek{\UKan}}
\and
\Infer
  {\ceqtm{A}{A'}{\Ux}}
  {\ceqtypex{A}{A'}}
\and
\Infer
  {\ceqtm{A}{A'}{\Ux[i]} \\
   i\leq j}
  {\ceqtm{A}{A'}{\Ux[j]}}
\and
\Infer
  {\ceqtm{A}{A'}{\UKan}}
  {\ceqtm{A}{A'}{\Upre}}
\and
\Infer
  {\ceqtm{A}{A'}{\Ux} \\
   \eqtm{\oft aA}{B}{B'}{\Ux}}
  {\ceqtm{\picl{a}{A}{B}}{\picl{a}{A'}{B'}}{\Ux}}
\and
\Infer
  {\ceqtm{A}{A'}{\Ux} \\
   \eqtm{\oft aA}{B}{B'}{\Ux}}
  {\ceqtm{\sigmacl{a}{A}{B}}{\sigmacl{a}{A'}{B'}}{\Ux}}
\and
\Infer
  {\ceqtm[\Psi,x]{A}{A'}{\Ux} \\
   (\forall\e)\ \ceqtm{P_\e}{P_\e'}{\dsubst{A}{\e}{x}}}
  {\ceqtm{\Path{x.A}{P_0}{P_1}}{\Path{x.A'}{P_0'}{P_1'}}{\Ux}}
\and
\Infer
  {\ceqtm{A}{A'}{\Upre} \\
   \ceqtm{M}{M'}{A} \\
   \ceqtm{N}{N'}{A}}
  {\ceqtm{\Eq{A}{M}{N}}{\Eq{A'}{M'}{N'}}{\Upre}}
\and
\Infer
  { }
  {\coftype{\void}{\Ux}}
\and
\Infer
  { }
  {\coftype{\nat}{\Ux}}
\and
\Infer
  { }
  {\coftype{\bool}{\Ux}}
\and
\Infer
  { }
  {\coftype{\wbool}{\Ux}}
\and
\Infer
  { }
  {\coftype{\C}{\Ux}}
\and
\Infer
  {\ceqtm<r=0>{A}{A'}{\Ux} \\
   \ceqtm{B}{B'}{\Ux} \\
   \ceqtm<r=0>{E}{E'}{\Equiv{A}{B}}}
  {\ceqtm{\ua{r}{A,B,E}}{\ua{r}{A',B',E'}}{\Ux}}
\and
\Infer
  {i<j}
  {\coftype{\Ux[i]}{\Upre[j]}}
\and
\Infer
  {i<j}
  {\coftype{\UKan[i]}{\UKan[j]}}
\and
\Infer
  { {\begin{array}{ll}
   &\wfshape{\etc{r_i = r_i'}} \\
   &\cwftypek{A} \\
   &\ceqtm{M}{M'}{A} \\
   (\forall i,j) &\ceqtypek[\Psi,y]<r_i = r_i',r_j = r_j'>{B_i}{B_j} \\
   (\forall i,j) &\ceqtm<r_i = r_i',r_j = r_j'>{N_i}{N_j'}{\dsubst{B_i}{r'}{y}} \\
   (\forall i)   &\ceqtypek<r_i = r_i'>{\dsubst{B_i}{r}{y}}{A} \\
   (\forall i)   &\ceqtm<r_i = r_i'>{\Coe{y.B_i}{r'}{r}{N_i}}{M}{A}
   \end{array}}}
  {\ceqtm{\Kbox{r}{r'}{M}{\sys{r_i=r_i'}{N_i}}}%
    {\Kbox{r}{r'}{M'}{\sys{r_i=r_i'}{N_i'}}}%
    {\Hcom{\UKan[j]}{r}{r'}{A}{\sys{r_i=r_i'}{y.B_i}}}}
\and
\Infer
  {\coftype{M}{A}}
  {\ceqtm{\Kbox{r}{r}{M}{\sys{r_i=r_i'}{N_i}}}{M}{A}}
\and
\Infer
  { {\begin{array}{ll}
   & r_i = r_i' \\
   &\cwftypek{A} \\
   &\coftype{M}{A} \\
   (\forall i,j) &\ceqtypek[\Psi,y]<r_i = r_i',r_j = r_j'>{B_i}{B_j} \\
   (\forall i,j) &\ceqtm<r_i = r_i',r_j = r_j'>{N_i}{N_j}{\dsubst{B_i}{r'}{y}} \\
   (\forall i)   &\ceqtypek<r_i = r_i'>{\dsubst{B_i}{r}{y}}{A} \\
   (\forall i)   &\ceqtm<r_i = r_i'>{\Coe{y.B_i}{r'}{r}{N_i}}{M}{A}
   \end{array}}}
  {\ceqtm{\Kbox*{r_i=r_i'}}{N_i}{\dsubst{B_i}{r'}{y}}}
\and
\Infer
  { {\begin{array}{ll}
   &\wfshape{\etc{r_i = r_i'}} \\
   &\cwftypek{A} \\
   (\forall i,j) &\ceqtypek[\Psi,y]<r_i = r_i',r_j = r_j'>{B_i}{B_j'} \\
   (\forall i)   &\ceqtypek<r_i = r_i'>{\dsubst{B_i}{r}{y}}{A} \\
   &\ceqtm{M}{M'}{\Hcom{\UKan[j]}{r}{r'}{A}{\sys{r_i=r_i'}{y.B_i}}} \\
   \end{array}}}
  {\ceqtm{\Kcap{r}{r'}{M}{\sys{r_i=r_i'}{y.B_i}}}%
    {\Kcap{r}{r'}{M'}{\sys{r_i=r_i'}{y.B_i'}}}{A}}
\and
\Infer
  {\coftype{M}{A}}
  {\ceqtm{\Kcap{r}{r}{M}{\sys{r_i=r_i'}{y.B_i}}}{M}{A}}
\and
\Infer
  { {\begin{array}{ll}
   &r_i = r_i' \\
   &\cwftypek{A} \\
   (\forall i,j) &\ceqtypek[\Psi,y]<r_i = r_i',r_j = r_j'>{B_i}{B_j'} \\
   (\forall i)   &\ceqtypek<r_i = r_i'>{\dsubst{B_i}{r}{y}}{A} \\
   &\ceqtm{M}{M'}{\Hcom{\UKan[j]}{r}{r'}{A}{\sys{r_i=r_i'}{y.B_i}}} \\
   \end{array}}}
  {\ceqtm{\Kcap{r}{r'}{M}{\sys{r_i=r_i'}{y.B_i}}}%
    {\Coe{y.B_i}{r'}{r}{M}}{A}}
\and
\Infer
  { {\begin{array}{ll}
   &\wfshape{\etc{r_i = r_i'}} \\
   &\cwftypek{A} \\
   &\ceqtm{M}{M'}{A} \\
   (\forall i,j) &\ceqtypek[\Psi,y]<r_i = r_i',r_j = r_j'>{B_i}{B_j} \\
   (\forall i,j) &\ceqtm<r_i = r_i',r_j = r_j'>{N_i}{N_j'}{\dsubst{B_i}{r'}{y}} \\
   (\forall i)   &\ceqtypek<r_i = r_i'>{\dsubst{B_i}{r}{y}}{A} \\
   (\forall i)   &\ceqtm<r_i = r_i'>{\Coe{y.B_i}{r'}{r}{N_i}}{M}{A}
   \end{array}}}
  {\ceqtm{\Kcap{r}{r'}{\Kbox{r}{r'}{M}{\sys{r_i=r_i'}{N_i}}}{\sys{r_i=r_i'}{y.B_i}}}%
    {M}{A}}
\and
\Infer
  { {\begin{array}{ll}
   &\wfshape{\etc{r_i = r_i'}} \\
   &\cwftypek{A} \\
   (\forall i,j) &\ceqtypek[\Psi,y]<r_i = r_i',r_j = r_j'>{B_i}{B_j} \\
   (\forall i)   &\ceqtypek<r_i = r_i'>{\dsubst{B_i}{r}{y}}{A} \\
   &\coftype{M}{\Hcom{\UKan[j]}{r}{r'}{A}{\sys{r_i=r_i'}{y.B_i}}}
   \end{array}}}
  {\ceqtm{\Kbox{r}{r'}{\Kcap{r}{r'}{M}{\sys{r_i=r_i'}{y.B_i}}}{\sys{r_i=r_i'}{M}}}%
   {M}{\Hcom{\UKan[j]}{r}{r'}{A}{\sys{r_i=r_i'}{y.B_i}}}}
\end{mathparpagebreakable}

\newpage
\section{Future work}
\label{sec:future}

\paragraph{Formal Cartesian cubical type theory}

With Guillaume Brunerie, Thierry Coquand, and Dan Licata, we have developed a
formal Cartesian cubical type theory with univalent universes, accompanied by a
constructive cubical set model, most of which has been formalized in Agda in the
style of \citet{ortonpitts16topos}. This forthcoming work explores the the Kan
operations described in this paper---in particular, with the addition of $x=z$
diagonal constraints---in a proof-theoretic and model-theoretic setting, rather
than the computational setting emphasized in this paper.

\paragraph{Cubical (higher) inductive types}

Evan Cavallo is currently extending this work to account for a general class of
inductive types with higher-dimensional recursive constructors. In the
cubical setting, such types are generated by dimension-parametrized constructors
with prescribed boundaries. (For example, $\C$ is generated by $\base$ and
$\lp{x}$, whose $x$-faces are $\base$.)

\paragraph{Discrete, $\Hcom$, and $\Coe$ types}

In this paper we divide types into pretypes and Kan types, but finer
distinctions are possible. Some types support $\Hcom$ but not necessarily
$\Coe$, or vice versa. Exact equality types always have $\Hcom$ structure
because $\ax$ is a suitable composite for every box, but not $\Coe$ in general.
Types with $\Hcom$ or
$\Coe$ structure are not themselves closed under all type formers,
but depend on each other; for example,
\begin{enumerate}[itemsep=.2ex,parsep=.2ex]
\item $\cwftype{hcom}{\picl{a}{A}{B}}$ when $\cwftypep{A}$ and
$\wftype{hcom}{\oft{a}{A}}{B}$,
\item $\cwftype{hcom}{\sigmacl{a}{A}{B}}$ when $\cwftype{hcom}{A}$ and
$\wftypek{\oft{a}{A}}{B}$,
\item $\cwftype{coe}{\picl{a}{A}{B}}$ when $\cwftype{coe}{A}$ and
$\wftype{coe}{\oft{a}{A}}{B}$, and
\item $\cwftype{coe}{\Path{x.A}{M}{N}}$ when $\cwftypek[\Psi,x]{A}$,
$\coftype{M}{\dsubst{A}{0}{x}}$, and $\coftype{N}{\dsubst{A}{1}{x}}$.
\end{enumerate}

\emph{Discrete Kan} types, such as $\nat$ and
$\bool$, are not only Kan but also strict sets, in the sense that all paths are
exactly equal to reflexivity. To be precise, we say $\ceqtype{disc}{A}{B}$ if
for any $\tds{\Psi_1}{\psi_1}{\Psi}$, $\tds{\Psi_2}{\psi_2,\psi_2'}{\Psi_1}$,
we have $\ceqtypek[\Psi_2]{\td{A}{\psi_1\psi_2}}{\td{B}{\psi_1\psi_2'}}$,
and for any $\coftype[\Psi_1]{M}{\td{A}{\psi_1}}$, we have
$\ceqtm[\Psi_2]{\td{M}{\psi_2}}{\td{M}{\psi_2'}}{\td{A}{\psi_1\psi_2}}$.
Discrete Kan types are closed under most type formers, including exact equality.
Exact equality types do not in general admit coercion, because
$\Coe{x.\Eq{A}{\dsubst{P}{0}{x}}{P}}{0}{1}{\ax}$ turns any line $P$ into an
exact equality $\Eq{A}{\dsubst{P}{0}{x}}{\dsubst{P}{1}{x}}$ between its end
points. However, if $\cwftype{disc}{A}$ then
$\wftype{disc}{\oft{a}{A},\oft{a'}{A}}{\Eq{A}{a}{a'}}$, because paths in $A$
\emph{are} exact equalities.

\paragraph{Further improvements in \RedPRL{}}

Implementing and using this type theory in \RedPRL{} has already led to several
minor improvements not described in this paper:
\begin{enumerate}[itemsep=.2ex,parsep=.2ex]
\item We have added \emph{line types} to \RedPRL{}, $(x{:}\mathsf{dim})\to A$,
path types whose end points are not fixed. Elements of line types are simply
terms with an abstracted dimension, which has proven cleaner in practice than
the iterated sigma type $\sigmacl{a}{A}{\sigmacl{a'}{A}{\Path{\_.A}{a}{a'}}}$.

\item We are experimenting with alternative implementations of the Kan
operations for $\Fcom$ and $\ua$ types in \RedPRL{}, some inspired by the work
in the forthcoming formal Cartesian cubical type theory mentioned above.

\item The \RedPRL{} proof theory includes discrete Kan, $\Hcom$, and $\Coe$
types as described above, in addition to the Kan types and pretypes described in
this paper.

\item The definitions of the $M \ssteps M'$ and $\sisval{M}$ judgments have been
extended to account for computations that are stable by virtue of taking place
under dimension binders.
\end{enumerate}

\newpage
\bibliographystyle{plainnat}
\bibliography{uni}

\begin{thebibliography}{27}
\providecommand{\natexlab}[1]{#1}
\providecommand{\url}[1]{\texttt{#1}}
\expandafter\ifx\csname urlstyle\endcsname\relax
  \providecommand{\doi}[1]{doi: #1}\else
  \providecommand{\doi}{doi: \begingroup \urlstyle{rm}\Url}\fi

\bibitem[uue()]{uuee}
{S}heremetyevo {I}nternational {A}irport.
\newblock URL \url{http://svo.aero/}.

\bibitem[Allen(1987)]{allen1987types}
Stuart~F. Allen.
\newblock A {N}on-type-theoretic {D}efinition of {M}artin-{L}{\"{o}}f's
  {T}ypes.
\newblock In D.~Gries, editor, \emph{Proceedings of the 2nd IEEE Symposium on
  Logic in Computer Science}, pages 215--224. IEEE Computer Society Press, June
  1987.

\bibitem[Altenkirch et~al.(2016)Altenkirch, Capriotti, and
  Kraus]{altenkirch16strict}
Thorsten Altenkirch, Paolo Capriotti, and Nicolai Kraus.
\newblock Extending homotopy type theory with strict equality.
\newblock In \emph{25th EACSL Annual Conference on Computer Science Logic (CSL
  2016)}, pages 21:1--21:17, Dagstuhl, Germany, 2016.
\newblock \doi{http://dx.doi.org/10.4230/LIPIcs.CSL.2016.21}.
\newblock URL \url{http://drops.dagstuhl.de/opus/volltexte/2016/6561}.

\bibitem[Angiuli and Harper(2016)]{ah2016cubicaldep}
Carlo Angiuli and Robert Harper.
\newblock Computational higher type theory {II}: Dependent cubical
  realizability.
\newblock Preprint, June 2016.
\newblock URL \url{http://arxiv.org/abs/1606.09638}.

\bibitem[Angiuli et~al.(2016)Angiuli, Harper, and Wilson]{ahw2016cubical}
Carlo Angiuli, Robert Harper, and Todd Wilson.
\newblock Computational higher type theory {I}: Abstract cubical realizability.
\newblock Preprint, April 2016.
\newblock URL \url{https://arxiv.org/abs/1604.08873}.

\bibitem[Angiuli et~al.(2017)Angiuli, Harper, and Wilson]{ahw2017cubical}
Carlo Angiuli, Robert Harper, and Todd Wilson.
\newblock Computational higher-dimensional type theory.
\newblock In \emph{Proceedings of the 44th ACM SIGPLAN Symposium on Principles
  of Programming Languages}, POPL 2017, pages 680--693, New York, NY, USA,
  2017. ACM.
\newblock ISBN 978-1-4503-4660-3.
\newblock \doi{10.1145/3009837.3009861}.
\newblock URL \url{http://doi.acm.org/10.1145/3009837.3009861}.

\bibitem[Awodey(2016)]{awodey16cartesian}
Steve Awodey.
\newblock A cubical model of homotopy type theory.
\newblock June 2016.
\newblock URL
  \url{https://www.andrew.cmu.edu/user/awodey/preprints/stockholm.pdf}.

\bibitem[Beki{\'{c}}(1984)]{Bekic1984}
Hans Beki{\'{c}}.
\newblock Definable operations in general algebras, and the theory of automata
  and flowcharts.
\newblock In C.~B. Jones, editor, \emph{Programming Languages and Their
  Definition: H. Beki{\v{c}} (1936--1982)}, pages 30--55. Springer Berlin
  Heidelberg, Berlin, Heidelberg, 1984.
\newblock ISBN 978-3-540-38933-0.
\newblock \doi{10.1007/BFb0048939}.
\newblock URL \url{https://doi.org/10.1007/BFb0048939}.

\bibitem[Bezem et~al.(2014)Bezem, Coquand, and Huber]{bch}
Marc Bezem, Thierry Coquand, and Simon Huber.
\newblock A model of type theory in cubical sets.
\newblock In \emph{19th International Conference on Types for Proofs and
  Programs (TYPES 2013)}, volume~26, pages 107--128, 2014.

\bibitem[Boulier and Tabareau(2017)]{boulier17twolevel}
Simon Boulier and Nicolas Tabareau.
\newblock Model structure on the universe in a two level type theory.
\newblock Preprint, 2017.
\newblock URL \url{https://hal.archives-ouvertes.fr/hal-01579822}.

\bibitem[Buchholtz and Morehouse(2017)]{buchholtz2017}
Ulrik Buchholtz and Edward Morehouse.
\newblock Varieties of cubical sets.
\newblock In \emph{Relational and Algebraic Methods in Computer Science: 16th
  International Conference, RAMiCS 2017, Lyon, France, May 15-18, 2017,
  Proceedings}, pages 77--92. Springer International Publishing, Cham, 2017.
\newblock ISBN 978-3-319-57418-9.
\newblock \doi{10.1007/978-3-319-57418-9_5}.
\newblock URL \url{https://doi.org/10.1007/978-3-319-57418-9_5}.

\bibitem[Cohen et~al.(2016)Cohen, Coquand, Huber, and
  M{\"{o}}rtberg]{cohen2016cubical}
Cyril Cohen, Thierry Coquand, Simon Huber, and Anders M{\"{o}}rtberg.
\newblock Cubical type theory: a constructive interpretation of the univalence
  axiom.
\newblock In \emph{21st International Conference on Types for Proofs and
  Programs (TYPES 2015)}, Dagstuhl, Germany, 2016.
\newblock To appear.

\bibitem[{Constable, et al.}(1985)]{constableetalnuprl}
Robert~L. {Constable, et al.}
\newblock \emph{Implementing Mathematics with the Nuprl Proof Development
  Environment}.
\newblock Prentice-Hall, 1985.

\bibitem[Davey and Priestley(2002)]{daveypriestleylattices}
B.~A. Davey and H.~A. Priestley.
\newblock \emph{Introduction to lattices and order}.
\newblock Cambridge University Press, Cambridge, UK, 2002.
\newblock ISBN 0-521-78451-4.
\newblock URL \url{http://opac.inria.fr/record=b1077513}.

\bibitem[Harper(1992)]{harper1992typesys}
Robert Harper.
\newblock Constructing type systems over an operational semantics.
\newblock \emph{J. Symb. Comput.}, 14\penalty0 (1):\penalty0 71--84, July 1992.
\newblock ISSN 0747-7171.
\newblock \doi{10.1016/0747-7171(92)90026-Z}.
\newblock URL \url{http://dx.doi.org/10.1016/0747-7171(92)90026-Z}.

\bibitem[Huber(2016)]{hubercanonicity}
Simon Huber.
\newblock \emph{Cubical Interpretations of Type Theory}.
\newblock PhD thesis, University of Gothenburg, November 2016.

\bibitem[Licata and Brunerie(2014)]{licata2014cubical}
Daniel~R. Licata and Guillaume Brunerie.
\newblock A cubical type theory, November 2014.
\newblock URL
  \url{http://dlicata.web.wesleyan.edu/pubs/lb14cubical/lb14cubes-oxford.pdf}.
\newblock Talk at Oxford Homotopy Type Theory Workshop.

\bibitem[Licata and Harper(2012)]{lh2dtt}
Daniel~R. Licata and Robert Harper.
\newblock Canonicity for 2-dimensional type theory.
\newblock In \emph{Proceedings of the 39th Annual ACM SIGPLAN-SIGACT Symposium
  on Principles of Programming Languages}, POPL '12, pages 337--348, New York,
  NY, USA, 2012. ACM.
\newblock ISBN 978-1-4503-1083-3.
\newblock \doi{10.1145/2103656.2103697}.
\newblock URL \url{http://doi.acm.org/10.1145/2103656.2103697}.

\bibitem[{Martin-L\"{o}f}(1984)]{cmcp}
P.~{Martin-L\"{o}f}.
\newblock Constructive mathematics and computer programming.
\newblock \emph{Philosophical Transactions of the Royal Society of London
  Series A}, 312:\penalty0 501--518, October 1984.
\newblock \doi{10.1098/rsta.1984.0073}.

\bibitem[Martin-{L\"{o}f}(1984)]{martin1984intuitionistic}
Per Martin-{L\"{o}f}.
\newblock Intuitionistic type theory.
\newblock \emph{Naples: Bibliopolis}, 1984.

\bibitem[Orton and Pitts(2016)]{ortonpitts16topos}
Ian Orton and Andrew~M. Pitts.
\newblock {Axioms for Modelling Cubical Type Theory in a Topos}.
\newblock In \emph{25th EACSL Annual Conference on Computer Science Logic (CSL
  2016)}, pages 24:1--24:19, Dagstuhl, Germany, 2016.
\newblock \doi{10.4230/LIPIcs.CSL.2016.24}.
\newblock URL \url{http://drops.dagstuhl.de/opus/volltexte/2016/6564}.

\bibitem[Pitts(2015)]{pittsnominal}
A.~M. Pitts.
\newblock {Nominal Presentation of Cubical Sets Models of Type Theory}.
\newblock In \emph{20th International Conference on Types for Proofs and
  Programs (TYPES 2014)}, pages 202--220, Dagstuhl, Germany, 2015.
\newblock \doi{http://dx.doi.org/10.4230/LIPIcs.TYPES.2014.202}.
\newblock URL \url{http://drops.dagstuhl.de/opus/volltexte/2015/5498}.

\bibitem[Spiwack(2011)]{spiwack2011}
Arnaud Spiwack.
\newblock \emph{Verified Computing in Homological Algebra, A Journey Exploring
  the Power and Limits of Dependent Type Theory}.
\newblock PhD thesis, \'Ecole Polytechnique, 2011.

\bibitem[Sterling and Harper(2017)]{sterling2017}
Jonathan Sterling and Robert Harper.
\newblock {Algebraic Foundations of Proof Refinement}.
\newblock \url{https://arxiv.org/abs/1703.05215}, 2017.

\bibitem[Sterling et~al.(2017)Sterling, {Hou (Favonia)}, Cavallo, Wilcox,
  Akentyev, Christiansen, Gratzer, and Morrison]{redprl}
Jonathan Sterling, Kuen-Bang {Hou (Favonia)}, Evan Cavallo, James Wilcox,
  Eugene Akentyev, David Christiansen, Daniel Gratzer, and Darin Morrison.
\newblock {RedPRL} -- the {P}eople's {R}efinement {L}ogic.
\newblock \url{http://www.redprl.org/}, 2017.

\bibitem[Voevodsky(2010)]{voevodskycmu}
Vladimir Voevodsky.
\newblock The equivalence axiom and univalent models of type theory, 2010.
\newblock URL \url{http://www.math.ias.edu/vladimir/files/CMU_talk.pdf}.
\newblock Notes from a talk at Carnegie Mellon University.

\bibitem[Voevodsky(2013)]{voevodsky13hts}
Vladimir Voevodsky.
\newblock A simple type system with two identity types.
\newblock Lecture notes, February 2013.
\newblock URL
  \url{https://www.math.ias.edu/vladimir/sites/math.ias.edu.vladimir/files/HTS.pdf}.

\end{thebibliography}

\end{document}